\documentclass[a4paper,12pt,twoside]{report}
\usepackage[top=3.8cm, bottom=3.8cm, left=3.8cm, right=2.7cm]{geometry} 

\usepackage{indentfirst}
\usepackage[nottoc]{tocbibind} 
\usepackage{verbatim}          
\usepackage[pdftex]{graphicx,color} 
\usepackage[suffix=]{epstopdf}      
\usepackage{epsfig}   
\usepackage{epsf}     
\usepackage{textcomp} 
\usepackage[colorlinks,hypertexnames=false]{hyperref} 
\usepackage[figure,table]{hypcap} 
\usepackage[innercaption]{sidecap}
\sidecaptionvpos{figure}{c}

\hypersetup{ 
    bookmarksnumbered,
    pdfstartview={FitH},        
    pdftitle={Evolution and Consequences of Coronal Mass Ejections in the Heliosphere}, 
    pdfauthor={Wageesh Mishra},   
    pdfsubject={Solar Physics}, 
    pdfnewwindow=true,          
    colorlinks=true,            
    linkcolor=blue,             
    citecolor=blue,             
    filecolor=blue,             
    urlcolor=red,               
    pdfpagemode={UseOutlines}
}
\usepackage{subfigure}      
\usepackage{amsmath}        
\usepackage{amssymb}        
\usepackage{amsthm}
\usepackage{float}          
\usepackage{array}          
\usepackage{pdfpages}       
\usepackage[small,bf,belowskip=9pt,aboveskip=9pt,labelsep=quad]{caption} 
\newcolumntype{x}[1]{{\centering\arraybackslash}p{#1}}%
\usepackage{natbib}         

\usepackage{psfrag}                 
\usepackage{wrapfig}                
\usepackage[figuresright]{rotating} 
\usepackage{setspace,cite}          
\usepackage{fancyhdr}               
\usepackage{lineno}                 
\usepackage{longtable}
\usepackage[titletoc,page]{appendix}
\usepackage{cancel}

\makeatletter

\newcommand{\Rmnum}[1]{\expandafter\@slowromancap\romannumeral #1@}

\newcommand\rightlast{\leftskip0ptplus1fil
\rightskip0ptplus-1fil\parfillskip0ptplus1fil}
\DeclareCaptionJustification{rightlast}{\rightlast}

\newcommand{\aap}{    {\it Astron. Astrophys.}}

\newcommand{\ag}{     {\it Ann. Geophys.}}

\newcommand{\apj}{    {\it Astrophys. J.}}
\newcommand{\apjl}{   {\it Astrophys. J.}}
\newcommand{\apjs}{   {\it Astrophys. J. Sup. Ser.}}

\newcommand{\araa}{   {\it Ann. Rev. Astron. Astrophys.}}

\newcommand{\grl}{    {\it Geophys. Res. Lett.}}

\newcommand{\jgr}{    {\it J. Geophys. Res.}}

\newcommand{\lrsp}{   {\it Liv. Rev. Solar Phys.}}
\newcommand{\mnras}{  {\it Mon. Not. Roy. Astron. Soc.}}
\newcommand{\nat}{    {\it Nature}}

\newcommand{\pasj}{   {\it Pub. Astron. Soc. Japan}}

\newcommand{\prl}{    {\it Phys. Rev. Lett.}}

\newcommand{\solphys}{{\it Solar Phys.}}

\newcommand{\ssr}{    {\it Space Sci. Rev.}}

\newcommand{\planss}{		{\it Planet. Space Sci.}}
\newcommand{\ao}{		{\it ApOpt}}
\newcommand{\zap}{		{\it ZA}}
\newcommand{\aig}{		{\it Ann. Int. Geophys. Year}}
\newcommand{\icarus}{		{\it Icarus}}
\newcommand{\TeMAE}{		{\it Terrestrial Magnetism and Atmospheric Electricity (J. Geophys. Res.)}}
\newcommand{\jgg}{		{\it Journal of Geomagnetism and Geoelectricity}}

\def\thanks{\xdef\@thefnmark{}\@footnotetext}

\newcommand{\arcdeg}{\hbox{$^\circ$}}

\def\earth{\hbox{$\oplus$}}
\def\la{\mathrel{\hbox{\rlap{\hbox{\lower4pt\hbox{$\sim$}}}\hbox{$<$}}}}
\def\ga{\mathrel{\hbox{\rlap{\hbox{\lower4pt\hbox{$\sim$}}}\hbox{$>$}}}}

\def\arcmin{\hbox{$^\prime$}}
\def\arcsec{\hbox{$^{\prime\prime}$}}

\def\lesssim{\mathrel{\hbox{\rlap{\hbox{\lower4pt\hbox{$\sim$}}}\hbox{$<$}}}}
\def\gtrsim{\mathrel{\hbox{\rlap{\hbox{\lower4pt\hbox{$\sim$}}}\hbox{$>$}}}}

\def\level #1 #2#3#4{$#1 \: ^{#2} \mbox{#3} ^{#4}$}

\makeatother 

\usepackage{titlesec}      

\begin{document}
\titleformat{\section}{\large\bfseries}{\thesection}{1em}{}
\titleformat{\subsection}{\large\bfseries}{\thesubsection}{1em}{}


\newpage
\pagenumbering{alph}

 \newpage
\thispagestyle{empty}


\begin{center}
\doublespacing
{\Large \bf Evolution and Consequences of Coronal Mass Ejections in the Heliosphere}\\
\vspace{0.5cm}
{\bf A Thesis}\\
\vspace{0.5cm}
{\bf submitted for the award of Ph.D. degree of} \\
{\large \bf MOHANLAL SUKHADIA UNIVERSITY}\\
{\bf in the} \\
{\bf Faculty of Science} \\
{\bf By} \\
{\Large \bf Wageesh Mishra} \\ 
\begin{figure*}[htb]
\hspace{6cm}
\includegraphics[width=0.2\textwidth,clip=]{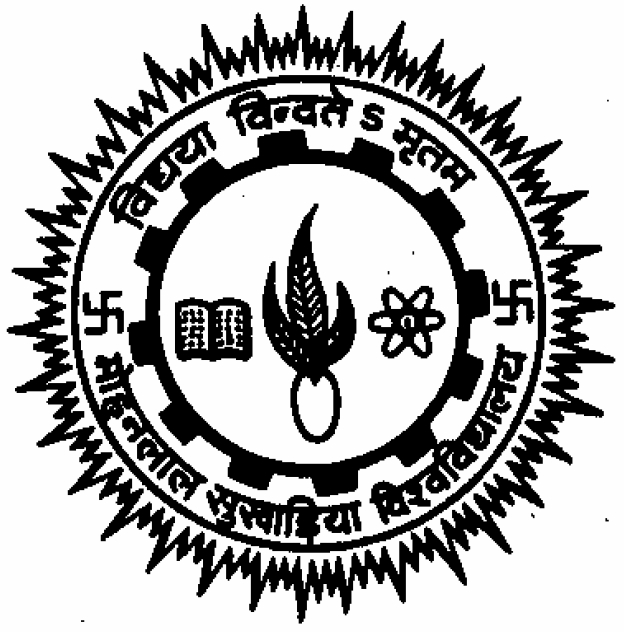}
\end{figure*}
{\bf Under the Supervision of} \\
{\large \bf Prof. Nandita Srivastava}\\
\vspace{0.05cm}
{\large \bf Udaipur Solar Observatory\\ Physical Research Laboratory, Udaipur}\\
\vspace{1cm}
{\bf Department of Physics}\\
{\bf Faculty of Science}\\
\vspace{0.2cm}
{\large \bf MOHANLAL SUKHADIA UNIVERSITY}\\
{\large \bf UDAIPUR}\\
\vspace{0.7cm}
{Year of submission: 2015}
\end{center}


%
%
%
%
%
%
%
%
%
%
%


\newpage
\thispagestyle{empty}

\begin{center}
\begin{Large}\textit{\underline{DECLARATION}}\end{Large}
\end{center}
\vspace{1cm}

\doublespacing
\par {I, Mr. {\bf Wageesh Mishra},  S/o  Mr. Nagesh Mishra,   resident of C-2, USO staff colony, Badi road, Udaipur-313 001, hereby declare that the  research work incorporated in the present thesis entitled, \textbf{\textquotedblleft Evolution and Consequences of Coronal Mass Ejections in the Heliosphere\textquotedblright} is my own work and is original. This work (in part or in full) has not been submitted to any University for the award of a Degree or a Diploma. I have properly acknowledged the material collected from secondary sources wherever required. I solely own the responsibility for the originality of the entire content.

\singlespacing
\vspace*{3cm}

\begin{tabular}{ll}
\begin{minipage}{0.3\textwidth}
 \begin{tabular}{l}
\vspace*{3cm}
Date:\\
\\
\\
\\
\end{tabular}
\end{minipage}

& \hspace*{2.1cm}
\begin{minipage}{0.7\textwidth}
\begin{tabular}{r}
\textbf{\large{Wageesh Mishra}}\\
\textbf{(Author)}\\
Udaipur Solar Observatory\\
Physical Research Laboratory\\
(Dept. of Space, Govt. of India)\\
P.B. No. 198, Dewali, Badi Road\\
Udaipur-313 001\\
Rajasthan, India.\\
\end{tabular}
\end{minipage}
\end{tabular}


\singlespacing
\newpage
\thispagestyle{empty}

\begin{center}
\begin{Large}\textit{\underline{CERTIFICATE}}\end{Large}
\end{center}

\vspace{1cm}

\begin{doublespace}
\hspace{0.15cm}I feel great pleasure in certifying that the thesis entitled \textbf{\textquotedblleft Evolution and Consequences of Coronal Mass Ejections in the Heliosphere\textquotedblright} embodies a record of the results of investigations carried out by \textbf{Mr. Wageesh Mishra} under my guidance. He has completed the following requirements as per Ph.D. regulations of the University.\\
    \hspace{2.5cm} (a) Course work as per the university rules. \\
    \hspace{2.5cm} (b) Residential requirements of the university. \\
    \hspace{2.5cm} (c) Regularly submitted six monthly progress report. \\
    \hspace{2.5cm} (d) Presented his work in the departmental committee. \\
    \hspace{2.5cm} (e) Published/accepted minimum of one research paper in a refereed research journal. \\

I recommend the submission of thesis.

\end{doublespace}

\singlespacing

\vspace*{1cm}
\begin{tabular}{ll}
\begin{minipage}{0.3\textwidth}
\begin{tabular}{l}
\vspace*{1cm}
Date:\\
\\
\\
\\
\\
Countersigned by\\
Head of the Department\\
\\
\\
\\
\end{tabular}
\end{minipage}

& \hspace*{2.1cm}
\begin{minipage}{0.7\textwidth}
 \begin{tabular}{r}
\textbf{\large{Prof. Nandita Srivastava}}\\
\textbf{(Thesis Supervisor)}\\
Udaipur Solar Observatory\\
Physical Research Laboratory\\
(Dept. of Space, Govt. of India)\\
P.B. No. 198, Dewali, Badi Road\\
Udaipur-313 001\\
Rajasthan, India.\\ 
\end{tabular}
\end{minipage}
\end{tabular}

\newpage
\thispagestyle{empty}
\vspace*{8cm}

\begin{center}
{\LARGE \textbf{This thesis is dedicated to my parents,}} \\
\vspace{0.2cm}
\normalsize{who have made sacrifices throughout to help me get where I am today.}
\end{center}




\newpage
\pagenumbering{roman}
\setcounter{page}{1}
\pdfbookmark[1]{Contents}{table} 
\pagestyle{fancy} 
\lhead{}          
\rhead{Contents}
\tableofcontents

\newpage
\rhead{List of Figures}
\listoffigures

\newpage
\rhead{List of Tables}
\listoftables

\newpage
\pagestyle{plain}
\setstretch{1.4}
\rhead{Acknowledgments}


\newpage
\phantomsection
\rhead{Acknowledgments}

{\begin{center}
\begin{Large}\textbf{Acknowledgments}               
\end{Large}                         
\end{center}}

\addcontentsline{toc}{chapter}
		 {\protect\numberline{Acknowledgments\hspace{-96pt}}}


I feel immense pleasure on the completion of my Ph.D. thesis work. I received help directly or indirectly from several people during  various stages of this work. First and foremost, I thank almighty God for giving me the determination, patience and environment required to pursue my research work successfully. I express my sincere gratitude to my Ph.D. supervisor Prof. Nandita Srivastava for her invaluable guidance, support and encouragement throughout to mould me as an independent researcher. I could participate in several international and national conferences/schools/workshops due to her backing and boost. This provided me an opportunity to interact with many scientists and students for realizing varieties of research problems. I am indeed proud to be her student and feel fortunate for having such an experienced and knowledgeable supervisor. I am grateful to her for allowing me to work under her guidance. I believe whatever I have learned working with her, I could not have learned anywhere else.

I would like to extend special thanks to Prof. P. Venkatakrishnan, Head Udaipur Solar Observatory (USO), who always encouraged and helped me. I  enjoyed discussing with him at his house on several subjects ranging from science to mythology. I remember the day when I gave my  first USO seminar after which Prof. Ashok Ambastha had literally told me `Well done', I sincerely thank him for his continuous encouragement. The informal discussion with Dr. Shibu K. Mathew has been beneficial in my work and I express my thanks to him. I am a great admirer of his supportive nature regarding genuine issues of students. His exemplary hard work has always prompted me to do  the same. I thank Dr. Brajesh Kumar and Dr. Bhuwan Joshi for their suggestions at various stages of my thesis. I have also been  benefited from the lectures given by Dr. Ramit Bhattacharyya at USO. His questions during my presentations often motivated me for further reading. I am delighted to acknowledge the instant help received from Mr. Raja Bayanna whenever required. He helped me in learning IDL programming in the initial phase at USO. I also thank Ms. Bireddy Ramya for her pleasant company. Her single minded dedication towards work is tremendously inspiring.

I express my gratitude to Prof. J. N. Goswami, Director, Physical Research Laboratory (PRL), and all faculty and staff of PRL, for giving me a nice opportunity to pursue my research work. I have immensely enjoyed the lectures by Dr. Dibyendu Chakrabarty and Dr. Bhuwan Joshi during my course work in PRL. I sincerely thank the academic committee for reviewing my progress from time to time and encouraging me. I also thank Prof. K. S. Baliyan and Dr. S. Naik for scientific discussions during my stay in PRL.

I would like to thank Dr. J. A. Davies (RAL, UK), Dr. T. A. Howard (SwRI, USA), Dr. Y. D. Liu (UC Berkeley, USA), Dr. N. Lugaz (UNH, USA) and Dr. C.  M{\"o}stl (UC Berkeley, USA) for their help and discussion via emails. Especially, I am indebted to Dr. Davies for her quick reply to my emails and guidance regarding HI images processing and data analysis. I also thank Prof. Yuming Wang (USTC, China), Dr. B. Vr\v{s}nak (Hvar Observatory, Croatia), Prof. C. Farrugia (UNH, USA), and Dr. M. Temmer (University of Graz, Austria) for the discussion, especially via emails. I thank Prof. R. Harrison (RAL, UK), Dr. Davies and Dr. Lugaz for giving me an opportunity to participate in `CME-CME Interaction Workshop' held in Oxford, UK. I am thankful to organizers of `Heliophysics Summer School-2013' for selecting me to attend the school held in HAO, Boulder, USA. I am grateful to Dr. Mausumi Dikpati (HAO) for arranging my talk in her institute during my visit to attend the summer school. I thank Prof. P. K. Manoharan (NCRA, Pune) and organizers of `International Space Weather Winter School-2013' for supporting me to attend the school held in NCU, Taiwan. I am extremely lucky to meet and discuss with several scientists in various conferences, namely Dr. Prasad Subramanian (IISER, Pune),  Prof. Manoharan, Dr. Durgesh Tripathi (IUCAA, Pune), Prof. B. N. Dwivedi (IIT, BHU), Dr. D. Banerjee (IIA, Bangalore), Prof. S. Ananthakrishnan (University of Pune), Dr. R. Ramesh (IIA, Bangalore), Dr. A. K. Srivastava (IIT, BHU) and many others.

I extend thanks to my friends at USO, Anand D. Joshi, Vemareddy Panditi, Suruchi Goel, Dinesh Kumar, Upendra Kushwaha, Sanjay Kumar, Alok Tiwari, Rahul Yadav, Hemant Saini, Ramya and Sudarshan. I have enjoyed their company whether it be playing badminton, outings, having party, taking meals or discussing over science, society, politics, mythology and many other non-scientific aspects of life. I am especially grateful to Anand from whom I have learnt a lot about IDL programing. I also learnt several concepts of MHD from Sanjay. I also thank my friends Arushi, Puneet and Tanuj who although spent a short time at USO during their projects, are still in contact. We learn and enjoy together among the true friends, and I believe to achieve all these throughout my academics.

I thank the administrative staff of USO, Mr. Raju Koshy, Mr. Rakesh Jaroli, and Mr. Pinakin S. Shikari for providing me good facilities at USO. The affection and support shown by Mr. Koshy is greatly acknowledged. I also thank Mr. Naresh Jain, Mr. Mukesh Saradava, Mr. Laxmi Lal Suthar, Mr. Jagdish Singh Chauhan and Mr. Dal Chand Purohit for providing me the cordial environment at USO.  My especial thanks go to late Mr. Sudhir Kumar Gupta who was very kind, helpful and always encouraged me. We had celebrated several festivals together and his absence is a grief to me. I sincerely acknowledge the friendly nature of USO library staff Mr. Nurul Alam and Ms. Hemlata Kumhar. I also thank the staff of Physics department and Ph.D. section of Mohanlal Sukhadia University for their kind cooperation throughout my thesis.

I have enjoyed the hospitality of  Mrs. Usha Venkatakrishnan, Mrs. Mahima Kanthalia, Mrs. Saraswathi, Mrs. Bharati and Mrs. Richa, I express my gratitude to all of them. I thank everyone living at USO colony for their help, arranging get-together, and homely environment provided. I am indebted to Usha madam for her affection and encouragement. The kind and friendly nature of Mahima is greatly acknowledged.

Although, I spent a short time at PRL but I enjoyed my stay with my friends. I would like to thank friends of my batch, Gaurav Tomar, Gaurav Sharma, Naveen Negi, Girish, Yashpal, Priyanka, Monojit, Anjali, Avdhesh, Bhavya, Lekshmy, Gangi Reddy, Tanmoy and Gulab. I would like to acknowledge the continuous help and encouragement from Sharma, Tomar and Negi which is beyond the description. I thank my other friends at PRL, Arun Awasthi, Susanta Bisoi, Sunil Chandra, Arun Pandey, Girish, Diptiranjan, Rukmani, Kuldeep, Navpreet and several others. I also thank my school friends Jainendra, Yogendra and Anandita for their help and inspiring company. I convey thanks to my college and university friends Arjun Tiwari, Anuj Verma, Shishir Pandey, Himanshu Rai, Abhishek, Rashmi, Jamwant, Siddhi, Suneeta, Narayan Datt and Jyotsana who have been always supportive and encouraging. I also thank Sargam Mulay, Bidya Binay, Sharad, Hitaishi, Sajal, Pandey ji, Yogita, Vasanth and Rahul for their support and suggestions at various stages of my thesis.

My especial thanks go to my teachers of the Gorakhpur university, Prof. U. S. Pandey, Prof. M. Mishra, Prof. S. N. Tiwari, Prof. S. Rastogi and Dr. R. S. Singh who have had an impact on me. I always admire Prof. Pandey for his helpfulness, simplicity and dedication that has helped me in my career. I also thank my college teachers Dr. S. C. Verma and Dr. B. K Verma, although with whom I am not associated now. My gratitude towards my teachers would remain incomplete without remembering my school teachers Mr. J. P. Pathak, Mr. K. Tiwari and Mr. O. P. Chaturvedi. 

I am in dearth of words to thank my parents (Mr. Nagesh Mishra and Mrs. Shyama Devi) for the encouragement and the support they have provided me throughout my life. Their sacrifices and way of instilling discipline, compassion, honesty and spirituality into me, for my true success, cannot be expressed in words. Therefore, I dedicate my thesis to them. I am deeply sad that my late mother, who used to ask me about my work and life and always gave her blessings, could not see the completion of my thesis. I miss her but feel that she is continuously watching and blessing me from the heaven. I also thank my dear brothers (Rajiv, Ashutosh and Gyaneesh) and sisters (Sadhana and Kamana) for their support, suggestions and love. I extend thanks to my grandmother, grandfather, aunty, uncle, bhabhi and cousins for their affection and support. Last and really not least, I thank all my friends, relatives, teachers, preachers (DJJS) and students from whom I have learned something to be where I am today.

\singlespacing
\vspace*{0.1cm}
\hspace*{9cm}
\textbf{\large Wageesh Mishra}\\

\newpage
\pagestyle{plain}
\rhead{Abstract}


\newpage
\phantomsection
{
\begin{center}
\begin{Large}
\textbf{Abstract}
\end{Large}
\end{center}
}
\addcontentsline{toc}{chapter}
  {\protect\numberline{Abstract\hspace{-96pt}}}

\setstretch{1.4}

Coronal mass ejections (CMEs), the most energetic eruptive phenomena occurring in the heliosphere, are recognized as the primary driver of many space weather events. Investigating their heliospheric evolution and consequences is critical to understanding the solar-terrestrial relationship. Prior to the development of wide-angle imaging of the heliosphere, the studies about propagation of CMEs was limited to analyzing their plane-of-sky projected remote observations within few solar radii of the Sun, and in situ observations in the vicinity of the Earth. Heliospheric Imagers (HIs) onboard \textit{STEREO} providing multiple views of CMEs in the heliosphere, for the first time, have filled the vast and crucial observational gap between near the Sun and the Earth. We notice that three-dimensional (3D) speeds of CMEs near the Sun, derived by implementing several stereoscopic reconstruction methods on coronagraphs (CORs) data, are not quite sufficient for understanding the propagation and accurate forecasting of the arrival time at the Earth of a majority of CMEs. This may be because of many factors that significantly change the CME kinematics beyond the CORs field of view, such as the interaction/collision of two or more CMEs or the interaction of CMEs with the ambient solar wind medium. In order to understand the heliospheric propagation of CMEs, several reconstruction methods, based on the use of time-elongation profiles of propagating CMEs viewed from single or multiple vantage points, are implemented to estimate the kinematics of the CMEs in the heliosphere. The time-elongation profiles of the tracked features of the Earth-directed CMEs, selected for our study, are derived from the \textit{J}-maps constructed from Heliospheric Imagers (HI1 and HI2) data. Using the kinematic properties as inputs to the Drag Based Model (DBM) for the distance beyond which the CMEs cannot be tracked unambiguously in the \textit{J}-maps, the arrival time of these CMEs have been estimated. These arrival times have also been compared with the actual arrival times as observed by in situ instruments located near the Earth.

We assess relative performance of a total of 10 existing reconstruction methods applicable on SECCHI/HI observations to derive the kinematic properties of the selected CMEs. The ambient solar wind into which these selected CMEs, traveling with different speeds, are launched, is different. Therefore, these CMEs evolve differently during their journey from the Sun to 1 AU. Our results show  that stereoscopic reconstruction methods perform better, especially those which take into account the global geometry of a CME and assume that the line of sight of both observers simultaneously images different parts of a CME. For understanding the association between remote and in situ observations of CMEs, we have continuously tracked different density enhanced features using \textit{J}-maps. Further, we compared their estimated heliospheric kinematics and arrival time, and then associated them with features observed in situ.

We have attempted to understand the evolution and consequences of the interacting/colliding CMEs in the heliosphere using 
SECCHI/HI, \textit{WIND} and \textit{ACE} observations. By estimating the true mass and 3D kinematics of these interacting CMEs in the inner heliosphere, we have studied their pre- and post-collision dynamics, momentum and energy exchange between them during the collision phase. We found a significant change in the dynamics of the CMEs after their collision and interaction. Relating heliospheric imaging observations with in situ measurements at L1, we find that the interacting CMEs move adjacent to each other after their collision in the heliosphere and are recognized as distinct structures in in situ observations. These observations also show heating and compression, formation of magnetic holes (MHs) and interaction region (IR) as signatures of CME-CME interaction. We also noticed that long-lasting IR, formed at the rear edge of preceding CME, is responsible for large geomagnetic perturbations. Our analysis shows an improvement in arrival time prediction of CMEs using their post-collision dynamics than using pre-collision dynamics. We also examined the differences in geometrical evolution of slow and fast CMEs during their propagation in the heliosphere.

Our study highlights the significance of using \textit{J}-maps constructed from \textit{STEREO}/HI observations in studying heliospheric evolution of CMEs, CME-CME collision, identifying and associating three-part structure of CMEs in their remote and in stiu observations, and hence for the purpose of improved space weather forecasting.

\newpage
\pagestyle{plain}

\newpage
\phantomsection
\pagestyle{plain}
\begin{center}
\begin{LARGE}\textbf{List of Publications}
\end{LARGE}          
\end{center}
\addcontentsline{toc}{chapter}
 		 {\protect\numberline{List of Publications\hspace{-96pt}}}

\begin{large}\textbf{I. Research Papers in Refereed Journals:}
\end{large}

\begin{enumerate}

\item \textbf{Wageesh Mishra} and Nandita Srivastava, 2014, 
\textit{Heliospheric tracking of enhanced density structures of the 2010 October 6 CME},
\textit{\textbf{J. Space Weather Space Clim.}}, (under revision)

\item \textbf{Wageesh Mishra}, Nandita Srivastava, and D. Chakrabarty, 2015, 
\textit{Evolution and consequences of interacting CMEs of 9-10 November 2012 using \textit{STEREO}/SECCHI and in situ observations},
\textit{\textbf{Solar Phys.}}, \textbf{290}, 527.

\item \textbf{Wageesh Mishra} and Nandita Srivastava, 2014, 
\textit{Morphological and kinematic evolution of three interacting coronal mass ejections of 2011 February 13-15},
\textit{\textbf{Astrophys. J.}}, \textbf{794}, 64.

\item \textbf{Wageesh Mishra}, Nandita Srivastava, and Jackie A. Davies, 2014, 
\textit{A comparison of reconstruction methods for the estimation of coronal mass ejections kinematics based on SECCHI/HI observations}, 
\textit{\textbf{Astrophys. J.}}, \textbf{784}, 135.

\item \textbf{Wageesh Mishra} and Nandita Srivastava, 2013, 
\textit{Estimating the arrival time of earth-directed coronal mass ejections at in situ spacecraft using COR and HI observations from \textit{STEREO}},
\textit{\textbf{Astrophys. J.}}, \textbf{772}, 70.

\end{enumerate}

\begin{large}\textbf{II. Research Papers in Proceedings:}
\end{large}

\begin{enumerate}

\item \textbf{Wageesh Mishra} and Nandita Srivastava, 2013, 
\textit{Estimating arrival time of 10 October 2010 CME using \textit{STEREO}/SECCHI and in situ observations}
\textit{\textbf{ASI Conference Series}}, \textbf{10}, 127.

\end{enumerate}


\newpage
\pagenumbering{arabic}
\setcounter{page}{1}
\pagestyle{fancy}
\lhead{}
\setstretch{1.4}
\chapter{Introduction}
\label{Chap1:IntMot}
\rhead{Chapter~\ref{Chap1:IntMot}. Introduction}

\section{The Sun}
\label{Sun}

The early humans of several countries and civilizations had considered the Sun as their God or Goddess and worshipped it. This is possibly because they realized from their daily experience that without the Sun there will be no light, warmth, life, change in seasons, and measurement of time on the Earth. Even in modern times, the Sun is worshipped in many countries and religions. Understanding the importance of the Sun, several observatories or observing sites were built and used by various civilizations around the globe, beginning from the middle neolithic period (\citealp{Mukerjee2003}, \citealp[ch. 1]{Bhatnagar2005}; \citealp{Boser2006}).

With the progress made in the field of astronomy, it is now well understood that the Sun is the nearest star from the Earth, and therefore it can be observed with good spatial resolution and studied in details. The Sun being in plasma state acts as a laboratory to test and understand various theories of plasma physics. In addition, the study of the Sun is important from the astrophysics perspective, in general. The modern space and ground based observations have unmasked the dynamic and active nature of the Sun in the form of sunspots, faculae, filaments, prominences, coronal holes, flares, and coronal mass ejections (CMEs). Further, the Sun is the driving factor for the terrestrial and space weather, therefore the study of solar-terrestrial relations is of great importance for our space and ground based technological systems as well as for human life and health.

The Sun is a main sequence star of spectral type G2V with mass \textit{M$_{\odot}$} $\approx$ 1.98 $\times$ 10$^{30}$ kg, luminosity \textit{L$_{\odot}$} $\approx$ 3.84 $\times$ 10$^{26}$ W and radius \textit{R$_{\odot}$} $\approx$ 6.96 $\times$ 10$^{8}$ m \citep[p. 24]{Lang2006}. The mass of the Sun is about 99\% of the total mass of the solar system. The solar system, to which the Sun, the 8 planets, asteroids, meteorites, comets and other small dust particles belong, is located in our Milky Way galaxy. Similar to other stars, the Sun was born from the gravitational collapse of a molecular cloud approximately around 4.6 $\times$ 10$^{9}$ years ago, and now is currently in a state of hydrostatic equilibrium. It is predicted that the Sun will enter in a red giant phase in another $\approx$ 5 billion years before ending its life as a white dwarf \citep[p. 432]{Foukal2004}. 

\section{Structure of the Sun} 
\label{StructSun}

\subsection{Solar interior}
The modern picture of the internal structure of the Sun has been built up over time. The three most important contributions to this have been the `standard solar model' (SSM; \citealt{Bahcall1982}), helioseismology \citep{Leibacher1985}, and solar neutrino observations \citep{Bahcall2001}. As the interior of the Sun cannot be directly observed, its structure is modeled and then compared to the observed properties by iteratively changing the model parameters, until they match the observations. The SSM is essentially several differential equations, constrained by boundary conditions (mass, radius, and luminosity of the Sun), which are derived from the principles of fundamental physics. Helioseismology allows us to probe the solar interior by studying the propagation of waves in the Sun, mainly the sound waves \citep{Leighton1962,Ulrich1970}.

The Sun's interior includes the core, the radiation zone, and the convection zone. The core, which extends out to about 
0.25 \textit{R$_\odot$} from the center, is at a temperature of about 1.5 $\times$ 10$^{7}$ K and has a density $\approx$ 1.5 $\times$ 10$^{5}$ kg m$^{-3}$ \citep[p. 24]{Lang2006}. The core of the Sun is the source of its energy via the process of thermonuclear fusion which results in the formation of heavier elements as well as the release of energy in the form of gamma ray photons. 

Outside the core is the radiative zone which extends out from 0.25 \textit{R$_\odot$} to 0.70 \textit{R$_\odot$}. The gamma photons produced in the core are absorbed and re-emitted repeatedly by nuclei in the radiative zone, with the re-emitted photons having successively lower energies and longer wavelengths. The temperature drops from about 7 $\times$ 10$^{6}$ K at the bottom of the radiative zone to 2 $\times$ 10$^{6}$ K just at the top of the radiative zone. Due to the high density ($\approx$ 2 $\times$ 10$^{4}$ kg m$^{-3}$) in the radiative zone, the mean free path of the photons is very small ($\approx$ 9.0 $\times$ 10$^{-2}$ cm), hence the photons take approximately tens to hundreds of thousands of years to travel through the radiative zone \citep{Mitalas1992}. Hence, if the energy generation processes in the core of the Sun suddenly stopped, the sun will continue to shine for millions of years.

Above the radiative zone is the convective zone extending from about 0.70 \textit{R$_\odot$} to 1 \textit{R$_\odot$} at the surface of the Sun. The convection zone rotates differentially and temperature in this zone decreases very rapidly with increasing height and becomes around 5700 K at its outer boundary. In this zone the energy is transported by convection. Hot regions at the bottom of this layer become buoyant and rise, cooler material from above descends, and giant convective cells are formed which can be seen on the surface of the Sun (i.e. photosphere) as granules. In the convection zone, the temperature gradient set up by radiative transport is larger than adiabatic gradient and hence a convection pattern is established \citep[p. 204]{Foukal2004}. Convective circulation of plasma (charged particles) generates large magnetic fields that play an important role in producing solar activity on the Sun. From the recent helioseismology results, it has been proposed that around 0.7 \textit{R$_\odot$} from the center of the Sun (i.e. transition layer between radiative and convective zone with thickness of 0.04 \textit{R$_\odot$}), the solar magnetic fields are generated through a dynamo mechanism. Here the sound speed and density profiles show a distinct sudden `bump' called the tachocline \citep{Spiegel1992}.

\subsection{Solar atmosphere}
Based on the density, temperature, and composition,  the solar atmosphere is subdivided into three regions, the photosphere, chromosphere and corona (Figure~\ref{atmlayers}). The density of the plasma generally decreases from the photosphere to the corona. However, the temperature decreases before reaching a minimum at the base of the chromosphere, then slowly increases until there is a rapid increase at the transition region which continues into the corona. This rapid increase in temperature is termed as `coronal heating problem' \citep{Grotrian1939,Davila1987,Gudiksen2004,Klimchuk2006,Parnell2012}.

\begin{figure*}[!htb]
	\centering
		\includegraphics[scale=3.1]{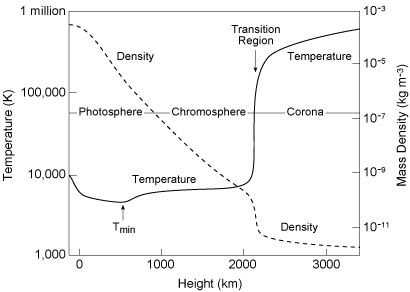}
\caption[Variations of temperature and density in the solar atmosphere]{The temperature of the solar atmosphere decreases from $\approx$ 5700 K at the visible photosphere to a minimum value of $\approx$ 4,400 K about 500 km higher up. The temperature increases with height, slowly at first, then extremely rapidly in the narrow transition region (less than 100 km thick, between the chromosphere and corona) from $\approx$ 10,000 K to about one million K. (Courtesy of Eugene Avrett, Smithsonian Astrophysical Observatory. \citealp[p. 115]{Lang2006})}
\label{atmlayers}
\end{figure*}

The Sun's corona extends millions of kilometers into space and is naturally visible during a total solar eclipse. The three components of the corona (i.e. K-corona, F-corona, and E-corona) are described based on the nature of radiation emitted by them. The Brightness variation of these three components of the solar corona as a function of radial distance is shown in Figure~\ref{Intcorona}. The K-corona dominates between 1.03 \textit{R$_\odot$}-2.5 \textit{R$_\odot$}. From this region, the emitted scattered light from the coronal plasma shows the continuous spectrum of the photosphere with no Fraunhofer lines and is found to be strongly polarised. The F-corona dominates beyond 2.5 \textit{R$_\odot$} displays the solar spectrum with Fraunhofer lines superimposed on the continuum. The outer part of the F-corona is observed to merge into the zodiacal light. The E-corona is due to spectral line emission from visible to EUV by several atoms and ions in the inner part of the corona, containing many forbidden line transitions, thus it contains many polarization states. These lines provide information on very low density and extremely high temperature of the corona. The density of the particles is only of the order of 10$^{6}$ to 10$^{8}$ cm$^{-3}$ as against the value of 10$^{10}$ to 10$^{12}$ cm$^{-3}$ for chromosphere and of 10$^{16}$ to 10$^{17}$ cm$^{-3}$ for the photosphere \citep[ch. 1]{Golub2009}.


\begin{figure*}[!htb]
	\centering
		\includegraphics[scale=0.8]{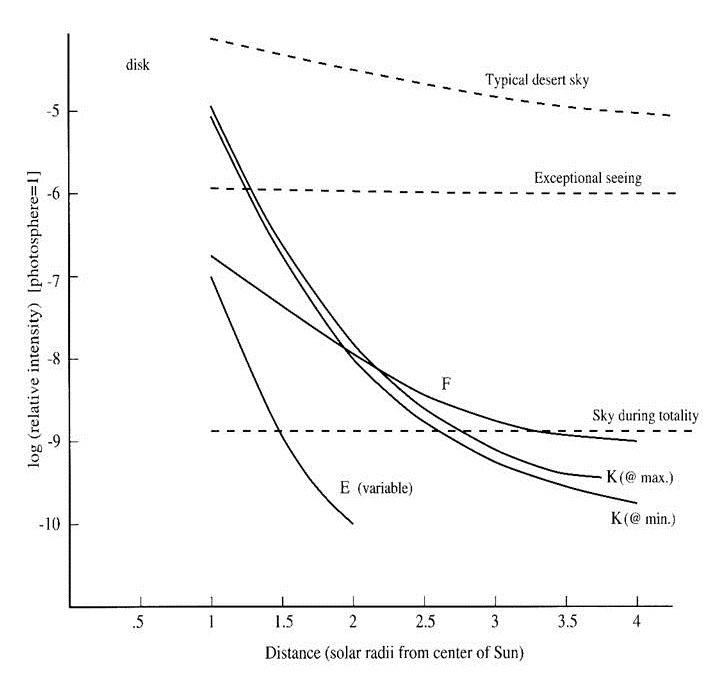}
\caption{Brightness variation of the different components of the solar corona with radial distance (reproduced from \citealp[Ch. 5]{Golub2009}).}
\label{Intcorona}
\end{figure*}

\subsection{Solar wind and the heliosphere}

The solar wind is constant out-stream of charged particles of plasma (mixture of ions and electrons) from the Sun's atmosphere and fills the space around the Sun \citep{Biermann1951,Parker1958}. The outer atmosphere of the Sun (i.e. corona) is so hot that even the gravity of the Sun can not prevent it from continuously evaporating. The escaping particles carry energies of $\approx$ 1 keV, and is observed in two states of fast and slow speed. The slow solar wind has speed of $\approx$ 400 km s$^{-1}$ with a typical proton density of $\approx$ 10 cm$^{-3}$ and the fast solar wind has speed of $\approx$ 800 km s$^{-1}$ with a typical proton density of $\approx$ 3 cm$^{-3}$ \citep{Schwenn1990}. In 1973, in the Skylab era, sources of fast speed solar wind was discovered as coronal holes  \citep{Krieger1973}. Coronal holes are usually found where ``open'' magnetic field lines prevail. 

As the solar wind runs into the interstellar medium (ISM) it becomes abruptly slow from being supersonic to sub-sonic speed at a certain location from the Sun. This location in ISM is called the termination shock. The Voyager 1 spacecraft in 2004 and Voyager 2 spacecraft in 2007 passed through the termination shock at $\approx$ 94 AU and 84 AU from the Sun, respectively  \citep{Richardson2008, Burlaga2008}. Beyond the termination shock there is heliosheath where the ISM and solar wind are in pressure balance. The outer boundary of heliosheath is called the heliopause also known as the edge of the heliosphere. The solar wind does not expand infinitely into space but it stops at heliopause. Voyager 1 had crossed the heliopause as of 2012 August 25 at a distance of 121 AU from the Sun \citep{Cowen2012}. The region around the Sun which is dominated by the solar wind is called the heliosphere.

\section{Coronal Mass Ejections}
\label{CMEs}

Observations of solar corona has been carried out earlier during solar eclipses. The earliest observation of a Coronal Mass Ejection (CME) probably dates back to the eclipse of 1860 which is clear from a drawing recorded by G. Temple. Then, in space era,  a CME was imaged on 1971 December 14 by a coronagraph on board the seventh Orbiting Solar Observatory (OSO-7) satellite \citep{Tousey1973}. A coronagraph is an instrument which blocks the photospheric light from the disk of the Sun and observes the corona by creating an artificial eclipse. After the discovery of CME from OSO-7, thousands of CMEs have been observed from a series of space-based coronagraphs e.g. Apollo Telescope Mount on board Skylab \citep{Gosling1974}, Solwind coronagraph on board P78-1 satellite \citep{Sheeley1980}, Coronagraph/Polarimeter on board Solar Maximum Mission (SMM) \citep{MacQueen1980}, Large Angle Spectrometric COronagraph (LASCO) on board SOlar and Heliospheric Observatory \citep{Brueckner1995}, and the coronagraphs (CORs) on Solar TErrestrial RElations Observatory (\textit{STEREO}) \citep{Howard2008}. These observations were complemented by white light data from the ground-based Mauna Loa Solar Observatory (MLSO) K-coronameter which had a FOV from 1.2 \textit{R$_\odot$}-2.9 \textit{R$_\odot$} \citep{Fisher1981} and emission line observations from the coronagraphs at Sacramento Peak, New Mexico \citep{Demastus1973} and Norikura, Japan \citep{Hirayama1974}. 

The name CME was initially coined for a feature which shows an observable change in coronal structure that occurs on a time scale of few minutes to several hours and involves the appearance (and outward motion) of a new, discrete, bright, white-light feature in the coronagraphic field of view (FOV) \citep{Hundhausen1984}. It is now well established that CMEs are frequent expulsions of magnetized plasma from the Sun into the heliosphere. In white light observations, it is noted that the frequency of occurrence of CMEs around solar maximum is $\approx$ 5 per day and at solar minimum is $\approx$ 1 per day 
\citep{StCyr2000,Webb2012}.

It is evident from a survey of literature that consequences of CME have been observed well before its discovery in 1971. For example, CMEs were observed at larger distances from the Sun in radio via interplanetary scintillation (IPS) observations from the 1960s. However, only around 1980s the association between IPS and CMEs could be established \citep{Hewish1964,Houminer1972,Tappin1983}. The IPS technique is based on measurements of the fluctuating intensity level of a large number of point-like distant meter-wavelength radio sources. Also, the zodiacal light photometers on the twin Helios spacecraft during 1975 to 1983 \citep{Richter1982} have observed the regions in the inner heliosphere from 0.3 AU to 1.0 AU but with an extremely limited FOV. In the present era of heliospheric imagers, e.g. Solar Mass Ejection Imager (SMEI) \citep{Eyles2003} on board the Coriolis spacecraft launched early in 2003 and Heliospheric Imagers (HIs) \citep{Eyles2009} launched on the twin \textit{STEREO} spacecraft in late 2006, several CMEs have been observed far from the Sun. \textit{SOHO}/LASCO has detected well over 10$^{4}$ CMEs till date and still continues (\citealt{Yashiro2004}; \url{http://cdaw.gsfc.nasa.gov/CME\_list/}). SMEI observed nearly 400 transients during its 8.5 year lifetime, it was switched off in September 2011. The number of CME ``events'' reported using the HIs on board \textit{STEREO} is now more than one  thousand (\url{http://www.stereo.rl.ac.uk/HIEventList.html}), although less than 100 have been discussed so far in the scientific literature \citep{Webb2012}.

\subsection{Observations of CMEs}
\label{CMEobs}

In white light images, CMEs are seen due to Thomson scattering of photospheric light from the free electrons of coronal and heliospheric plasma. The intensity of Thomson scattered light has an angular dependence which must be accounted for the measured brightness of CMEs \citep{Billings1966,Vourlidas2006,Howard2009}. They are faint relative to the background corona, but much more transient, therefore a suitable coronal background subtraction is applied to identify them. The advantage of white light observations over radio, IR or UV observations is that Thomson scattering only depends on the observed electron density and is independent of the wavelength and temperature \citep{Hundhausen1993}. The coronagraphs record two-dimensional (2D) image of a three-dimensional (3D) CME projected onto the plane of sky. Therefore, the morphology of CME in coronagraphic observations depends on the location of the observing instruments (e.g. coronagraphs) and launch direction of CME from the Sun. The CMEs launched from the Sun toward or away from the Earth, when observed from the near Earth coronagraphs (e.g. LASCO coronagraph on board \textit{SOHO} located at L1 point of Sun-Earth system), will appear as `halos' surrounding the occulting disk of coronagraphs \citep{Howard1982}. Such a CME is called a ``halo'' CME  (Figure~\ref{haloCME}). A CME having 360$\arcdeg$ apparent angular width is called ``full halo'' CME and with apparent angular width greater than 120$\arcdeg$ but less than 360$\arcdeg$ is called as ``partial halo''. Thus, the nomenclature of a CME is restricted by its viewing perspective. The observations of solar activity on the solar disk, associated with CME, are necessary to help distinguish whether a halo CME was launched from the front or backside of the Sun relative to the observer. Front side halo CMEs observed by \textit{SOHO}/LASCO are important as they are the key link between solar eruptions and major space weather phenomena such as geomagnetic storms and solar energetic particle events.  Such CMEs that are launched from near the disk center tend to be more geoeffective while those closer to solar limb are less so. It is important to note that among all the CMEs, only about 10\% are partial halo type (i.e. width greater than 120$\arcdeg$)  and about 4\% are full halo type \citep{Webb2000}.

\begin{figure}[!htb]
\centering
\includegraphics[scale=0.4]{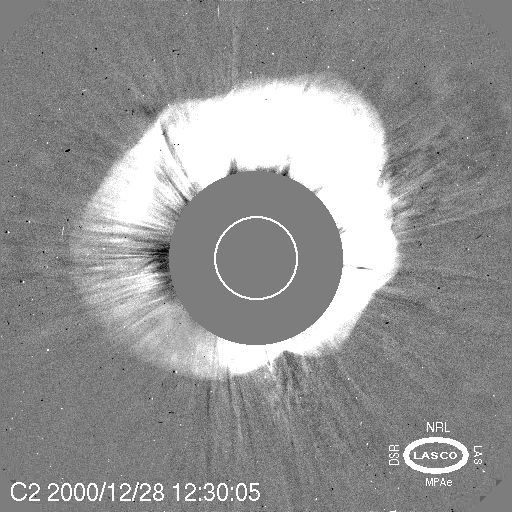}
\caption[A ``halo'' CME observed by LASCO-C2 coronagraph on \textit{SOHO}]{An image of a ``halo'' CME observed by LASCO-C2 coronagraph on \textit{SOHO}. The CME was launched from the Sun on 2000 December 28. The white circle in the center is the size and location of the solar disk, which is obscured by the occulting circular disk of radius 1.7 \textit{R$_\odot$}. (\textit{Image credit: \url{http://lasco-www.nrl.navy.mil}})}
\label{haloCME}
\end{figure}

A typical CME observed near the Sun often appears as  ``three-part'' structure comprising of an outer bright frontal loop (i.e. leading edge), and a darker underlying cavity within which is embedded a brighter core as shown in Figure~\ref{threepartCME} \citep{Illing1985}. The front may contain swept-up material by erupting flux ropes or the presence of pre-existing material in the overlying fields \citep{Illing1985,Riley2008}. The cavity is a region of lower plasma density but probably higher magnetic field strength. The core of CME can often be identified as prominence material based on their visibility in chromospheric emission lines  \citep{Schmieder2002} and often appear to have helical structure.

\begin{figure*}[!htb]
	\centering
		\includegraphics[trim = 10mm 76mm 165mm 34mm, clip]{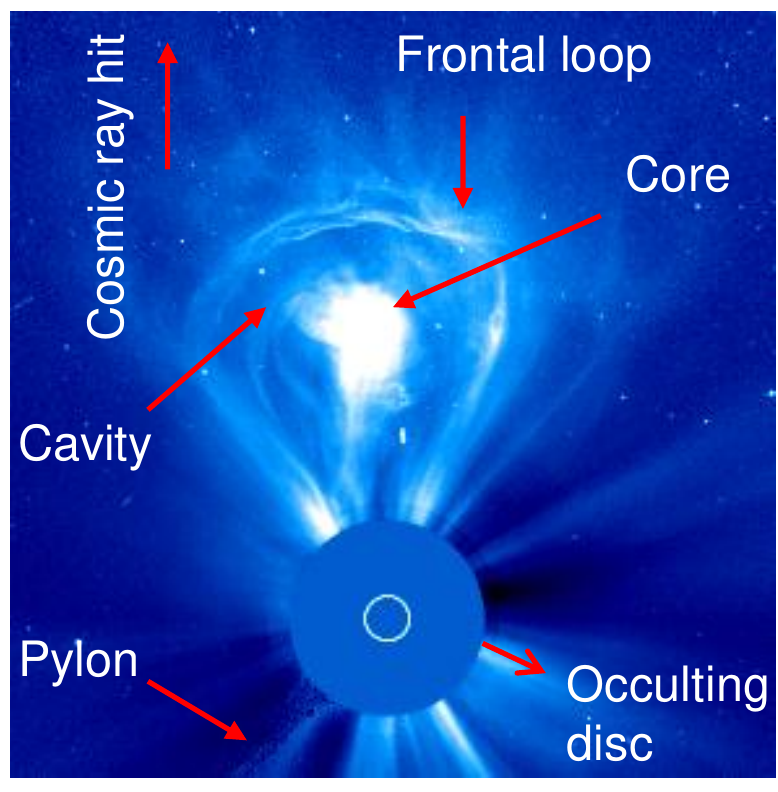}
\caption[A classical CME with 3-part structure]{A classical 3-part CME seen in the LASCO-C3 field of view on 2000 February 27 at 07:42 UT, showing a bright frontal loop surrounding a dark cavity with a bright core. (\textit{Image credit: \url{http://soho.nascom.nasa.gov}})}
\label{threepartCME}
\end{figure*}

The onset of CMEs has been associated with many solar disk phenomena such as flares \citep{Feynman1994}, prominence eruptions 
\citep{Hundhausen1999}, coronal dimming \citep{Sterling1997}, arcade formation \citep{Hanaoka1994}. Several CMEs have also been observed which can not be associated with any obvious solar surface activity \citep{Robbrecht2009,Ma2010,Vourlidas2011} and therefore have no easily identifiable signature to locate their source region on the Sun, and these are called the ``problem or stealth CMEs''. 

\subsection{Thomson scattering}

The scattering of white light in the solar corona and the solar wind is an example of Thomson scattering, which is a special case of the general theory of the scattering of electromagnetic waves by charged particles. Since the wavelength of white light is lesser than the separation between the charge particles in the corona, and the energy of the white light photons is less than the rest mass energy of the particles in the corona, therefore the photospheric light gets Thomson scattered by electrons in the corona and solar wind. The details of Thomson scattering is given in earlier studies \citep{Minnaert1930,Billings1966,Howard2009,Howard2012,Howard2013}. These studies have shown that the received intensity of the scattered light by an observer depends on its location relative to the scattering source and incident beam. If scattered light is decomposed into two components, then for an observer, the intensity of the component seen as transverse to the incident beam is isotropic, while the intensity of the component seen as parallel to the projected direction of the incident beam varies as square of cosine of scattering angle ($\chi$). Hence, the efficiency of Thomson scattering measured by an observer is minimum at $\chi$ = 90$\arcdeg$, i.e. on Thomson surface (TS), however, this is the point where incident light and electron density is found to be maximum. The combined  effect of all the three factors is that the TS is the locus of points where the scattering intensity is maximized however a spread of the observed intensity to larger distances from the TS is noted. This spreading is called `Thomson plateau' which is greater at larger distances (elongations) from the Sun, where elongation is the Sun-observer-scattering feature angle. Figure~\ref{TS} shows the relevant angles in the context of the Thomson scattering geometry.

\begin{figure*}[!htb]
	\centering
		\includegraphics[scale=5.0]{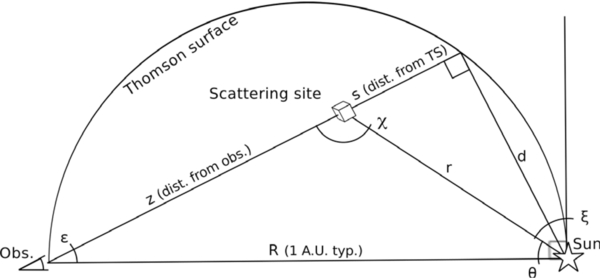}
\caption[Relevant angles in the context of the Thomson scattering geometry]{The Thomson surface, the line of sight with elongation $\varepsilon$ passes through the scattering site and marks an angle $\chi$ with the radial direction from the Sun, sky angle $\xi$ and observer-Sun-scattering point angle $\theta$, at a given heliocentric distance $r$ from the Sun is shown. (reproduced from \citealt{Howard2012})}
\label{TS}
\end{figure*}

It is also noted that when a feature is observed in terms of the observer-Sun-scattering point angle ($\theta$), the width of the peak scattering intensity is almost independent of the elongation but the location of the peak varies in accordance with the relative location of the TS. This width is large and shows only a small variation in intensity with a change of 30$\arcdeg$-40$\arcdeg$ from the TS \citep{Howard2009}. Recently, \citet{Howard2012} have shown that the sensitivity of unpolarized heliospheric imagers is not strongly affected by the geometry relative to the TS, and in fact, heliospheric imagers have observed the CMEs very far from the TS.

Conclusively, the observed brightness of a CME can change corresponding to its changing location from the TS and hence corresponding to observers at different locations, especially at larger elongation. This concept has implications for understanding how the kinematics and morphology of CMEs may be influenced from different locations of the observers.

\subsection{Properties of CMEs}

CMEs are characterized by their speed, angular width, source location relative to solar disk, mass and energies.
It is noted that speeds of CMEs near the Sun range from few km s$^{-1}$ to 3000 km s$^{-1}$ \citep{StCyr2000}. After a distance of about 2 \textit{R$_\odot$} from the Sun, CMEs accelerate or decelerate slowly in the FOV of coronagraphs 
\citep{StCyr2000,Yashiro2004}. It is also found that a typical CME shows three-phase kinematic profile, first, a slow rise over tens of minutes, then a rapid acceleration between 1.4 \textit{R$_\odot$}-4.5 \textit{R$_\odot$} during the main phase of a flare, and finally a propagation phase with constant or decreasing speed \citep{Zhang2006}. These three distinct phases of a CME are shown in Figure~\ref{CMEpropimg}. Excluding the partial and full halo CMEs, the apparent angular width of CMEs is found to vary from few degrees to more than 120$\arcdeg$, with an average value of about $\approx$ 50$\arcdeg$ 
\citep{Yashiro2004}. The source locations of a majority of CMEs are near the solar equator between 25$\arcdeg$ N and S, around the solar minimum, however, few CMEs are seen at higher latitudes also \citep{StCyr2000}. The estimated total mass of CMEs range from 10$^{10}$ kg to 10$^{13}$ kg, and the total energy from 10$^{20}$ J to 10$^{26}$ J. The average mass and energy of a CME is 1.4 $\times$ 10$^{12}$ kg and 2.6 $\times$ 10$^{23}$ J, respectively \citep{Vourlidas2002}.

\begin{figure*}[!htb]
	\centering
		\includegraphics[scale=2.5]{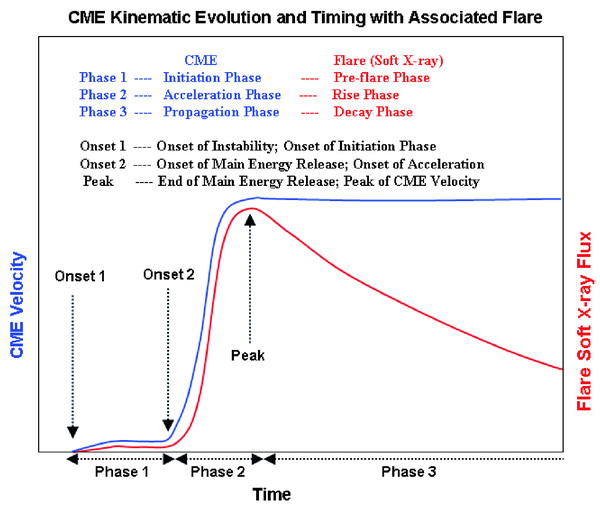}
\caption[Three phase kinematic profile of a CME]{The three different phases of CME kinematics and its relation with temporal evolution of GOES soft X-ray flux is shown. The initiation, acceleration, and propagation phase of the CME kinematics correspond to the preflare, rise, and decay phase of the associated flare, respectively. (reproduced from \citealt{Zhang2006})}
\label{CMEpropimg}
\end{figure*}

It is noted that aforementioned properties of CMEs are estimated from the 2D coronagraphic images of CMEs and therefore are subject to the problem of projection and perspective. These studies are based on the plane of sky assumption, i.e. CMEs are propagating orthogonal to observer, therefore if this assumption is not fulfilled, the speed, mass, and energies of CMEs will be underestimated \citep{Vourlidas2010} while the angular width will be severely overestimated \citep{Burkepile2004}. These properties derived from a statistical study, will also depend on the sensitivity of the coronagraphs and the selection of sample of CMEs.

\section{Heliospheric Evolution of CMEs}

A CME, after its launch, propagates in the heliosphere filled with ambient solar wind medium. It undergoes different morphological and kinematic evolution throughout its propagation in the heliosphere. When a CME enters in the heliosphere, its interplanetary counterpart is termed as ICME \citep{Dryer1994,Zhao2003}. The ICMEs have been identified in the in situ observations and is found that their plasma and magnetic field parameters are different from that of the ambient solar wind medium. Although, it is possible to record a CME near the Sun and to identify the same in in situ observations, a one to one association between remote and in situ observations of the CMEs is not always easy to establish. There may be several factors responsible for this which are discussed below.

It is understood that fast CMEs often generate large-scale density waves out into space which finally steepen to form collisionless shock waves \citep{Gopalswamy1998a}. This shock wave is similar to the bow shock formed in front of the Earth's magnetosphere. Following the shock there is a sheath structure which has signatures of significant heating and compression of the ambient solar wind \citep{Manchester2005}. After the shock and the sheath, the ICME structure is found. The main problem in understanding the evolution of CMEs is our limited knowledge about their physical properties. In addition, remote sensing observations (e.g. coronagraphs) do not provide plasma and magnetic field parameters of CMEs. Several attempts in the recent past have been made for associating near Earth in situ observed structures of ICME by Advanced Composition Explorer (\textit{ACE}) \citep{Stone1998} and \textit{WIND} \citep{Ogilvie1995} spacecraft with observed Earth-directed front-side halos CMEs in LASCO coronagraph images. Such studies have suffered severely because of difficulties in determining 3D speed of Earth-directed CMEs. Another problem is that an in situ spacecraft takes measurements along a certain trajectory through the ICME, therefore does not provide the global characteristics of CME plasma.

Different ICME structures may have strong, out-of the ecliptic components and therefore a southward pointing interplanetary magnetic field (IMF). Such southward \textit{B$_{z}$}, different than usual Parker spiral, have potential to produce severe consequences on the Earth's geomagnetism \citep{Dungey1961,Tsurutani1988,Gonzalez1994}. Keeping in mind the goal of understanding the Sun-Earth connection, several studies have been undertaken to estimate the arrival time of CMEs at 1 AU near the Earth.

\subsection{CME studies before \textit{STEREO} era}

Before the launch of \textit{STEREO} in 2006, several studies (\citealp{Schwenn2006}, and references therein) were carried out using imaging observations from several space based instruments mentioned in Section~\ref{CMEs}. Among all the space based instruments dedicated to observe the CMEs, the \textit{SOHO}/LASCO launched in 1995 can be considered as the most successful mission in observing several thousand CMEs which led to several research papers. \textit{SOHO}/LASCO with three coronagraphs C1 (no longer operating since June 1998), C2, and C3 could observe the solar corona from 1.1 \textit{R$_\odot$} to 30 \textit{R$_\odot$}, with overlapping FOV. Using these observations, studies were carried out to estimate the source location, mass, kinematics, morphology and arrival time of CMEs (\citealp{StCyr2000,Xie2004,Schwenn2005}, and references therein). To explain the initiation and propagation of CMEs, several theoretical models have also been developed  (\citealp{Chen2011}, and references therein). However, these models differ from one another considerably in the involved mechanism of progenitor, triggering, and eruption of CME. Realizing the consequences of CMEs on our modern high-tech society, several studies were dedicated to find a correlation between intensity of magnetic disturbances on the Earth's surface with the characteristics of CMEs estimated near the Sun \citep{Gosling1990,Srivastava2002,Srivastava2004}. Based on the angular width, CMEs were classified as halo, symmetric halo, asymmetric halo, partial halo, limb, and narrow CMEs. Further, based on the acceleration profile, the CMEs were classified as gradual and impulsive CMEs \citep{Sheeley1999,Srivastava1999}. However, it is believed that all the CMEs belong to a dynamical continuum with a single physical initiation process \citep{Crooker2002}. With the availability of complementary disk observations of solar active regions and prominences, statistical studies on association of different types of CMEs with flares and prominences were carried out in details (\citealp{Kahler1992,Gopalswamy2003}, and references therein).

\subsubsection{CME kinematics and their arrival time at L1}

Several studies of evolution of CMEs were carried out using \textit{SOHO}/LASCO observations, in situ observations near the Earth by \textit{ACE} and \textit{WIND} combined with modeling efforts \citep{Gopalswamy2000,Gopalswamy2001,Gopalswamy2005,Yashiro2004,Wood1999,Andrews1999}. These studies were based on understanding of the kinematics of CMEs using two point measurements, one near the Sun up to a distance of 30 \textit{R$_\odot$} using coronagraph (LASCO/C2 and C3) images, and the other near the Earth using in situ instruments. Using the LASCO images, one could estimate the projected speeds of CMEs, although we lacked information about the 3D speed and direction of the Earth-directed CMEs. These studies, carried out to calculate the kinematics and the travel time of CMEs from the Sun to the Earth, suffered from a lot of assumptions regarding the geometry and evolution of a CME in the interplanetary medium \citep{Howard2009,Vrsnak2010}.  

Various models have been developed to forecast the CME arrival time at 1 AU, based on an empirical relationship between measured projected speeds and arrival time characteristics of various events \citep{Gopalswamy2001b,Vrsnak2002,Schwenn2005}. However, the empirical models have inherent difficulties and individual CMEs must be studied separately in order to derive their kinematics and morphology to be compared with theoretical models. Also few attempts have been made to fit the observed kinematics profiles of CMEs using an appropriate mathematical function \citep{Gallagher2003}. The above mentioned studies are subject to large uncertainties due to projection effects. To overcome the projection effects, the methods such as forward modeling, which approximates a CME as a cone \citep{Zhao2002,Xie2004,Xue2005} and varies the model parameters to best fit the 2D observations of CME, have been used to estimate the CME kinematics. However, this derived kinematics is also subject to several new sources of errors due to presumed geometry of the CME. Another method known as polarimetric technique, using the ratio of unpolarised to polarised brightness of the Thomson-scattered K-corona, has been applied to estimate the average line of sight distance of CME from the instrument plane of sky \citep{Moran2004}. This technique is only applicable up to $\approx$ 5 \textit{R$_{\odot}$} because beyond this the F-corona cannot be considered as unpolarised. 

Analytical drag-based models (DBM; e.g., \citealp{Vrsnak2007,Lara2009,Vrsnak2010}) and numerical MHD simulation models (e.g., \citealp{Odstrcil2004,Manchester2004,Smith2009}) have been developed and used to predict CME arrival times \citep{Dryer2004,Feng2009,Smith2009}. These studies show that the predicted arrival time is usually within an error of $\pm$ 10 hr but sometimes the errors can be larger than 24 hr. Many studies have also shown that CMEs interact significantly with the ambient solar wind as they propagate in the interplanetary medium, resulting in acceleration of slow CMEs and deceleration of fast CMEs toward the ambient solar wind speed \citep{Lindsay1999,Gopalswamy2000,Gopalswamy2001b,Yashiro2004,Manoharan2006,Vrsnak2007}. The interaction between the solar wind and the CME is understood in terms of a `drag force' \citep{Cargill1996,Vrsnak2002}. However, even during the propagation phase of a CME the role of Lorentz force is acknowledged in some earlier studies \citep{Kumar1996,Subramanian2005,Subramanian2007,Subramanian2014}. Despite several studies on CME propagation, very little is known about the exact nature of the forces governing the propagation of CME. However, it is obvious from the observations that there must be some forces acting on the CME which tend to equalize the CME velocity to that of the background solar wind speed.

\subsubsection{In situ observations of CMEs}
\label{insitu}

As previously mentioned, a CME in the interplanetary medium is known as ICME. Various plasma, magnetic field and compositional parameters of an ICME are measured by in situ spacecraft at the instant when it intersects the ICME. The identification of ICME in in situ data is not very straightforward and is based on several signatures which are summarized below.

\paragraph*{Magnetic field signatures in the plasma} \hspace{0pt}\\
 The ICME in in situ observations is identified based on the increased magnetic field strength and reduced variability in magnetic field \citep{Klein1982}. A subset of ICMEs is known as Magnetic Cloud (MC) which shows additional signatures such as enhanced magnetic field greater than $\approx$ 10 nT, smooth rotation of magnetic field vector by greater than $\approx$ 30$\arcdeg$, and plasma $\beta$ (ratio of thermal and magnetic field energies) less than unity \citep{Lepping1990,Burlaga1991}.

\paragraph*{Dynamics signatures in the plasma} \hspace{0pt}\\
The ICME in in situ data is identified by its characteristics of expansion in the ambient solar wind. Due to expansion, CMEs also show depressed proton temperature inside the ICME in contrast to ambient solar wind. ICME leading edge, i.e. front has speed greater than its trailing edge and the difference of speeds at boundaries is equal to two times the expansion speed of CME. Hence, a monotonic decrease in the plasma velocity inside a ICME is noticed \citep{Klein1982}. According to \citet{Lopez1987}, for the normal solar wind there is an empirical relation between proton temperature and solar wind speed as  
follow:
\begin{subequations}
\label{eqvswt}
\begin{align}
T_{exp} &= (0.031 V_{sw} - 5.1)^{2} \times 10^{3} , \hspace{8pt} when~V_{sw} < 500~km~s^{-1} \\ 
T_{exp} &= (0.51 V_{sw} - 142) \times 10^{3},  \hspace{8pt} when~V_{sw} > 500~km~s^{-1}         
\end{align}
\end{subequations}

However, \citet{Richardson1995} found that ICMEs typically have  T$_{p}$ $<$ 0.5 T$_{exp}$ , where T$_{exp}$ is ``expected'' temperature determined from the equation~\ref{eqvswt}. Also, \citet{Neugebauer1997} defined a thermal index (I$_{th}$) using the observed proton temperature and proton velocity and found that I$_{th}$ $>$ 1 for the plasma  associated with an ICME, while this may or may not be the case when I$_{th}$ $<$ 1. The equation for thermal index is as below:

\begin{equation}
\label{eqvswindex}
I_{th}  = (500 V_{p} + 1.75 \times 10^{5})/ T_{p} 
\end{equation}

It is also noted that in an ICME, the electron temperature (T$_{e}$) is greater than proton temperature (T$_{p}$). \citet{Richardson1997} proposed the ratio of electron to proton temperature, i.e. T$_{e}$/T$_{p}$ $>$ 2 is a good indicator of an ICME.

\paragraph*{Compositional signatures in the plasma} \hspace{0pt}\\
The composition of an ICME is different than the ambient solar wind medium. In situ observations have shown that alpha to proton ratio (He$^{+2}$/H) is higher ($>$ 6\%) inside an ICME than its values in normal solar wind. This suggested that an ICME also contains material from the solar atmosphere below the corona \citep{Hirshberg1971,Zurbuchen2003}. It is observed that relative to the solar wind, an ICME shows an enhancement in value of $^{3}$He$^{+2}$/$^{4}$He$^{+2}$ \citep{HO2000}, heavy ion abundances (especially iron) and its enhanced charge states \citep{Lepri2001,Lepri2004}. ICME associated plasma with enhanced charge states of iron suggest that CME source is ``hot'' relative to the ambient solar wind. It is also noted that ICME shows relative enhancement of O$^{+7}$/O$^{+6}$ \citep{Richardson2004,Rodriguez2004}. However, few CMEs have been identified with unusual low ion charge states such as presence of singly-charged helium abundances well above solar wind values \citep{Schwenn1980,Burlaga1998,Skoug1999}. Such low charge states suggest that the plasma is associated with possibly the cool, and dense prominence material \citep{Gopalswamy1998,Lepri2010,Sharma2012}.

\paragraph*{Energetic particles signatures in the plasma} \hspace{0pt}\\
Since, ICMEs have loops structures rooted at the Sun, therefore the presence of bidirectional beams of suprathermal ($\geq$ 100 eV) electrons (BDEs) is considered as a typical ICME signature \citep{Gosling1987a}. Sometimes such BDEs are absent when the ICME field lines in the legs of the loops reconnect with open interplanetary magnetic field lines. In addition, the short-term (few days duration) depressions in the galactic cosmic ray intensity and the onset of solar energetic particles are well associated with ICMEs (\citealp{Zurbuchen2006}, and references therein).

\paragraph*{Association with interplanetary shock and sheath} \hspace{0pt}\\
It is understood that some of the fast CMEs generate a forward shock ahead of them. Such shocks are wide and span several tens of heliospheric longitude approximately two times the value of angular width of related driver ICME \citep{Richardson1993}. In in situ observations, a forward shock is identified based on a simultaneous increase in the density, temperature, speed and magnetic field in the plasma. The shock is followed by a sheath region before the ICME/MC. These sheaths are identified as turbulent and compressed regions of solar wind having strong fluctuations in magnetic fields which last for several hours 
(\citealp{Zurbuchen2006}, and references therein). The magnetic fields in the compressed sheath region may be deflected out of the ecliptic by draping around the ICME \citep{McComas1989}. The compressed and  deflected magnetic field in the sheath result in geo-effectiveness. If the shock is perpendicular, the compression of the magnetic field is especially strong and sheath can lead to more intense geomagnetic storms in comparison to that by parallel shocks \citep{Jurac2002}.

Several studies have shown that different ICMEs show different signatures \citep{Jian2006,Richardson2010}. For example, few ICMEs show signatures of flux ropes while others do not. However, it is still not well understood why few ICMEs are not observed as flux-ropes in in situ data. Similarly, cold filament materiel which is often observed in COR images as a `bright core' following the cavity is rarely observed in in situ observations near 1 AU \citep{Skoug1999,Lepri2010}.

Before the \textit{STEREO} era, the biggest limitation of CME study was that most of the in situ data analysis was restricted to a single point observations at 1 AU while ICMEs are large 3D structures. Figure~\ref{insitutraje} shows how a single point in situ instrument can measure different structures and hence show different signatures of an ICME.

\begin{figure*}[!htb]
	\centering
		\includegraphics[scale=0.60]{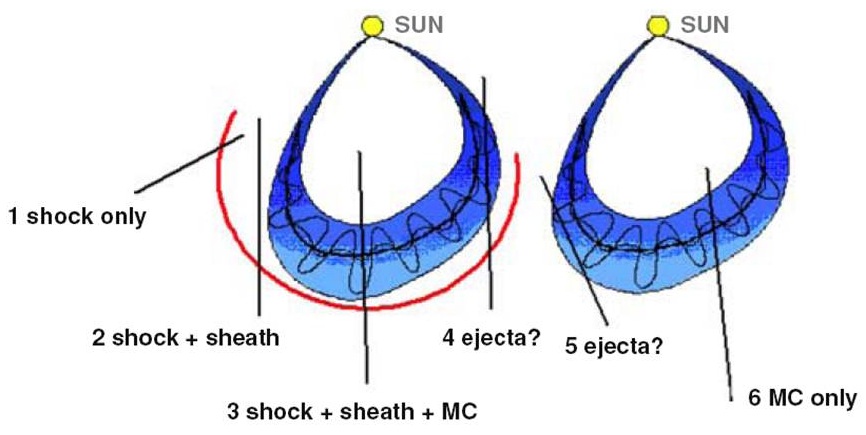}
\caption[Four possible tracks of an in situ spacecraft through a CME]{Four possible tracks of an observing spacecraft through a CME with a leading shock (left) and another two more tracks through the CME without leading shock (right). Track 1 passes through the shock only and track 2 passes through the shock and the sheath of the CME. Track 3 corresponds to a situation when the CME from the Sun is directed exactly towards the in situ spacecraft. In this case, the spacecraft measures the shock, sheath, and the magnetic cloud. Tracks 5 and 6 are similar to 4 and 3, respectively, where there is no shock possibly due to slow speed of CME. Trajectories 4 and 5 will not observe the MC structure. 
(reproduced from \citealt{Gopalswamy2006})}
\label{insitutraje}
\end{figure*}

Such single point in situ spacecraft will also measure different dynamics of an ICME based on its relative location with ICME. Hence, in the absence of information about the part of ICME being sampled by in situ spacecraft, finding an association between the speed derived in COR FOV and the one measured in situ may be erroneous. Hence, multi-point in situ observations and investigation of thermodynamic state of CMEs must be carried out.

\subsection{CME studies in \textit{STEREO} era}

\textit{STEREO} \citep{Kaiser2008}, launched late in 2006, have the capability to continuously image a CME from its lift-off in the corona out to 1 AU and beyond using its Sun Earth Connection Coronal and Heliospheric Investigation (SECCHI; \citealp{Howard2008}) coronagraph (COR) and Heliospheric Imager (HI) data. The twin \textit{STEREO} spacecraft move ahead and behind the Earth in its orbit with their angular separation increasing by 45$\arcdeg$ per year. \textit{STEREO} observations enable us to perform three-dimensional (3D) reconstruction of selected features of a CME based on suitable assumptions. The \textit{STEREO} mission overcomes a large observational gap between near Sun remote observations and near Earth in situ observations and provide information on the 3D kinematics of CMEs due to multiple viewpoints on the solar corona. The twin \textit{STEREO} spacecraft also carry in situ instruments called \textit{In situ Measurements of PArticles and CME Transients} 
(IMPACT: \citealp{Luhmann2008}) and \textit{PLAsma and SupraThermal Ion Composition} (PLASTIC: \citealp{Galvin2008}) and hence provide a chance to measure the in situ signatures of CMEs at 1 AU from multiple vantage points.

In  the \textit{STEREO} era, 3D aspects of CMEs could be studied for the first time. This is because of angular separation between the \textit{STEREO} spacecraft and the Sun-Earth line which provides two different viewpoints for the COR and HI observations. Such unique observations led to the development of various 3D reconstruction techniques (viz., tie-pointing: \citealp{Inhester2006}; forward modeling: \citealp{Thernisien2009}). Also, several other techniques were developed that are derivatives of the tie-pointing technique: the 3D height-time technique \citep{Mierla2008}, local correlation tracking and triangulation \citep{Mierla2009}, and triangulation of the center of mass \citep{Boursier2009}. These methods have been devised to obtain the 3D heliographic coordinates of CME features in the COR FOV. These kinematics of CMEs may change beyond the COR FOV either due to drag forces acting on them or due to CME-CME interaction in the heliosphere. Also, a CME may be deflected by another CME and by nearby coronal holes \citep{Gopalswamy2009a}.

The kinematics of CMEs in 3D over a range of heliocentric distances and their heliospheric interaction have been investigated by exploiting \textit{STEREO}/HI observations \citep{Davis2009,Temmer2011,Harrison2012,Liu2012,Lugaz2012}. \citet{Byrne2010} applied the elliptical tie-pointing technique on the COR and HI observations and determined the angular width and deflection of a CME of 2008 December 12. They used the derived kinematics as inputs in the ENLIL \citep{Odstrcil1999} model to predict the arrival time of a CME at the L1 near the Earth. The image of 2008 December 12 CME from \textit{STEREO}/SECCHI-A and B is shown in Figure~\ref{Byrne2010}.

\begin{figure*}[!htb]
	\centering
		\includegraphics[scale=2.7]{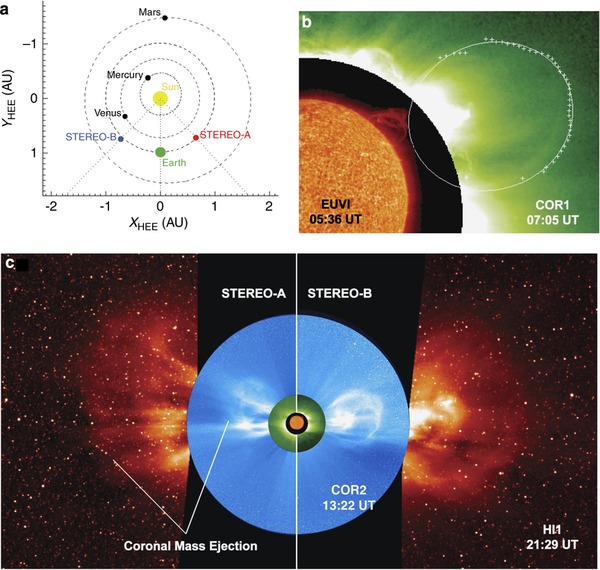}
\caption[A CME observed in \textit{STEREO}/SECCHI images]{\textit{STEREO} spacecraft separated by an angle of 86$\arcdeg$.7 at the time of launch of the 2008 December 12 CME and their locations are shown in panel `a'. Panel `b' shows a prominence, which is assumed to be the core of the CME, and erupted in EUVI-B off the north-west limb around 03:00 UT on December 12. The Earth-directed CME, being observed off the east limb in \textit{STEREO-A} and the west limb in \textit{STEREO-B}, in COR and HI FOV is shown in panel `c'. (reproduced from \citealt{Byrne2010})}
\label{Byrne2010}
\end{figure*}

\citet{Maloney2010} estimated 3D kinematics of CMEs in the inner heliosphere exploiting \textit{STEREO} observations and pointed out different forms of drag force for fast and slow CMEs. The aerodynamic drag force acting on different CMEs will be different and its magnitude will change as the CME propagate in the heliosphere. Therefore, the estimation of the CME arrival time using only the 3D speed estimated from the 3D reconstruction method in COR FOV may not be accurate \citep{Kilpua2012}.

\citet{Kilpua2009} analyzed the in situ and remote observations of CME by \textit{STEREO} and suggested that high latitude CMEs can be guided by the polar coronal fields and can be observed as ICME close to the ecliptic plane. In another study, \citet{Kilpua2011} emphasize that an ICME cannot be explained in terms of simple flux ropes models. They also noted different in situ structures at \textit{STEREO} spacecraft even when they were separated by only few degrees in longitude. Despite the advantage of multi-point in situ observations, it is still unclear whether all CMEs have flux ropes or in other words whether all interplanetary CMEs are magnetic clouds. Also, it is not well understood how a remotely observed CME evolves into a CME observed in situ in the solar wind.

\section{Heliospheric Consequences of CMEs}
\label{HelioConse}

Until the early 1990s, the observed geomagnetic activity was primarily attributed to solar flares, and CMEs were believed to be the result of flare-driven shock waves. However, the paper on `The solar flare myth' published by \citet{Gosling1993}, for the first time, highlighted that geomagnetic activity is mainly related to the CMEs and not flares. CMEs can lead to several consequences at the various locations in the heliosphere, e.g. interplanetary shocks, radio bursts, intense geomagnetic storms, large solar energetic particles (SEPs) events and Forbush decrease.

CMEs drive interplanetary shock when their speeds exceed the Alfv\'en speed in the heliosphere \citep{Gopalswamy2000a,Gopalswamy2006a}. Such shocks can cause SEPs near the Sun and in the interplanetary medium. When this shock arrives at the Earth's magnetosphere it causes the storm sudden commencement (SSC) \citep{Chao1974}. SSC is because of compression of the day-side Earth's magnetosphere by the shock and therefore the horizontal component of Earth's magnetic field which can be measured by ground based magnetometers, is found to be increased \citep{Dessler1960,Tsunomura1998}. It is known that type II radio bursts are excellent indicators of CME-driven shocks \citep{Robinson1985,Gopalswamy2001}. Type II radio bursts are produced via a plasma emission mechanism by the non-thermal electrons accelerated at the shock front. Such electrons generate plasma waves which is converted to electromagnetic radiation at the fundamental and harmonic of the plasma frequency. CMEs which give rise to type II bursts with emission wavelength from metric (m) to decameter-hectometric (DH) to the kilometric (Km) are the most energetic with a typical speed of $ \approx$ 1500 km s$^{-1}$ \citep{Gopalswamy2006a}. It is found that about two-third of CME-driven shocks observed at in situ spacecraft near 1 AU are radio-loud, i.e., they generate type II radio bursts near or far from the Sun. However, about one-third of interplanetary shocks observed at 1 AU are related to radio-quiet CMEs. Such CMEs have slow speeds near the Sun and accelerate away from the Sun, but their driven shocks are too weak to produce radio bursts \citep{Gopalswamy2010a}.

The magnetic nature of the planet decides the effect of CMEs on a planet. Mars has no magnetic field, and therefore there is no magnetic storm on Mars, but its atmosphere has been considerably eroded by the solar wind \citep{Haff1978,Luhmann1992}. The Earth has a magnetic field, and hence Earth-directed CMEs with a southward interplanetary magnetic field component interact with the Earth's magnetosphere at the dayside magnetopause. In this interaction, solar wind energy is transferred to the magnetosphere, primarily via magnetic reconnection that produces non-recurrent geomagnetic storms \citep{Dungey1961}. The rate of transfer of solar wind energy into the magnetosphere depends on the magnitude of the interplanetary convective electric field (E = V $\times$ B; \citealp{Gosling1991}). The main phase of a geomagnetic storm is characterized by a depression in the horizontal component of the Earth's magnetic field measured on the surface of the Earth. The main phase of a geomagnetic storm is the enhancement in the trapped magnetospheric particle population.
The gradient and curvature in the Earth's magnetic field and gyration of trapped particles, lead to ions moving from midnight toward dusk (i.e., westward) and electrons from midnight toward dawn (i.e., eastward) , giving an overall current in the westward direction around the Earth. This current can be visualized as a toroidal shaped electric current that is centered at the equatorial plane, with variable density at an altitude between 2 R$_{\earth}$ to 9 R$_{\earth}$.  During geomagnetic storms, the main carriers of the ring current are positive ions with energies from 1 keV to a few hundred keV having their origin from the solar wind and the ionosphere (\citealp{Gonzalez1994}, and references therein). It is found that during the quiescent time, ring current ions are basically of solar wind origin. However, during the storm, the relative and absolute abundance of O$^{+}$ ions are found to be increased, which are of ionospheric origin. The abundances of O$^{+}$ ions increase rapidly during the onset of geomagnetic storms, which causes a sharp depression in the Earth's magnetic field. After the main phase of a geomagnetic storm, the ring current weakens via various mechanisms, and Earth's horizontal component of magnetic field starts to recover to its quiescent value and is known as the `recovery phase' of a geomagnetic storm. The origin, evolution, dynamics, and losses of ring current are reviewed by \citet{Daglis1999}. A typical geomagnetic storm with its different phases is shown in Figure~\ref{geomag}.

\begin{figure*}[!htb]
	\centering
		\includegraphics[scale=0.7]{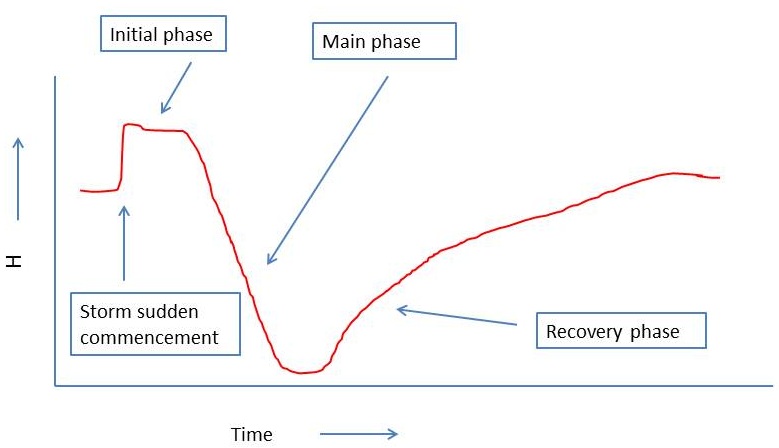}
\caption[Different phases of a typical geomagnetic storm]{Different phases of a typical geomagnetic storm are shown. The sudden commencement, initial, main and recovery phases are characterized by a sudden rise, constant, fast decrease, and slow recovery in the horizontal components of Earth's magnetic field, respectively.}
\label{geomag}
\end{figure*}

CMEs naturally have a southward component of magnetic field (negative B$_{z}$) because of their flux ropes structures. If the CME drives a shock, then the sheath region lying between the shock and flux ropes may also have negative B$_{z}$. Also, it has been shown that 83\% intense geomagnetic storms are due to CMEs \citep{Zhang2007}. Few intense storms may occur because of corotating interactions regions (CIRs). CIRs form when the fast speed solar wind overtakes the slow speed solar wind ahead and leads to the formation of a interface region with increased temperature, density, and magnetic field. The negative B$_{z}$ in the CIRs responsible for geomagnetic storms are due to enhancement in the pre-existing negative B$_{z}$ in the Alfv\'enic fluctuations in the stream interface region. From a space weather perspective, it is essential to estimate the arrival time and transit speeds of CMEs near the Earth well in advance to predict the severity of these events. However, the prediction of negative B$_{z}$ at the Earth is most important for predicting the occurrence of geomagnetic storms \citep{Gonzalez1989, Srivastava2002}. Determination of the negative B$_{z}$ component in the CMEs by exploiting the solar observations is far from reality. By examining the neutral line in the source region of a CME, one can attempt to guess the inclination of the flux rope, the expected direction of rotation, and the portion of flux rope where negative B$_{z}$ may occur \citep{Yurchyshyn2005}. Our understanding of the flux rope structure of a CME is very limited, and is still debated whether such flux ropes are formed during the eruption or exist before the eruption \citep{Chen2011}. 

In the \textit{STEREO} era, by exploiting the Sun to Earth remote observations of CMEs from twin viewpoints, we expect to have better success in predicting the speed and direction of CME near the 1 AU at the Earth. However, without the knowledge about the negative B$_{z}$ component of CME from remote observations, prediction of the intensity of resulting geomagnetic storms well in advance is difficult.

\citet{Kahler1978} found an association between the SEPs events and the CMEs and concluded that the protons must be accelerated at the front of the CME-driven shocks, which was confirmed later in several observations (\citealp{Reames1999}, and references therein). Direct evidence of particles' acceleration by the shock comes from observing energetic storm particle (ESP) events. In ESP events, the locally accelerated particles by the shocks and the shocks themselves are detected at in situ spacecraft at 1 AU. If a shock is generated near the Sun, it is expected to be strong enough to accelerate SEPs and inject them back into the heliosphere. \citet{Gopalswamy2003a} have shown that SEP events are associated with fast and wide CMEs. It is also well proven that CMEs are responsible for the periodic 11-year variation in the intensity of galactic cosmic rays (GCRs).
Further, CMEs are found to be responsible for Forbush decreases (FDs) (\citealp{Forbush1937, Cane2000}, and references therein). Non-recurrent FDs are defined as sudden shorter-term decrease of the recorded intensity of GCRs when the ICME passes through the Earth. In FDs, the depression in the intensity of GCRs typically lasts for less than one day, while its recovery to normal level takes place in several days. FDs are due to the exclusion of GCRs because of their inability to diffuse ``across" the relatively strong and ordered IMF in the vicinity of interplanetary shock, in the sheath and/or a MC region.

\section{Motivation and Organization of the Thesis} 
\label{MotOrg}

As mentioned earlier, until recently, understanding of the nature and propagation of the CMEs has been limited for many reasons. The most significant are projection effects, the lack of continuous high spatial resolution, and high cadence CME observations through the inner heliosphere. Also, many CME arrival time prediction models utilize the projected kinematics from single viewpoint. Very few studies have been carried out to continuously track the CMEs in the heliosphere and predict their arrival time at 1 AU based on their 3D kinematics estimated by exploiting multiple viewpoints observations for them. Therefore, the success in finding an association between remote observations and in situ observations of CME at 1 AU is still limited.

With the availability of stereoscopic observations for CMEs, several 3D reconstruction techniques have been developed recently; however a detailed study to understand the relative importance and validity of ideal assumptions used in these techniques has not been carried out. Hence, a comparison of existing 3D reconstruction techniques, which use the data from COR and HI to accurately predict the arrival time of CMEs is required. Such studies are essential from space weather prediction perspective and for understanding the bulk of plasma motion in an ambient medium.

A majority of work done so far has been focused on the propagation of a single CME from the Sun to Earth and its consequences on hitting the Earth's magnetosphere. However, the interaction or collision of successive CMEs can, in some cases, produce an extended period of southward B$_{z}$ and cause strong geomagnetic storms. Also, during the interaction of CMEs, their kinematics is expected to change. Hence, any scheme to estimate the arrival time of interacting CMEs must take their post-interaction kinematics into account. Further, it is essential to understand the nature of the CME-CME collision. In addition, various plasma processes during the interaction of CMEs that can change the initial identity and properties of CME plasma must be investigated in detail.

The specific objectives of my thesis related to the evolution and consequences of CMEs are outlined below:

\begin{enumerate}

\item{Estimating three-dimensional kinematics using \textit{STEREO} observations on the Earth-directed CMEs and compare with kinematics estimated using data from a single coronagraph alone.} 

\item{To continuously track the Earth-directed CMEs from the Sun to Earth and estimate their kinematic evolution using various 3D reconstruction methods. This approach has been used to estimate the arrival time of CMEs on the Earth. As different reconstruction techniques are based on specific assumptions, we also examine the relative performance of these reconstruction methods for various CMEs.}

\item{To understand the role of Drag Based Model (DBM) in the prediction of arrival time, particularly for the CMEs that cannot be tracked up to the Earth.}

\item{To associate the CME features continuously observed far from the Sun in heliospheric imaging observations with that observed in in situ observations.}

\item{A prime objective of the thesis is to understand the heliospheric consequences of interacting CMEs. In this context, one goal is to understand the consequences of a CME on another CME kinematics if many CMEs follow each other, especially when they collide en route from the Sun to the Earth. We also plan to investigate the geomagnetic consequences of interacting CMEs and examine if it is different than the geomagnetic response of an isolated CME.}
 
\end{enumerate}

Based on the work carried out to accomplish the aforementioned objectives, this thesis is organized into six chapters. A brief description of each chapter is given below. 

\paragraph*{Chapter 1: Introduction} \hspace{0pt}\\
In this chapter, the basics of the Sun and the heliosphere is presented briefly, followed by a discussion on CMEs and their properties. A review of work carried out on CMEs propagation, arrival time estimation, and their heliospheric consequences are presented based on pre-and post \textit{STEREO} observations.

\paragraph*{Chapter 2: Observational Data and Analysis Methodology} \hspace{0pt}\\
In this chapter remote sensing observations of CME from \textit{STEREO}/SECCHI and in situ observations from \textit{ACE} and \textit{WIND} spacecraft is given briefly. Then the analysis methodology used to carry out the objectives is presented in detail. The various 3D reconstruction techniques used for COR and HI observations of CMEs have been described, along with their possible advantages and limitations. The theory of drag-based model (DBM) implemented for selected CMEs in our study to understand the role of solar wind medium on the propagation of CMEs is also presented.

\paragraph*{Chapter 3: Estimation of Arrival Time} \hspace{0pt}\\
In this chapter, the application of 3D reconstruction techniques on selected features of a few Earth-directed CMEs is carried out in COR2 FOV. Further, the CMEs are tracked by constructing the time-elongation map (\textit{J}-map) using the HI observations taken from twin \textit{STEREO} spacecraft. Using the derived elongation-time profiles for the tracked features of CMEs from the \textit{J}-map and implementing several reconstruction techniques, the kinematics of all the CMEs are estimated. The kinematics is either extrapolated or used as input in the DBM to estimate the arrival time of a CME at 1 AU. We assess the performance of using various approaches, e.g., 3D speed near the Sun, extrapolation of heliospheric kinematics, and DBM combined with kinematics, in arrival time prediction of CME at 1 AU. An assessment of the relative performance of several reconstruction methods is also presented in this chapter. This chapter is based on the work published in \citet{Mishra2013} and \citet{Mishra2014}.   

\paragraph*{Chapter 4: Association Between Remote and In Situ Observations} \hspace{0pt}\\
In this chapter, we present the analysis of the in situ observations of selected Earth-directed CMEs and attempt to identify different interplanetary structures of CMEs, i.e., shock, sheath, and magnetic cloud. We also identified in in situ data the remotely tracked features of CMEs. The actual arrival time of tracked features of CMEs (described in Chapter 3) for assessing the performance of different reconstruction methods are explained. This chapter is based on the work published in \citet{Mishra2013} and \citet{Mishra2014}.

\paragraph*{Chapter 5: Interplanetary Consequences of CMEs} \hspace{0pt}\\ 
In this chapter, the interaction of CMEs observed in HI images is demonstrated as a heliospheric consequence of CME. We also describe the morphological and kinematic evolution of interacting CMEs selected for our study. By measuring the energy, and momentum exchange during collision/interaction of CMEs, we attempt to understand the nature of collision of CMEs. Also, the arrival time and in situ signatures of interacting CMEs at 1 AU is shown. Finally, the geomagnetic consequences of interacting CMEs are presented. This chapter is based on the published work in \citet{Mishra2014a} and \citet{Mishra2014b}.

\paragraph*{Chapter 6: Conclusions and Future Work } \hspace{0pt}\\ 
In this chapter, the main results on the propagation and consequences of CMEs, in particular the Earth-directed ones, are highlighted. The role of interaction between CMEs in defining the geomagnetic consequences is also emphasized. A short discussion on the limitation of the applied approaches in the thesis is also included in this chapter. Finally, a future research plan on this topic is mentioned to address a few unanswered questions.

\chapter{Observational Data and Analysis Methodology }
\label{Chap2:DataMthd} 
\rhead{Chapter~\ref{Chap2:DataMthd}. Observational Data and Analysis Methodology}

\section{Introduction}

The objective of the present thesis is to study the propagation and consequences of CMEs in the heliosphere. For this purpose, we have analyzed two sets of observations of Earth-directed CMEs, viz. remote and in situ. In remote observations, a CME is observed in visible photospheric light scattered by electrons of plasma. In in situ observations, the plasma parameters of a CME are measured when it passes through the instruments. The speed, direction, mass, and morphology of a CME at a particular location in the heliosphere can be studied by exploiting its remote observations, while the temperature, speed, density, magnetic field, composition, and charge states of CME plasma/solar wind can be measured from in situ observations. By the time a CME reaches the in situ spacecraft, it is already evolved, and therefore the plasma parameters are different than those measured remotely  \citep{Crooker2006}. However, if the physics of evolution of CME is known, then its properties estimated remotely can be extrapolated up to in situ spacecraft, and then a comparison between both sets of observations can be made with reasonable accuracy. In the absence of a complete understanding of the true nature of the evolution of CMEs, it is often difficult to predict their arrival time near the Earth based on their initial characteristics estimated from remote observations near the Sun. The CME characteristics estimated from remote observations suffer from the line of sight integration and projection effects, while CME/solar wind parameters can be measured along a specific trajectory through the CME by the in situ spacecraft. Few of these limitations can be overcome to some extent if there are remote observations of a CME from multiple viewpoints providing continuous spatial coverage from the Sun to in situ spacecraft. Such observations are possible by the instruments onboard twin \textit{STEREO} spacecraft. This chapter describes the details of remote observations, near-Earth in situ observations of CMEs, and the methodology to study their kinematics and propagation. The details of remote sensing and in situ instruments are described in Sections~\ref{RemoteObs} and ~\ref{ACEWIND}, respectively.

\subsection{Remote observations from \textit{STEREO}/SECCHI}
\label{RemoteObs}

NASA's twin \textit{STEREO} spacecraft was launched to understand the initiation of CMEs and their propagation in the inner heliosphere. The location of \textit{STEREO} on 2010 April 3 is shown in Figure~\ref{STEREO}. Each \textit{STEREO} spacecraft has identical optical, radio, and in situ particles and fields instruments. These instruments are in four different measurement packages named SECCHI, IMPACT, PLASTIC and S/WAVES. The suite of instruments in the SECCHI package consists of two white-light coronagraphs 
(COR1 and COR2), an Extreme Ultraviolet Imager (EUVI) and two white light heliospheric imagers (HI1 and HI2), which jointly can image a CME from its lift-off in the corona out to 1 AU.  

\begin{figure*}[htb]
	\centering
		\includegraphics[scale=0.8]{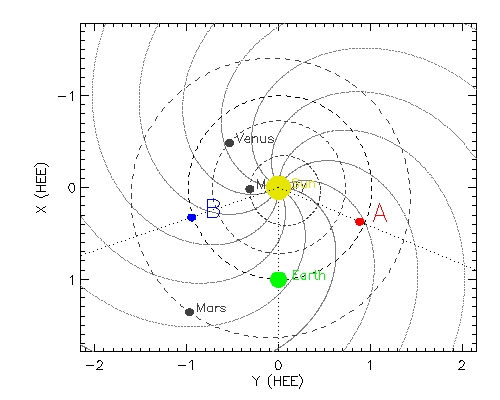}
\caption[The location of \textit{STEREO}, inner solar system planets and the Parker spiral]{The location of \textit{STEREO-A} (red) and \textit{STEREO-B} (blue) spacecraft, Sun (yellow) and Earth (green) in the Heliocentric Earth Ecliptic (HEE) coordinates with X and Y axis in unit of AU is shown. The separation angle of \textit{STEREO-B} and \textit{STEREO-A} with Earth is 
71.2$\arcdeg$ and 67.4$\arcdeg$, respectively. The location of inner solar system planets and the Parker spiral is also shown.}
\label{STEREO}
\end{figure*}

\subsection{SECCHI/COR observations}
A coronagraph is an instrument that images the faint solar corona visible due to scattered light from the much brighter solar photosphere. As mentioned in the previous section SECCHI has two white-light coronagraphs, COR1 is a Lyot internally occulting refractive coronagraph \citep{Lyot1939} and its field of view (FOV) is from 1.4 \textit{R}$_{\odot}$ to 4.0 \textit{R}$_{\odot}$. The internal occultation enables better spatial resolution closer to the limb. The 2 $\times$ 2 binned images size of COR1 are 1024 $\times$ 1024 pixel$^{2}$ and have a resolution of 7.5$\arcsec$ per pixel with a cadence of 8 min. 

COR2 is an externally occulted Lyot coronagraph similar to LASCO-C2 and C3 coronagraphs onboard SOHO spacecraft with a FOV from 
2.5 \textit{R}$_{\odot}$ to 15 \textit{R}$_{\odot}$. In an externally occulted coronagraph, the objective lens is shielded from direct sunlight, and therefore, there is a lower stray light level than COR1, and observations are possible to farther distances from the Sun. COR2 observes with an image sequence cadence of 15 min. COR2 acquires only polarized images of the corona since the polarizer is always in the beam. The standard sequence is three polarized images at -60$\arcdeg$, 0$\arcdeg$, and +60$\arcdeg$,
similar to COR1. The COR2 image size is 2048 $\times$ 2048 pixel$^{2}$ with a resolution of 14.7$\arcsec$ per pixel. The brightness sensitivity of COR1 and COR2 is $\approx$ 10$^{-10}$ B$_\odot$ and 10$^{-12}$ B$_\odot$, respectively. The calibration, operation, mechanical and thermal design of COR1 and COR2 coronagraphs are described in \citet{Howard2008}.

\subsection{SECCHI/HI observations}

SECCHI/Heliospheric Imagers (HIs) detect photospheric light scattered from free electrons in K-corona and interplanetary dust around the Sun (F-corona), similar to CORs. HI also detects the light from the stars and planets within its FOV. The F-corona is stable on a timescale far longer than the nominal image cadence of 40 and 120 min for the HI1 and HI2 cameras. The HI1 and HI2 telescopes have angular FOV of 20$\arcdeg$ and 70$\arcdeg$ and are directed at solar elongation angles of about 14$\arcdeg$ and $\approx$ 54$\arcdeg$ in the ecliptic plane. The HI-A telescopes are pointed at elongation angles to the east of the Sun, while HI-B axes are pointed to the west. HI1 and HI2 observe the heliosphere from 4$\arcdeg$-24$\arcdeg$ and 18.7$\arcdeg$-88.7$\arcdeg$ solar  elongation, respectively \citep{Eyles2009}. Hence, HI1 and HI2 have an overlap of about 5$\arcdeg$ in their FOVs and therefore permit photometric cross-calibration of the instruments. The  2 $\times$ 2 binned image size for HI1 and HI2 are 1024 $\times$ 1024 pixel$^{2}$ with a resolution of 70$\arcsec$ per pixel and 4$\arcmin$ per pixel, respectively. The brightness sensitivity of HI1 and HI2 is 3 $\times$ 10$^{-15}$ B$_\odot$ and 3 $\times$ 10$^{-16}$ B$_\odot$, respectively \citep{Eyles2009}.

The geometrical layout of the FOVs of the HI1, HI2, and the COR2-A and B FOV is shown in Figure~\ref{FOVsHI}.  The HI1 and 
HI2 FOVs provide an opening angle from the solar ecliptic of about 45$\arcdeg$, which is chosen to match the typical size of a CME. The optical designs of HI1 and HI2 are optimized for the circular FOVs of diameter 20$\arcdeg$ and 70$\arcdeg$ respectively however the CCD detectors with square format result in some response in the regions indicated by dotted lines in 
Figure~\ref{FOVsHI}. The details about the optical, baffle, and electronic design of HI1 and HI2 are given in \citet{Eyles2009}.

\begin{figure*}[htb]
	\centering
		\includegraphics[scale=0.8]{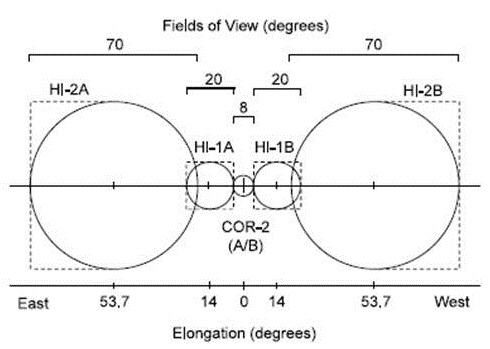}
\caption[Field of view (FOV) of CORs and HIs]{The field of view of the HIs and the COR2 is shown. The dotted lines represent the square format of the CCD detectors. The Sun-centred CORs observe all solar latitudes, while the HIs observe a maximum of $\pm$ 35$\arcdeg$ perpendicular to the ecliptic (reproduced from \citealp{Eyles2009}).}
\label{FOVsHI}
\end{figure*}

It must be emphasized that HI-A and HI-B view from two widely separated spacecraft at similar planetary angles (Earth-Sun-spacecraft), thus providing a stereographic view. Figure~\ref{FOVsHIsep}(a) shows the overall FOVs of HI instruments projected onto the ecliptic plane.  The two lines of sight drawn with arrows from both \textit{STEREO-A} (red) and \textit{STEREO-B} (blue) spacecraft represent the inner and outer edges of FOVs of HI. The region of the heliosphere observed in the common FOV of HI-A, and HI-B only will have a stereoscopic view from \textit{STEREO}. It is also clear from this figure that a CME directed towards the Earth can be observed continuously from the Sun to Earth and beyond from both HI-A and HI-B telescopes. In this scenario, a CME directed eastward from the Earth, and \textit{STEREO-B} can only be observed in HI-A FOV but not in HI-B FOV. Similarly, a CME directed westward from the Earth, and \textit{STEREO-B} will be observed only in HI-B FOV but not in HI-A FOV.

\begin{figure*}[!htb]
	\centering
		\includegraphics[scale=0.55]{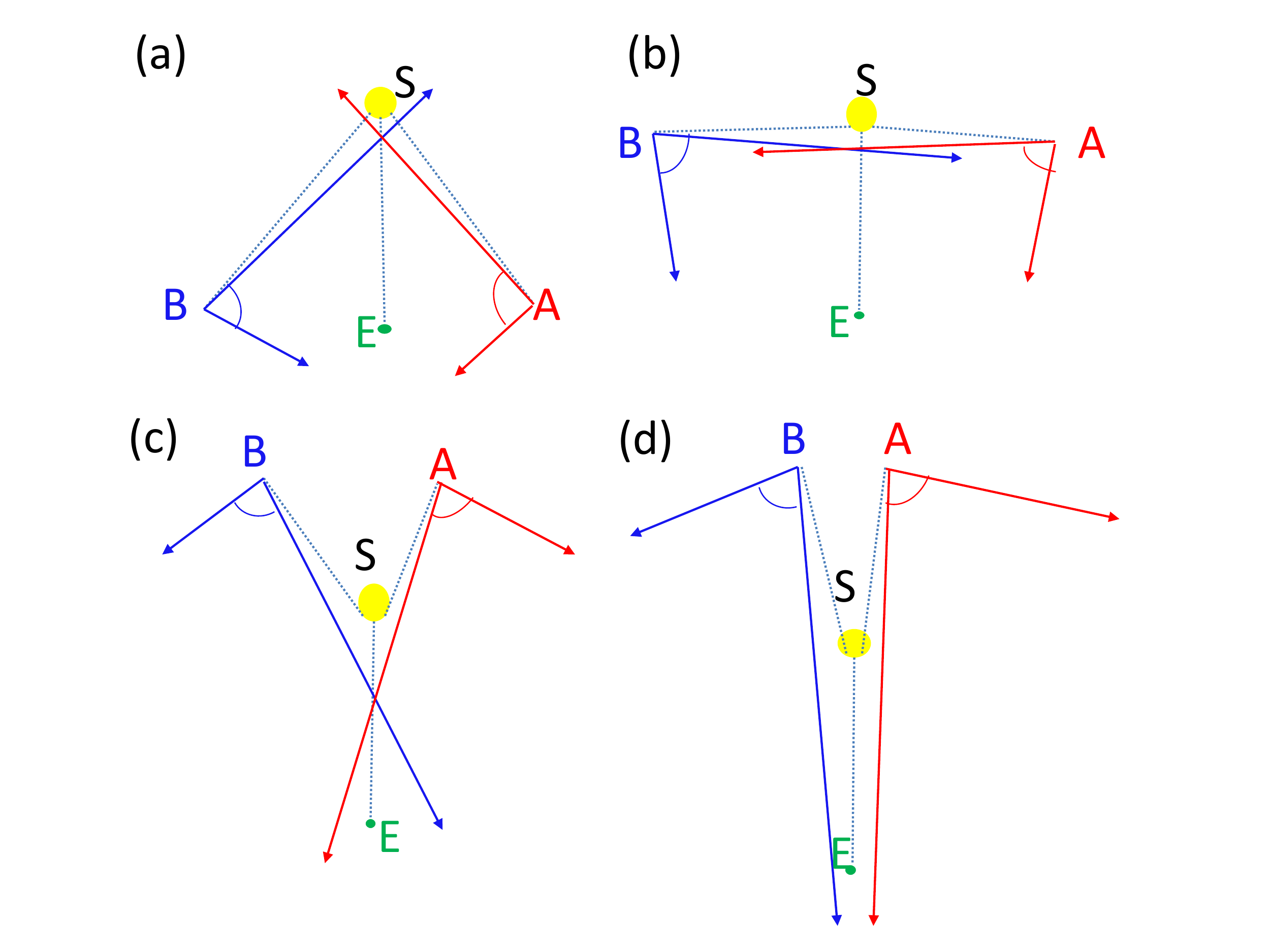}
\caption[Inner and outer edges of HI FOV corresponding to different separation angle of \textit{STEREO-A} and \textit{B}]{The locations of Sun (yellow), Earth (green), \textit{STEREO-A} (red) and \textit{STEREO-B} (blue) is shown corresponding to different separation angles of twin \textit{STEREO}. The arrows from the \textit{STEREO-A} and \textit{STEREO-B} locations represent the inner (near the Sun) and outer edges of the HI FOV.}
\label{FOVsHIsep}
\end{figure*}

It must be highlighted that as the separation (summation of longitude of both \textit{STEREO}) between the \textit{STEREO-A}, and \textit{STEREO-B} increases with time, the region of the heliosphere observed simultaneously by both HI-A and HI-B also changes.  From Figure~\ref{FOVsHIsep}(b), it is clear that the separation between \textit{STEREO-A} and \textit{B} is approximately 175$\arcdeg$ around December 2010 therefore Earth-directed CMEs near the Sun can not be observed. They can be observed only a little far from the Sun by HI-A and HI-B. Thus, the continuous (Sun to Earth) tracking of CMEs is not possible in this case. Figure~\ref{FOVsHIsep}(c) shows that the \textit{STEREO} spacecraft are behind the Sun from the Earth's perspective, i.e., separation between them is greater than 180$\arcdeg$, HI-A and HI-B will not provide continuous coverage between the Sun and Earth along the ecliptic. Hence, in this scenario, an Earth-directed CME will not be observed for a significant distance close to the Sun. The other issue of `detectability' of a CME arises when the \textit{STEREO} spacecraft are behind the Sun. In this case, if the CME is directed toward the Earth, then it is substantially far-sided for both the \textit{STEREO} spacecraft. Hence, the distance between the CME and \textit{STEREO} increases with time, and also as the CME diffuses with time; therefore its detection is difficult but not impossible. Even in such a scenario, some of the Earth-directed CMEs have been detected well in HI FOV \citep{Liu2013}. In Figure~\ref{FOVsHIsep}(d), the \textit{STEREO} spacecraft are the completely far side of the Earth. In this scenario, the Earth does not appear in HI FOV, which implies that any CME propagating toward the Earth will not be observed during its journey from the Sun to the Earth.

\section{Observations Near 1 AU}

The term `Interplanetary Coronal Mass Ejection' (ICME) was originally coined for near-Earth observations of CME by an in situ spacecraft to distinguish it from the coronagraphic observations near the Sun. In the present \textit{STEREO} era, where CMEs can be tracked from near the Sun to the Earth and beyond, the term ICME has become redundant \citep{Webb2012}. Therefore, hereafter, throughout this thesis, we will use the term CME to refer to both CME and ICME. The details about signatures and identification of CMEs in in situ observations have been described in Section~\ref{insitu} of Chapter~\ref{Chap1:IntMot}. In the present section, we briefly state the principles on which the in situ instruments work. When a CME arrives at the Earth's magnetosphere, it produces magnetic disturbance on the surface of the Earth and inside the magnetosphere, which can be observed by ground-based magnetometers (See, Section~\ref{GroundObs}).        

\subsection{In situ observations from \textit{ACE} \& \textit{WIND}}
\label{ACEWIND}

\textit{ACE} spacecraft was launched in 1995 and is located close to the L1 Lagrangian point. It carries nine high-resolution instruments which measure the elemental, isotopic, and ionic charge-state composition of energetic nuclei by monitoring the state of the solar wind medium. For CME identification, we have mainly used the observations from three instruments, namely Solar Wind Ion Composition Spectrometer (SWICS) \citep{Gloeckler1998}, Solar Wind Electron, Protn Alpha Monitor (SWEPAM) \citep{McComas1998} and Magnetometer (MAG) \citep{Smith1998}. SWICS measures the elemental and ionic charge-state composition of solar wind ions (H to Fe) using a combination of electrostatic deflection, post-acceleration, time-of-flight, and energy measurements. SWEPAM contains an electrostatic analyzer and provides the electron and ion distribution functions in 3D over all the velocity space to characterize the solar wind's bulk flow and kinetic properties. MAG consists of twin boom-mounted, triaxial flux-gate sensors that measure the local vector magnetic field in the interplanetary medium.

\textit{WIND} spacecraft was launched in 1994 and is located near the L1 point. This spacecraft carries a total of nine instruments and continuously monitors the solar wind plasma, energetic particles, magnetic fields, radio and plasma waves, and cosmic gamma-ray bursts. In the present thesis, we have used the observations from three instruments, namely Magnetic Field Instrument (MFI) \citep{Lepping1995}, Solar Wind Experiment (SWE) \citep{Ogilvie1995}, Three-Dimensional Plasma and Energetic Particle Investigation (3DP) \citep{Lin1995}. MFI consists of boom-mounted double triaxial sensors and measures the interplanetary magnetic field. SWE instrument is a suite of two Faraday cup (FC) sensors, Vector Electron and Ion Spectrometer (VEIS), and a strahl sensor. FC sensors measure the density, temperature, and velocity of the solar wind. VEIS measures the foreshock (region upstream of the bow shock) ions and electrons reflected from the bow shock. Strahl sensor can measure the electron velocity distribution function near the direction of the magnetic field. 3DP measures the 3D distribution of suprathermal electrons and ions in the solar wind. It must be noted that \textit{WIND} spacecraft lacks an instrument like SWICS on board \textit{ACE}. Therefore chemical abundance and charge state composition of the solar wind can only be studied using \textit{ACE} data.

\subsection{Ground-based magnetometer observations}
\label{GroundObs}
As a CME having a southward orientation of the magnetic field passes through the Earth, various electric current systems in Earth's magnetosphere are intensified, which lead to perturbations in the geomagnetic field. These perturbations at the surface of the Earth can be measured by a series of ground-based magnetometers. Based on these measurements, various geomagnetic indices are calculated, which monitor different geomagnetic responses, e.g., geomagnetic storms and substorms \citep{Akasofu1964}. The perturbations measured in the geomagnetic field can be resolved in horizontal (north-south dipole) direction and orthogonal (east-west) direction and are denoted by H and D components, respectively. The formation of the magnetospheric ring current and its effect on the horizontal component (H) of Earth's magnetic field is explained in Section~\ref{HelioConse} of Chapter~\ref{Chap1:IntMot}. A series of magnetometers around the equator (mid-latitudes) of the Earth measure the perturbations in the horizontal component of Earth's magnetic field and monitor the ring current intensity and hence geomagnetic storms \citep{Dessler1959, Sckopke1966}. The hourly average of such measured perturbations is known as Disturbance storm time (Dst) index \citep{Sugiura1964, Mayaud1980}. The negative values of the Dst index represent the intensification of the ring current. Also, high resolution (1 min) measurements of symmetric (SYM) and asymmetric (ASY) disturbance are introduced for both H and D components. Sym-H index is very similar to the Dst index, but with high resolution and derived from different sets of  stations using a slightly different coordinate systems and base values 
\citep{Iyemori1990}.

There are a series of magnetometers in the northern hemisphere auroral zone of the Earth. During the perturbations in the magnetosphere, these magnetometers can measure the positive (upper bound) or negative (lower bound) fluctuations in the H-component of the geomagnetic field. Such positive or negative fluctuations are averaged for several stations and are known as the AU or AL index. The difference between maximum negative (AL) and positive (AU) fluctuations is known as AE index \citep{Davis1966}. AL and AU index monitors the westward and eastward electrojet, respectively. AE index represents the magnetospheric substorms and has durations of tens of minutes to several hours \citep{Akasofu1969, McPherron1970}. It must be noted that although a geomagnetic index primarily monitors a single current system but a small contribution to the index by several other current systems formed during magnetospheric perturbations cannot be neglected. Also, an index known as the PC index is derived from a magnetic station in the northern and southern hemispheres designed to monitor the perturbation in the geomagnetic field due to ionospheric and field-aligned currents \citep{Troshichev2000}. To examine the geomagnetic effects of the CMEs, we have mainly used the Dst index data; however, for a few events the AL and PC indices are also analyzed.

\section{Analysis Methodology}
\label{anameth}

To achieve our objective of understanding the evolution and consequences of CMEs, we followed three main steps. In the first step, we selected those CMEs that could be continuously observed from the Sun to the Earth in remote sensing observations. Their plasma parameters could also be measured in in situ observations. As we also aimed to understand the geomagnetic consequences of CMEs, the selected CMEs for our study are directed approximately towards the Earth. These selected CMEs have been broadly classified into two groups, one consists of isolated CMEs to study their heliospheric evolution, and other comprises interacting CMEs to study the nature of their collision and geomagnetic responses. In the second step, CMEs features are tracked through the heliosphere by constructing \textit{J}-maps \citep{Sheeley1999}. Using \textit{J}-maps, the variation in elongation (angle of the feature concerning the Sun-spacecraft line) of a moving CME feature is estimated. In the third step, the kinematics of tracked features of CMEs is calculated using stereoscopic reconstruction techniques on COR2 observations.
Further, to understand the heliospheric propagation of selected CMEs, various stereoscopic or single spacecraft methods are used on their time-elongation profiles (derived from \textit{J}-maps) observed in HI FOV. The stereoscopic methods based on certain assumptions and used on COR2 observations are generally not suitable for HI observations because these assumptions break down at large distances from the Sun. In the fourth step, we extrapolate the kinematics of CME features and use them as input to the drag-based model (DBM) of propagation of CMEs for estimating their arrival time at 1 AU. In the fifth step, in situ observations of CME at 1 AU are analyzed, and actual arrival time of a CME is marked. Finally, we attempt to associate remotely tracked features of CME to the features observed in in situ observations. The aforementioned methodology has been adopted for several cases of geo-effective CMEs and interacting/colliding CMEs to address the specific objectives.

\subsection{Reconstruction methods using COR2 observations}
Various 3D reconstruction methods have been developed which can be used on SECCHI/COR observations, i.e., for a CME feature close to the Sun. These have been reviewed in \citet{Mierla2010}. In our study, we have used the tie-pointing method of 3D reconstruction \citep{Thompson2009} and forward modeling method \citep{Thernisien2009} on the SECCHI/COR observations of CMEs. These methods have been used to estimate the kinematics of selected features of CMEs before they enter into the HI FOV.

\subsubsection{Tie-point (TP) reconstruction}
\label{tiepoint}

The tie-pointing method of stereoscopic reconstruction is based on the concept of epipolar geometry. The position of two \textit{STEREO} spacecraft and the point to be triangulated defines a plane called epipolar plane \citep{Inhester2006}. Since every epipolar plane is seen head-on from both spacecraft, it is reduced to a line in the respective image projection. This line is called the epipolar line. Epipolar lines in each image can easily be determined from the observer's position and the direction of the observer's optical axes.

Any object which lies on a certain epipolar line in one image must lie on the same epipolar line in the other image. This straightforward geometrical consequence is known as epipolar constraint. Due to the epipolar constraint, finding the correspondence of an object in the contemporaneous image of both spacecraft reduces to finding out correspondence along the same epipolar lines in both images. Once the correspondence between the pixels is found, the 3D reconstruction is achieved by calculating the line of sight rays corresponding to those pixels and back tracking them in 3D space. Since the rays are constrained to lie in the same epipolar plane, they intersect at a point on tracking backward. This procedure is called tie-pointing \citep{Inhester2006}. The point of intersection of both lines of sight gives the 3D coordinates of the identified object or feature in both sets of images.

In our study, we have used the tie-pointing method (scc\_measure: \citealp{Thompson2009}) of 3D reconstruction. Before implementing this method, the processing of SECCHI/COR2 images and creation of minimum intensity images, and then its subtraction from the sequence of processed COR2 images were carried out as described in \citet{Mierla2008} and \citet{Srivastava2009}. This method has a graphical user interface (GUI) in the Interactive Data Language (IDL) and has been widely used by several authors to estimate the 
3D coordinates of a selected feature of the CME.

\subsubsection{Forward modeling method}

In this method, a specific parametric shape of CME is assumed and iteratively fit until it matches with its actual image. This method produces a physical solution based on the model assumptions but presumes that the solution only fits that model. 
We have used the 3D reconstruction method of \citet{Thernisien2009} where they have applied the Graduated Cylindrical Shell (GCS) model on CMEs observed by SECCHI/COR2-A and B. This model represents the flux rope structure of CMEs with two shapes; the conical legs and the curved (tubular) fronts (Figure~\ref{modelGCS}). The resulting shape is like a ``hollow croissant''. The model also assumes that the GCS structure moves in a self-similar way. In principle, this technique can be applied to images from all other telescopes of the SECCHI package. However we applied this technique only to COR2 images. This is because, in COR2 FOV, the flux-rope structure of CMEs is well identified, while it is not fully developed in COR1 FOV and is too faint in the  HI FOV.

\begin{figure*}[!htb]
	\centering
		\includegraphics[scale=0.60]{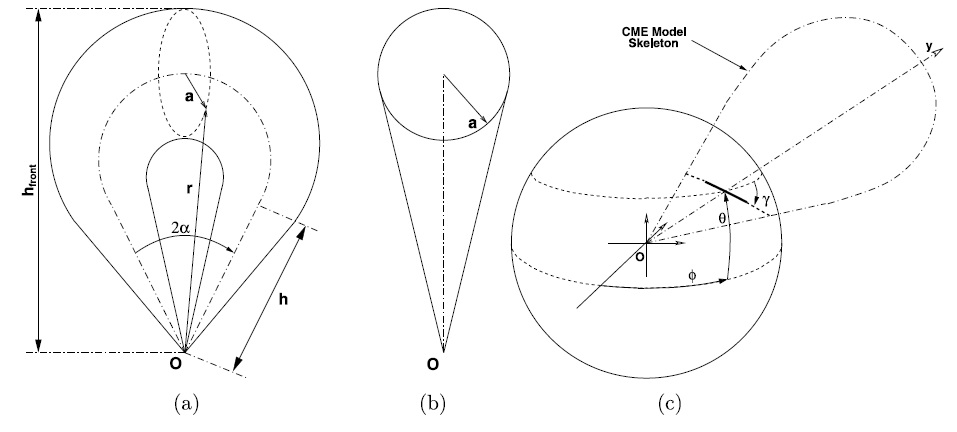}
\caption[GCS model representation]{The Graduated Cylindrical Shell (GCS) model, as seen in the face-on (a) and edge-on (b) view, is shown with several positioning parameters (c). The dash-dotted and the solid line represents the axis and a planar cut through the cylindrical shell, respectively.  The $\phi$ and $\theta$ are the longitude and latitude of the axis through the centre of the shell, respectively, and $\gamma$ is the tilt angle around the axis of symmetry of the model (reproduced from \citealp{Thernisien2009}.)}
\label{modelGCS}
\end{figure*}

GCS model fitting tool in IDL involves simultaneous adjusting six model parameters so that the resulting GCS flux structure matches well with the observed flux rope structure of the CME. These six parameters, including the longitude, latitude, tilt angle of the flux ropes with a height of the legs, half-angle between the legs, and aspect ratio of the curved front, are adjusted to match the spatial extent of the CME. These have been discussed in detail in \citet{Thernisien2011}. The best fit six parameters obtained for the observed CME with GCS modeled CME used to calculate various geometrical dimensions of a CME.

From a space weather perspective, the main advantage of using SECCHI/COR data and the 3D reconstruction methods described above is that it enables estimation of true speed and hence forecasting of the arrival time of CME near the Earth in advance with better accuracy. However, information on the deceleration, acceleration, or deflection experienced by a CME beyond COR2 FOV cannot be obtained. This may lead to erroneous arrival time estimation of CMEs.

\subsection{Reconstruction methods using COR \& HI observations}
\label{Recnsmthd}

It is often observed that when CMEs leave the coronagraphic FOV, the Thomson scattered signal becomes too low to identify a particular feature in both sets of images obtained by \textit{STEREO-A} and \textit{STEREO-B}. Therefore, a method of time- elongation map (\textit{J}-map), initially developed by \citet{Sheeley1999} for \textit{SOHO}/LASCO images, is used to track a CME feature in the interplanetary medium. \citet{Rouillard2009} implemented the same technique on HI images to reveal the outward motion of plasma blobs in the interplanetary medium. \textit{J}-maps are now considered necessary for the best exploitation of \textit{STEREO}/HI observations. 

\subsubsection{Construction of \textit{J}-maps}
\label{Jmapsmthd}

We have constructed \textit{J}-maps along the ecliptic plane using long-term background-subtracted running difference images taken from COR2, HI1, and HI2 on \textit{STEREO-A} and \textit{STEREO-B} spacecraft. A previous image is subtracted from the current image in the running difference procedure. This reveals the changes in electron density between consecutive images. Before taking running difference, the HI image pair is aligned to prevent the stellar contribution in the difference images. This alignment requires precise pointing information of the HI instruments \citep{Davies2009}. For this purpose, we use the Level 2 HI data that were corrected for cosmic rays, shutterless readout, saturation effects, flat fields, and instrument offsets from spacecraft pointing. Also, a long-term background image is subtracted to prepare Level 2 HI data. We use the Level 0 data for the COR images and process these images to Level 1 using IDL before taking their running difference. We calculate the elongation and position angles for each pixel of the difference images from COR and HI and extract a strip of constant position angle interval along the position angle of the Earth. The position angle tolerance considered for the COR2 images is 5$\arcdeg$ and 2.5$\arcdeg$ for both HI1 and HI2. We bin the pixels of the extracted strip over a specific elongation angle bin size, viz., 0.01$\arcdeg$ for COR2 and 0.075$\arcdeg$ for both HI1 and HI2.
Furthermore, we take the resistant mean of all pixels over a position angle tolerance in each bin to represent the intensity at a corresponding elongation angle. The resistant mean for each elongation bin is scaled to reveal a significant elongation bin. These scaled resistant means are stacked as a function of time and elongation to produce a time-elongation map 
(\textit{J}-map). Figure~\ref{JmapMay} shows a typical \textit{J}-map in which the bright curves with positive inclination reveal the propagation of a CME feature.

\begin{figure*}[!htb]
	\centering
		\includegraphics[scale=0.60,angle=90]{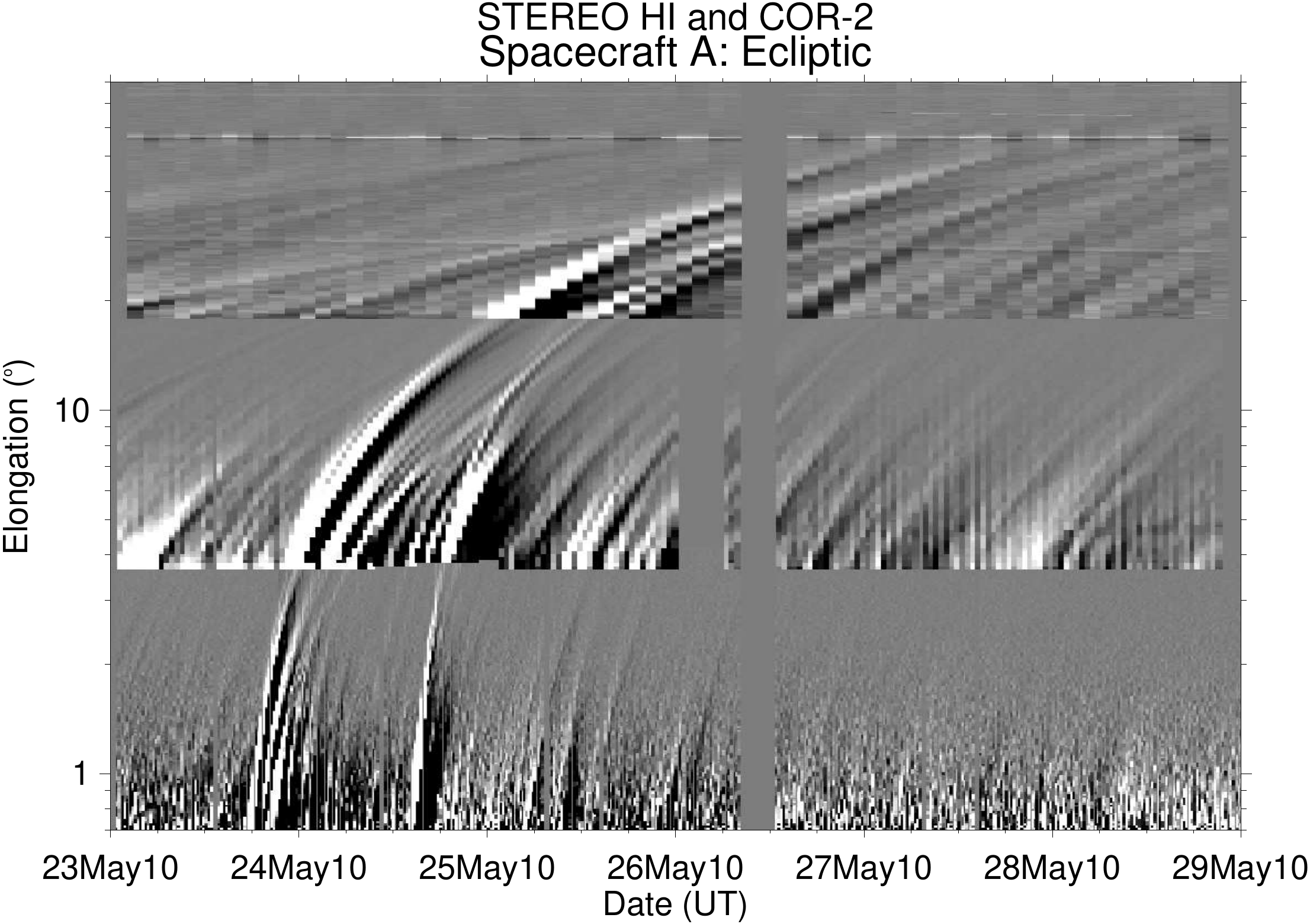}
\caption[\textit{J}-maps using COR2, HI1 and HI2 images]{\textit{J}-map constructed from running difference images of COR2, HI1 and HI2 instruments along the ecliptic plane for \textit{STEREO-A} spacecraft is shown. The Y-axis shows the elongation angles plotted in logarithmic scale while the X-axis shows the time in UT. Two bright tracks starting on 2010 May 23 at 19:00 UT and May 24 at 14:30 UT represent features of two CMEs and can be tracked up to 50$\arcdeg$ elongation angles.}
\label{JmapMay}
\end{figure*}

Once the elongation-time profile of a tracked CME feature in the heliosphere is derived from the \textit{J}-map, it can be used as input in different reconstruction methods to estimate its heliospheric kinematics. Several reconstructions methods based on various assumptions are described below and have been applied to selected CMEs in this thesis.

\subsubsection{Single spacecraft reconstruction methods}
\label{SinRcnsMthd}
Single viewpoint observations can determine 3D kinematics of CMEs at large distances from the Sun. This is because when CMEs are very far from the Sun, the `linearity' condition imposed on the CMEs near the Sun breaks down. In other words, the `linear assumptions' imposed on an observed CME feature in coronagraphic FOV to convert its measured elongation into the distance are no longer valid. Near the Sun, the plane of sky assumption is used, i.e. distance of a feature d =tan$\alpha$, and further for small 
$\alpha$; d = $\alpha$ can be used. Far from the Sun, it is difficult to assume that the same feature of a CME can be observed from different viewpoints or even at different locations in the heliosphere. This increases the complexity in the stereoscopic reconstruction techniques. However, if the images of the CMEs are taken at large distances from the Sun and across a large FOV, then, with proper treatment of Thomson scattering and simplistic assumptions about the geometry and trajectory of CMEs, some 
3D parameters of CMEs can be estimated by exploiting the images from a single viewpoint alone. However, such an approach cannot be applied to images obtained from coronagraphs, as they observe across a small angular extent, and therefore the geometric effects of the CME structure are not detectable. As HIs have large FOV and can observe the CMEs at farther distances from the Sun, therefore, 
3D parameters of the CMEs can be derived using the elongation-time profile of a tracked feature from only \textit{STEREO-A} or \textit{STEREO-B} location. The single spacecraft reconstruction methods require observations from a single viewpoint and are described below.

\paragraph{Point-P (PP) method} \hspace{0pt}\\
\label{PP} 
The Point-P (PP) method was developed by \citep{Howard2006} to convert the elongation angle to distance from the Sun center. This method was developed soon after the launch of SMEI \citep{Eyles2003}, which can measure the elongation angle of a moving feature of a CME. The accuracy of this conversion is constrained by the effects of the Thomson scattering process and the geometry of CMEs, which govern their projection in the images. In this method, to remove the plane of sky approximation, especially for HIs, it is assumed that a CME is a wide circular structure centered on the sun, and an observer looks and tracks the point where the CME intersects the Thomson surface \citep{Vourlidas2006}—under these assumptions derived radial distance (R$_{PP}$) of CME from the Sun center is,  R$_{PP}$ = $d_{0}$ $\sin\epsilon$, where $\epsilon$ is the measured elongation of a moving feature and $d_{0}$ is the distance of the observer from the Sun. This method has been used earlier by \citet{Howard2007, Wood2009, Wood2010, Mishra2014}. If small (elongation) angle approximation can be applied, the PP method is close to the plane of sky approximation.

However, recently \citet{Howard2012} and \citet{Howard2013} have de-emphasized this concept by showing that the maximum intensity of scattered light per unit density is spread over a broad range of scattering angles (called Thomson plateau). They conclude that CME features can be observed far from the Thomson surface and that their detectability is governed by the location of the feature relative to the plateau rather than the Thomson surface. The existence of this Thomson plateau and the oversimplified CME geometry assumed in the PP method is likely to lead to significant errors in the estimated kinematics of CMEs.

\paragraph{Fixed-phi (FP) method} \hspace{0pt}\\
\label{FP}
\citet{Sheeley1999}, while analyzing the LASCO data, introduced the concept that the time-elongation map shows an apparent acceleration and deceleration of a CME due to imposed projective geometry on it. But this effect of apparent acceleration/deceleration was not significant in the LASCO FOV, which covers a narrow elongation range. After the advent of truly wide-angle imaging with SMEI, \citet{Kahler2007} developed a  method to convert elongation to radial distance by assuming that a CME feature can be considered as a point source moving radially outward in a fixed direction ($\phi_{FP}$) relative to an observer located at a distance $d_{0}$ from the Sun. Using this concept, elongation ($\epsilon(t)$) variation of a moving CME feature can be converted to distance ($R_{FP}(t)$) from the Sun. With these assumptions, the following expression can be derived. 

\begin{equation}
R_{FP}(t) = \frac{d_{0}\; \sin(\epsilon(t))}{\sin(\epsilon(t)+\phi_{FP})}
\label{FPeqn}
\end{equation}

The CME's fixed radial direction of propagation can be determined using the source region of the CME. Also, the initial direction of propagation of a CME can be derived from the 3D reconstruction techniques applicable to COR observations, and can be used in Equation~\ref{FPeqn}. One major drawback of the FP method is that it does not take into account the finite cross-sectional extent of a CME. Figure~\ref{Davies2012}(a) shows that open black dots moving along a fixed radial direction are being observed in the FP method.

\begin{figure*}[!htb]
	\centering
		\includegraphics[scale=8.0]{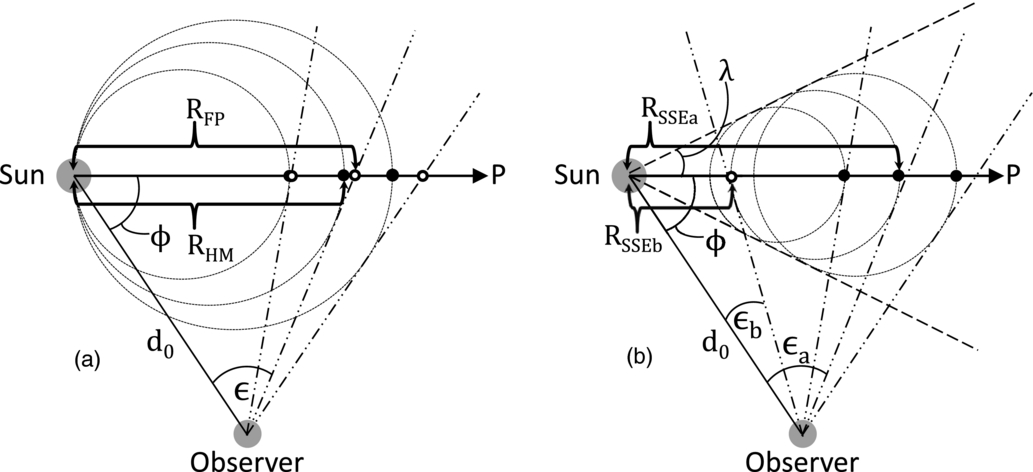}
\caption[CME feature tracking in FP, HM and SSE model geometry]{The tracked CME features in FP (open black dots) and HM (circles/filled black dots) model geometries are shown in panel (a). The tracked feature corresponding to the geometry of the SSE model is shown in panel (b) (reproduced from \citealp{Davies2012}).}
\label{Davies2012}
\end{figure*}

\paragraph{Harmonic mean (HM) method} \hspace{0pt}\\
\label{HM}
To convert elongation angle to radial distance from the center of the Sun, \citet{Lugaz2009} assumed that a CME could be
represented as a self-similarly expanding sphere attached to Sun-center, with its apex traveling in a fixed radial direction. They further assumed that an observer measures the scattered emission from that portion of the sphere where the line of sight
intersects tangentially. Based on these assumptions, they derived the distance (R$_{HM}$) of the apex of the CME from the Sun-center
as a function of elongation. They found that this distance is the harmonic mean of the distances estimated using the FP and PP
methods. Hence, the method is referred to as the HM method.  The distance ($R_{HM}$) of the apex of the sphere from the Sun can be estimated by,

\begin{equation}
R_{HM}(t) = \frac{2d_{0}\; \sin(\epsilon(t))}{1 + \sin(\epsilon(t)+\phi_{HM})}
\label{HMeqn}
\end{equation}

where $\phi_{HM}$ is the radial direction of propagation of CME from the Sun-observer line and $\epsilon$ is the elongation angle, and $d$ is the distance of the observer from the Sun. Although, in this method, the geometry of CMEs is taken into account, the assumption of such geometry may not be valid due to possible flattening of CME front on interaction with the solar wind. Figure~\ref{Davies2012}(a) shows that the filled black dots moving along a fixed radial direction will be observed for a circular model geometry considered for the HM method.

\paragraph{Self-similar expansion (SSE) method} \hspace{0pt}\\
\label{SSE}
Davies et al. (2012) derived an expression for the elongation variation as a function of the time of a CME viewed from a single vantage point and termed as Self-Similar Expansion (SSE) method. In this method, a CME is considered to have a circular cross-section, in the plane corresponding to the position angle (PA) of interest, is not anchored to the Sun, and, during its propagation away from the Sun, its radius increases such that it always subtends a fixed angle to the Sun center. They also showed that the SSE geometry could be characterized by an angular half-width ($\lambda$), and in its extreme forms, the SSE geometry is equivalent to the FP ($\lambda$ = 0$\arcdeg$) and HM methods ($\lambda$ = 90$\arcdeg$). It must be noted that $\lambda$ can also be considered a parameter related to the curvature of the CME front. Figure~\ref{Davies2012}(b) shows the feature marked with filled black dots that will be observed under the SSE method. The distance ($R_{SSE}$)  of a feature using this method at a certain elongation measured from \textit{STEREO-A} or \textit{STEREO-B} can be calculated from the Equation~\ref{SSEeqn}.

\begin{equation}
R_{SSE}(t) = \frac{d_{0}\; \sin(\epsilon(t)) (1+\sin(\lambda))}{\sin(\epsilon(t)+ \phi_{SSE}) +\sin(\lambda)}
\label{SSEeqn}
\end{equation}

In all the single spacecraft methods described above, i.e., FP, HM, and SSE, it is assumed that a CME propagates along a fixed radial trajectory (in particular, estimated in COR FOV), ignoring real or ``artificial'' heliospheric deflections. This assumption is likely to introduce errors. As a CME moves away from the Sun, not only the direction of propagation but also the geometry plays a role \citep{Howard2011}. Such a geometrical effect comes to the picture because distances are estimated, taking into account that part of the CME which makes a tangent with the line of sight. Therefore, as the CME is far from the Sun, the observer from a certain location cannot estimate the kinematics of the same part of the leading edge of a CME in subsequent consecutive images. This is because of geometrical effect, which produces a situation similar to the deflection of CME and is called `artificial deflection.' This effect leads to overestimating the distance of the CME estimated from the FP method, which is more severe when the CME approaches longer elongation angles. 

\subsubsection{Single spacecraft fitting methods}

\paragraph{Fixed-phi fitting (FPF) method} \hspace{0pt}\\
\label{FPF}
The original concept of \citet{Sheeley1999} about deceptive acceleration or deceleration of a CME moving with constant speed in the imager (SMEI $\&$ HI) at large elongation angles from the Sun is used widely to assess the direction of propagation and speed of CME \citep{Rouillard2008, Sheeley2008, Davis2009, Mostl2009, Mostl2010, Howard2009, Mostl2011}. Under the assumption that a CME is traveling at a constant speed, the shape of the observed elongation-time profile of CME will be different for observers at different locations. Solving the Equation~\ref{FPeqn} for the elongation ($\epsilon(t)$) with assumption of constant velocity (v$_{FP}$) of CME along the fixed radial direction ($\phi_{FP}$), we get,

\begin{equation}
\epsilon(t) = \arctan \Big(\frac{v_{FP}(t)\; \sin(\phi_{FP})}{d_{0} - v_{FP}(t) \; \cos(\phi_{FP})} \Big)
\label{FPFeqn}
\end{equation}

From this Equation~\ref{FPFeqn}, the launch time of CME from the sun center, i.e. t$_{0FP}$ can also be calculated, and for this 
$\epsilon$(t$_{0FP}$) = 0 will be satisfied. In practice, we should calculate the launch time of a CME in the corona, i.e., at an elongation corresponding to height in the corona. But to make the calculation simpler, we consider the launch time on the Sun's center \citep{Mostl2011}. Theoretical elongation variation obtained from Equation~\ref{FPFeqn} can be fitted to match most closely with the observed elongation variation for a real CME by finding the most suitable physically realistic combinations of v$_{FP}$, $\phi_{FP}$ and t$_{0FP}$ values. This approach to finding the direction of CME propagation and its speed is called the Fixed-Phi-Fitting (FPF) method. This method has been applied to transients like CIRs \citep{Rouillard2008} and also on CMEs \citep{Davis2009, Davies2009, Rouillard2009, Mishra2014}.

\paragraph{Harmonic mean fitting (HMF) method} \hspace{0pt}\\
\label{HMF}
Based on HM approximation \citep{Lugaz2009} for CMEs, \citet{Lugaz2010} solved the equation of elongation angle and obtained the Harmonic mean fitting (HMF) relation. Further, following the fitting version of the FP method, i.e. FPF, \citet{Mostl2011} derived a new fitting version of the HM method. They write the $\epsilon$ for Equation~\ref{HMeqn} as below, assuming a constant speed (v$_{HM}$) of CME propagating along a fixed radial direction ($\phi_{HM}$).

\begin{equation}
\epsilon(t) = \arccos \Big(\frac{-b + a\; \sqrt{a^{2}+b^{2}-1}}{a^{2}+ b^{2}} \Big) 
\label{HMFeqn}
\end{equation}
 
In this equation, $a$ and $b$ are represented as below. 
	
\begin{equation*}
  a = \frac{2d_{0}} {v_{HM}t} - \cos(\phi_{HM}) \qquad\text{and}\qquad b = \sin(\phi_{HM})
\end{equation*}

It must be noted that in the case of a limb CME, its flank will be observed in HI FOV because of the Thomson scattering surface. The flank of a CME is relatively closer to the Sun than its apex. HMF method accounts for this effect and estimates the propagation direction is always farther away from the observer compared to the direction derived by the FPF method.    

\paragraph{Self-similar expansion fitting (SSEF) method} \hspace{0pt}\\
\label{SSEF}
\citet{Davies2012} derived a method to convert the measured elongation of an outward moving feature into distance based on the selection of an intermediate geometry for the CMEs. They derived the fitting version of the SSE method described in Section~\ref{SSE}. They inverted the Equation~\ref{SSEeqn} for elongation as given below.   

\begin{equation}
\epsilon(t) = \arccos \Big(\frac{-bc + a\; \sqrt{a^{2}+b^{2}-c^{2}}}{a^{2}+ b^{2}} \Big) 
\label{SSEFeqn}
\end{equation}

In this equation, $a$, $b$ and $c$ are represented as below. 
\begin{equation*}
a = \frac{d_{0}(1+c)} {v_{SSE}t} - \cos(\phi_{SSE}) \quad\text{;}\quad b = \sin(\phi_{SSE}) \qquad\text{and}\qquad 
c = \pm\sin(\lambda_{SSE})
\end{equation*}

It must be highlighted that FPF and HMF techniques can be used to estimate only the propagation direction, speed, and launch time of the CMEs, while SSEF can estimate the additional angular half-width ($\lambda_{SSE}$) of CMEs. Thus, implementation of the SSEF
technique requires a four-parameter curve fitting procedure with the assumptions that $\phi_{SSE}$, v$_{SSE}$ and $\lambda_{SSE}$
are constant over the complete duration of the time-elongation profile. The $\lambda_{SSE}$ measures the angular extent of the CME in a plane orthogonal to the observer's FOV.  It must be noted that the positive root of the quadratic in Equation~\ref{SSEFeqn} is allowed for the regimes relevant to currently operational HI on board \textit{STEREO}. If the SSEF is applied to the front, i.e., apex of CMEs, then the positive form of $c$ is used, while for the trailing edge of the CMEs, its negative form is used. Hence, for CMEs propagating in certain directions, identifying the correct form of the equation is very important. \citet{Davies2012} have pointed out that in the case where SSEF can be applied to time-elongation profiles of both features at the front and rear of a CME, then their fitted radial speed would differ while other fitted parameters would be the same.  In the SSEF method, the uncertainties arising from the degrees of freedom associated with the four-parameter fit could also be solved by putting constraints on the other parameters, like $\phi_{SSE}$, $\lambda_{SSE}$, and v$_{SSE}$ to reduce the number of free parameters in the fit. Again, we must emphasize that FPF and HMF methods are the special cases of the SSEF method corresponding to $\lambda$ = 0$\arcdeg$ and $\lambda$ = 90$\arcdeg$, respectively.

The main advantage of using FPF, HMF, and SSEF methods is that these fitting methods are quick and straightforward to apply in real-time \citep{Mostl2014}. In addition, these methods can be used for single spacecraft HI observations, i.e., when any one of \textit{STEREO} spacecraft suffers from a data gap. A significant disadvantage is that these methods assume constant speed and direction of propagation of the CMEs.

\subsubsection{Twin spacecraft methods}
Twin spacecraft reconstruction methods require simultaneous observations from two viewpoints of \textit{STEREO}. Hence, time-elongation profiles of the features of a CME derived from observations of both \textit{STEREO-A} and \textit{STEREO-B} viewpoints are used to determine the 3D characteristics of CMEs.    

\paragraph{Geometric triangulation (GT) method} \hspace{0pt}\\
\label{GTFP}

The geometric triangulation (GT) method was developed by \citet{Liu2010}. It assumes that the same feature of a CME can be observed from two different viewpoints and that the difference in measured elongation angles for the tracked feature from \textit{STEREO-A} and \textit{STEREO-B} is entirely due to two viewing directions. Using imaging observations and a Sun-centered coordinate system, the elongation angle of a moving feature can be calculated in each consecutive image. The details of the Geometric Triangulation (GT) method in an ecliptic plane applicable for a feature propagating between the two spacecraft have been explained in detail in \citet{Liu2010, Liu2010a}. A schematic diagram for the location of the twin spacecraft and the tracked feature is shown in Figure~\ref{GT}. Using this geometry, \citet{Liu2010} derived the following sets of equations.

\begin{equation}
d_{A} =  \frac{r \sin(\alpha_{A} + \beta_{A})}{\sin\alpha_{A}} 
\label{GTeqn1}
\end{equation}
\begin{equation}
d_{B} =  \frac{r \sin(\alpha_{B} + \beta_{B})}{\sin\alpha_{B}} 
\label{GTeqn2}
\end{equation}
\begin{equation}
\beta_{A}+\beta_{B} = \gamma
\label{GTeqn3}
\end{equation}
\\
In the above Equations~\ref{GTeqn1} to \ref{GTeqn3}, $r$ is the radial distance of the feature from the Sun, $\beta_{A}$ and $\beta_{B}$ are the propagation angles of the feature relative to the Sun-spacecraft line. The $d_{A}$ and $d_{B}$ are the distances of the spacecraft from the Sun, and $\gamma$ is the longitudinal separation between the two spacecraft that are known. Once the elongation angles ($\alpha_{A}$ and $\alpha_{B}$) are derived from imaging observations, the above equations can be solved for $\beta_{A}$.

\begin{equation}
\beta_{A} = \arctan \Big(\frac{\sin(\alpha_{A}) \sin(\alpha_{B}+\gamma) - f \sin(\alpha_{A}\sin(\alpha_{B})} 
{\sin(\alpha_{A}) \cos(\alpha_{B}+\gamma) + f \cos(\alpha_{A}\sin(\alpha_{B})} \Big)
\label{GTeqn4}
\end{equation} 
\\
where $f$ = $d_{B}$/$d_{A}$ ($f$ varies between 1.04 and 1.13 during a full orbit of the \textit{STEREO}
spacecraft around the Sun). Using Equation~\ref{GTeqn4}, the propagation direction of a CME can be estimated. Once the propagation direction has been estimated, the distance of the moving CME feature can be estimated using the  
Equation~\ref{GTeqn1}.

\begin{figure*}[!htb]
	\centering
		\includegraphics[scale=0.55]{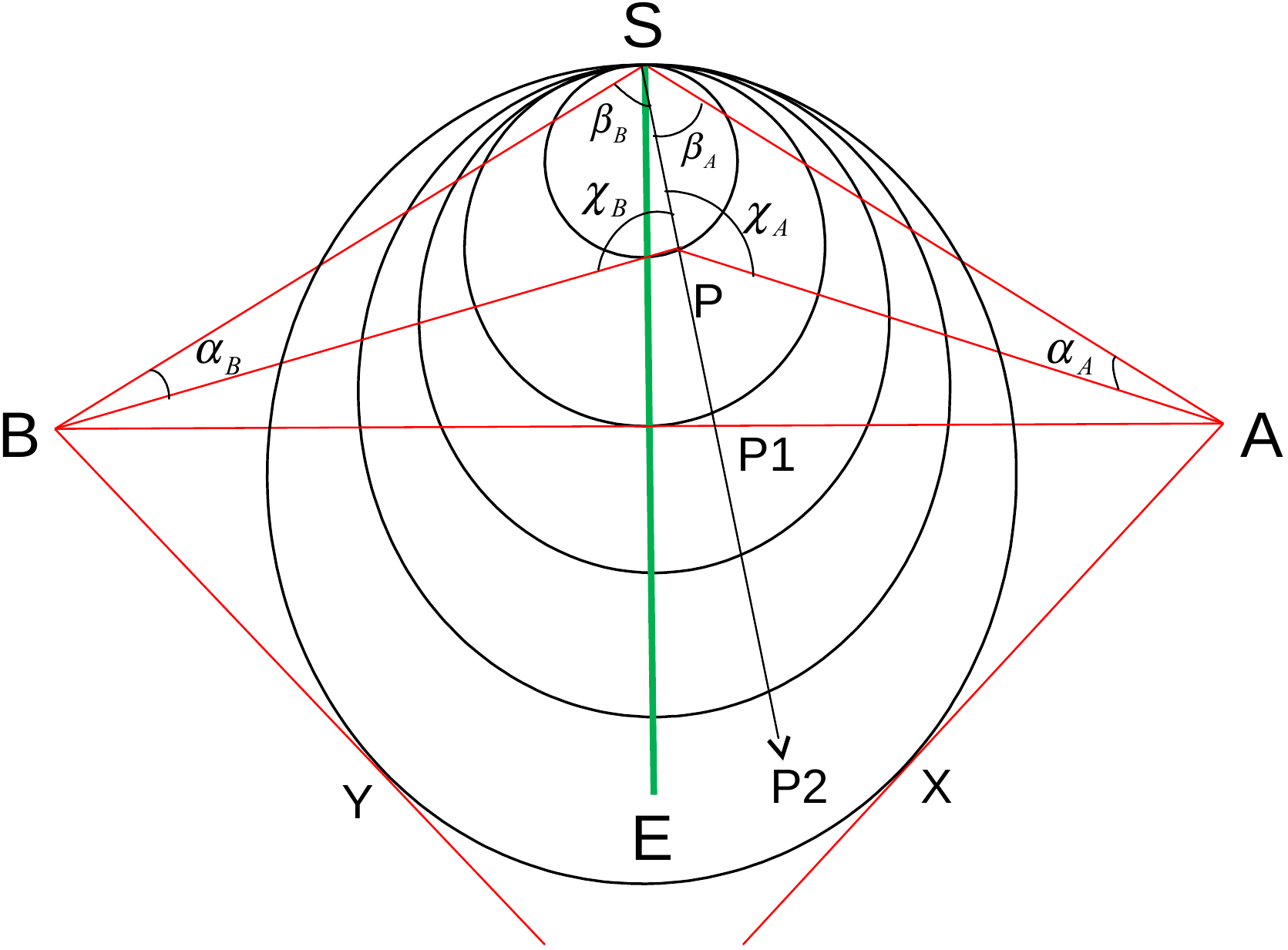}
\caption[Geometric triangulation for a moving CME feature]{Schematic diagram of geometric triangulation for a moving CME feature between the two spacecraft \textit{STEREO-A} and \textit{B}, in the direction of the arrow. Line SE represents the Sun-Earth line and $\alpha$, $\beta$, and $\chi$ denote the elongation, propagation, and scattering angles, respectively. Subscripts A and B represent angles measured from the \textit{STEREO-A} and \textit{STEREO-B} viewpoints.}
\label{GT}
\end{figure*}

In this reconstruction method, \citet{Liu2010} did not take into account the effects of Thomson scattering and the geometry of CMEs. The total scattered intensity received by an observer at a specific location depends on the radial and tangential component of the scattered radiation from the scattering source \citep{Howard2009}. The tangential component of the scattered radiation does not depend on the scattering angle, but the radial component does. However, for Earth-directed events, both view directions
(line-of-sight AP and BP as shown in Figure~\ref{GT}) will be nearly symmetrically located from the Sun-Earth line. Therefore, the scattering angles ($\chi_{A}$ and $\chi_{B}$) for both the observers will only be slightly different, and the resulting difference in the received radial intensity for both the observers (\textit{STEREO-A} and \textit{STEREO-B}) will be small. The approximation that both observers view the same part of CME may not be accurate when Earth-directed CMEs are at a considerable distance from the Sun (for view directions AX and BY as shown in Figure~\ref{GT}) and also near the Sun for very wide or rapidly expanding CMEs. It is also rather unlikely that the same feature of a CME will be tracked in each successive image. In light of the aforementioned points, it is clear that the geometry of the CME should be taken into account in any of the reconstruction methods. However, the breakdown of idealistic assumptions about the geometry can result in new errors in the estimated kinematics.

\paragraph{Tangent to a sphere (TAS) method} \hspace{0pt}\\
\label{TAS}

Following the development of the GT method \citep{Liu2010}, \citet{Lugaz2010.apj} proposed another method for stereoscopic reconstruction of the CMEs using HIs observations. They assume that CME has a circular cross-section anchored at the Sun, and twin \textit{STEREO} observe the tangent to the circular CME front in contrast to the assumption that CME is a point by \citet{Liu2010}. Hence the observers from two viewing locations of \textit{STEREO} do not observe the same CME feature. Using HM approximation of 
\citet{Lugaz2009}, the diameter $R_{A}$ and $R_{B}$ of CME sphere for both viewpoints respectively, was obtained by \citet{Lugaz2010.apj} and they solved it for the condition $R_{A}$ = R$_{B}$, where
 
\begin{equation} 
R_{A}= \frac{2d_{A}\; \sin(\alpha_{A})} {1+\sin(\alpha_{A} + \beta_{A} -\phi_{TAS})} 
\label{TAS1}
\end{equation}

\begin{equation}
R_{B}= \frac{2d_{B}\; \sin(\alpha_{B})} {1+\sin(\alpha_{B}+ \beta_{B} +\phi_{TAS})}
\label{TAS2}
\end{equation} 
  
In the above Equations~\ref{TAS1} and ~\ref{TAS2}, the parameters $d$, $\alpha$, $\beta$, and $\phi_{TAS}$ are the distance of the observer from the Sun, elongation angle, separation angle of the observer from the Sun-Earth line, and propagation direction of CME from the Sun-Earth line respectively. The $\phi_{TAS}$ is considered positive in the westward direction from the Sun-Earth line. The solution of these equations for $\phi_{TAS}$ can be used to estimate the propagation direction of the CMEs. This method to calculate the kinematics of the CME was referred to as the tangent-to-a-sphere (TAS) method. This method assumes that the measured elongation angle refers to the point where the observer line of sight intersects tangentially with the spherical front of the CME.   

\paragraph{Stereoscopic self-similar expansion (SSSE) method} \hspace{0pt}\\

Both GT and TAS methods described in the Section~\ref{GTFP} and ~\ref{TAS} are based on extreme geometrical descriptions of solar wind transients (a point source for GT and an expanding circle attached to the Sun for TAS). Therefore, \citet{Davies2013} proposed a stereoscopic reconstruction method based on a more generalized SSE geometry of \citet{Lugaz2010.apj} and \citet{Davies2012} and  named it the Stereoscopic Self-Similar Expansion (SSSE) method. They showed that the GT and TAS methods could be considered as the limiting cases of the SSSE method. Such a stereoscopic reconstruction is illustrated in Figure~\ref{SSSE}. In this figure, the propagation direction of a CME is shown as $\phi_{A}$ relative to observer \textit{STEREO-A}, $\phi_{B}$ relative to \textit{STEREO-B}, and $\phi_{E}$ relative to Earth (E) and $\gamma$ is separation angle between both observer located at distances $d_{A}$ and $d_{B}$ from the Sun. At each instance, $\epsilon_{A}$ and $\epsilon_{B}$ is the elongation measured from line of sight from \textit{STEREO-A} and \textit{STEREO-B}, respectively.

\begin{figure*}[!htb]
	\centering
		\includegraphics[scale=8.5]{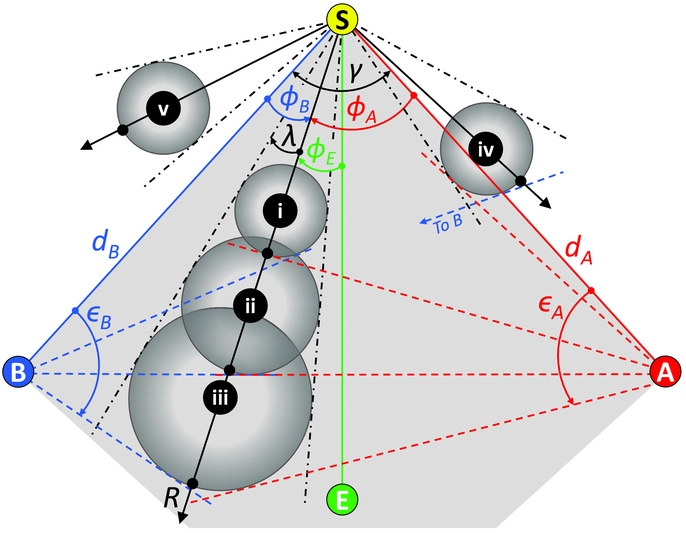}
\caption[The SSE modeled circular CME]{The SSE modeled circular CME, with a constant $\lambda$ is labeled as (i), (ii) and (iii), show three instances of propagation away from the Sun (S) in the common FOV of two observers. The shaded region with gray color represents the common FOV of \textit{STEREO-A} and \textit{STEREO-B}. Geometry marked with (iv) is outside the common FOV however both observers can observe it while geometry (v) is outside the FOV of \textit{STEREO-B} and therefore can only be observed by \textit{STEREO-A} (reproduced from \citealp{Davies2013}).}
\label{SSSE}
\end{figure*}

\citet{Davies2012} used the Equation~\ref{SSEeqn} corresponding to both viewpoints of \textit{STEREO-A} and \textit{STEREO-B} and derived an expression for calculating the propagation direction ($\phi_{A}$ or $\phi_{B}$) of the CME. The propagation direction can be used in Equation~\ref{SSEeqn} to estimate the distance of tracked CME feature. This is the same methodology utilized in the GT and TAS methods. SSSE method is special as we can take a reasonable angular extent ($\lambda$) of CME  geometry contrary to extreme geometrical description taken in both GT and TAS methods. The details of the SSSE method and important considerations for implementation of this method have been discussed in \citet{Davies2013}.

\subsection{Estimation of the arrival time of CMEs}
In our study, the kinematics of CMEs estimated by implementing 3D reconstruction methods in COR2 and HI FOV is used to calculate their arrival time at the Earth. In our study, we use the estimated kinematics either in the drag-based model or extrapolate it to find the arrival time of CMEs near 1 AU at the Earth. 

\subsubsection{Drag based model for propagation of CMEs}
\label{DBM}
When a CME is far from the Sun, the Lorentz and gravity forces decrease such that drag can be considered to govern CME dynamics. Although it is not proven that drag is the only force that shapes  CME dynamics in the interplanetary medium, the observed deceleration/acceleration of some CMEs have been closely reproduced by considering only the drag force acting between the CME and the ambient  solar  wind medium \citep{Lindsay1999, Cargill2004, Manoharan2006, Vrsnak2009, Lara2009}. In our study, we have used the drag-based model (DBM) \citep{Vrsnak2013} to derive the kinematic properties for the distance range beyond which a CME cannot be tracked in the \textit{J}-maps. The DBM model assumes that, after 20 \textit{R}$_\odot$, the dynamics of CMEs is solely governed by the drag force and that the drag acceleration has the form, $a_{d}$ = -$\gamma$ $(v-w)$ $|(v-w)|$, (see e.g. \citealt{Cargill1996, Cargill2004,Vrsnak2010}), where $v$ is the speed of the CME, $w$ is the ambient solar wind speed and $\gamma$ is the drag parameter.

The drag parameter is given by $\gamma$ = $\frac{c_{d}A \rho_{w}}{M + M_{v}}$, where c$_{d}$ is the dimensional drag coefficient, $A$ is the cross-sectional area of the CME perpendicular to its propagation direction (which depends on the CME-cone angular width), $\rho_{w}$ is the ambient solar wind density, $M$  is the CME mass, and $M_{v}$  is the virtual CME mass. The latter is written as, $M_{v}$ = $\rho_{w} V/2$, where $V$ is the CME volume. A statistical study has shown that the drag parameter generally lies between 0.2 $\times$ 10$^{-7}$ and 2.0 $\times$ 10$^{-7}$ km$^{-1}$ \citep{Vrsnak2013}. They assumed that the mass and angular width of CMEs do not vary beyond 20 \textit{R}$_{\odot}$ and also showed that the solar wind speed lies between 300 and 400 km s$^{-1}$ for slow solar wind conditions. For the case where a CME propagates in high-speed solar wind or if a coronal hole is present in the vicinity of the CME source region, the ambient solar wind speed should be chosen to lie between 500 and 600 km s$^{-1}$, along with a lower value of the drag parameter.  

\section{Identification of CMEs and Their Consequences Near the Earth} 

Once the CMEs arrive near the 1 AU at the Earth, they can cause severe geomagnetic consequences in the Earth's magnetosphere. Analyzing the observation of in situ spacecraft (e.g., \textit{ACE}, \textit{WIND}), we identify various parts of CME structures and mark their boundaries. The identification of CME boundaries is carried out using multiple signatures of CMEs (described in Section~\ref{insitu} of Chapter~\ref{Chap1:IntMot}). It is essential to mention here that no CMEs show all the signatures, and therefore there is no unique scheme to identify them in in situ observations. Also, different signatures may appear for different intervals of time, and hence, CMEs may have different boundaries in plasma, magnetic field, and other signatures. This is possible as different signatures have their origin due to different physical processes. If we identify CMEs based on only a few signatures, then they may be falsely identified.
On the other hand, if several in situ signatures are considered, then such an attempt may lead to the omission of a few CMEs. Therefore, our approach is to identify as many signatures as possible. Such a procedure helps for reliable identification of the CMEs in in situ observations; however, marking their boundaries may still be ambiguous. Various lists of CMEs observed near the Earth have been compiled based on different criteria, e.g. by \citet{Richardson1995, Cane2003,Richardson2010}. \citet{Richardson2010} have identified approximately 300 CMEs near the Earth during the complete solar cycle 23, i.e., between year 1996 to 2009.

Identifying CMEs in in situ observations is extremely difficult when they arrive as structures formed due to the interaction or collision of several CMEs. The collision of CMEs is expected during their evolution between the Sun and the Earth when they are launched in quick succession in the same direction from the Sun. As they interact, they experience a change in their plasma, dynamic and magnetic field parameters. Hence, the collision of CMEs may lead to a new type of solar wind structure which is expected to show different in situ signatures than isolated normal CMEs. In addition, such new structures might have a different geomagnetic response than isolated CMEs. We have investigated the geomagnetic responses of interacting CMEs as an example of the heliospheric consequences of CMEs. The actual arrival times of remotely tracked features of the CMEs are marked in in situ observations.  We compare the estimated arrival times of tracked features based on the kinematics and actual arrival times. In this way, we attempt to make an association between remote and in situ observations of the CMEs and assess the performance of several 3D reconstruction methods used to estimate CME kinematics.

In summary, continuous tracking of CMEs in the heliosphere and implementation of suitable 3D reconstruction methods to estimate and understand their kinematics is an important step in our study. Finding an association of remote observations to in situ and ground-based magnetometer observations is also carried out. The heliospheric consequences and geomagnetic responses of interacting CMEs have been studied.

\chapter{Estimation of Arrival Time of CMEs}
\label{Chap3:ArrTim}
\rhead{Chapter~\ref{Chap3:ArrTim}. Estimation of Arrival Time of CMEs}

\section{Introduction}
Due to the significant role of CMEs and interplanetary shocks in the context of space weather, understanding their heliospheric evolution and predicting their arrival times at the Earth is a major objective of various forecast centers. The prediction of CME/shock arrival time means that forecasters utilize the observables of solar disturbance obtained prior to arrival as inputs to predict whether/when they will arrive. A longer lead time in prediction is yielded if the solar observables are used. The arrival time of CMEs at 1 AU can be related to their characteristics (velocity, acceleration) near the Sun in order to develop the prediction methods for CME’s arrival time. Different models of CME/shock arrival time prediction have been developed, e.g., empirical models, expansion speed models, drag-based models, physics-based models, and MHD models.

The models that adopt relatively simple equations to fit the relations between the arrival time of the CME disturbance at the Earth and their observables near the Sun (such as initial velocity) are called empirical prediction models. \citet{Vandas1996} found that the transit time (in hr) to 1 AU for the CME flux rope (cloud/driver) leading edge is T$_{driver}$ = 85-0.014V$_{i}$ for a slow background solar wind speed (say, 361 km s$^{-1}$), and T$_{driver}$ = 42-0.0041V$_{i}$ for a faster background solar wind speed (say, 794 km s$^{-1}$). Here V$_{i}$ (km s$^{-1}$) is the propagation speed of the leading edge of CME at 18 R$_{\odot}$. Then the transit time of the shock preceding the magnetic cloud is T$_{shock}$ = 74 - 0.015V$_{i}$  for slow solar wind and T$_{shock}$ = 43-0.006V$_{i}$ for fast solar wind. \citet{Brueckner1998} found that the difference in time between the CME launch on the Sun and the time when the associated geomagnetic storm reaches its peak is about 80 hr. One of the most typical and widely used empirical prediction models is empirical CME arrival (ECA) and empirical shock arrival (ESA) models. The empirical CME arrival (ECA) model developed by \citet{Gopalswamy2001b} considers that a CME has an average acceleration up to a distance of 0.7 AU-0.95 AU.
After the cessation of acceleration, a CME is assumed to move at a constant speed. They found that the average acceleration has a linear relationship with the initial plane-of-sky speed of the CME. The ECA model has been able to predict the arrival time of CMEs within an error of approximately $\pm$ 35 hr with an average error of 11 hr. Later, an empirical shock arrival (ESA) model was developed that was able to predict the arrival time of CMEs to within an error of approximately $\pm$ 30 hr with an average error of 12 hr \citep{Gopalswamy2005}. The ESA model is a modified version of the ECA model in which a CME is considered to be the driver of magnetohydrodynamic (MHD) shocks. The other assumption is that fast mode MHD shocks are similar to gas dynamic shocks. The gas-dynamic piston-shock relationship is thus utilized in this model. Various efforts have been made to derive an empirical formula for CME arrival time based on the projected speed of a large number of CMEs \citep{Wang2002, Zhang2003,Srivastava2004, Manoharan2004}. \citet{Vrsnak2007}  found that CME transit time depends on both the CME take-off speed and the background solar wind speed. In the majority of these models, the initial speeds of CMEs used were measured from the plane of sky LASCO/\textit{SOHO} observations, and therefore the measured kinematics are not representative of the true CME motion.

To overcome plane-of-sky effects, \citet{DalLago2003} used a sample of 57 limb CMEs to derive an empirical relationship between their radial and expansion speeds as V$_{rad}$ = 0.88V$_{exp}$. This result led to lateral expansion speed as a proxy for the radial speed of halo CMEs that could not be measured. Also, \citet{Schwenn2005} analyzed 75 events to derive an empirical  formula for transit time of CMEs to Earth T$_{tr}$ = 203 - 20.77 ln(V$_{exp}$). Their results show that the formula can predict ICME arrivals, with a 95\% error margin of about 24 hr.

Several observations have revealed that the dynamics of CMEs are governed mainly by their interaction with the background solar wind beyond a certain helio-distance. On this basis, several analytical models have been developed to depict the propagation of CMEs and predict their arrival times. These analytical models are based on the equation of motion of CMEs, where the drag acceleration/deceleration has a quadratic dependence on the relative speed between CME and the background solar wind. It was found that the measured deceleration rates are proportional to the relative velocity between CME and the background solar wind, as well as a dimensionless drag coefficient (c$_{d}$) \citep{Vrsnak2001, Vrsnak2002,Cargill2004}. Recently, \citet{Subramanian2012} have discussed the variation of the drag coefficient (c$_{d}$) with heliocentric distance for the first time. They adopt a microphysical prescription for viscosity in the turbulent solar wind to obtain an analytical model for the drag coefficient. \citet{Vrsnak2013} have simplified the drag-based model and presented an explicit solution for the Sun-Earth transit time of CMEs and their impact speed at 1 AU.

One physics-based prediction model is the ``Shock Time of Arrival'' (STOA) model, which is based on the theory of similar blast waves from point explosions. This concept was revised by introducing the piston-driven concept \citep{Dryer1974, Smart1985}. Another such model is the ``Interplanetary Shock Propagation Model'' (ISPM), which is based on a 2.5D MHD parametric study of numerically simulated shocks. The model demonstrates that the organizing parameter for the shock is the net energy released into the solar wind \citep{Smith1990}. The ``Hakamada-Akasofu-Fry version 2'' (HAFv.2) model is a ``modified kinematic'' solar wind model that calculates the solar wind speed, density, magnetic field, and dynamic pressure as a function of time and location \citep{Dryer2001, Dryer2004,Fry2001, Fry2007,Smith2009}. This model provides a global description of the propagation of multiple and interacting shocks in nonuniform, stream-stream interacting solar wind flows in the ecliptic plane. The STOA, ISPM, and HAFv.2 models use similar input solar parameters (i.e., the source location of the associated flare, the start time of the metric Type II radio burst, the proxy piston driving time duration, and the background solar wind speed).

Also, several physics-based magnetohydrodynamics (MHD) numerical models have been developed. The coupled Wang-Sheeley-Arge (WSA) + ENLIL + Cone model \citep{Odstrcil2004} is widely used to simulate the propagation and evolution of CMEs in interplanetary space and provides a 1-2 day lead time forecasting for major CMEs \citep{Taktakishvili2009, Pizzo2011}. WSA is a quasi-steady global solar wind model that uses synoptic magnetograms as inputs to predict ambient solar wind speed and interplanetary magnetic field polarity at Earth \citep{Wang1995, Arge2000}. The ENLIL model is a time-dependent, 3D ideal MHD model of the solar wind in the heliosphere \citep{Odstrcil2002, Odstrcil2004}. The cone model assumes a CME as a cone with constant angular width in the heliosphere \citep{Zhao2002, Xie2004}. The input of ENLIL at its inner boundary of 21.5 \textit{R}$_{\odot}$ is taken from the output of WSA to get the background solar wind flows and interplanetary magnetic field.

Some of the aforementioned models are complicated, while others are relatively simple and easy. However, no significant differences are found between their prediction capabilities of CME arrival time. The predictions yield a root-mean-square error of $\approx$ 12 hr and a mean absolute error of $\approx$ 10 hr for many CMEs. There are many factors that are responsible for the limited accuracies of these models, e.g. (1) The inputs parameters (kinematics and morphology) of the model have their uncertainties. (2) The real-time background solar wind condition into which CME travels is difficult to observe or simulate from MHD. (3) The change in the kinematics of the CME due to its interaction with other large or small scale solar wind structures. These factors are challenging to be taken into account in a single model. Improvement in the accuracy of these arrival time models requires a better understanding of CME's heliospheric evolution and the ambient solar wind medium. Using the observations of CMEs, such as by SECCHI instruments onboard \textit{STEREO}, their heliospheric evolution can be investigated and compared with results from the models. Such observations can help to impose several constraints on the aforementioned models.

\textit{STEREO} observations have greatly enhanced our ability to track the CMEs continuously. This is because of \textit{STEREO's} two viewpoints which allow the 3D reconstruction of CMEs. Also, the large field of view (FOV) of its imaging telescopes (such as HI1 and HI2) enables the tracking of CMEs to a much larger distance in the heliosphere. Using \textit{STEREO} observations, several attempts have been made to understand the 3D propagation of CMEs and estimate their arrival time \citep{Mierla2009,Srivastava2009,Kahler2007,Liu2010,Mostl2011,Davies2012,Davies2013}. In a recent study, a CME was tracked beyond the Earth's distance, and was shown that a proper treatment of CME geometry must be performed in estimating CME kinematics, especially when a CME is directed away from the observer \citep{Liu2013}. Using different reconstruction methods on HI observations, \citet{Mostl2014} show an absolute difference between predicted and observed CME arrival times as 8.1 $\pm$ 6.3 hr. The HI observations have also revealed imaging of a few cases of interacting CMEs. The interaction of CMEs complicates the problem of estimating their arrival time. Several attempts have been made to understand the propagation of interacting CMEs using SECCHI/HI observations \citep{Harrison2012, Temmer2012,Mostl2012, Lugaz2012}. In an attempt to combine the CME kinematics with a model, \citet{Kilpua2012} estimated the 3D speed of CMEs in the COR FOV (close to the Sun) from the forward modeling \citep{Thernisien2009} method and used in CME travel-time prediction models of \citet{Gopalswamy2000, Gopalswamy2001b}. They compared the estimated travel time with the actual travel time of CME from the Sun to \textit{STEREO}, \textit{ACE} and \textit{WIND} spacecraft. They also compared the estimated travel time with that estimated using the projected CME speed in the models. Their study shows that CME 3D speeds give slightly ($\approx$ 4 hr) better predictions than projected CME speeds. However, in their study, a large average error of 11 hr is noted between the predicted and observed travel times.

For understanding the heliospheric evolution of CMEs from the Sun to Earth, we have estimated the kinematics of various selected CMEs by implementing suitable 3D reconstruction methods on the remote sensing observations of the CMEs. This chapter consists of two major studies:
\begin{enumerate}
\item{Estimating the arrival time of CMEs at L1 by tracking them into heliosphere (HI FOV) and applying geometric triangulation (GT) method \citep{Liu2010} of 3D reconstruction.} 
 \item{Assessment of the relative performance of several 3D reconstruction methods, applicable on HI observations, for estimating the arrival time of CMEs.}
\end{enumerate}

\section{Estimating the Arrival Time of CMEs Using GT Method}
\label{ArrtimGT}
In this study, we attempt to understand the 3D propagation of CMEs by tracking them continuously throughout the interplanetary medium. For this purpose, we use the GT reconstruction method on the time-elongation maps (\textit{J}-maps) \citep{Sheeley1999}, constructed from COR2 and HI observations, to estimate the 3D kinematics of CMEs. These estimated values of kinematics are used as inputs in the drag-based model \citep{Vrsnak2013} beyond the distance where a CME could not be tracked unambiguously, and its arrival time, as well as its transit velocity at the L1 point, are predicted. The predicted arrival time and transit velocity of the CME at L1 is then compared with the actual arrival time and transit velocity as observed by in situ instruments, e.g., \textit{ACE} and \textit{WIND}. The predicted arrival time is also compared with the arrival time estimated using the 3D speed obtained by the 3D reconstruction methods applicable on COR2 observation alone, e.g., tie-pointing procedure (scc\_measure: \citealp{Thompson2009}) from SECCHI/COR2 data alone. Hence, in this study, we attempt to examine the improvement in arrival time prediction of CMEs by using the GT method on heliospheric observations of CMEs (HI FOV) over using only coronagraph (COR) observations. The in situ observations of selected CMEs and their interpretation are described separately in detail in Chapter~\ref{Chap4:Associa}.   

\paragraph*{Selection of CMEs} \hspace{0pt}\\
Keeping in mind our goal to track a CME in the heliosphere and then implement the 3D reconstruction method, we selected only the Earth-directed CMEs in our study. This is because the HI instruments on \textit{STEREO} have been designed in such a way that only a CME is directed towards the Earth (i.e., between \textit{STEREO-A} and \textit{STEREO-B}) can be observed in both FOV of HI-A and HI-B. In addition, the in situ observations of these CMEs can be used as a reference for their actual arrival time to constrain the kinematics of the CMEs. Further, these CMEs are important to study because of their consequences on the Earth. We selected eight Earth-directed CMEs observed on different dates after the launch of the \textit{STEREO} spacecraft. These CMEs observed with \textit{STEREO} at different separation angles between them, the arrival time of CME shock/sheath, leading and trailing edge at L1 are listed in Table~\ref{SelectCMEs}. These CMEs have been observed from their birth in the corona through the inner heliosphere by coronagraphs and HIs. It must be noted that during the time of favorable separation between \textit{STEREO} spacecraft, for more appropriate 3D reconstruction of Earth-directed CMEs in HI FOV, very few Earth-directed CMEs occurred during 2006 to 2010.

\begin{table}
  \centering

 \begin{tabular}{|p{2.3cm}| p{2.0cm}| p{2.5cm}|p{2.7cm}|p{2.7cm}|}
    \hline
	 &  & \multicolumn{3}{ c| }{Arrival time of CME at L1 (UT)} \\ \cline{3-5}
\hspace{1.5cm}CME dates & \textit{STEREO} separation ($^\circ$) & \hspace{1.5cm}Shock/sheath & \hspace{1.7cm}Leading edge & 
\hspace{1.8cm}Trailing edge  \\ \hline

12 Dec 2008 & 86 & 16 Dec 11:55 & 17 Dec 04:39 & 17 Dec 15:48 \\ \hline
07 Feb 2010 & 135 & 11 Feb 01:00 & 11 Feb 12:47 & 11 Feb 23:13 \\ \hline
12 Feb 2010 & 135 & 15 Feb 18:42 & 16 Feb 04:32 & 16 Feb 12:38 \\ \hline
14 Mar 2010 & 138 &      \hspace{1cm}--       & 17 Mar 21:19 & 18 Mar 11:26   \\ \hline
03 Apr 2010 & 139 & 05 Apr 08:28  & 05 Apr 13:43 & 06 Apr 16:05  \\ \hline
08 Apr 2010 & 139 & 11 Apr 12:44  & 12 Apr 02:10 &  12 Apr 13:52  \\ \hline
10 Oct 2010 & 161 & 15 Oct 04:30 & \hspace{1cm}-- & 16 Oct 01:38  \\ \hline 
26 Oct 2010 & 164 & 30 Oct 10:32 & 31 Oct 06:30 & 01 Nov 21:35 \\ \hline 

 \end{tabular}
\caption[List of selected CMEs with their arrival times at L1]{Selected CMEs for our study, separation angle between \textit{STEREO} on launch time of CMEs, the arrival time of their shock/sheath, leading and trailing edge at L1 are listed. The blank shell in the table is because of the lack of identification of that particular feature in in situ data.}
\label{SelectCMEs}
\end{table}

Remote sensing observational data of CMEs by the twin \textit{STEREO} spacecraft are taken from UKSSDC (http://www.ukssdc.ac.uk/solar/stereo/data.html). In situ observations of CMEs were obtained from the \textit{ACE} and \textit{WIND} spacecraft situated at the L1 point, upstream from the Earth. We used the OMNI data with a 1 min time resolution for solar wind parameters, e.g., magnetic field, proton velocity, proton density, proton temperature, and plasma beta. We also used the latitude and longitude of magnetic field vector data with a time resolution of 1 hr. Combined OMNI data were taken from NASA CDAWeb (http://cdaweb.gsfc.nasa.gov). We present our analysis for each event sequentially; the event of 2008 December 12 has been described in-depth to explain the implemented techniques. Research for the other events has been carried out by adopting the same methodology as explained for the event of 2008 December 12 and is summarized only briefly.

\subsection{2008 December 12}
\label{Rmt12Dec08}

This CME was observed in SECCHI/COR1-A images at 04:35 UT in the NE quadrant and in SECCHI/COR1-B at 04:55 UT in the NW quadrant. SOHO/LASCO also observed this as a partial halo CME with an angular width of 184$\arcdeg$ and a linear speed of 203 km s$^{-1}$ (\url{http://cdaw.gsfc.nasa.gov/CME\_list/}; see \citealp{Yashiro2004}). This CME was tracked in LASCO-C3 images out to 12 \textit{R}$_\odot$ where its quadratic speed was measured as 322 km s$^{-1}$. The CME was associated with a filament eruption that started at 03:00 UT in the NE quadrant observed in SECCHI/EUVI-A 304 {\AA} images. The appearance of the CME in the SECCHI-A COR2, HI1, and HI2 data is displayed in Figure~\ref{12Dec08CME}.

\begin{figure}[!htb]
\includegraphics[angle=0,scale=.38]{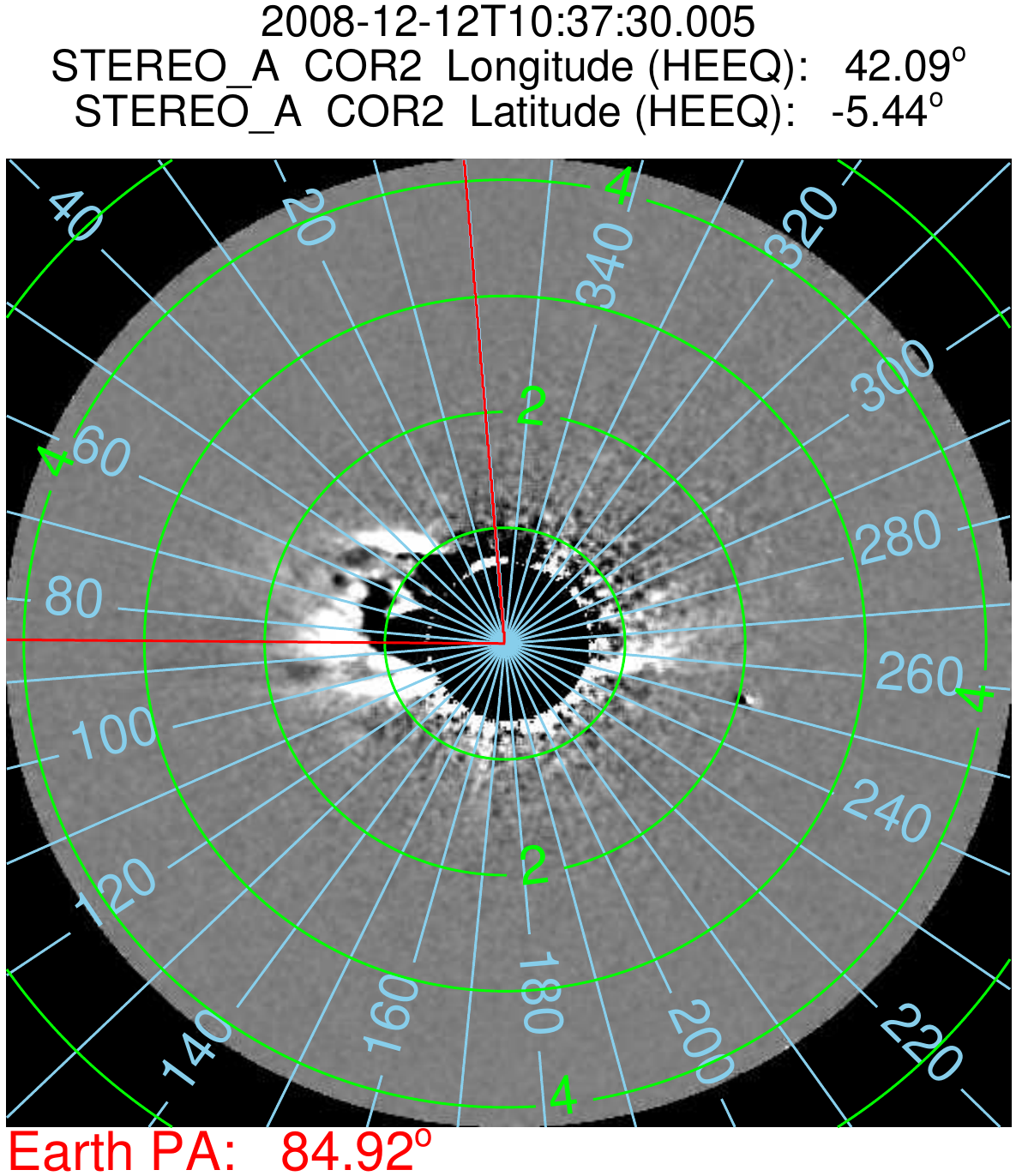}
\includegraphics[angle=0,scale=.38]{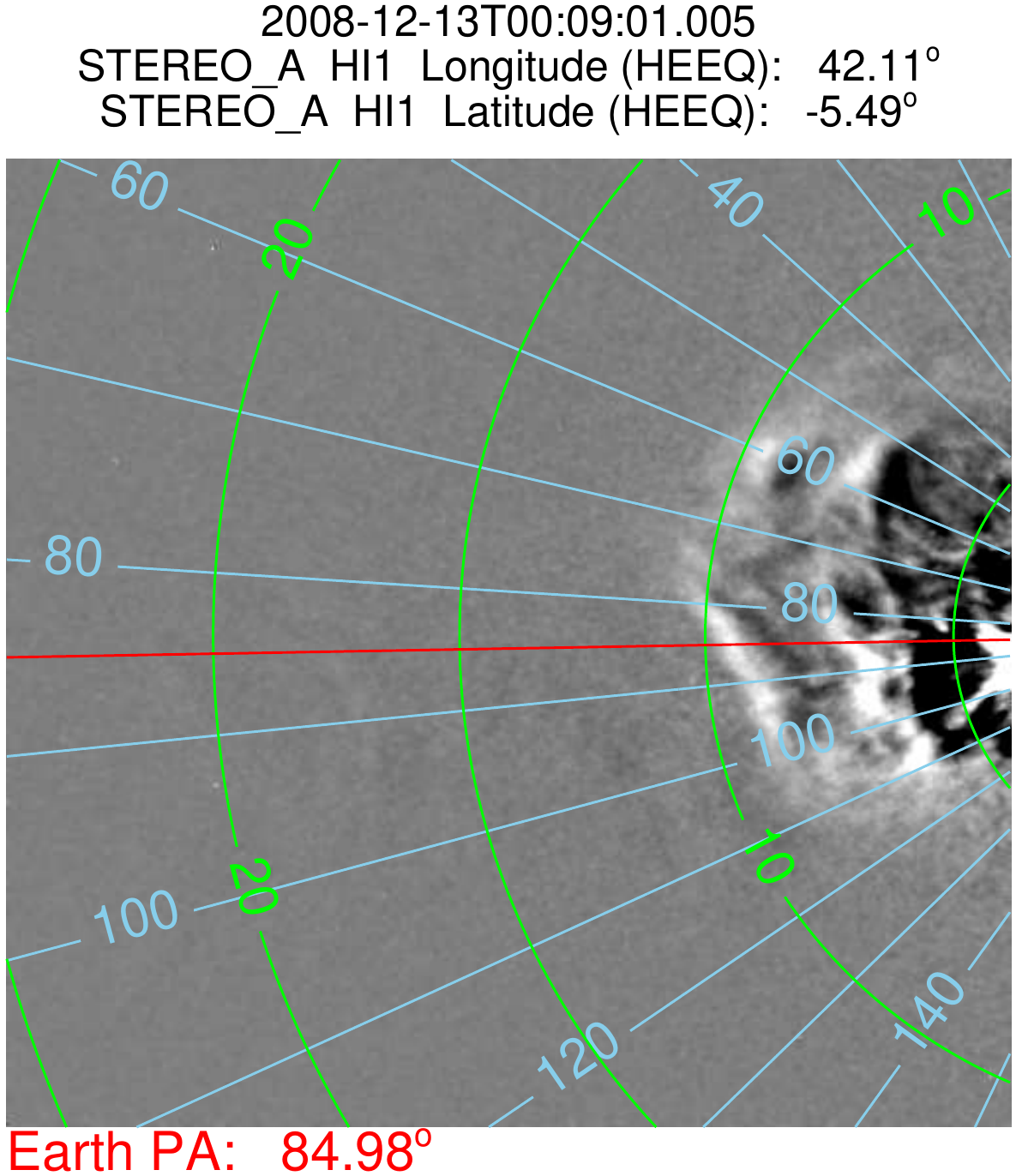}
\includegraphics[angle=0,scale=.38]{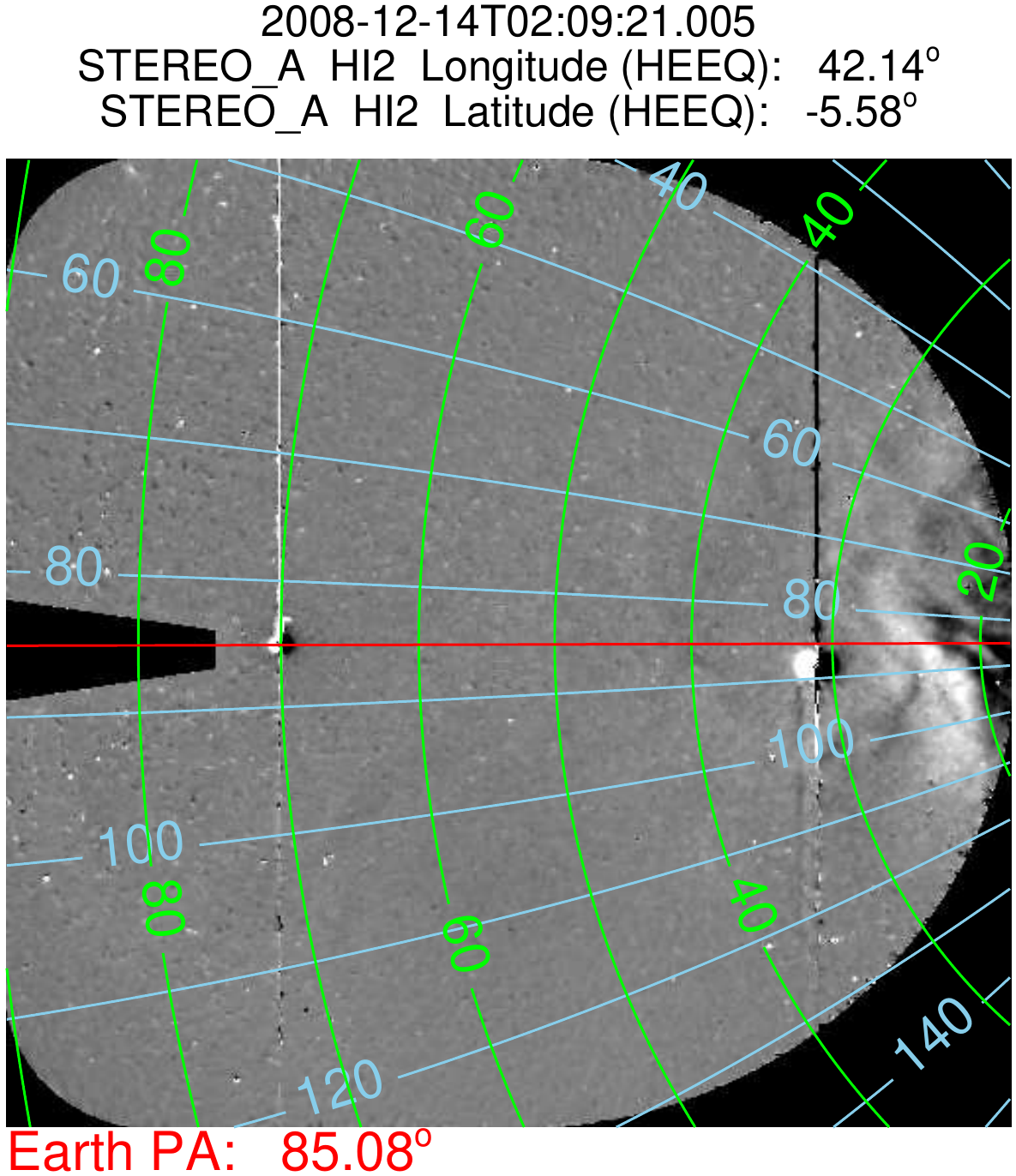}
\caption[Evolution of the 2008 December 12 CME as seen in COR and HI FOV]{Left, middle and right figures show the running difference image of COR2, HI1 and HI2 respectively taken by \textit{STEREO}/SECCHI-A with contours of elongation angle (green) and position angle (blue) overplotted. The red line is along the ecliptic at the position angle of Earth.}
\label{12Dec08CME}
\end{figure}

We constructed the \textit{J}-maps along the ecliptic plane using long-term background-subtracted running difference images
from COR2, HI1, and HI2 taken by the \textit{STEREO-A} and \textit{B} spacecraft (Figure~\ref{Jmaps12Dec08}). The \textit{J}-maps were constructed as per the procedure described in Section~\ref{Jmapsmthd} of Chapter~\ref{Chap2:DataMthd}. In this \textit{J}-map, the bright curve with positive inclination reveals the evolution of the CME. In the \textit{J}-map constructed from images taken by the \textit{STEREO-A} spacecraft, two nearly horizontal lines in the HI2 FOV start at an elongation angle of 30.7$\arcdeg$ and 70.1$\arcdeg$, respectively. These lines are due to the appearance of the planets Venus and Earth in the HI2-A FOV. In the HI2-A images taken on 2008 December 12 at 00:09:21 UT, Venus is seen at a position angle (helioprojective radial longitude) of 88.1$\arcdeg$ and an elongation angle (helioprojective radial latitude) of 30.7$\arcdeg$. These measurements correspond to the pixels 830 $\times$ 487 in the image of 1024 $\times$ 1024 size. At this time, the Earth is observed corresponding to pixels 277 $\times$ 509 in the HI2-A image and to pixels 655 $\times$ 510 in the HI2-B image. The appearance of planets in the HI FOV saturates the pixels, and also their signal bleeds up and down in the CCD, creating vertical columns of saturated pixels in the HI images 
(Figure~\ref{12Dec08CME}). In the \textit{J}-map constructed from the \textit{STEREO-B} images, one horizontal line is observed that starts at an elongation angle of 63.9$\arcdeg$ and is due to the appearance of the Earth in the HI2-B FOV. Vertical columns of saturated pixels correspondingly appear in the HI2-B images. 

\begin{figure}[!htb]
\centering
\includegraphics[angle=0,scale=.44]{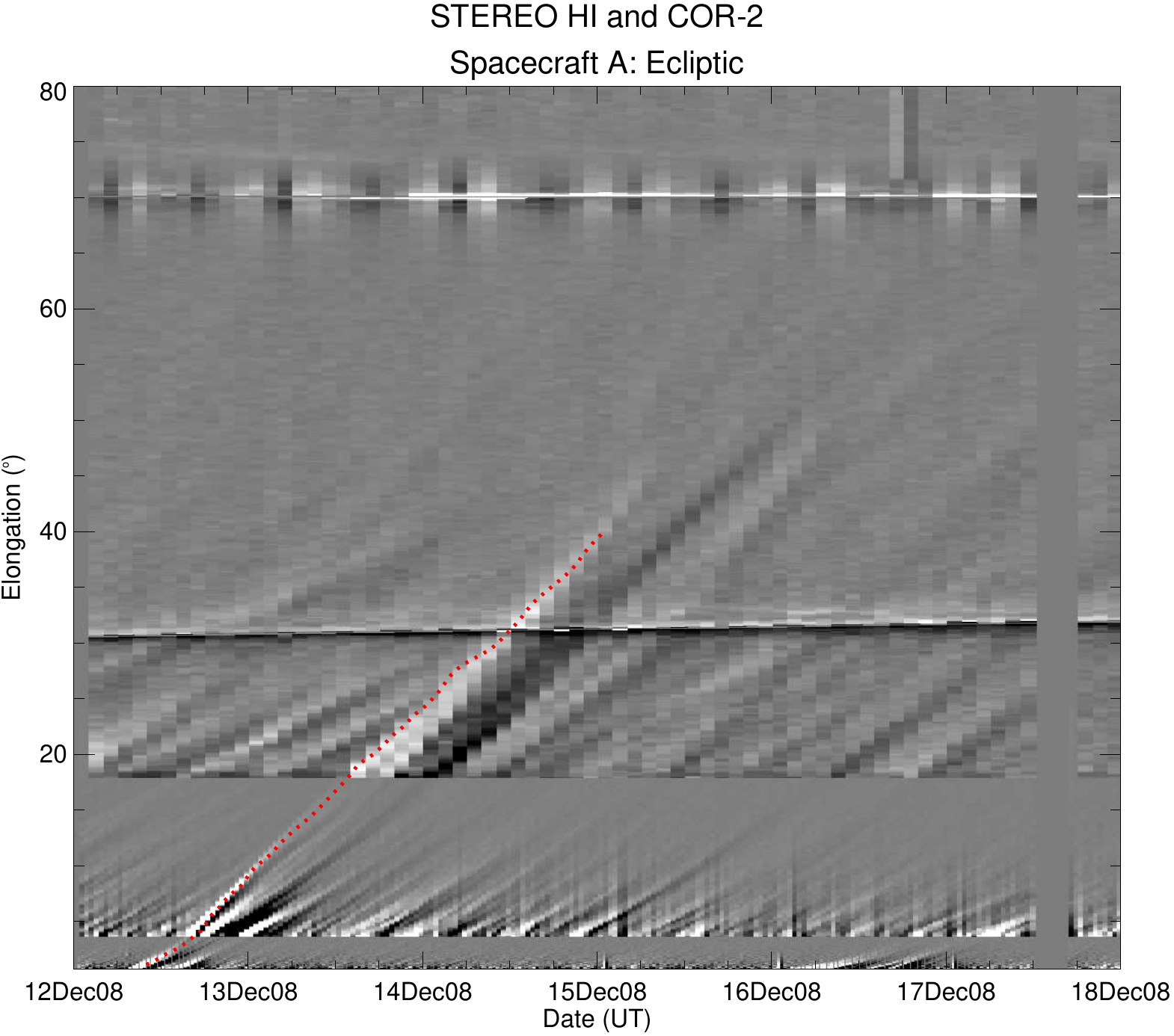}
\includegraphics[angle=0,scale=.44]{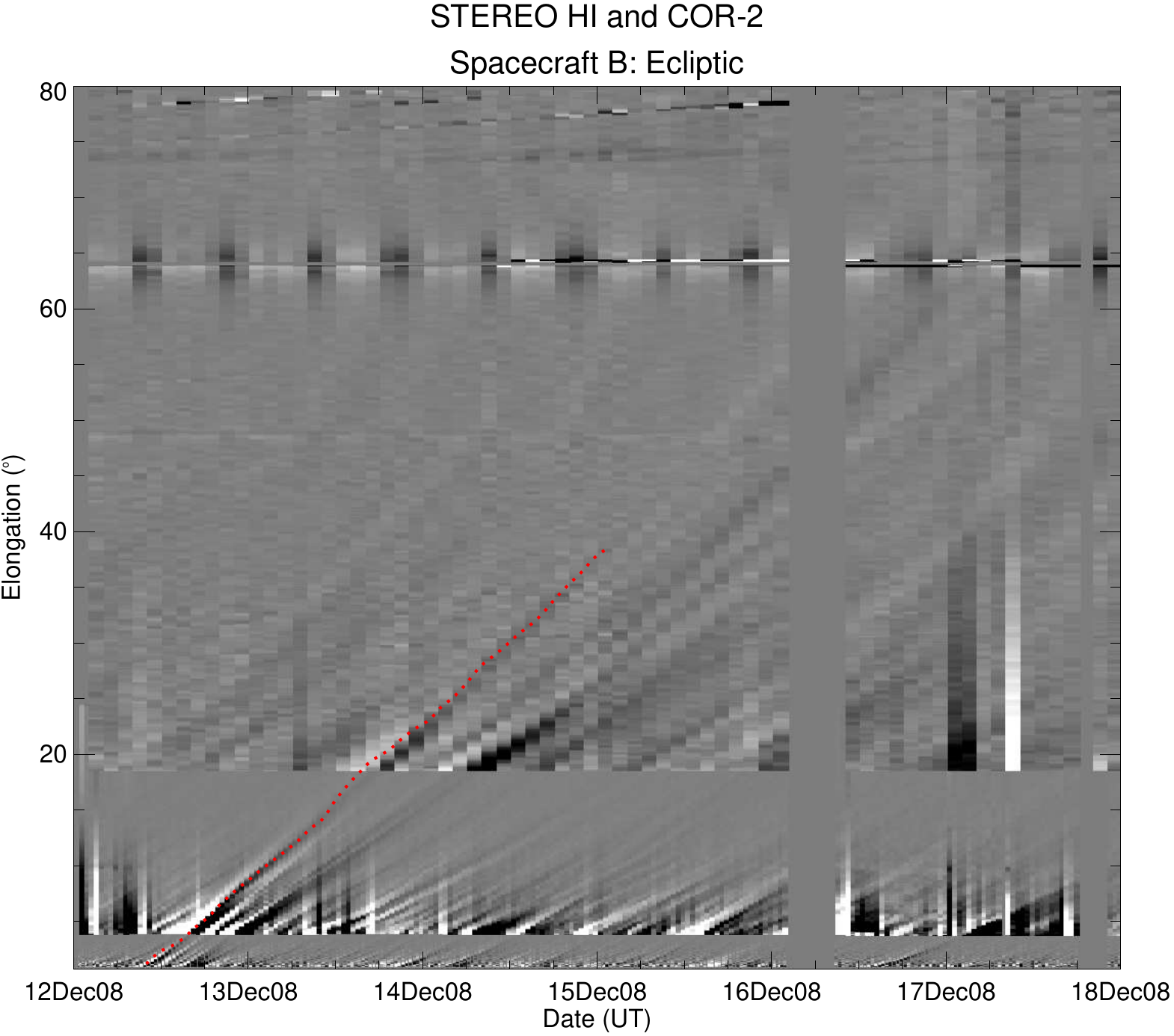}
\caption[\textit{J}-maps for the 2008 December 12 CME]{Time elongation map (\textit{J}-map) for \textit{STEREO-A} (left) and \textit{STEREO-B} (right) for the interval of 2008 December 12 to 18 is shown. The red dot shows the tracked feature corresponding to CME. Dots overplotted on the \textit{J}-map mark the elongation variation with time.}
\label{Jmaps12Dec08}
\end{figure}

\subsubsection{Application of 3D reconstruction method in COR FOV}

To estimate the 3D kinematics of the 2008 December 12 CME in the COR2 (2.5-15 \textit{R}$_\odot$) FOV, we carried out a 3D reconstruction
of its selected features. For this, we applied the tie-pointing method (scc\_measure: \citealp{Thompson2009}) on both sets of images taken by COR2-A and B. The details of the method are described in Section~\ref{tiepoint} of Chapter~\ref{Chap2:DataMthd}. The kinematics obtained after 3D reconstruction is shown in Figure~\ref{KinTP12Dec08}. This plot shows the 3D height, speed, acceleration, Stonyhurst heliographic latitude, and longitude of the CME leading edge. The values of latitude and longitude show that this CME was Earth-directed.

\begin{figure}[!htb]
    \centering
    \includegraphics[scale=0.8]{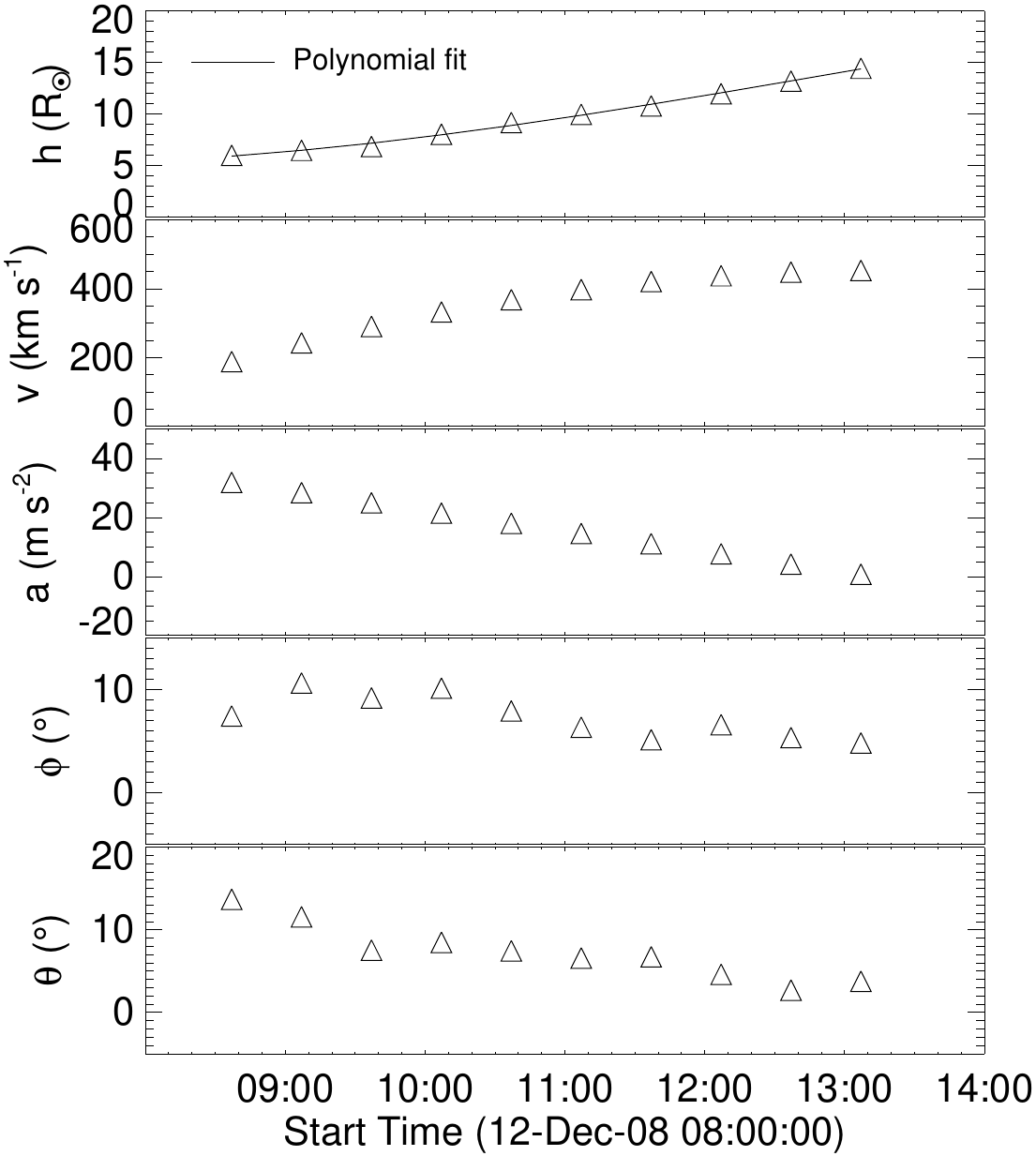}
    \caption[Estimated 3D height, radial velocity, acceleration, longitude and latitude of the 2008 December 12 CME using tie-pointing method]{From top to bottom, the panels show the estimated 3D height, radial velocity, acceleration, longitude and latitude and time on X-axis using the tie-pointing method. Velocity and acceleration are calculated by first and second order differentiation of the fitted polynomial of third-order for 3D height.}
		\label{KinTP12Dec08}
\end{figure}

\subsubsection{Application of 3D reconstruction method in HI FOV}

In this section, we reconstruct the CME using the Geometric Triangulation (GT) method \citep{Liu2010} on HI observations. The geometry and complex treatment of Thomson scattering physics are not incorporated in this GT method. Optimistically, even after neglecting many real effects, we expect that errors arising from implementing the GT method will be minimized, particularly for Earth-directed CME events.

By tracking a CME continuously in the \textit{J}-maps, separately for the \textit{STEREO-A} and \textit{B} images, independent elongation angles of a moving CME feature are estimated. Using precise separation angle between \textit{STEREO-A} and \textit{B}, their heliocentric distances, and elongation angles as inputs in the GT method \citep{Liu2010, Liu2010a}, we obtain the distance and propagation direction of the moving CME feature. The estimated propagation direction of the CME is converted to an angle with respect to the Sun-Earth line in the ecliptic plane. Its positive value implies that the CME was moving to the west from the Sun-Earth line, while a negative value would mean that the CME was propagating to the east. The velocity of the CME is calculated from the estimated distance profile by using numerical differentiation with a three-point Lagrange interpolation method. Figure~\ref{KinGT12Dec08} shows the kinematics of the CME on 2008 December 12. Red vertical lines show the calculated error bars, taking into account the uncertainty in the measurements of the elongation angles. We have considered an uncertainty of 5 pixels in measurements of elongation angles, which is equivalent to uncertainties of 0.02$\arcdeg$, 0.1$\arcdeg$, and 0.35$\arcdeg$ in elongation for the COR2, HI1, and HI2 images, respectively.

\begin{figure}[!htb]
    \centering
		\includegraphics[angle=0,scale=0.8]{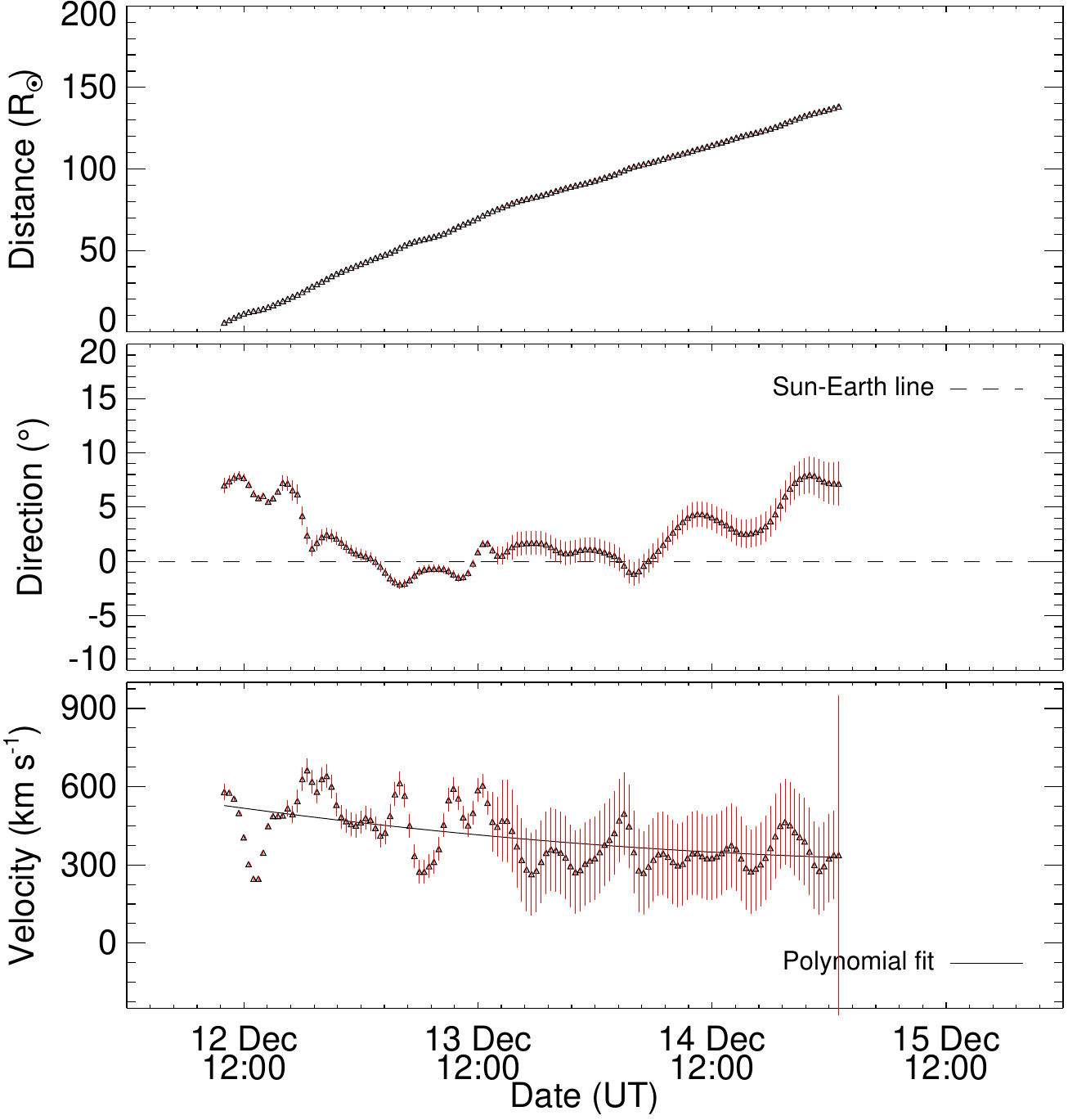}
    \caption[Estimated distance, propagation direction, and velocity of the 2008 December 12 CME using GT method]{From top to bottom, panels show the distance, propagation direction, and velocity respectively of 12 December 2008 CME. In the second panel, the horizontal dashed line represents the Sun-Earth line. The red vertical lines show the error bars. The bottom panel velocity is estimated from adjacent distances using three-point Lagrange interpolation, therefore having large error bars. The solid line in the third panel is the polynomial fit of actual velocity data points.}
		\label{KinGT12Dec08}
  \end{figure}

\subsubsection{Estimation of arrival time and transit speed of 2008 December 12 CME at L1}
\label{arrtim12Dec08}

Using the GT method on HI data, we have estimated the kinematics of the CME of 2008 December 12 out to approximately 138 \textit{R}$_\odot$. We assume that after traversing such a large distance, the speed of the CME will depend solely on aerodynamic drag.
Therefore, we use the Drag-based model (DBM) developed by \citet{Vrsnak2013} to estimate the arrival time of the 2008 December 12 CME. The details of the DBM are described in Section~\ref{DBM} of Chapter~\ref{Chap2:DataMthd}. 

Although there is limited accuracy in reliable estimation of the variables (mass, cone angle of the CME, and solar wind density) on which the drag parameter depends, we used the extreme values of the range of the drag parameter for CMEs in our study. The limitation
on the accurate and reliable estimation of the CME mass using single coronagraph images has been discussed in 
\citet{Vourlidas2000}. \citet{Colaninno2009} discussed the limitations on the reliable estimation of CME mass using multiple viewpoints COR images. Choosing extreme values of the range of the drag parameter, we estimate the maximum possible error in arrival time that can occur due to CMEs having different characteristics.

We used the estimated kinematics as inputs for three different approaches in order to predict the arrival time of CMEs at L1
($\approx$ 1 AU). These three approaches are described as follows.

\begin{enumerate}

\item{\textbf{Using 3D speed estimated in the COR2 FOV}}\\
From the 3D reconstruction of the CME leading edge using the tie-pointing (scc\_measure) method on SECCHI/COR2 data, the 3D velocity was found to be 453 km s$^{-1}$ at 14.4 \textit{R}$_\odot$ at 13:07 UT on 2008 December 12. With a simple assumption
that the speed of the CME remains constant beyond the COR2 FOV, the predicted arrival time at L1 is 01:45 UT on 2008 December 16.

\item{\textbf{Using a polynomial fit of distance estimated from the GT technique with COR2 and HI data}}\\
The radial distance of a moving feature, estimated by implementing the GT method, is fitted to a second-order polynomial.  We obtained the arrival time of the CME at L1 by extrapolation for the distance beyond which the moving feature could not be tracked. The predicted arrival time is 06:23 UT on 2008 December 17. In an earlier study by \citet{Liu2010}, the predicted arrival time for this CME was found to be 16:00-18:00 UT on 2008 December 16. The difference in the two arrival times may arise due to different extrapolation techniques but is most likely due to manual tracking of different features in \textit{J}-maps. Since fitting the estimated distance using a second-order polynomial includes all the points with different velocity phases, this technique may result in more significant uncertainties in the extrapolated arrival times.

\item{\textbf{From DBM with inputs from GT on COR2 and HI}}\\
We used the DBM, combined with the inputs obtained by implementing the GT technique on COR2 and HI observations. For the inputs in the DBM, we used the initial radial distance and CME take-off date and time as obtained from the last data points estimated in the GT method. The initial take-off velocity is taken as the average of the last few velocity points of the fitted polynomial for the estimated velocity in the ecliptic plane. In this particular event of 2008 December 12, we used the DBM with a take-off date
and time of 2008 December 15 at 01:00 UT, a take-off distance of 138.3 \textit{R}$_\odot$, a take-off velocity of 330 km s$^{-1}$, an
ambient solar wind speed of 350 km s$^{-1}$, and a drag parameter of 0.2 $\times$ 10$^{-7}$ km$^{-1}$. Using these values, we obtained its arrival time as 20:25 UT on 2008 December 16, with a transit speed of 331 km s$^{-1}$ at the L1 point. Using the
maximum value of the average range of the drag parameter (2 $\times$ 10$^{-7}$ km$^{-1}$) and keeping other parameters the same in the DBM, the predicted arrival time of the CME was found to be 19:55 UT on 2008 December 16 with a transit speed of 338 km s$^{-1}$.

\end{enumerate}

An in-depth analysis of this CME was carried out earlier by \citet{Byrne2010} using \textit{STEREO} COR and HI observations. They predicted the arrival time of this CME using 3D reconstruction of the CME front and the ENLIL simulation. In their study, they found that the predicted arrival time of the CME agreed well (within $\pm$ 10 hr) with the CME front plasma pileup ahead of a magnetic cloud observed in situ by the \textit{ACE} and \textit{WIND} spacecraft.

In our study, we have tracked the positively inclined bright features in the \textit{J}-maps that are considered to be enhanced density regions of CMEs moving along the ecliptic. The arrival time of these bright features is expected to match with the arrival of the enhanced density features in in situ observations. Therefore, in the present study, we define the actual arrival time of a CME as the time when the first density peak is observed in in situ measurements taken at the L1 point. The in situ observations of 2008 December 12 CME are described in Chapter~\ref{Chap4:Associa}. The predicted arrival time and transit speed of the tracked feature of the CME are marked in Figure~\ref{insitu12Dec08}.

\subsection{2010 February 7 CME}
\label{Rmt7Feb10}

\textit{SOHO}/LASCO recorded this CME on 2010 February 7 at 03:54 UT as a halo CME with a linear speed of 421 km s$^{-1}$.
The speed of the CME was nearly constant in the LASCO FOV. It appeared in the COR1 FOV of \textit{STEREO-A} and \textit{B} at east
and west limbs, respectively, at 02:45 UT. We constructed the \textit{J}-map for this CME as described for the 2008 December 12
CME. At the time of this CME, the distances of \textit{STEREO-A} and \textit{B} from the Sun were 0.96 and 1.01 AU, respectively. In the HI2-A and HI2-B FOVs, the planets Earth and Mars were visible. As a result, a vertical column of saturated pixels in the images appears, and parallel lines in the \textit{J}-maps corresponding to the planet's elongation angles are visible. This event was tracked in the heliosphere, and an independent elongation of a moving feature of the CME from two vantage points was estimated using the \textit{J}-maps. These estimated independent elongation angles and the positional inputs of the twin \textit{STEREO} spacecraft were used in the GT scheme to estimate the kinematics of the CME. The obtained kinematics are shown in Figure~\ref{KinGT7Feb10}, in which the gap in the estimated parameters is due to the existence of a singularity in the GT method. The estimated points in this range have nonphysical variations; these have therefore been removed \citep{Liu2011}.

\begin{figure}[!htb]
    \centering
		\includegraphics[angle=0,scale=0.7]{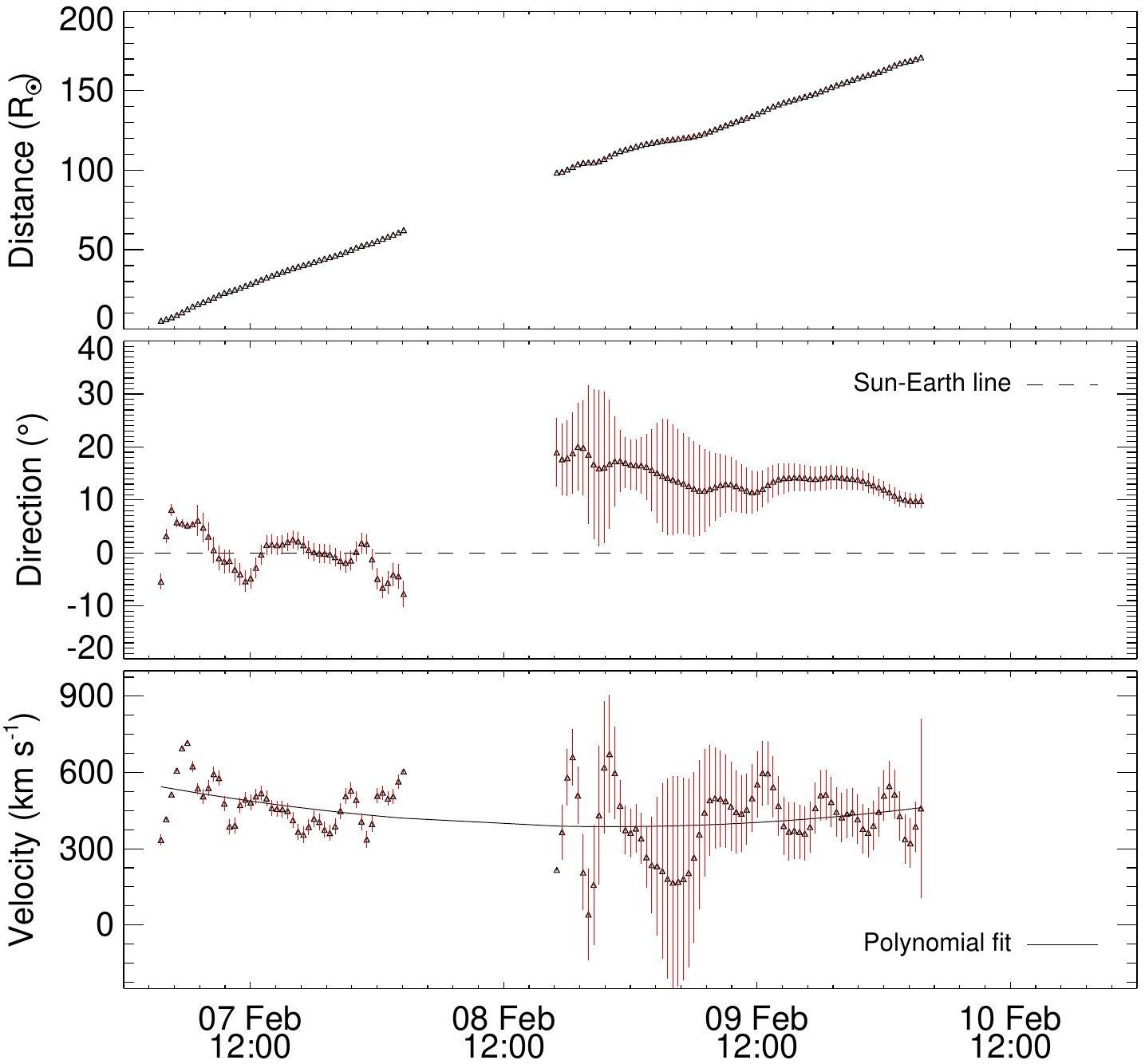}
    \caption[As Figure~\ref{KinGT12Dec08}, for the 2010 February 7 CME]{From top to bottom, panels show the distance, propagation direction, and velocity, respectively, of the 2010 February 7 event. The details are given in the Figure~\ref{KinGT12Dec08} caption.}
		\label{KinGT7Feb10}
  \end{figure}

We found that the singularity in the GT method occurs when the sum of both independent elongation angles measured from two spacecraft and the separation angle between two spacecraft becomes nearly 180$\arcdeg$. In this situation, the line of sight from the two viewpoints of the spacecraft coincides along their entire length. Therefore, a single value of longitude cannot be obtained in this scenario. In Figure~\ref{GT} of Chapter~\ref{Chap2:DataMthd}, we show that the line of sight from the two locations of \textit{STEREO-A} and \textit{B} (AP1 and BP1) will be parallel for point P1 to be triangulated. Therefore, in this case, a singularity will occur in a triangulation scheme. If the separation angle between \textit{STEREO}  spacecraft is larger, the occurrence of a singularity is an issue at smaller elongation angles.

We used the drag model with inputs from the last data point of estimated kinematics to calculate the arrival time of the CME
at the L1 point. Using a take-off distance of 171 \textit{R}$_\odot$, a take-off time of 2010, February 10 at 03:30 UT, a take-off velocity of 455 km s$^{-1}$, a drag parameter of 0.2 $\times$ 10$^{-7}$ km$^{-1}$, and an ambient solar wind speed of 350 km s$^{-1}$, we obtained a predicted arrival time of 21:40 UT on February 10 and a transit speed 442 km s$^{-1}$ at the L1 point. Keeping all the input parameters the same but changing the drag parameter to 2.0 $\times$ 10$^{-7}$ km$^{-1}$, the predicted arrival time is 22:50 UT on February 10, and the transit speed is 393 km s$^{-1}$. Using a second-order polynomial fit for the estimated distance points and extrapolating it, we obtained an arrival time of 00:50 UT on February 11 at the L1 point. We also carried out the 3D reconstruction of this CME using SECCHI/COR2 data and estimated the 3D heliographic coordinates of a selected feature along the leading edge. The 3D velocity was estimated as 480 km s$^{-1}$ at 13.5 \textit{R}$_\odot$ at 06:39 UT on 2010 February 7. Assuming that the 3D velocity of the CME is constant beyond the COR2 FOV, the predicted arrival time is 14:55 UT on 2010 February 10 at L1.  The in situ observations of the CME of 2010 February 7 are given in Figure~\ref{insitu7Feb10} of Chapter~\ref{Chap4:Associa}.

\subsection{2010 February 12 CME}
\label{Rmt12Feb10}

This CME was observed in the NE quadrant in SECCHI/COR1-A observations and in the NW quadrant in SECCHI/COR1-B images at 11:50 UT on February 12. It was also observed by \textit{SOHO}/LASCO coronagraph at 13:42 UT as a halo CME with a linear speed of 509 km s$^{-1}$. The CME decelerated in the LASCO FOV, and its speed at final height ($\approx$ 25 \textit{R}$_\odot$) was measured to be 358 km s$^{-1}$. We constructed the \textit{J}-maps using COR2, HI1, and HI2 observations for the CME. In the SECCHI/HI2-A \& B FOV, planets Earth and Mars were visible as parallel lines corresponding to their elongation angles in the \textit{J}-maps. The independent elongation angle of a moving feature from two viewpoints was estimated by tracking the bright, positively inclined feature in the \textit{J}-maps corresponding to this CME. Implementing the GT technique, the kinematics of the CME were estimated and are shown in Figure~\ref{KinGT12Feb10}. The data gap in this plot is due to the occurrence of a singularity in the GT method; details are explained in Section~\ref{Rmt7Feb10}.

\begin{figure}[!htb]
    \centering
		\includegraphics[angle=0,scale=0.7]{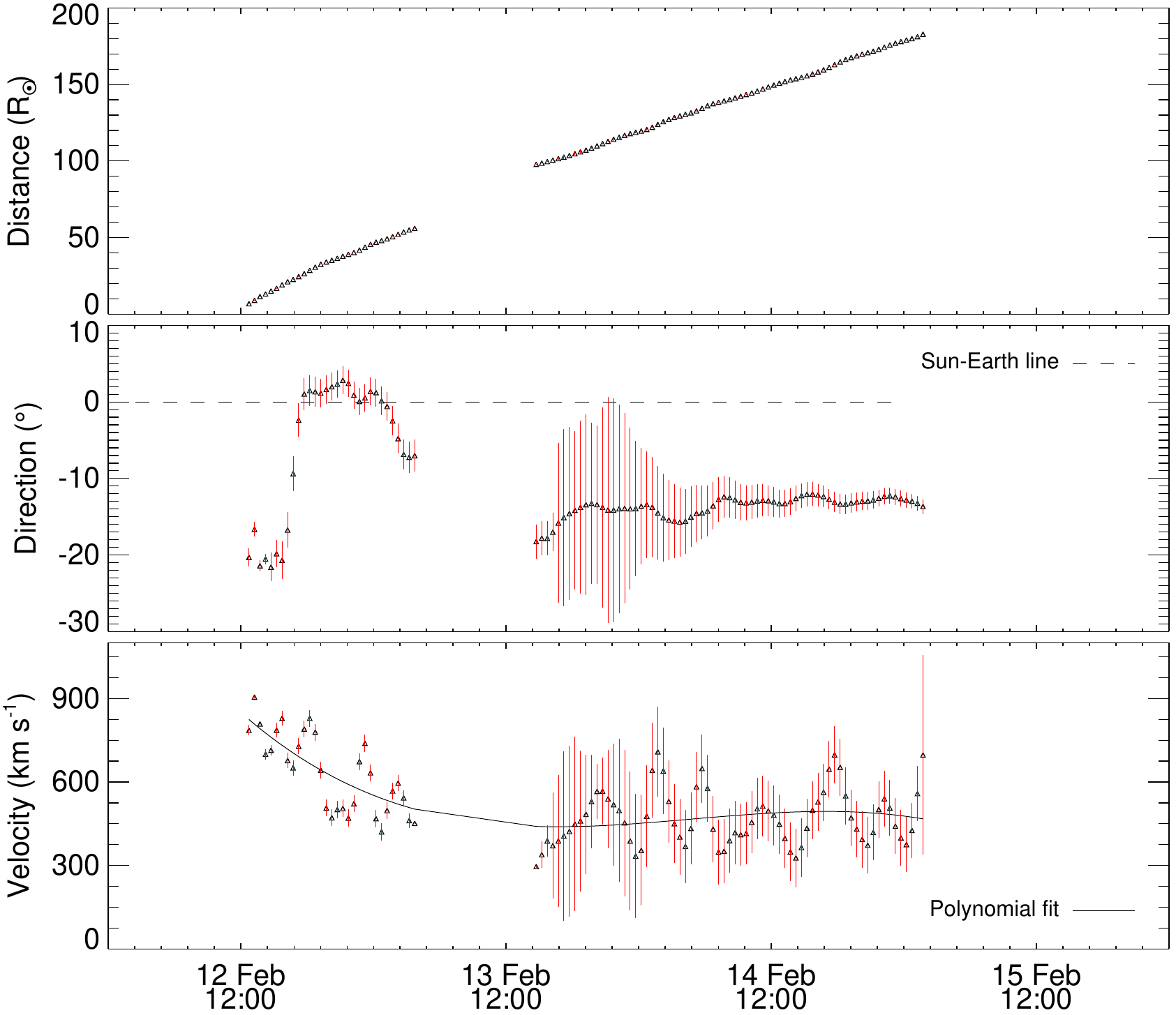}
    \caption[As Figure~\ref{KinGT12Dec08}, for the 2010 February 12 CME]{From top to bottom, panels show the distance, propagation direction, and velocity, respectively, of the 2010 February 12 CME. The details are given in the  Figure~\ref{KinGT12Dec08} caption.}
		\label{KinGT12Feb10}
  \end{figure}

To predict the arrival time of this CME, we combined the estimated kinematics with the DBM. Input parameters used in
the DBM were calculated in the same way as explained earlier in Section~\ref{arrtim12Dec08}. Using the inputs, e.g., an initial 
take-off distance of 183 \textit{R}$_\odot$, a take-off time of 2010 February 15 at 01:44 UT, a take-off velocity of 450 km s$^{-1}$, an ambient solar wind speed of 350 km s$^{-1}$, and a drag parameter of 0.2 $\times$ 10$^{-7}$ km$^{-1}$ in the DBM, the predicted arrival time and transit speed of the CME at L1 are 14:35 UT on 2010 February 15 and 442 km s$^{-1}$, respectively. If the drag parameter is taken as 2.0 $\times$ 10$^{-7}$ km$^{-1}$ in the DBM, keeping the rest of the input parameters the same, the
predicted arrival time of the CME is found to be 15:20 UT on 2010 February 15, and the transit speed is found to be 401 km s$^{-1}$. 
Using the second-order polynomial fit for distance, we obtained the predicted arrival time of CME at L1 as 16:07 UT on February
15. Furthermore, from 3D reconstruction in the COR2 FOV, the 3D velocity of the leading edge of CME was estimated to be
867 km s$^{-1}$ at a distance of 14.8 \textit{R}$_\odot$ at 14:54 UT on February 12. Beyond this distance, considering that the CME speed was constant up to L1, the predicted arrival time at L1 is estimated as 11:02 UT on February 14.

\subsection{2010 March 14 CME}
\label{Rmt14Mar10}

This CME was observed on 2010 March 14 by SECCHI/ COR1-A in the NE quadrant and by SECCHI/COR1-B in
the NW quadrant of the coronagraphic images. \textit{SOHO}/LASCO coronagraph recorded this CME as a partial halo (angular width $\approx$ 260$\arcdeg$) at 00:30 UT on March 14 with a linear speed of 351 km s$^{-1}$. In the LASCO FOV, a nearly constant speed of the CME was observed. In the \textit{J}-maps constructed from COR2-A and HI-A images, this CME could be tracked nearly up to 35$\arcdeg$ while in the \textit{J}-maps constructed from COR2-B and HI-B images, tracking was possible up to 50$\arcdeg$. In the SECCHI/HI2-A FOV, the planets Earth and Mars were seen, while in the HI2-B FOV, only the Earth could be seen. We tracked the CME in the heliosphere and estimated its independent elongation from two \textit{STEREO} locations. These elongation angles and the separation angle between the twin \textit{STEREO} spacecraft were used as inputs in the GT method to obtain the propagation direction and the distance of the CME. Velocity was calculated from the adjacent distances using three-point Lagrange interpolation. The kinematics of this CME is shown in Figure~\ref{KinGT14Mar10}. Singularity (described in Section~\ref{Rmt7Feb10}) occurred in this case also, and therefore, the estimated kinematics in the time range of the singularity is not shown.

\begin{figure}[!htb]
    \centering
		\includegraphics[angle=0,scale=0.7]{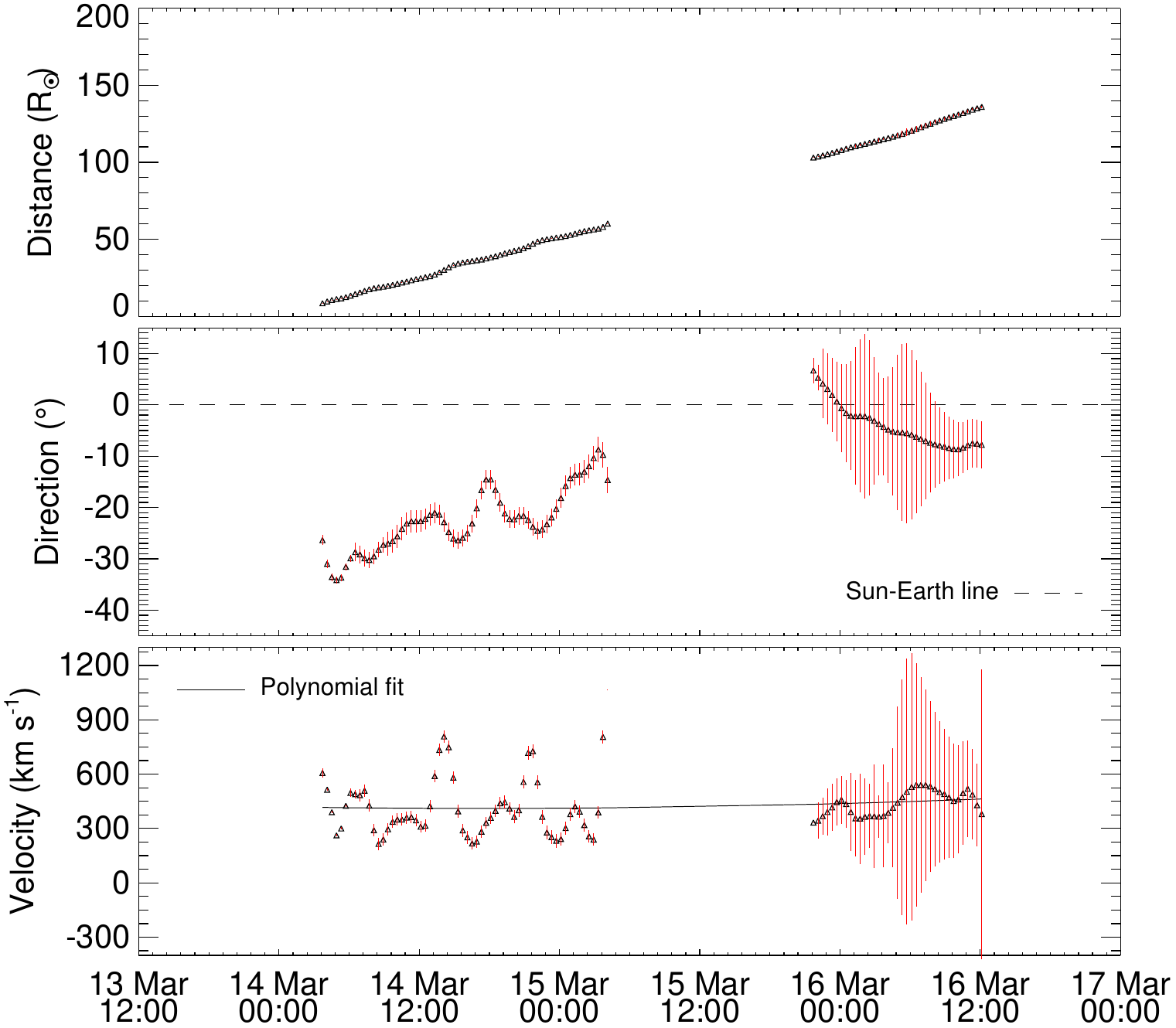}
    \caption[As Figure~\ref{KinGT12Dec08}, for the 2010 March 14 CME]{From top to bottom, panels show the distance, propagation direction, and velocity, respectively, of the 2010 March 14  CME. The details are given in the Figure~\ref{KinGT12Dec08} caption.}
		\label{KinGT14Mar10}
  \end{figure}

The estimated kinematics are used as inputs in the DBM to predict the arrival time of the CME at the L1 point. We used the DBM with a take-off distance of 135.9 \textit{R}$_\odot$, a take-off velocity of 460 km s$^{-1}$, a take-off date and time of 2010 March 16 at
12:07 UT, an ambient solar wind speed of 350 km s$^{-1}$, and a drag parameter of 0.2 $\times$ 10$^{-7}$ km$^{-1}$ as inputs and obtained its predicted arrival time at 21:10 UT on March 17 with a transit velocity of 437 km s$^{-1}$ at L1. Keeping all these input parameters the same and using the maximum value of the statistical range of the drag parameter (2 $\times$ 10$^{-7}$ km$^{-1}$) in the DBM, we obtained a predicted arrival time of 01:00 UT on March 18 and a transit velocity of 378 km s$^{-1}$ at L1. The predicted arrival time of the CME at L1, using the second-order polynomial fit for distance is 16:21 UT on March 17. We also implemented the tie-pointing method of 3D reconstruction on the leading edge of the CME in the COR2 FOV and estimated the 3D kinematics of a CME feature. Assuming that the 3D speed (335 km s$^{-1}$) estimated at 3D height (11 \textit{R}$_\odot$) at 03:54 UT on March 14 is constant beyond the COR2 FOV, the predicted arrival time of the CME is 00:17 UT on March 19.

\subsection{2010 April 3 CME}
\label{Rmt3Apr10}

This geo-effective (D$_{st}$ = -72 nT) CME was detected by \textit{SOHO}/LASCO on 2010 April 3 at 10:33 UT as a halo. It had a projected plane-of-sky linear speed of around 668 km s$^{-1}$ as measured from the LASCO images. It was observed at 09:05 UT by SECCHI/COR1-A in the SE quadrant and by SECCHI/COR1 B in the SW quadrant. The source region of the CME was NOAA AR 11059. The CME was accompanied by a filament disappearance, a post eruption arcade, coronal dimming, an EIT wave, and a B7.4 long-duration flare peaking at 09:54 UT \citep{Liu2011}.

We constructed the \textit{J}-maps to track the CME in the heliosphere. Due to the appearance of the Milky Way galaxy in the SECCHI/HI2-B images, the signal of this CME is not well pronounced. Therefore, it could not be tracked beyond 27$\arcdeg$ elongation
in the \textit{J}-map constructed from \textit{STEREO-B} images. Planets Earth and Mars are visible in the HI2-A images at 58.1$\arcdeg$ and 50.6$\arcdeg$, respectively. The Earth is visible in the HI2-B images at 54.3$\arcdeg$ elongation. Independent elongation angles are extracted from the leading edge of the track of the CME in the \textit{J}-maps. Then, the GT method is implemented to estimate the distance and propagation direction of the CME. The estimated kinematics are displayed in Figure~\ref{KinGT3Apr10}; a discontinuity in the plot for approximately 6 hr in the estimated kinematics is due to the occurrence of a singularity in the implemented triangulation method.

\begin{figure}[!htb]
    \centering
		\includegraphics[angle=0,scale=0.7]{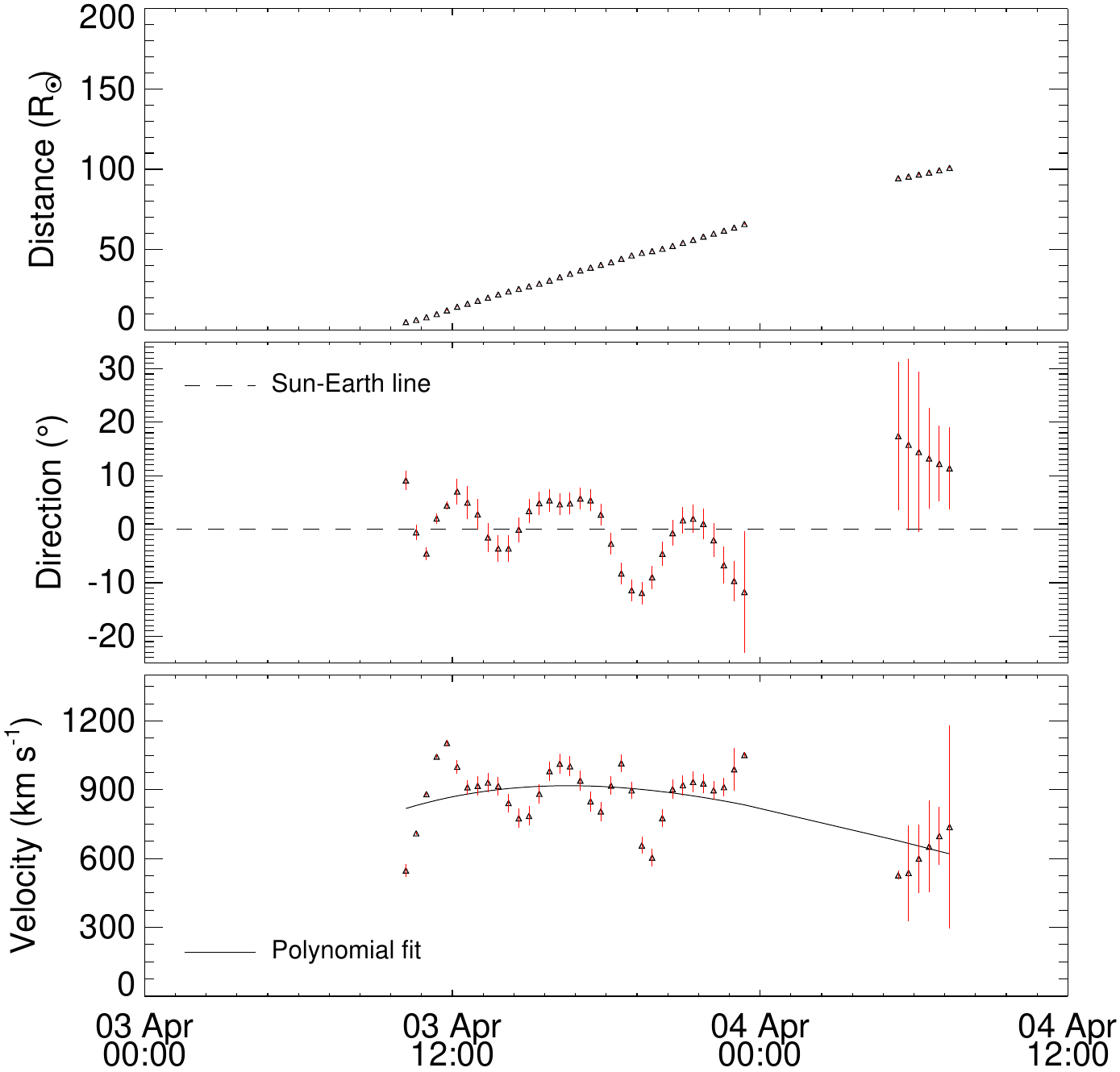}
    \caption[As Figure~\ref{KinGT12Dec08}, for the 2010 April 3 CME]{From top to bottom, panels show the distance, propagation direction, and velocity, respectively, of the 2010 April 3   CME. The details are given in the Figure~\ref{KinGT12Dec08} caption.}
		\label{KinGT3Apr10}
  \end{figure}

We used the distance and velocity of the CME at the last estimated data point as inputs in the DBM to predict its arrival
time at L1. At the time of this CME, the Earth was blown over by a high-speed solar wind stream emanating from the northern
polar coronal hole. Due to the presence of this coronal hole and the generally large spatial scale of the CME, its kinematics will
be partly governed by the high-speed stream as shown by \citet{Vrsnak2013}. They also showed that an ambient solar wind speed
of 550 km s$^{-1}$ and a low value of the drag parameter should be considered as inputs in the DBM for such CMEs traveling in a fast solar wind medium. We used a CME take-off speed of 640 km s$^{-1}$, a take-off date and time of April 4 at 07:23 UT, a take-off distance of 100.8 \textit{R}$_\odot$, and a drag parameter of 0.2 $\times$ 10$^{-7}$ km$^{-1}$ in the DBM as inputs. We obtained the predicted arrival time of the CME at 17:35 UT on April 5 and a transit speed of 624 km s$^{-1}$ at the L1 point. The predicted arrival time by extrapolating the second-order polynomial fit of distance is 09:00 UT on April 5. This CME has been studied in detail by \citet{Liu2011}. They predicted the arrival by extrapolating the estimated distance at 12:00 UT on April 5, which is approximately the same as we predicted using a polynomial fit, with an error of 3 hr. We also performed 3D reconstruction of the CME leading edge using SECCHI/COR2 data and obtained its 3D kinematics. Assuming that this 3D estimated velocity (816 km s$^{-1}$) at 12:24 UT on April 3 is constant beyond the COR2 FOV, the predicted arrival time of
the CME at L1 is 11:25 UT on April 5.

\subsection{2010 April 8 CME}
\label{Rmt8Apr10}

In the \textit{SOHO}/LASCO observations, this CME was detected at 04:54 UT on April 8 as a partial halo. The plane-of-sky speed of
this CME was 264 km s$^{-1}$; the CME decelerated in the LASCO FOV. SECCHI/COR1-A observed this CME in the NE quadrant
, and COR1-B observed it in the NW quadrant at 03:25 UT on April 8. The CME was accompanied by a B3.7 flare in NOAA
AR 11060.

A time-elongation plot (\textit{J}-map) in the ecliptic plane was constructed for this CME. The CME was tracked in the heliosphere
up to 54$\arcdeg$ elongation angle in \textit{J}-maps constructed from SECCHI-A images. In \textit{J}-maps constructed from 
SECCHI-B images, could be tracked up to 44$\arcdeg$ only. For comparison, planets Earth and Mars are seen at 58$\arcdeg$ and 48.4$\arcdeg$ elongation, respectively, in the HI2-A FOV on April 8. In the HI2-B FOV at this time, the Earth is seen at 54.5$\arcdeg$ elongation. The GT method is implemented to estimate the distance and propagation direction of the CME in the heliosphere. The obtained kinematics are shown in Figure~\ref{KinGT8Apr10}, and the gap in the estimated kinematics for nearly 12 hr is due to the occurrence of a singularity in the GT method.

\begin{figure}[!htb]
    \centering
		\includegraphics[angle=0,scale=0.7]{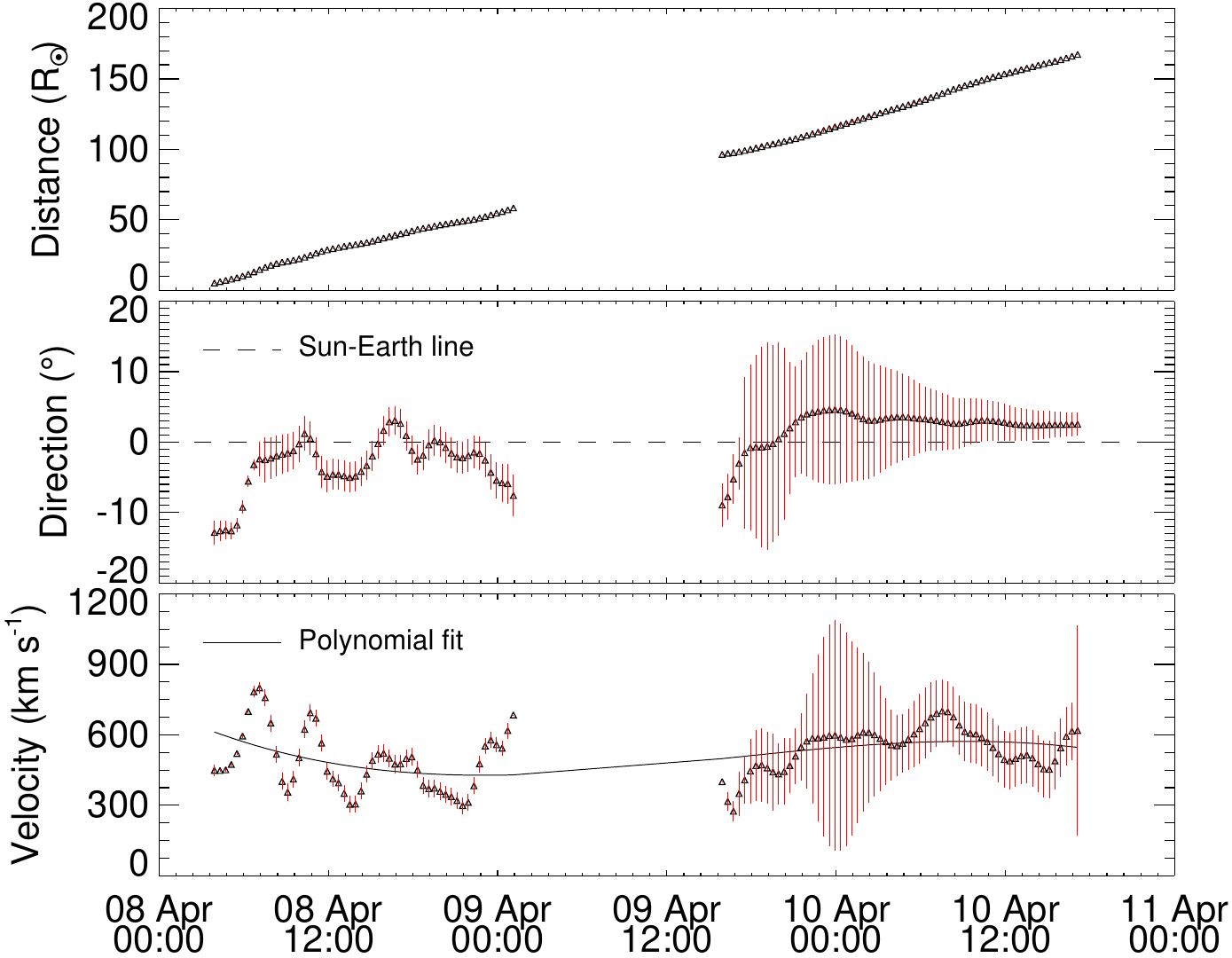}
    \caption[As Figure~\ref{KinGT12Dec08}, for the 2010 April 8 CME]{From top to bottom, panels show the distance, propagation direction, and velocity, respectively, of the 2010 April 8   CME. The details are given in the Figure~\ref{KinGT12Dec08} caption.}
		\label{KinGT8Apr10}
  \end{figure}

We used the drag model with the inputs of a take-off velocity of 550 km s$^{-1}$, a take-off distance of 167.3 \textit{R}$_\odot$, a take-off date and time of April 10 at 17:07 UT, an ambient solar wind speed of 350 km s$^{-1}$, and a drag parameter of 0.2 $\times$ 10$^{-7}$ km$^{-1}$. Using these inputs in the DBM, the predicted arrival time of the CME is 09:45 UT on April 11, and the transit velocity is 511 km s$^{-1}$ at L1. Keeping all the input parameters the same and taking the drag parameter value as 2 $\times$ 10$^{-7}$ km$^{-1}$ in the DBM, the predicted arrival time of the CME is 12:55 UT on April 11 with a transit speed of 402 km s$^{-1}$. By extrapolating the second-order polynomial fit of distance, the predicted arrival time of the CME at L1 is 06:32 UT on April 11. Using the tie-pointing technique on SECCHI/COR2 data, the 3D velocity of the CME leading edge at the height of 12 \textit{R}$_\odot$ was found to be 478 km s$^{-1}$ at 07:24 UT on April 8. Assuming the CME 3D speed is constant beyond the COR2 FOV, its predicted arrival time at L1 is 16:35 UT on April 11.

\subsection{2010 October 10 CME}
\label{Rmt10Oct10}

This CME was accompanied by a filament eruption in the SE quadrant of the solar disk. \textit{SOHO}/LASCO coronagraph observed this event at 22:12 UT on October 10 as a slow (projected linear speed $\approx$ 262 km s$^{-1}$) and partial halo (angular
width $\approx$ 150$\arcdeg$) CME. The projected speed calculated from SECCHI/COR1-A images was 297 km s$^{-1}$. The projected
speed calculated from SECCHI/COR1-B images was estimated as 328 km s$^{-1}$. The CME first appeared in the SECCHI/COR1-A FOV at 
19:25 UT and then in the SECCHI/COR1-B FOV at 20:05 UT.

In the \textit{J}-map, tracking of a feature was feasible at elongations up to 35$\arcdeg$ and 30$\arcdeg$ for \textit{STEREO-A} and \textit{STEREO-B}, respectively. In the \textit{J}-map constructed from images taken by the \textit{STEREO-A}, there are two nearly horizontal lines and one slanted line in the HI2 FOV starting at elongation angles of 35.2$\arcdeg$, 49.4$\arcdeg$,
and 69.6$\arcdeg$, respectively. These lines are due to planets Venus, Earth, and Jupiter, respectively, in the HI2-A FOV. In the \textit{J}-map constructed from \textit{STEREO-B} images, two horizontal lines that start at elongation angles of 39.8$\arcdeg$ and 47.9$\arcdeg$, respectively, are due to the appearance of planets Venus and Earth in the HI2-B FOV. The kinematics obtained for this CME is shown in Figure~\ref{KinGT10Oct10}, where the discontinuity in the plot occurs due to the existence of a singularity in the method.

\begin{figure}[!htb]
    \centering
		\includegraphics[angle=0,scale=0.7]{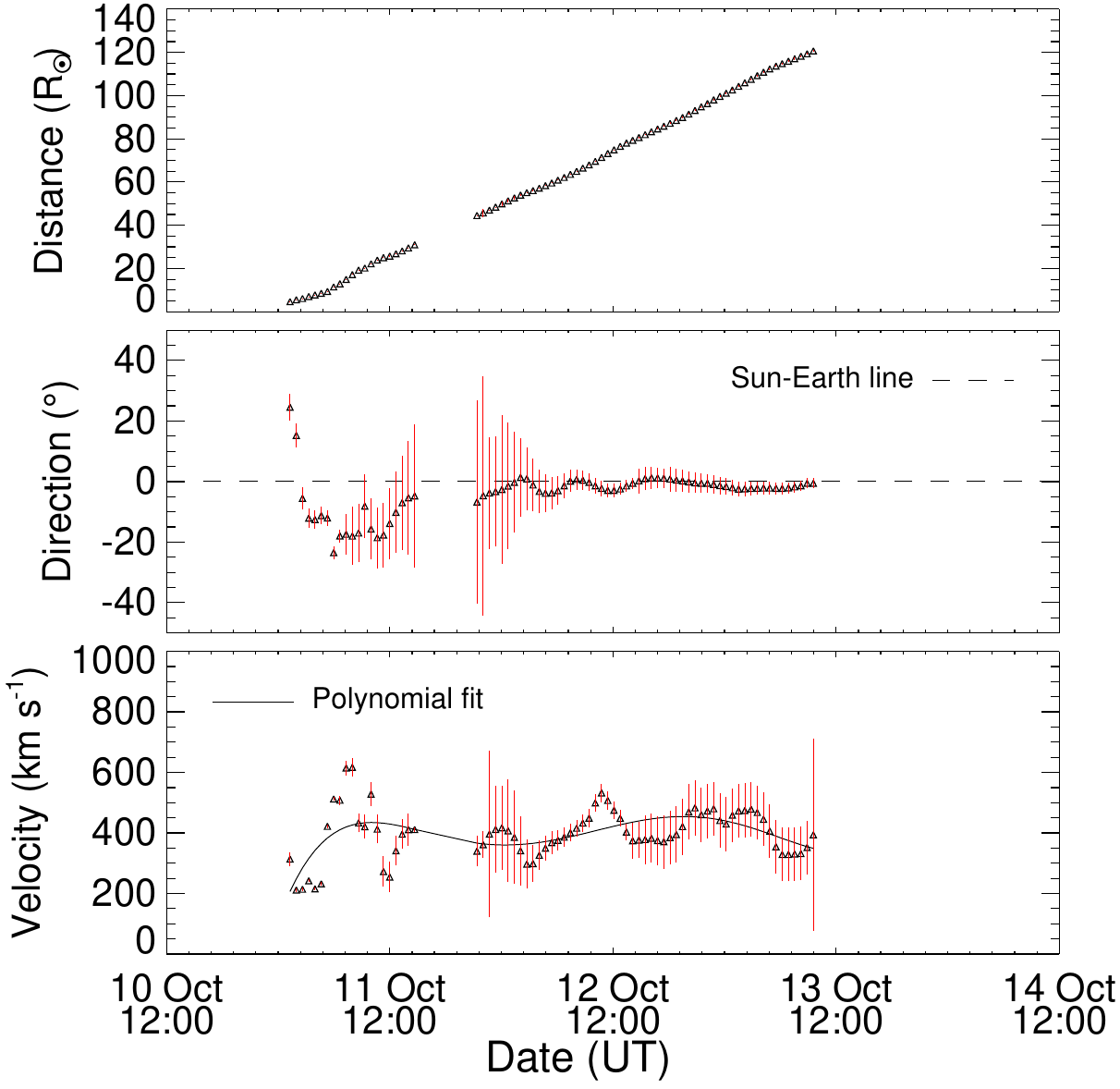}
    \caption[As Figure~\ref{KinGT12Dec08}, for the 2010 October 10 CME]{From top to bottom, panels show the distance, propagation direction, and velocity, respectively, of the 2010 October 10  CME. The details are given in the Figure~\ref{KinGT12Dec08} caption.}
		\label{KinGT10Oct10}
  \end{figure}

We used the DBM developed by \citet{Vrsnak2013} to estimate the arrival time of the CME. For this event, we used the DBM with a 
take-off distance of 120.65 \textit{R}$_\odot$, a take-off date and time of 2010 October 13 at 09:33 UT, a take-off velocity of 354 
km s$^{-1}$, an ambient solar wind speed of 350 km s$^{-1}$, and a drag parameter of 0.2 $\times$ 10$^{-7}$ km$^{-1}$ to 2 $\times$  10$^{-7}$ km$^{-1}$. We obtained its arrival time as 11:40-11:45 UT on October 15 and its transit velocity as 354-353 km s$^{-1}$  at L1. By extrapolating the second-order polynomial fit for distance, the predicted arrival time of the CME at L1 is October 14 at 22:53 UT. From the 3D reconstruction of the CME leading edge using the tie-pointing method on SECCHI/COR2 data, the estimated 3D velocity was found to be 565.8 km s$^{-1}$ at 14.5 \textit{R}$_\odot$ at 06:50 UT on 2010 October 11. Assuming that the speed of CME is constant beyond the COR2 FOV, the predicted arrival time is 02:33 UT on 2010 October 14 at L1.

\subsection{2010 October 26 CME}
\label{Rmt26Oct10}

This CME was observed in \textit{SOHO}/LASCO images around 01:36 UT on October 26 with an angular width of 83$\arcdeg$ and had a projected linear speed of 214 km s$^{-1}$ at a position angle of 210$\arcdeg$. It was also observed by both the \textit{STEREO} spacecraft in COR1 images on October 26.

The independent elongation angle of a tracked feature from two different vantage points is extracted using \textit{J}-maps. In both
\textit{J}-maps, the leading edge of the bright feature has a positive slope, revealing that the CME propagation could be tracked up to
28$\arcdeg$ without ambiguity. For comparison, on October 26 in the HI2-A FOV, planets Venus, Earth, and Jupiter are observed
at elongation angles of 38.3$\arcdeg$, 48.9$\arcdeg$, and 55.8$\arcdeg$, respectively. In the HI2-B FOV, planets Venus and Earth are observed at elongation angles of 36.9$\arcdeg$ and 46.7$\arcdeg$, respectively. Figure~\ref{KinGT26Oct10} shows the kinematics of the CME of 2010 October 26, which were obtained by implementing the GT technique \citep{Liu2010} with the independent elongations estimated from two different viewpoints as inputs. There is a gap in the estimated parameters (Figure~\ref{KinGT26Oct10}) due to the existence of a singularity in the triangulation scheme.

\begin{figure}[!htb]
    \centering
		\includegraphics[angle=0,scale=0.7]{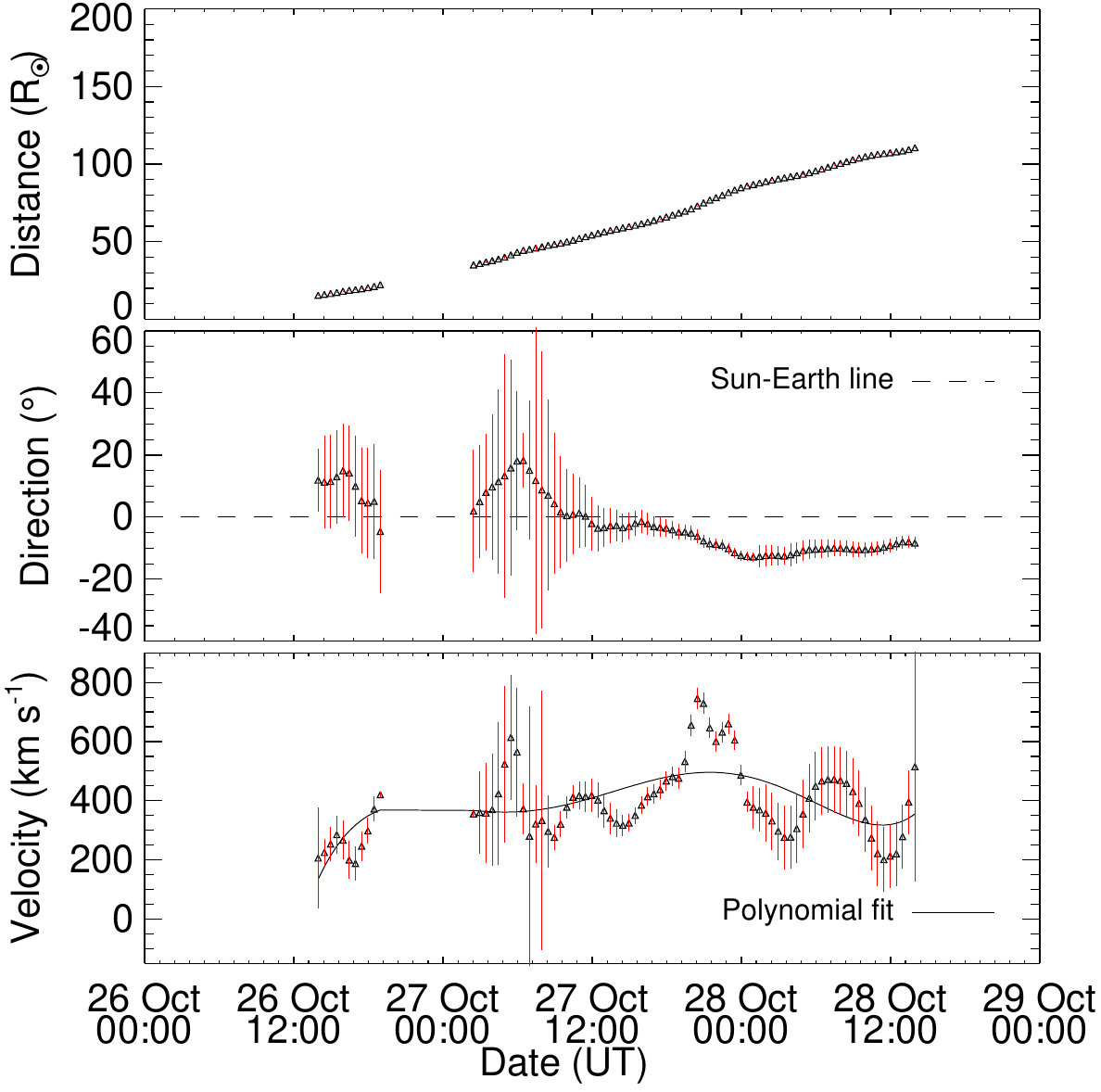}
    \caption[As Figure~\ref{KinGT12Dec08}, for the 2010 October 26 CME]{From top to bottom, panels show the distance, propagation direction, and velocity, respectively, of the 2010 October 26 CME. The details are given in the Figure~\ref{KinGT12Dec08} caption.}
		\label{KinGT26Oct10}
  \end{figure}

We applied the tie-pointing method for the 3D reconstruction of a feature along with the identified bright blob behind the leading edge in both sets of calibrated and background-subtracted images obtained by SECCHI/COR2. Our goal was to estimate its 3D kinematics in the COR2 FOV. Estimated latitude ($\approx$ -25$^{\circ}$) and longitude ($\approx$ 15$^{\circ}$) show that the CME is Earth-directed. The speed of the CME using 3D reconstruction at the outer edge of the COR2 FOV is about 600 km s$^{-1}$ with an acceleration of 30 m s$^{-2}$. Assuming that the speed of the CME is constant beyond this distance, the predicted arrival time of the CME at L1 is 07:45 UT on 2010 October 29. We used the DBM with a take-off date and time of 2010 October 28 at 13:57 UT, a take-off distance of 110.5 \textit{R}$_\odot$ a take-off velocity of 341 km s$^{-1}$, an ambient solar wind speed of 350 km s$^{-1}$, and a drag parameter value of 0.2 $\times$ 10$^{-7}$ km$^{-1}$. The resulting CME arrival time is 23:45 UT on 2010 October 30, and the transit speed is 341 km s$^{-1}$ at L1. Keeping all the input parameters the same but using a different drag parameter value, 2.0 $\times$ 10$^{-7}$ km$^{-1}$, the CME arrival time is 23:35 UT on 2010 October 30, and the transit speed is 343 km s$^{-1}$ at L1. The predicted arrival time of the CME at L1, obtained by extrapolating a second-order polynomial fit for distance, is
08:32 UT on October 30.

\section{Arrival Time Estimation of the CMEs Using GT Method: Results and Discussion}
\label{ResDis31}

We estimated the kinematics for eight selected CMEs by implementing the GT method \citep{Liu2010} that uses elongation (derived from \textit{J}-maps) as input. We predicted the arrival time (T$_{arr}$) of tracked density enhanced feature of CMEs at the L1 point using three different approaches, described in Section~\ref{arrtim12Dec08}. The arrival time predicted using the two different approaches (DBM and polynomial fit) that use inputs of derived parameters from GT with \textit{J}-maps were compared with the actual arrival time of the enhanced density feature at L1. For each CME studied here, the actual arrival time and errors in the predicted arrival time using these two approaches are shown in Table~\ref{ErrDBMpoly}. The predicted arrival time of a CME using DBM is shown in the third column of Table~\ref{ErrDBMpoly}, corresponding to the two extreme values of the range of the drag parameter. Predicted transit speed (v$_{1}$) of the CME at L1 is also compared with the actual measured speed (fifth column) at the time of the arrival of the enhanced density feature. Negative (positive) values of errors in arrival time indicate that the predicted arrival is earlier (later) than the actual arrival time. Negative (positive) values of error in the transit speed of the CME at L1 indicate that the predicted transit velocity of the CME at L1 is lesser (greater) than the transit velocity measured in situ.

To examine the efficacy of the third approach, one must be certain that the arrival time of the tracked feature at
L1 is correctly marked. It is generally difficult to mark the actual (reference) arrival time of a remotely observed (in COR2)
feature at L1 by analyzing the in situ data. This difficulty arises due to uncertainty in the one-to-one identification of remotely observed structures with in situ observed structures. For all the CMEs studied in our case, except 2010 October 26, we could estimate the 3D speed of a feature along its leading edge in the COR2 FOV by implementing the tie-pointing method. Therefore, we take the arrival of the leading edge of the CME in in situ data at L1 as a reference for the CME arrival time (Table~\ref{SelectCMEs}). The details of in situ observations of selected CMEs are presented in Chapter~\ref{Chap4:Associa}. The error in the predicted arrival time of the CME using 3D speed (estimated in the COR2 FOV) is shown in Table~\ref{ErrCOR2}. This table shows the 3D speed in the COR2 FOV and the measured speed of the CME leading edge at L1. The CME of 2010 October 10, is not included here because the CME boundaries could not be identified in in situ data; it seems probable that only the CME flank was encountered by the spacecraft. It would have been possible to confirm the above by discussing multi-point observations of this CME if the \textit{STEREO} spacecraft had a smaller separation angle at this time (similar to the study carried out by \citet{Kilpua2011}.

From Tables~\ref{ErrDBMpoly} and ~\ref{ErrCOR2}, it is clear that, in general, a more accurate prediction of CME arrival time is possible using DBM combined with kinematics estimated from the GT method. It is also evident from Table~\ref{ErrCOR2} that, for the 2010 April 3 and April 8 CMEs, the implementation of the tie-pointing technique provides a more accurate arrival time prediction. In
these cases, CME transit speeds at L1 are approximately equal to the measured CME speeds in the COR2 FOV. This result shows that the speeds of the CMEs did not change significantly during their propagation in the heliosphere. Therefore, in these particular cases, the observed CME speeds in the COR2 FOV are sufficient to predict their arrival times near 1AU with reasonably good accuracy. Our analysis of the April 3 CME shows that CME speed is partly governed by the high-speed stream from a coronal hole located in a geo-effective location on the Sun. The estimated speed ($\approx$ 816 km s$^{-1}$) of the April 3 CME in the COR2 FOV and its in situ speed ($\approx$ 800 km s$^{-1}$) highlight that it experienced a weak drag force throughout its journey up to 1 AU. These findings are in good agreement with the results of \citet{Temmer2011}.

\begin{sidewaystable}
  \centering
 \begin{tabular}{|p{2.7cm}| p{2.8cm}| p{4.0cm} |p{2.0cm} |p{2.0cm} |p{4.0cm}|}
    \hline
 &  & \multicolumn{2}{ c| }{Error in predicted T$_{arr}$ at L1 (hr) } &  &  \\ \cline{3-4} 
CME dates & Actual T$_{arr}$ (Peak density time) (UT) & Kinematics + Drag Based Model [$\gamma$ = 0.2 - 2.0 (10$^{-7}$ km$^{-1}$)]  & Distance + Polynomial fit & Actual v$_{1}$ at L1 (km s$^{-1}$) & Error in predicted v$_{1}$ at L1 (km s$^{-1}$) [$\gamma$ = 0.2 - 2.0 (10$^{-7}$ km$^{-1}$)]  \\ \hline 

12 Dec 2008 & 16 Dec 23:50 & -3.4 to -3.9 & +6.5  & 356  & -25 to -18  \\ \hline
07 Feb 2010 & 11 Feb 02:05 & -4.3 to -3.2 & -1.2  & 370  & +72 to +23  \\ \hline
12 Feb 2010 & 15 Feb 23:15 & -8.7 to -7.9 & -7.1  & 320  & +122 to +81 \\ \hline
14 Mar 2010 & 17 Mar 21:45 & -0.6 to +3.2 & -5.4  & 453  & -16 to -75  \\ \hline
03 Apr 2010 & 05 Apr 12:00 & +5.5       & -3.0    & 720  & -96         \\ \hline
08 Apr 2010 & 11 Apr 14:10 & -4.4 to -1.2 & -7.6  & 426  & +85 to -24  \\ \hline
10 Oct 2010 & 15 Oct 06:05 & +5.5 to +5.6 & -7.2  & 300  & +54 to +53  \\ \hline
26 Oct 2010 & 31 Oct 03:30 & -3.7 to -4.0 & -18.9 & 365  & -24 to -22  \\ \hline 

 \end{tabular}
\caption[Errors in the predicted arrival time of CMEs using two different approaches]{The table shows the errors in the predicted arrival time using two different approaches. The actual arrival time and the transit velocity of the tracked feature at the L1 point are shown in Columns 2 and 5, respectively. Errors in the predicted arrival time using two approaches are shown in Columns 3 and 4 and errors in the predicted transit velocity in Column 6.}
\label{ErrDBMpoly}
\end{sidewaystable}

\begin{table}
  \centering

 \begin{tabular}{|p{2.5cm}| p{2.5cm}| p{2.0cm}|p{2.2cm}|p{1.8cm}|}
    \hline
CME dates &  Actual arrival time (UT) of CME leading edge at L1 &  Error in predicted arrival time & Measured velocity of CME leading edge at L1 & Velocity (km s$^{-1}$) in COR2 FOV  \\ \hline

12 Dec 2008 & 17 Dec 04:39 & -26.9 & 365 & 453 \\ \hline
07 Feb 2010 & 11 Feb 12:47 & -21.8 & 360 & 480 \\ \hline
12 Feb 2010 & 16 Feb 04:32 & -41.5 & 310 & 867  \\ \hline
14 Mar 2010 & 17 Mar 21:19 & +27 & 450 & 335  \\ \hline
03 Apr 2010 & 05 Apr 13:43 & -2.3 & 800 & 816 \\ \hline
08 Apr 2010 & 12 Apr 02:10 & -9.5 & 410 & 478  \\ \hline
26 Oct 2010 & 31 Oct 06:30 & -46.7 & 365 & 600 \\ \hline 

 \end{tabular}

\caption[Errors in the predicted arrival time estimated using the 3D velocity in the COR2 FOV]{The table shows errors in the predicted arrival time estimated using the 3D velocity in COR2 FOV. The second column shows the actual arrival time at L1, the third column shows errors in the predicted arrival time, the fourth column shows the CME leading edge velocity measured at the L1 point by in situ spacecraft, and the fifth column shows the 3D velocity in the COR2 FOV.}
\label{ErrCOR2}
\end{table}

Predicting arrival time by extrapolating the fitted second-order polynomial for estimated distance is better than the
prediction made using only the 3D speed estimated in the COR2 FOV. The decelerating trend of the 2008 December 12
CME in the inner heliosphere is in good agreement with the results of \citet{Liu2010}. Using extrapolation, the error in the predicted arrival time is also less if a CME is tracked up to large distances in the heliosphere (using HI), as in the case of
the 2010 February 7 CME where the \textit{J}-map allowed tracking of CME features at elongations up to nearly 50$\arcdeg$ 
($\approx$ 170 \textit{R}$_\odot$). The high-speed ($\approx$ 867 km s$^{-1}$) CME of 2010 February 12 shows a significant, continuous deceleration in the heliosphere. This is in agreement with the results of previous studies that demonstrate that the drag force plays an important role in shaping CME dynamics \citep{Lindsay1999, Gopalswamy2001,Manoharan2006, Vrsnak2007}.

In our study, CMEs could be tracked in HI images up to a large elongation angle ($\approx$ 35$\arcdeg$). However, relating the tracked feature to features observed in situ is often challenging. Though in the GT method, we assume that the same enhanced
density structure is tracked using \textit{J}-maps from both \textit{STEREO} spacecraft. We also assume that the same structure appears in each consecutive image. These assumptions are not always truly valid. Even if these assumptions hold good, a single in situ spacecraft will likely be unable to sample the tracked feature. In such a situation, relating a remotely observed tracked feature with in situ observations will lead to an incorrect interpretation. Since the location of an in situ spacecraft concerning the CME structure determines which part of the CME will be intercepted by the in situ spacecraft, it may be more appropriate to also take into account the geometry of CMEs in the GT method. In the present work, we made the \textit{J}-map along the ecliptic plane as the estimated velocity in this plane is more suitable than the radial velocity at other position angles for estimating the arrival time of CMEs at the L1 point. But, even in the ecliptic plane, the \textit{J}-map gives information about only a part of the CME structure that has different velocities at different longitudes. Different parts of a CME traverse a fixed radial distance from the Sun at different times. The time is minimized for the apex of a CME \citep{Schwenn2005}. If the tracked feature happens to be at the apex of a CME moving in a different direction (longitude) from the Sun-L1 line, then the predicted arrival time of this CME using the estimated speed of the tracked feature will be earlier than the actual arrival time of the CME at the L1 point. Therefore, to obtain the actual arrival time and transit velocity of a CME passing the in situ spacecraft, it is necessary to take into account the propagation direction of the tracked feature.  \citet{Mostl2011} showed that the speed of a CME flank measured at a given angle $\theta$ to the CME apex is reduced by a factor of $\cos(\theta)$ for a circular geometry of the CME. Therefore, it seems reasonable that the transit speed measured in situ and the arrival time of the CME at L1 should be compared with the corresponding corrected speed of the tracked feature along the Sun-L1 line and the arrival time, respectively. However, we note that if the apex of the CME moves with a linear speed of 400 km s$^{-1}$ at 10$\arcdeg$ to the Sun-L1 line, the correction in the actual arrival time of the CME at L1 is only approximately 2 hr for a fixed distance of nearly 1 AU. It seems appropriate to also take the geometry of CMEs into account in the triangulation method. However, idealistic assumptions about geometry also are far from the real structure. Further, CME shapes can be distorted in the heliosphere by the solar wind, interplanetary shocks, and CME interactions. Therefore, one needs to be cautious, as assumptions about the CME geometry may result in new sources of errors.

Despite various assumptions made in the DBM \citep{Vrsnak2013} and the GT method \citep{Liu2010}, the predicted arrival times of CMEs are found within acceptable error. It must be mentioned here that the estimated velocity profiles of the selected CMEs show small apparent accelerations and decelerations for a few hr that do not seem to be real. We believe that these may be due to errors in the manual tracking of a CME feature using \textit{J}-maps and the extraction of the elongation angles. We have shown the error bars with vertical red lines in the kinematics plot. However, these do not denote the actual errors in the triangulation method but are representative of the sensitivity of the technique to elongation uncertainties \citep{Liu2010}. Considering that the elongation angles determined for each pixel in level 0 data for COR2 and level 2 data for HI1
and HI2 is quite accurate; the resolution will determine the minimum uncertainty in the elongation in elongation
that is used to construct the \textit{J}-maps. However, as mentioned above, the actual error in this GT method owes to the manual 
error in tracking the bright points using \textit{J}-maps. This error is assessed by repeating the manual tracking of a feature several times and comparing the derived parameters. One should also ensure that the same feature is tracked continuously in the \textit{J}-maps. This is often difficult due to the different sensitivities of the COR2, HI1, and HI2 imaging instruments, which cover a wide range of elongation angles. The actual sources of error in the GT technique are inherent in the assumption that the CME feature is a single point and the same point is being tracked continuously from both twin \textit{STEREO} spacecraft. If the assumption fails (likely at larger elongations) at some segment of the journey in the interplanetary medium, the CME kinematics will not be correctly estimated.

We also note that the density distribution along the line of sight is not well known, as we project the 3D structure of CME
features onto a 2D image. Relating remote sensing observations to in situ measurements of CMEs is often uncertain because the bright feature observed in \textit{J}-maps are due to the contribution of intensity along with the entire depth of the line of sight, but \textit{ACE} or \textit{WIND} density measurements near 1 AU are only along the Sun-Earth line. Despite these constraints on observations and assumptions in implemented techniques, our study shows a fair agreement in relating remotely observed features with in situ measurements. We track the CMEs up to large distances in the heliosphere. The efficacy of any forecasting scheme for
CME arrival time must be validated with real-time data so that the results are unbiased.

\section{Assessing the Relative Performance of Ten Reconstruction Methods for Estimating the Arrival Time of CMEs Using SECCHI/HI Observations}
\label{AssMthd}

Several reconstruction methods have been developed, based on a variety of assumptions regarding their geometry, propagation direction, and speed, to derive CME kinematics by exploiting heliospheric imagers (i.e., SMEI and HI) observations (e.g., \citealt{Howard2006, Jackson2006, Jackson2007, Jackson2008, Kahler2007, Sheeley2008, Howard2009, Tappin2009, Lugaz2009, Lugaz2010, Liu2010, Lugaz2010.apj, Davies2012, Davies2013}). These methods are based on a different set of assumptions which make them independent of each other to some degree. Several of the methods treat CMEs as a point, while other methods consider CMEs to have a larger-scale geometry as explained in Section~\ref{Recnsmthd} of Chapter~\ref{Chap2:DataMthd}. In some of the methods, CMEs are assumed to propagate with a constant speed, while other methods can estimate the time variations of the speed of a CME as it propagates through the heliosphere.

Since some of the critical questions in CME and space weather research relate to the propagation of CMEs, it is an obvious next step to attempt to ascertain the relative merits of various reconstruction methods for estimating the kinematic properties of CMEs, including their arrival time at Earth. Several such studies, based mainly on HI observations, have been performed previously. For example, \citet{Davis2010} applied the Fixed-Phi Fitting method \citep{Sheeley2008.apjl, Rouillard2008,Davies2009} to HI observations to estimate the propagation direction and speed of CMEs and compared their results with those obtained by \citet{Thernisien2009} using the Forward Modeling method for the same CMEs observed in the COR FOV. The authors found that their retrieved CME directions were in good agreement with those from forward modeling. At the same time, the discrepancy in speed between the two methods could be explained in terms of the acceleration of slow CMEs and the deceleration of fast CMEs in the HI FOV. It is also worth noting that \citet{Thernisien2009} had compared their estimated CME propagation directions with  those  obtained by \citet{Colaninno2009} using an entirely different technique. The method of \citet{Colaninno2009} is based on the constraint that both COR2-A and COR2-B should record the same true mass for any given CME. \citet{Wood2009} applied the Point-P \citep{Howard2006} and Fixed-Phi methods to the CME of 2008 February 4; their results suggested that the Fixed-Phi method performs better, which they argued for the studied CME having a small angular extent.
Similarly, \citet{Wood2010} implemented the Point-P, Fixed-Phi, and Harmonic Mean \citep{Lugaz2009} methods on the 2008 June 1 CME and showed that different methods could give significantly different kinematic profiles, especially in the HI2 FOV. Recently, \citet{Lugaz2010} has assessed the accuracy and limitations of two fitting methods (Fixed-Phi Fitting and Harmonic Mean Fitting) and two stereoscopic methods, Geometric Triangulation \citep{Liu2010} and Tangent to A Sphere  \citep{Lugaz2010.apj}, to estimate the propagation direction of 12 CMEs launched during 2008 and 2009. Their results showed that the Fixed-Phi Fitting approach could result in significant errors in the CME direction when the CME is propagating outside 60$\arcdeg$ $\pm$ 20$\arcdeg$ of the Sun-spacecraft direction, and Geometric Triangulation can yield large errors if the CME is propagating outside $\pm$ 20$\arcdeg$ of the Sun-Earth line. More recently, \citet{Colaninno2013} derived the deprojected height-time profiles of CMEs by applying the graduated cylindrical shell model (Forward Modeling: \citealt{Thernisien2009}) using SECCHI and LASCO images. They fitted the derived height-time data with six different methods to estimate the CME arrival time and speed at Earth.

The studies above are mainly limited to comparing CME propagation directions and sometimes speeds retrieved using different methods. These studies do not consider the characteristics of either the individual CMEs or the ambient medium into which they are launched, nor do they (except \citealt{Colaninno2013}) use the estimated kinematics of CMEs to predict their arrival time near 1 AU. They do not compare the derived CME arrival time with the actual CME arrival time identified from in situ observations. However, to understand the validity of these methods for space weather forecasting, we use different reconstruction methods to predict the arrival time and speed of CMEs launched at different speeds and launched  into different ambient solar wind conditions. Our selection of three CMEs, launched with different speeds into different solar wind conditions (on 2010 October 6, April 3, and February  12), satisfy such criteria. On the other hand, previous studies have been carried out by implementing several methods to an individual CME.

We implemented a total of 10 different reconstruction methods, ranging from single spacecraft methods and their fitting analogues to stereoscopic methods, to ascertain the performance of specific methods under specific conditions. Such extensive analysis undertaken in our study has not previously been reported. Moreover, we have not only compared the estimated direction and speed of the three selected CMEs but have also assessed the relative performance of these techniques in estimating CME arrival time at L1. We compare the estimated arrival time and speed of each CME to the actual arrival time and speed based on in situ signatures at L1 \citep{Zurbuchen2006}. Such a study is a useful step toward identifying the most appropriate techniques for the practical purpose of forecasting CME arrival time at Earth in the near future.

\section{Reconstruction Methods and Their Application to the Selected CMEs}
\label{Mthds3CMEs}

We applied ten reconstruction methods to the following three CMEs observed by \textit{STEREO}:

\begin{itemize}
\item{2010 October 6 CME, slow (3D speed in COR2 FOV = 340 km s$^{-1}$), traversed through a slow speed ambient solar wind medium.}

\item{2010 April 3 CME, fast (3D speed in COR2 FOV = 867 km s$^{-1}$), traversed through a fast speed ambient solar wind medium.}

\item{2010 February 12 CME, fast, (3D speed in COR2 FOV = 816 km s$^{-1}$), traversed through a slow speed ambient solar wind medium.}
\end{itemize}

The fast 2010 April 3 CME experienced only a modest deceleration during its journey and was the fastest CME at 1 AU since the 2006 December 13 CME. We note that, during this time, the Earth was in the throes of high-speed solar wind, which perhaps governed the dynamics of this CME. The 2010 April 3 CME has been investigated extensively in earlier studies \citep{Mostl2010,Liu2011,Wood2011,Temmer2011,Mishra2013}. 

The methods used to derive the heliospheric kinematics of CMEs can be classified into two groups, one which uses single spacecraft observations and the other which requires simultaneous observations from two viewpoints. We have implemented seven single spacecraft methods and three stereoscopic methods.  

\begin{itemize}

\item{\textbf{Single spacecraft reconstruction methods:} (i) Point-P (PP: \citealt{Howard2006}) (ii) Fixed-Phi (FP: \citealt{Kahler2007}) (iii) Harmonic Mean (HM: \citealt{Lugaz2009}) (iv) Self-Similar Expansion (SSE: \citealt{Davies2012}) (v) Fixed-Phi Fitting (FPF: \citealt{Rouillard2008}) (vi) Harmonic Mean Fitting (HMF: \citealt{Lugaz2010}) (vii) Self-Similar Expansion Fitting (SSEF: \citealt{Davies2012})}

\item{\textbf{Stereoscopic reconstruction methods:} (i) Geometric Triangulation (GT: \citealt{Liu2010}) (ii) Tangent to A Sphere (TAS: \citealt{Lugaz2010.apj}) (iii) Stereoscopic Self-Similar Expansion (SSSE: \citealt{Davies2013})}

\end{itemize}

Using the FP, HM, and SSE methods, one can estimate the kinematics of a CME, provided an estimate of its 3D propagation direction is known. The  3D propagation direction of a CME close to the Sun can be estimated by applying the scc$\_$measure.pro routine \citep{Thompson2009} to COR2 observations.  We derive the kinematics of these three CMEs in the heliosphere using the methods above and use those kinematics (averaged over the last few data points) as inputs to the Drag Based Model (DBM: \citealt{Vrsnak2013}) to estimate the arrival times of the CMEs at L1. If, for any method, the derived kinematics show implausible (unphysical) variations over the last few points, then the kinematics prior to that time are used instead. The details about the nature of the drag forces and DBM developed by \citet{Vrsnak2013} have been described in Section~\ref{DBM} of Chapter~\ref{Chap2:DataMthd}. The details of these reconstruction methods implemented in our study have been summarized in Section~\ref{Recnsmthd} of Chapter~\ref{Chap2:DataMthd}.

In the following sections, we discuss the application of various reconstruction methods to the three selected CMEs. The analysis of the CME of 2010 October 6 is presented in Section~\ref{Mthds6Oct10} in detail. Results of similar analyses of the 2010 April 3 and February 12 CMEs are presented briefly in Sections~\ref{Mthds3Apr10} and ~\ref{Mthds12Feb10}, respectively.

\subsection{2010 October 6 CME}
\label{Mthds6Oct10}

The CME of 2010 October  6 was first observed by SOHO/LASCO C2 at 07:12 UT as a partial halo with a linear plane of sky speed of 282 km s$^{-1}$ (online LASCO CME catalog: http://cdaw.gsfc.nasa.gov/CME$\_$list/; see \citealt{Yashiro2004}). This CME was associated with a filament eruption from the north-east (NE) quadrant of the solar disc. In the LASCO C3 FOV, this CME was observed to accelerate at 7 m s$^{-2}$. COR1-A and COR1-B, with an angular separation of approximately 161$\arcdeg$, first observed the CME at 04:05 UT in the NE and north-west (NW) quadrants, respectively. The CME was subsequently observed by COR2, and HI1 and HI2, on both \textit{STEREO-A} and \textit{B} (Figure~\ref{Evolution6Oct10}). The arrival of this CME at the Earth caused a moderate geomagnetic storm with a peak Dst index of approximately -80 nT on 2010 October  11 at 19:00 UT. 

We applied the COR2 data processing scheme as described by \citet{Mierla2010} before implementing the tie-pointing technique of 3D reconstruction \citep{Thompson2009}. The 3D radial speed of the 2010 October 6 CME is estimated to be 340 km s$^{-1}$  at a 3D height of nearly ten \textit{R}$_\odot$ from the Sun. The central latitude of the CME feature was estimated to be $\approx$ 20$\arcdeg$ North, and its longitude  $\approx$ 10$\arcdeg$ East of the Earth. As the CME was propagating slightly north-eastward of the Sun-Earth line, it was expected to impact the Earth.  Assuming a constant speed of 340 km s$^{-1}$ beyond the COR2 FOV, the predicted CME arrival time at L1 is estimated to be on 2010 October 11, at 06:10 UT.

\begin{figure}[!htb]
\begin{center}
\includegraphics[angle=0,scale=.42]{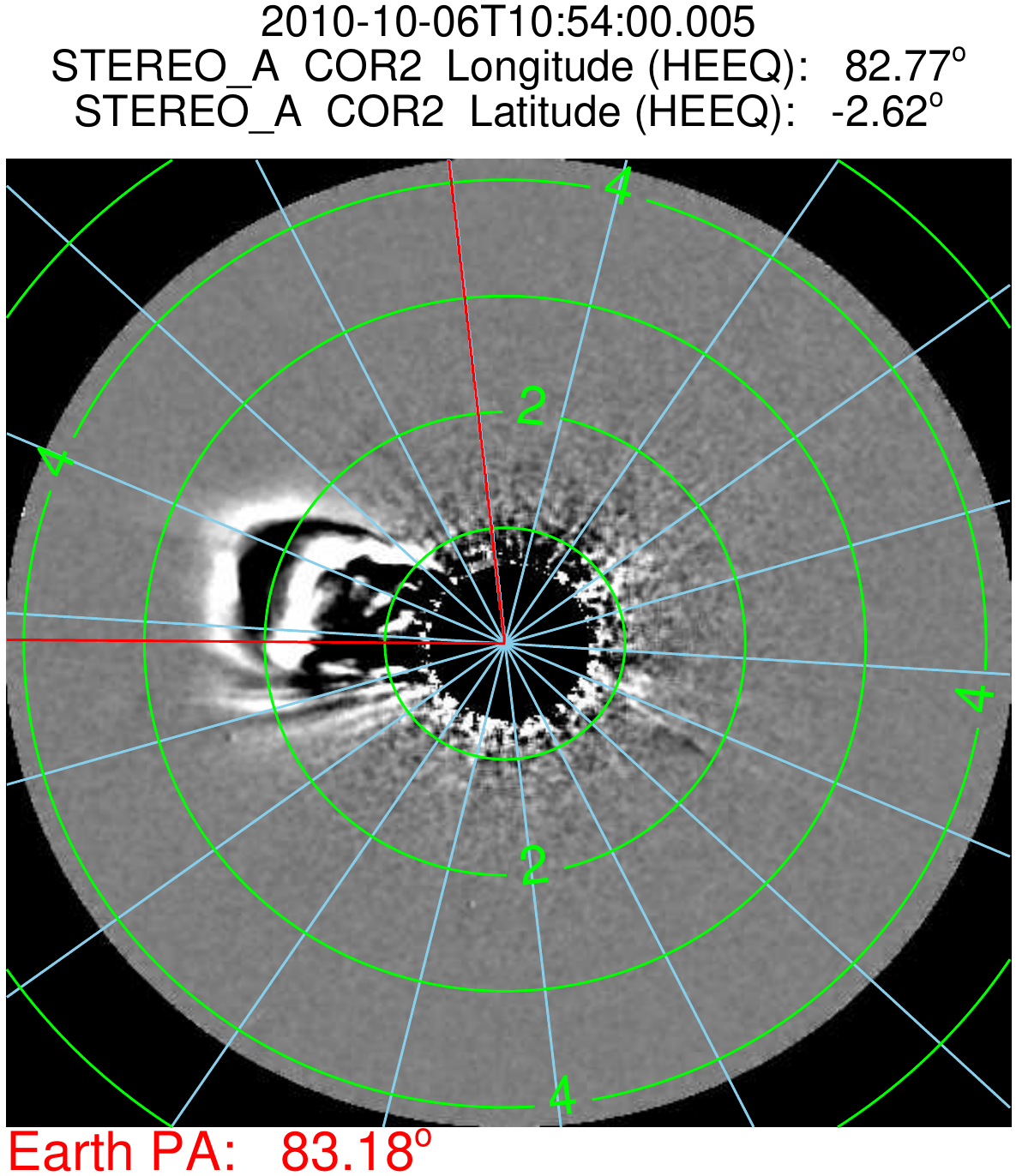}
\includegraphics[angle=0,scale=.42]{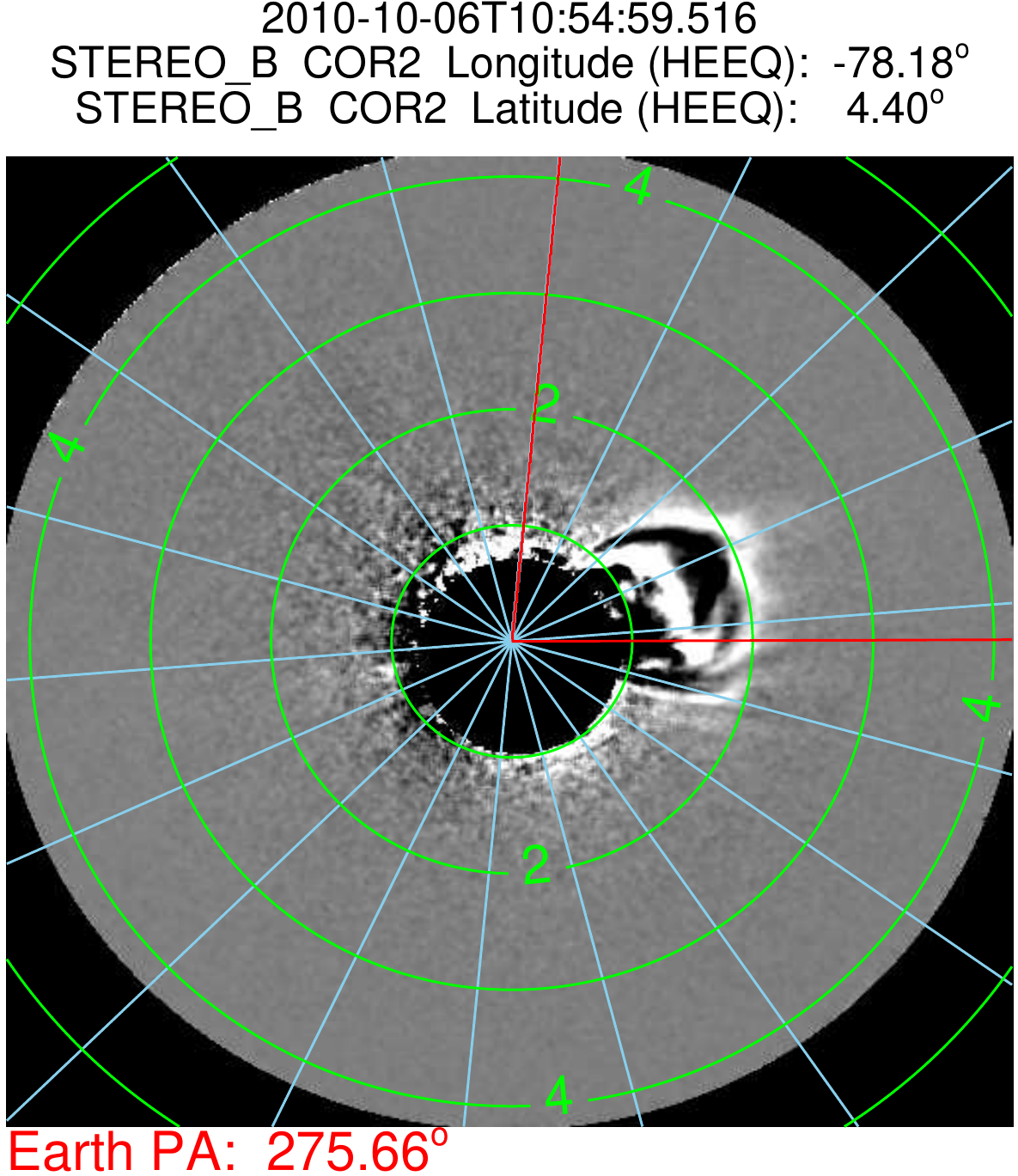}  \\
\includegraphics[angle=0,scale=.42]{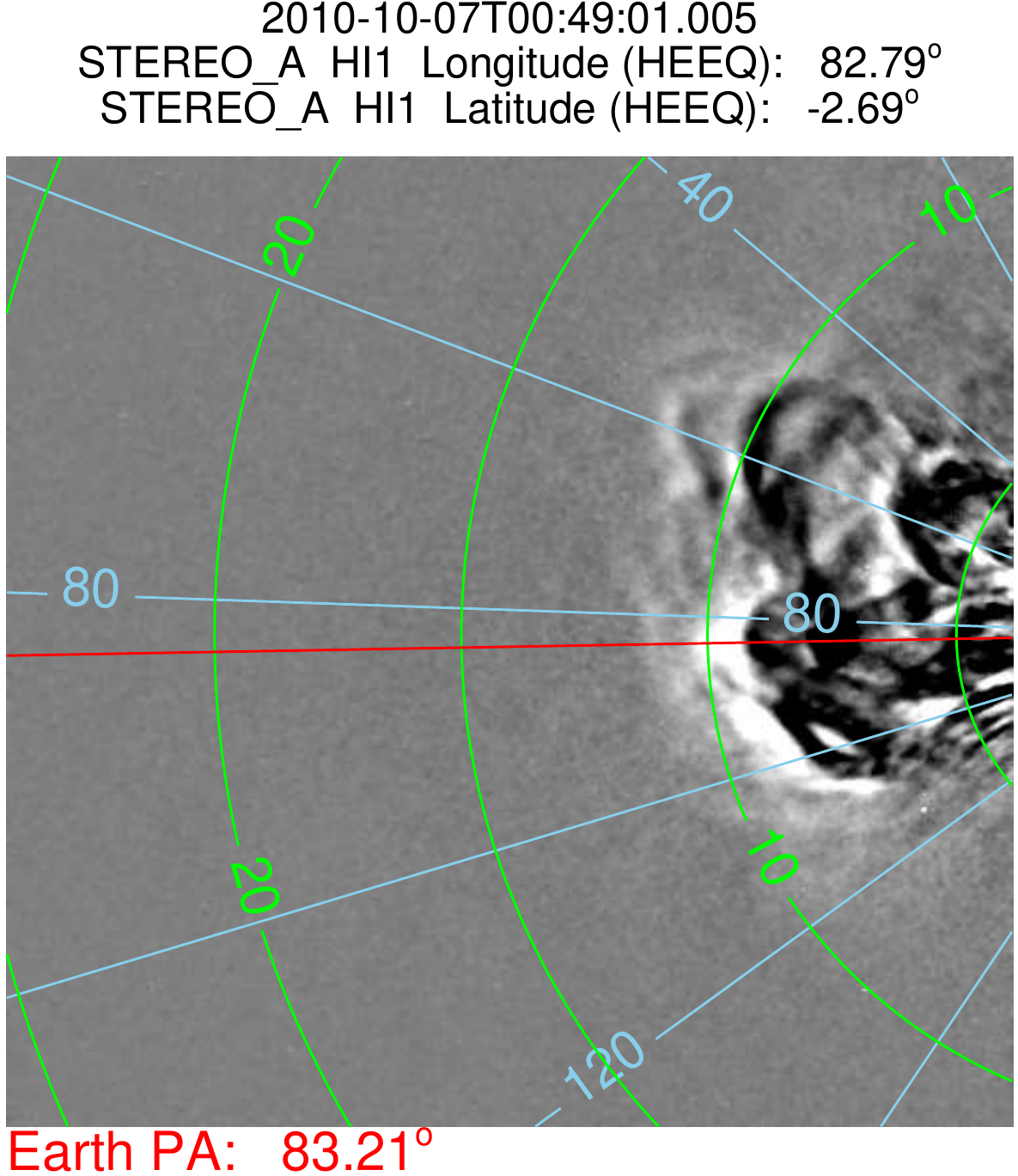}
\includegraphics[angle=0,scale=.42]{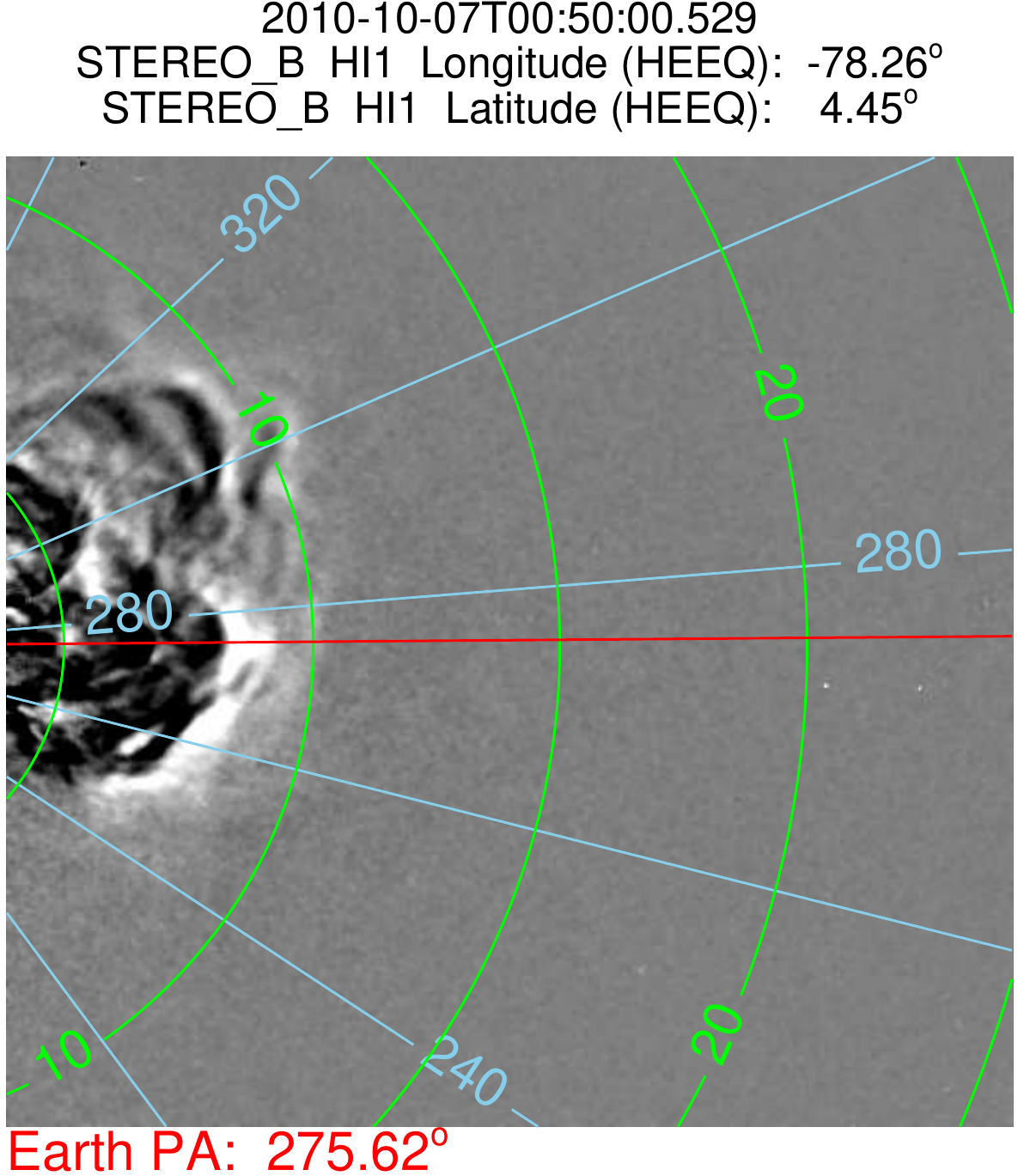}  \\
\includegraphics[angle=0,scale=.42]{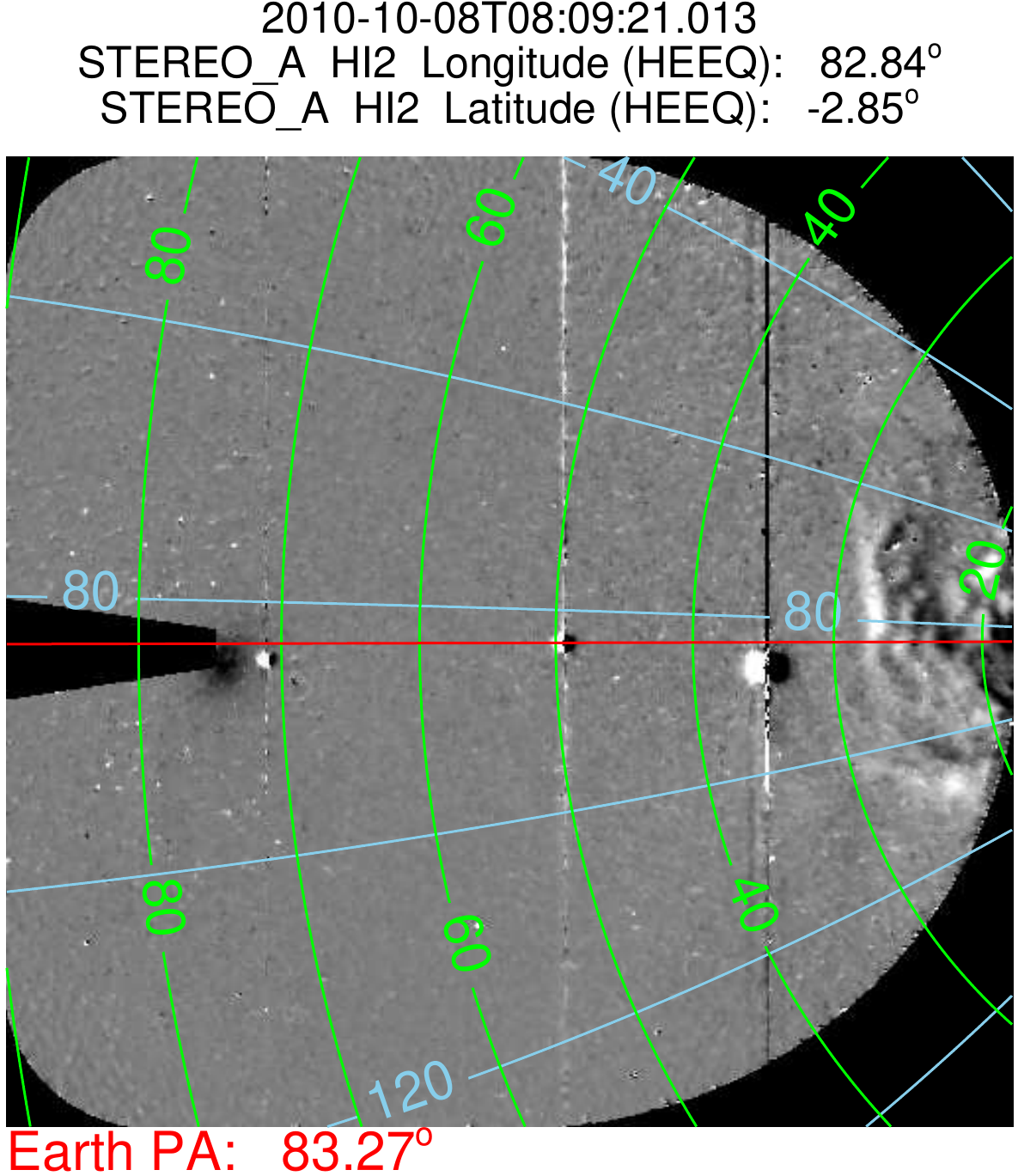}
\includegraphics[angle=0,scale=.42]{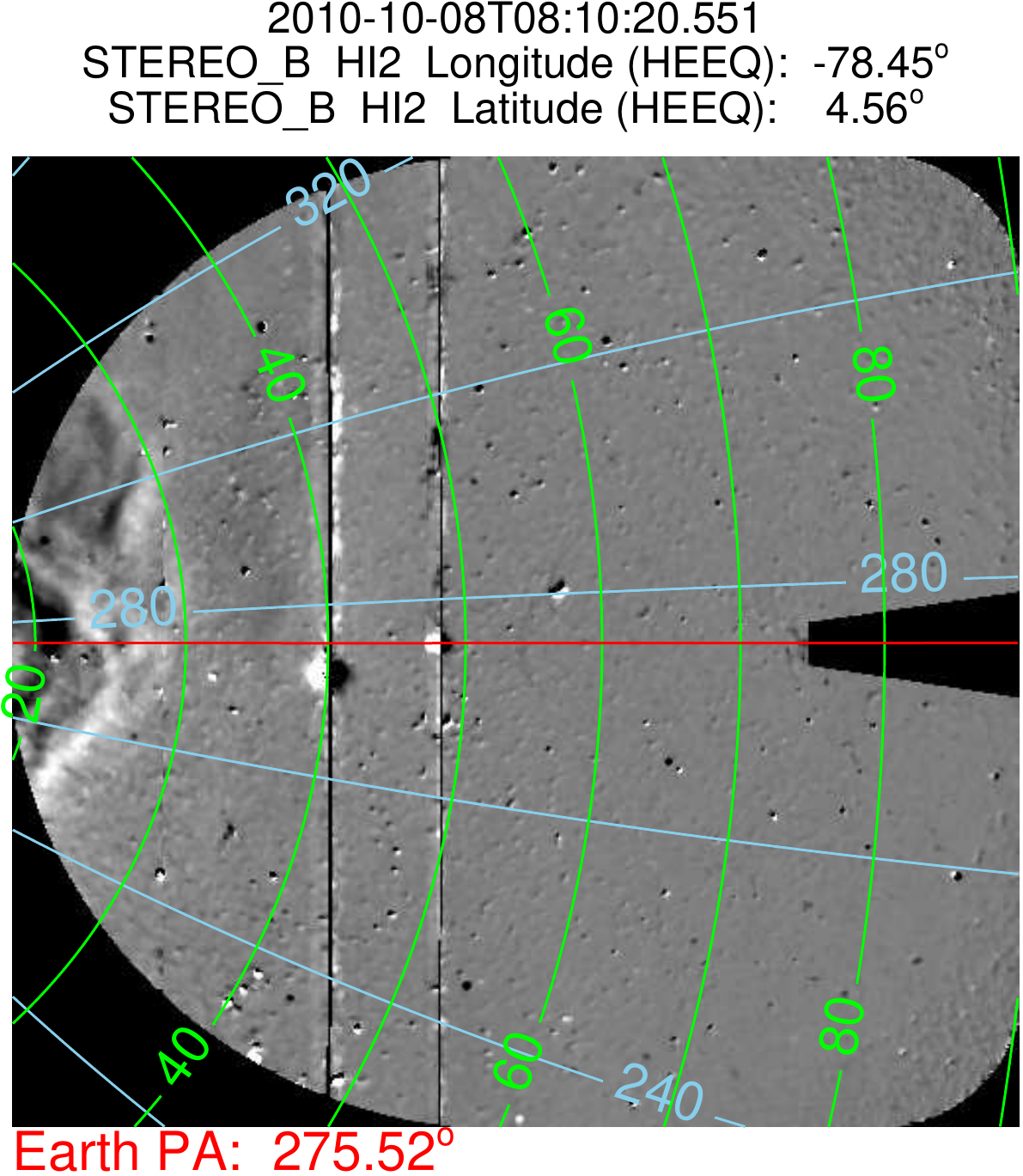}
\caption[Evolution of the 2010 October 6 CME observed in COR2, HI1 and HI2 images]{Evolution of the 2010 October 6 CME observed in COR2, HI1 and HI2 images from \textit{STEREO-A} (left column) and \textit{B} (right column). Contours  of elongation angle (green) and position angle (blue) are overplotted. The vertical red line in the COR2 images marks the 0$\arcdeg$ position angle contour. The horizontal lines (red) on all panels indicate the position angle of the Earth.}
\label{Evolution6Oct10}
\end{center}
\end{figure}

\subsubsection{Reconstruction methods using single spacecraft observations}

Using heliospheric imagers, which image at and across large distances from the Sun, 3D information about CMEs can be extracted without multiple viewpoint observations. In this section, we apply reconstruction methods based on single viewpoint observations of CMEs in the heliosphere, i.e., the PP, FP, HM, and SSE methods described in Section~\ref{SinRcnsMthd} of Chapter~\ref{Chap2:DataMthd}. 

\paragraph{Point-P (PP) method} \hspace{0pt}\\
\label{PP6Oct10}
It becomes very diffuse when a CME is in the FOV of a heliospheric imager (Coriolis/SMEI or SECCHI/HI). Therefore to track and extract the time-elongation profile of a moving solar wind structure, a technique developed originally by \citet{Sheeley1999}, involves the generation of time-elongation maps (\textit{J}-maps: \citealt{Rouillard2008, Davies2009, Mostl2010, Liu2010, Harrison2012}) is often applied. For the current study, we constructed \textit{J}-maps using running difference images from the HI1 and HI2 instruments, as explained in Section~\ref{Jmapsmthd} of Chapter~\ref{Chap2:DataMthd} (see also, \citealp{Mishra2013}). Ecliptic \textit{J}-maps covering the passage of the 2010 October 6 CME, from the viewpoints of both \textit{STEREO-A} and  \textit{B}, are shown in Figure~\ref{Jmaps6Oct10}. In the HI2-A FOV on 2010 October 6, Venus, Earth, Jupiter, and Neptune were identified at position angles of 86.5$\arcdeg$, 83.1$\arcdeg$, 84.4$\arcdeg$, and 83.8$\arcdeg$ with elongations of 34.3$\arcdeg$, 49.5$\arcdeg$, 73.1$\arcdeg$, and 50.3$\arcdeg$ respectively (see Figure~\ref{Evolution6Oct10}). In the HI2-B FOV, Venus and Earth were identified at elongations of 40.4$\arcdeg$ and 48.2$\arcdeg$ respectively. In each \textit{J}-map (Figure~\ref{Jmaps6Oct10}), two horizontal lines are due to the presence of Venus and Earth in the HI2 FOV. A slanted line that appears in the \textit{STEREO-A} \textit{J}-map on October 10 is due to the entrance of Jupiter into the HI2-A FOV.

\begin{figure}[!htb]
\begin{center}
\includegraphics[angle=0,scale=.43]{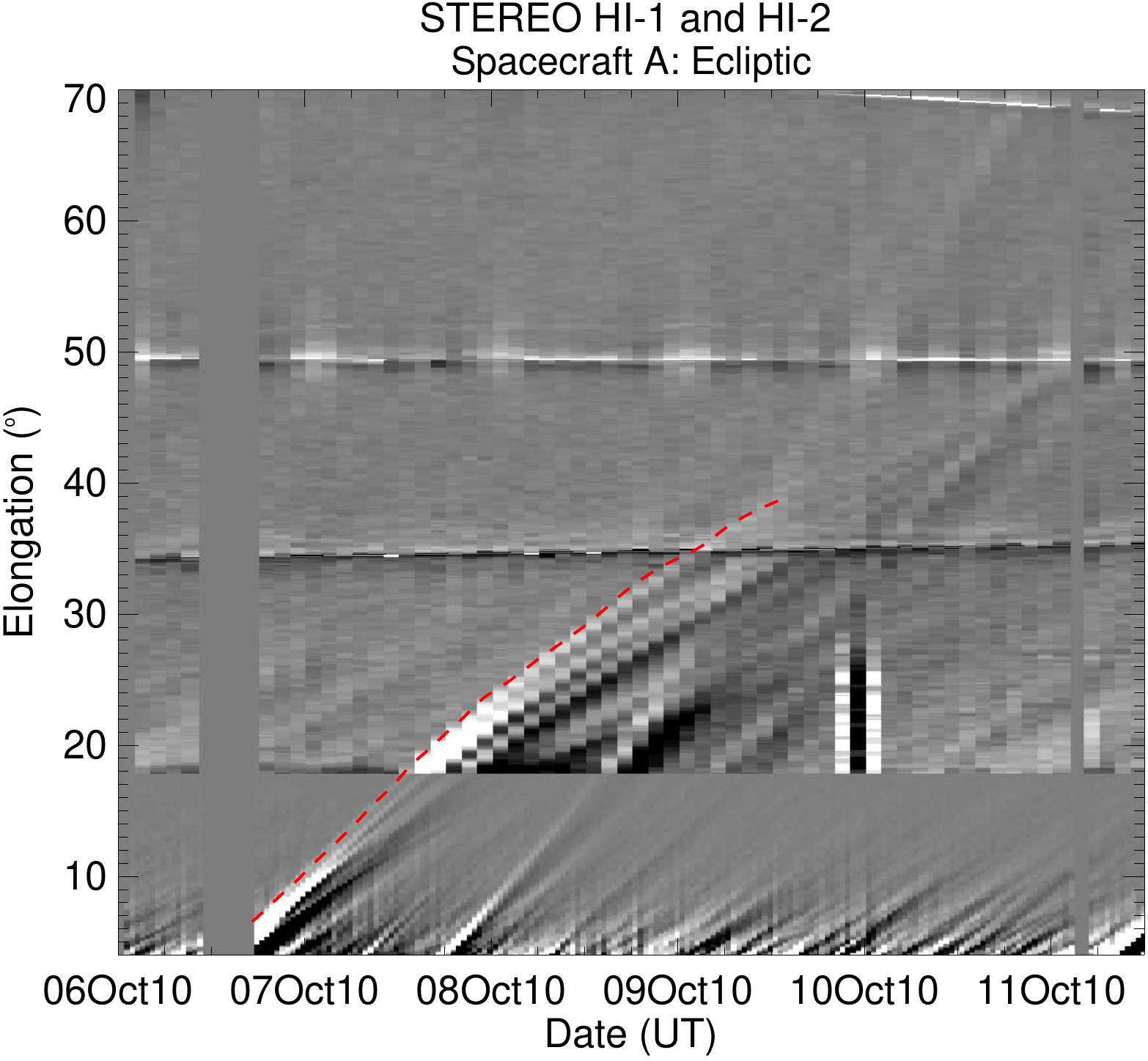}
\includegraphics[angle=0,scale=.43]{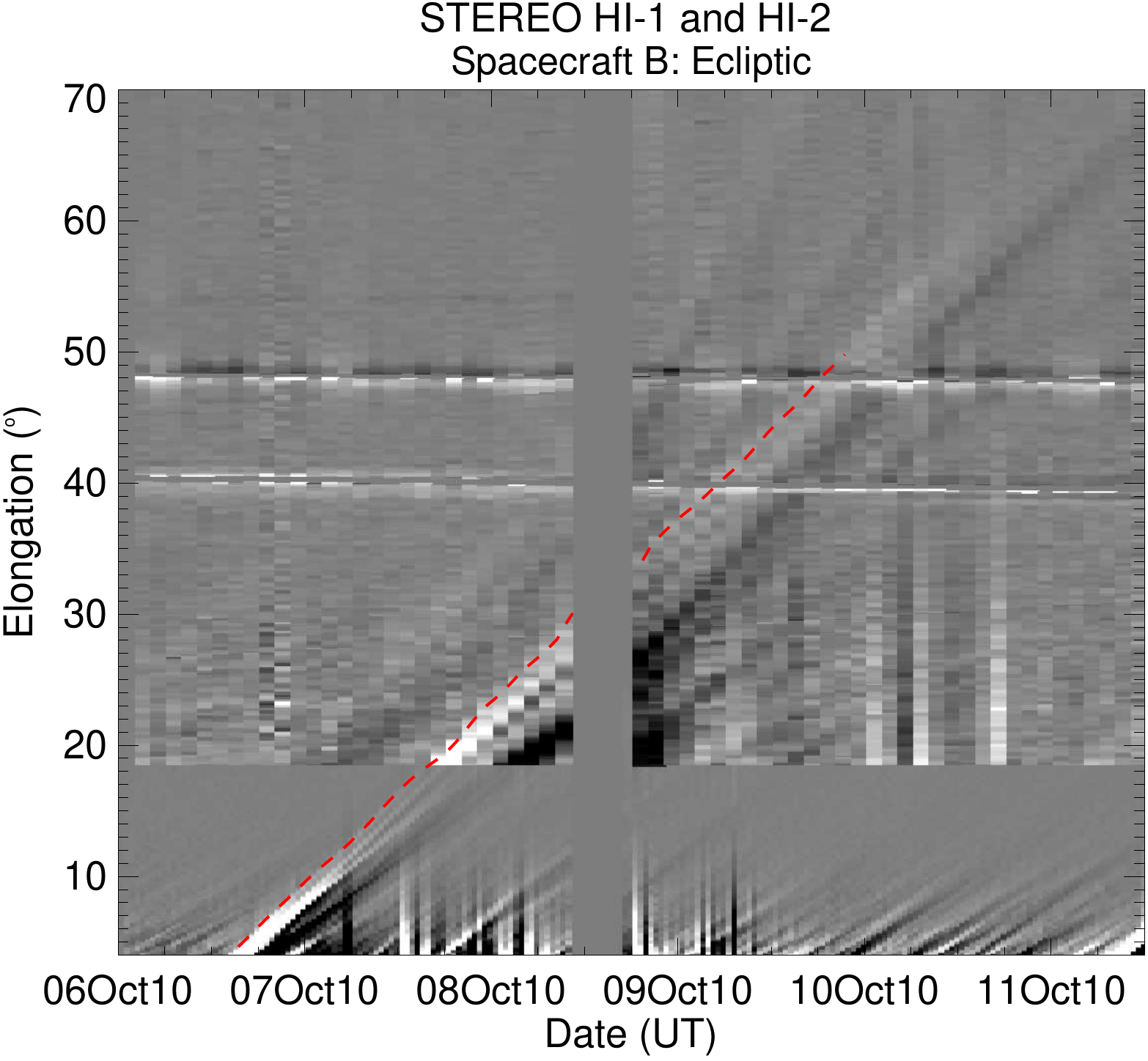}
\caption[\textit{J}-maps for the 2010 October 6 CME]{Ecliptic time-elongation maps (\textit{J}-maps) for \textit{STEREO-A} (left) and \textit{STEREO-B} (right) constructed from running differences images from HI1 and HI2, for the time interval extending from 06 Oct 2010 to 11 Oct 12:00 UT, 2010. The leading edge of the bright feature (corresponding to the leading edge of the initial CME front) is tracked in the \textit{J}-maps (red lines).}
\label{Jmaps6Oct10}
\end{center}
\end{figure}

In each \textit{J}-map, a set of the positively inclined bright features corresponds to the 2010 October 6 CME. We tracked the leading edge of the first bright track, corresponding to the initial CME front, in the \textit{STEREO-A} \textit{J}-map (red dashed line in the left panel in Figure~\ref{Jmaps6Oct10}). We manually extracted the time-elongation profile of this outward-moving feature and applied the PP approximation to the elongation data. The CME front can be tracked out to 39$\arcdeg$ elongation in the \textit{STEREO-A} \textit{J}-map. The tracked feature's derived radial distance and speed are plotted in Figure~\ref{STAA6Oct10} in black. It is to be noted that, using this method, we obtain time variations of CME radial distance and speed; this is not possible using the single spacecraft fitting techniques (FPF, HMF, and SSEF), which give only a single value of the radial speed unless we apply the triangulation approach. Note, however, that an estimation of the CME propagation direction is not retrievable using the PP approach. The speed is calculated from adjacent estimates of distance using the Interactive Data Language (IDL) $deriv$ function, which performs a three-point Lagrange interpolation on the data points to be differentiated. The estimated speed variation in Figure~\ref{STAA6Oct10} suggests that the tracked feature undergoes deceleration.

\begin{figure}[!htb]
\begin{center}
\includegraphics[angle=0,scale=.50]{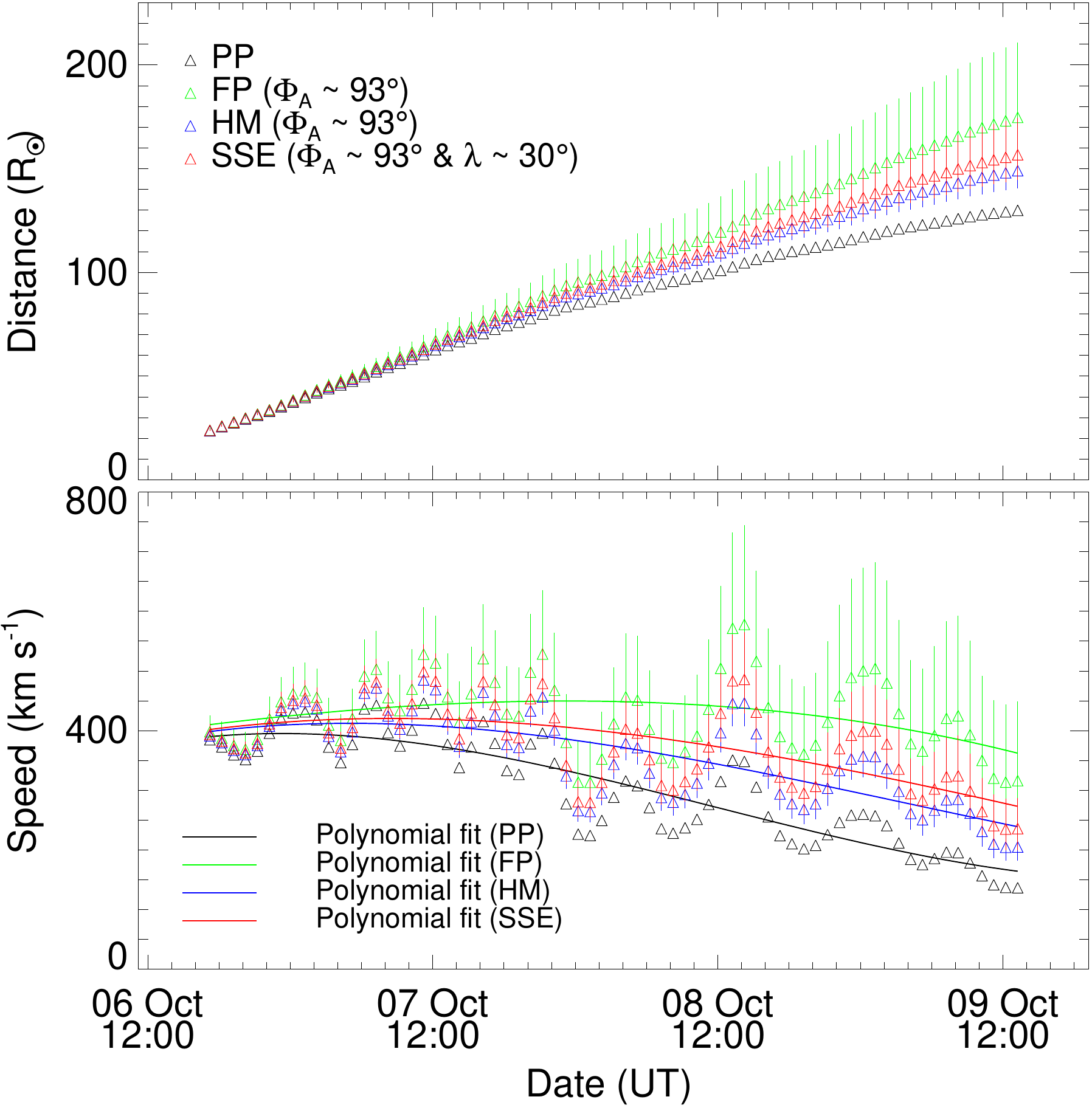}
\caption[Estimated distance and speed profiles for the tracked feature of the 2010 October 6 CME using the PP, FP, HM, and SSE methods]{The derived distance profiles based on the application of the PP, FP, HM, and SSE methods for the tracked feature are shown in the top panel. The bottom panel presents speed profiles derived from the adjacent distances using three-point Lagrange interpolation (solid line shows the polynomial fit). Vertical lines show the errors bars, calculated using  propagation  directions that are +10$\arcdeg$ and -10$\arcdeg$ different from the  value ($\phi$) estimated  using  tie-pointing, as described in Section~\ref{ErrSing6Oct10}.}
\label{STAA6Oct10}
\end{center}
\end{figure}

We input the CME kinematics, estimated by implementing the PP method, into the DBM to predict the arrival time of the CME at L1. The CME front is tracked out to a heliocentric distance of 129.9 \textit{R}$_\odot$ (0.6 AU) on 2010 October 9 at 09:14 UT, where it has a radial speed of 180 km s$^{-1}$.  The kinematics at the furthest distances to which this feature can be tracked are used as inputs to the DBM, along with the minimum and maximum values of the drag parameter estimated by \citet{Vrsnak2013} and an ambient solar wind speed of 350 km s$^{-1}$. Table~\ref shows the predicted arrival time and transit speed of the tracked CME front at L1, corresponding to the extreme values of the drag parameter, and Table~\ref{Tab6Oct10}.

\begin{sidewaystable}
  \centering
{\scriptsize
 \begin{tabular}{ p{3.0cm}|p{2.0cm}| p{3.0cm}| p{2.5cm}|p{2.5cm}|p{2.5cm}}
    \hline
		
 Method & Kinematics as inputs in DBM [t$_{0}$, R$_{0}$ (\textit{R}$_\odot$), v$_{0}$ (km s$^{-1})$] & Predicted arrival time using kinematics + DBM (UT) [$\gamma$ = 0.2 to 2.0 (10$^{-7}$ km$^{-1}$)] & Predicted transit speed at L1 (km s$^{-1}$)  [$\gamma$ = 0.2 to 2.0 (10$^{-7}$ km$^{-1}$)] & Error in predicted arrival time (hrs) [$\gamma$ = 0.2 to 2.0 (10$^{-7}$ km$^{-1}$)] &  Error in predicted speed (km s$^{-1}$)  [$\gamma$ = 0.2 to 2.0 (10$^{-7}$ km$^{-1}$)]  \\  \hline

PP (\textit{STEREO-A}) & 09 Oct 09:14, 130, 180  & 12 Oct 07:29 to 11 Oct 14:54  & 259 to 328 	& 25.6 to 9.05 	& -96 to -28  \\ \hline

PP (\textit{STEREO-B}) & 09 Oct 21:27, 158, 300  & 11 Oct 08:15 to 11 Oct 06:41  &   306 to 327  &  2.4 to 0.8  &  -50 to -28  \\  \hline

FP (\textit{STEREO-A}) & 09 Oct 13:14, 174, 390  & 10 Oct 08:23 to 10 Oct 08:45	& 388 to 376 	& -21.4 to -21	& 33 to 21  \\ \hline

FP (\textit{STEREO-B}) & 09 Oct 07:27, 162, 435 &  10 Oct 06:11 to 10 Oct 07:45 & 425 to 384  & -23.6 to -22.1  & 70 to 30 \\ \hline

HM (\textit{STEREO-A}) 	& 09 Oct 13:14, 149, 230 & 11 Oct 14:22 to 11 Oct 06:21	& 266 to 324 	& 8.5 to 0.55	& -89 to -31   \\ \hline

HM (\textit{STEREO-B})  & 09 Oct 21:27, 189, 410 &  10 Oct 08:35 to 10 Oct 08:52 & 407 to 390  & -21.3 to -21 & 52 to 35   \\ \hline

SSE (\textit{STEREO-A}) & 09 Oct 13:14, 157, 255  & 11 Oct 05:34 to 10 Oct 23:48	 & 276 to 322  &	0.1 to -6.0	 &  -79 to -33  \\ \hline

SSE (\textit{STEREO-B}) & 09 Oct 21:27, 193, 470  & 10 Oct 05:33 to 10 Oct 06:03	 & 462 to 419  &	-24.3 to -23.8	&  107 to 64  \\ \hline

GT	& 09 Oct 13:14, 177, 470 &  10 Oct 04:04 to 10 Oct 05:24	& 456 to 400 	& -25.8 to -24.4	& 101 to 45 \\  \hline

TAS &  09 Oct 13:14, 166, 385 &  10 Oct 12:39 to 10 Oct 13:04	 & 383 to 372 	& -17.1 to -16.7 & 28 to 17    \\  \hline
SSSE &  09 Oct 13:14, 169, 410 &  10 Oct 09:53 to 10 Oct 10:41	& 405 to 381 	& -19.9 to -19.2	 & 50 to 26    \\  \hline		
			
 \end{tabular}

\begin{tabular}{p{3.0cm}| p{3.0cm}| p{2.5cm}|p{2.5cm}|p{2.5cm}| p{2.0cm}}
   
\multicolumn{6}{c}{Time-elongation track fitting methods} \\  \hline
   
 Methods   & Best fit parameters [t$_{(\alpha = 0)}$, $\Phi$ ($\arcdeg$), v (km s$^{-1}$)]  &   Predicted arrival time at L1 (UT)   & Error in predicted arrival time  & Error in predicted speed at L1 (km s$^{-1}$) & Longitude ($\arcdeg$) 
   \\ \hline
	
	FPF (\textit{STEREO-A})   & 06 Oct 07:38, 96.2, 462 & 10 Oct 00:39 & -29.1  &  107      &  -13   \\  \hline
	FPF  (\textit{STEREO-B})  & 06 Oct 05:05,61, 414  & 10 Oct 08:25  & -21.3  &    59     &  -17 \\  \hline
	 
	HMF (\textit{STEREO-A})   & 06 Oct 08:40, 136, 610 & 11 Oct 00:19 & -5.5  &    12     &   -53  \\  \hline
	HMF  (\textit{STEREO-B})  & 06 Oct 06:53, 74.5, 434 & 10 Oct 05:50 &  -24  &    78    &   -4 \\  \hline
	
	SSEF (\textit{STEREO-A})  & 06 Oct 08:16, 115, 525  & ---   &   --- &   --- &   -32  \\  \hline
	SSEF (\textit{STEREO-B})  & 06 Oct 06:19, 68.1, 426  & 10 Oct 09:57  & -19.8  &  55    &   -10   \\  \hline

\end{tabular}
}
\caption[Results of different reconstruction techniques applied to the 2010 October 6 CME]{\scriptsize{Results of the applied methods for the 2010 October 6 CME (first column). 
Upper section -- Second column: the kinematics parameters obtained directly by the techniques and used as input to the  DBM. Third and fourth columns: predicted arrival time  and speed of the CME at  L1 corresponding to the extreme range of the drag parameter used in the DBM.  Fifth and sixth columns: errors in predicted arrival time and speed compared with in situ arrival time and speeds. 
Lower section (fitting techniques) -- Second column: best-fit launch time, longitude measured from observer, and speed. Third column: predicted arrival time at L1. Fourth and fifth columns: errors in predicted arrival time and speed (computed as above). Sixth column: longitude obtained from Sun-Earth line. 
The \textit{STEREO-A} and \textit{B} shown in parentheses for each method denote the spacecraft from which the derived elongation is used. Negative (positive) errors in predicted arrival time correspond to a predicted arrival time before (after) the actual CME  arrival time determined from in situ measurements.  Negative (positive) errors in predicted speed correspond to a predicted speed that is less (more)  than the actual speed of the CME  at L1.}}

\label{Tab6Oct10}
\end{sidewaystable}

From Figure~\ref{Jmaps6Oct10}, we notice a data gap in the \textit{STEREO} HI-B observations and that the \textit{J}-map quality is slightly better for \textit{STEREO} HI-A. An important reason for the poorer quality of the HI-B \textit{J}-maps is that the HI2-B images are out of focus compared to HI2-A images \citep{Brown2009}. Another reason for the poorer quality of HI-B images is that HI on \textit{STEREO-B} suffers small pointing discontinuities due to dust impact. Since HI-B is facing the direction of travel of the \textit{STEREO-B} spacecraft, it gets impacted directly by interplanetary dust; HI-A is on the opposite side to the direction of travel of \textit{STEREO-A}, so it does not suffer direct impact \citep{Davis2012}. Due to the large gradient inherent in the F-coronal signal, even a small pointing offset can result in an inaccurate F-corona subtraction, which results in degraded image quality. Despite this, we tracked the leading edge of the October 6 CME in HI-B \textit{J}-maps even beyond the data gap (right panel in Figure~\ref{Jmaps6Oct10}). Adopting the same procedure described for the feature tracked by \textit{STEREO-A}, we also estimated the kinematics and the arrival times at L1 of the CME based on its elongation profile extracted from the \textit{STEREO-B} \textit{J}-map. Its estimated kinematics over the last few tracked points, which are used as inputs to the DBM, and predicted arrival time at L1 are noted in Table~\ref{Tab6Oct10}.

\paragraph{Fixed-phi (FP) method} \hspace{0pt}\\
\label{FP6Oct10}
 The distance profile of the CME can be derived using this so-called FP approximation by assuming a propagation direction that can be determined by identifying the CME source region. However, in our analysis of the CME of 2010 October 6, we use the propagation direction ($\phi_{FP}$) derived from the 3D reconstruction of COR2 data as discussed in Section~\ref{Mthds6Oct10}. We assume that, beyond the COR2 FOV, the CME will continue to travel in the same direction. The estimated longitude of the CME is $\approx$ 10$\arcdeg$ east of the Earth, which corresponds to a longitude difference of $\approx$ 93 $\arcdeg$ from the \textit{STEREO-A} spacecraft; the separation angle between \textit{STEREO-A} and the Earth was $\approx$ 83$\arcdeg$ at that time. Using the elongation variation of the tracked feature, extracted manually from the ecliptic \textit{J}-maps constructed from HI-A images as shown in Figure~\ref{Jmaps6Oct10} (left), and $\phi_{FP}$ = 93$\arcdeg$, we calculated the distance profile of the leading bright front of the CME. The obtained time variations of the radial distance and speed are shown in green in Figure~\ref{STAA6Oct10}. The unphysical deceleration of the CME beyond 100 R$_{\odot}$ (suggested by a speed less than ambient solar wind speed) may be due to the erroneous fixing of the propagation angle in the case of a real deflection or, indeed, the inaccurate characterization of the propagation angle. However, it is most likely due to the breakdown of the simple assumption that the observer is always looking at the same point-like feature of the CME. This can lead to large errors in the estimated height of the CME leading edge, particularly at greater elongations where the expanding CME geometry plays a significant role and the observer will be unlikely to record the intensity from the same part of CME's leading edge (artificial deflection: \citealp{Howard2009, Howard2011}). The limitations of this and other methods will be discussed in Section~\ref{ResDis32}.

As with the PP analysis, we apply the DBM based on the kinematics of the tracked CME feature estimated using the FP method. Running the DBM with the derived CME kinematics as inputs, along with an ambient solar wind speed of 350 km s$^{-1}$ and the two extreme values of the drag parameter, we obtained the L1 arrival times and transit speeds given in Table~\ref{Tab6Oct10}.

We also applied the FP method to the elongation profile derived from the \textit{STEREO-B} \textit{J}-map. 
(Figure~\ref{Jmaps6Oct10}, right). In \textit{STEREO-B}, the feature can be tracked out to 162 \textit{R}$_\odot$ (0.75 AU) on 09 October 07:27 UT, where its speed is approximately 435 km s$^{-1}$.1 Again, these kinematics are used as inputs to the DBM to estimate the arrival times and transit speeds at L1 (see Table~\ref{Tab6Oct10}).

\paragraph{Harmonic mean (HM) method} \hspace{0pt}\\
\label{HM6Oct10}
 We used the CME longitude estimated from the 3D reconstruction of COR2 data (as described earlier) and the elongation profile extracted from the \textit{STEREO-A} \textit{J}-map to estimate the distance and speed profiles of the CME front using the HM approximation (blue lines in Figure~\ref{STAA6Oct10}). A polynomial fit to the speed profile (solid blue line in the bottom panel of Figure~\ref{STAA6Oct10}) suggests an overall deceleration of the tracked CME feature. The non-physical deceleration of this feature at large distances could result from the real deflection of this feature or inaccuracy in the assumed `fixed' direction, or be because the observer (in this case \textit{STEREO-A}) detects scattered light from a different part of the CME than that assumed (artificial deflection). The limitations of the HM method are discussed in Section~\ref{ResDis32}.

Again, the estimated kinematics derived over the last segment of the tracked time-elongation profile at around 13:14 UT on 2010 October 9 (distance: 149 \textit{R}$_\odot$, i.e., 0.69 AU and speed 230 km s$^{-1}$) are used as inputs into the DBM, to predict the CME arrival time and transit speed at L1 (Table~\ref{Tab6Oct10}). As above, the DBM assumes an ambient solar wind speed of 350 km s$^{-1}$ and is run for the two extreme values of the drag parameter.

The same methodology is applied to the CME track observed by \textit{STEREO-B}. At the end of its observed track, on October 9 at 21:27 UT, the CME speed is estimated to be 410 km s$^{-1}$ at a distance of 189 \textit{R}$_\odot$ (0.87 AU). These values are used as inputs in the DBM to derive arrival times and transit speeds at L1 (Table~\ref{Tab6Oct10}).

\paragraph{Self-similar expansion (SSE) method}
\label{SSE6Oct10}
Assuming $\lambda$ = 30$\arcdeg$, we use the longitude estimated from 3D reconstruction in the COR2 FOV to estimate the distance and speed profiles of the CME tracked in the \textit{STEREO-A} \textit{J}-map, using the SSE method (shown in red in Figure~\ref{STAA6Oct10}). The DBM was run in the same manner as described earlier, based on these kinematics inputs, to obtain the arrival times and transit speeds (Table~\ref{Tab6Oct10}1). The same methodology was applied to \textit{STEREO-B} observations (results are included in Table~\ref{Tab6Oct10}). 

\paragraph{Error analysis for FP, HM and SSE methods} \hspace{0pt}\\
\label{ErrSing6Oct10}
As described above, as input to the FP, HM, and SSE method, we used the propagation direction ($\phi$) of the CME  estimated from the tie-pointing  method  of 3D  reconstruction. To examine the uncertainties arising from using the tie-pointing method, we compared the CME kinematics derived using propagation directions obtained from other methods.  We note that, for all CMEs that form part of this study, the propagation directions estimated using tie-pointing and forward modeling \citep{Thernisien2009} are within  10$\arcdeg$, which is  in agreement with  results of  \citet{Mierla2010}. Propagation directions estimated from CME source locations are within 10$\arcdeg$ of the values obtained from tie-pointing.  We repeated FP, HM, and SSE analysis, as described above, using propagation  directions  that are +10$\arcdeg$ and -10$\arcdeg$ different from the value ($\phi$) estimated using  tie-pointing to estimate uncertainties in the distance (vertical error bars in Figure~\ref{STAA6Oct10}). The speed's standard deviation (uncertainty) is calculated using the IDL $derivsig$ function.  These error bars do not denote all errors in these single spacecraft methods but simply represent the method's sensitivity to uncertainties in direction.

\subsubsection{Single spacecraft fitting methods for SECCHI/HI observations}
\label{Fitting6Oct10}
\paragraph{Fixed-phi fitting (FPF), Harmonic mean fitting (HMF) and Self-similar expansion fitting (SSEF) method} \hspace{0pt}\\
We used Equation~\ref{FPFeqn} (Chapter~\ref{Chap2:DataMthd}) to get the most physically realistic combinations of speed ($v_{FP}$), direction ($\phi_{FP}$), and launch time from Sun-centre ($t_{0FP}$), where $\alpha(t_{0FP})$ = 0, which can closely reproduce the elongation-time profiles of CMEs derived from the \textit{J}-maps.

We applied the FPF technique to the elongation variation of the 2010 October 6 CME derived from the \textit{STEREO-A} HI \textit{J}-map. We implemented the IDL routine MPFITFUN \citep{Markwardt2009} to find the set of $v_{FP}$, $\phi_{FP}$, and $t_{0FP}$ parameters that best reproduced the observed elongation variation. In the upper panel of Figure~\ref{FF_HMF6Oct10} (red line), we show how well the observed variation is produced by equation (1) of \citet{Rouillard2008}. We estimated the fitting residuals (bottom panels) by following the approach of \citet{Mostl2011}. Applying FPF to the initial front of the October 6 CME tracked in the 
\textit{STEREO-A} \textit{J}-map yields a propagation direction of 96.3$\arcdeg$ from the spacecraft (i.e., 13.3$\arcdeg$ East of the Sun-Earth line), a speed of 462 km s$^{-1}$, and a launch time of 07:38 UT on 2010 October 6. Assuming this speed is constant from the Sun to L1, the CME is predicted to arrive at L1 at 00:39 UT on 2010 October 10.

\begin{figure}[!htb]
\begin{center}
\includegraphics[angle=0,scale=.70]{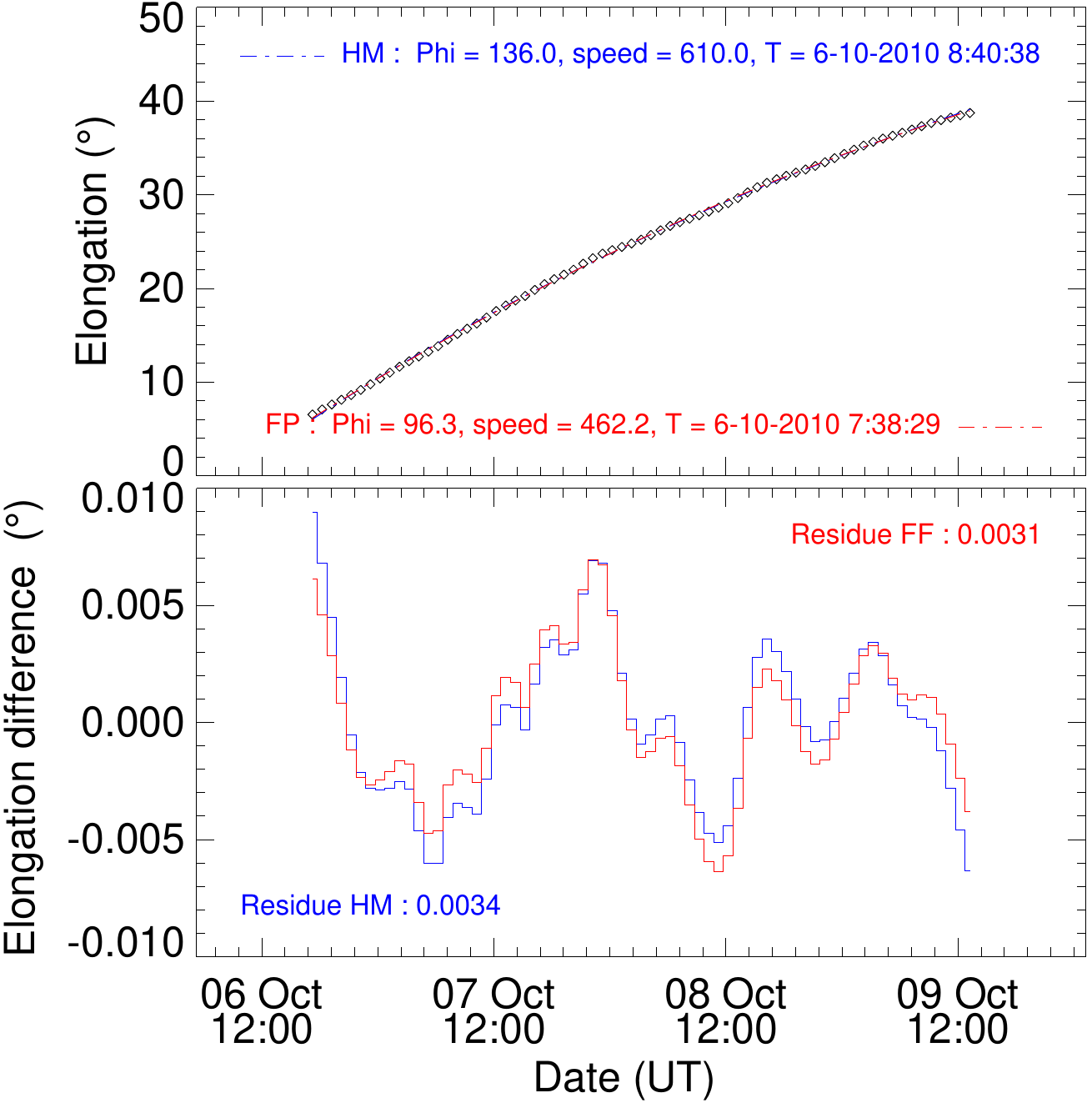} 
\caption[Best fit FPF and HMF results for the 2010 October 6 CME]{Best-fit FPF and HMF results for the 2010 October 6 CME are shown with red and blue colors, respectively, for the tracked CME feature. In the top panel, best-fit theoretically obtained elongation variations are shown. The bottom panel shows residuals between the best-fit theoretical elongation variations and the observed elongation variations.}

\label{FF_HMF6Oct10}
\end{center}
\end{figure}

Following the same basic procedure as for FPF, the best fit elongation variation, derived using the HMF approach, is shown in Figure~\ref{FF_HMF6Oct10} (upper panel) in blue. Applying the HMF technique to the time-elongation profile of the CME tracked in \textit{STEREO-A} gives a propagation direction of 136$\arcdeg$ from the spacecraft, i.e., 53$\arcdeg$  East of the Sun-Earth line, a speed of 610 km s$^{-1}$, and a launch time of 08:40 UT on 2010 October 6. This speed (when corrected for off-axis propagation, see below) gives a predicted L1 arrival time of 00:19 UT on October 11.

We fixed the value of half angular width (in our case to 30$\arcdeg$) and applied the SSEF technique to retrieve the best-fit speed, direction, and launch time in a similar manner to FPF and HMF. The best-fit parameters, and the estimated predicted L1 arrival time, for the October 6 CME based on SSEF are given in Table~\ref{Tab6Oct10}.

The FPF, HMF, and SSEF methods are also applied to the elongation variation of the CME extracted from the \textit{STEREO-B} \textit{J}-map. The retrieved best fit parameters (launch time, propagation direction from the observer, and speed) and arrival time at L1 are also given in Table~\ref{Tab6Oct10}.

We emphasize that the HMF and SSEF methods estimate the propagation speed of the CME apex and, to estimate its speed in an off-apex direction, a geometrical correction must be applied. The off-apex corrections applicable to the HM and SSE geometries are given by equation (8) of \citet{Mostl2011} and equation (18) of \citet{Mostl2013}, respectively. The speed of the CME in the off-apex direction is less than its speed derived in the apex direction. As the CME apex directions derived from both the HMF and SSEF technique are offset from the Sun-Earth line, we used the geometrically corrected speed to obtain the predicted arrival time of the CME at the L1 point. This corrected speed is compared later to the speed measured in situ at L1. Such a geometrical correction does not apply to the FPF technique, as the CME is assumed to be a point in this case.

\subsubsection{Stereoscopic reconstruction methods using SECCHI/HI observations}
\label{StrRecn6Oct10}
Several stereoscopic techniques have also been developed to determine the distance, speed, and direction profiles of CMEs based on simultaneous observations from the two viewpoints of \textit{STEREO}. In this section, we apply three such methods to determine the kinematics of the 2010 October 6 CME, namely the GT method \citep{Liu2010}, TAS method \citep{Lugaz2010.apj} and the SSSE method \citep{Davies2013}.

\paragraph{Geometric triangulation (GT) method} \hspace{0pt}\\
\label{GT6Oct10}
The use of stereoscopic HI (combined with COR2) observations to estimate the kinematics of CMEs was pioneered by \citet{Liu2010, Liu2010a}, who proposed the GT method. The details about this method have been explained in Section~\ref{GT} of Chapter~\ref{Chap2:DataMthd}. This triangulation technique has been applied to CMEs at different \textit{STEREO} spacecraft separation angles in studies that relate the imaging observations to near-Earth in situ measurements \citep{Liu2010, Liu2011, Liu2012, Liu2013, Mostl2010, Harrison2012, Temmer2012}. The elongation angle profiles for the October 6 CME,  extracted from the \textit{STEREO-A} and \textit{STEREO-B} ecliptic \textit{J}-maps, were interpolated onto a common time grid. We implemented the appropriate triangulation equations from \citet{Liu2010a} to obtain the kinematics of the CME. The derived distance, propagation direction, and speed profiles of the CME (the latter derived from adjacent distance points) are shown in Figure~\ref{STAABB6Oct10} in blue. The kinematic parameters obtained using the GT method at the sunward edge of the HI1 FOV are not shown in Figure~\ref{STAABB6Oct10} due to the occurrence of a singularity at those elongations (see \citealp{Liu2010, Mishra2013}).

\begin{figure}[!htb]
\begin{center}
\includegraphics[angle=0,scale=.50]{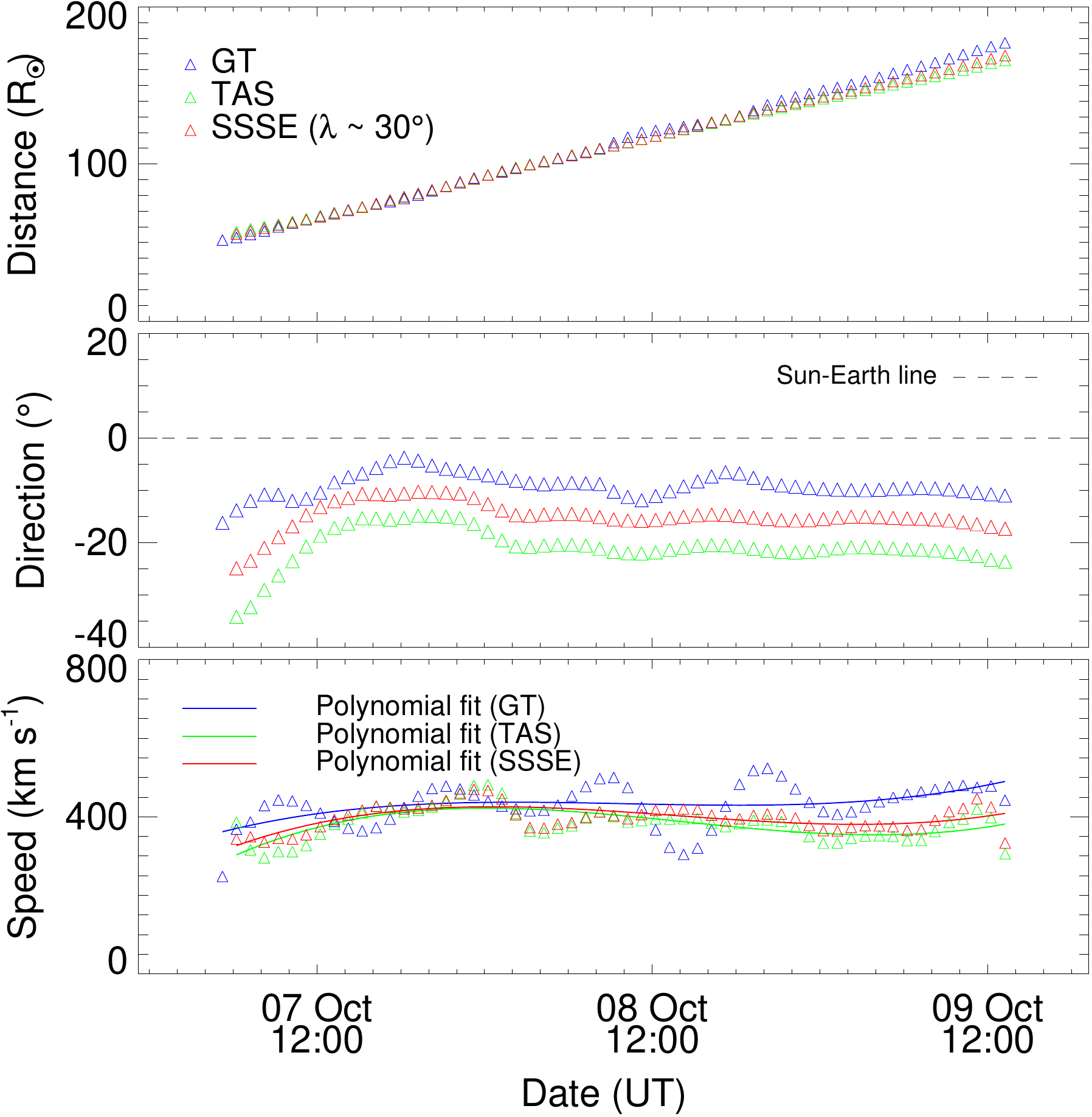} 
\caption[Distance, propagation direction, and speed profiles for the tracked feature of the 2010 October 6 CME using GT, TAS, and SSSE methods]{From top to bottom, panels show distance, propagation direction (relative to the Sun-Earth line), and speed profiles are shown for tracked CME feature as derived using the stereoscopic GT, TAS and SSSE methods. The horizontal line in the middle panel marks the Sun-Earth line.}

\label{STAABB6Oct10}
\end{center}
\end{figure}

We used the kinematics derived using the GT method as input to the DBM to predict the CME arrival time and speed at L1. To initiate the DBM, we used a CME speed of 470 km s$^{-1}$ (the average of the last few values) at a distance of 177 \textit{R}$_\odot$ (0.82 AU) at 13:14 UT on 2010 October 9. As before, the ambient solar wind speed was set to 350 km s$^{-1}$. The resultant L1 arrival times and speeds, corresponding to the extreme range of the drag parameter, are given in Table~\ref{Tab6Oct10}.

\paragraph{Tangent to a sphere (TAS) method} \hspace{0pt}\\
The details about the TAS method have been explained in Section~\ref{TAS} of Chapter~\ref{Chap2:DataMthd}. We apply equation (2) of \citet{Lugaz2010.apj} to estimate the propagation direction of the tracked CME using this technique. As in the previous section, we use the derived time-direction profiles to estimate the distance, hence the speed, profiles of the CME based on the expression for the radial distance of the transient's apex. The results are shown in Figure~\ref{STAABB6Oct10} in green.

The central panel of Figure~\ref{STAABB6Oct10} (green line) suggests that the CME is propagating slightly eastward of the Sun-Earth line. At the last point of its track (13:14 UT on October 9), the estimated CME distance and speed are 166 \textit{R}$_\odot$ (0.77 AU) and 385 km s$^{-1}$, respectively. The kinematics at the sunward edge of the HI1 FOV are not shown due to the occurrence of a singularity, as in the GT method. We applied the DBM exactly as discussed earlier; the results are shown in Table~\ref{Tab6Oct10}.

\paragraph{Stereoscopic self-similar expansion (SSSE) method} \hspace{0pt}\\
 The details about the SSSE method have been explained in Section~\ref{SSSE} of Chapter~\ref{Chap2:DataMthd}. We used the equations. 
(23), (24), and (4a) of \citet{Davies2013} to estimate the propagation direction from the observer and distance profiles for the tracked CME. Although this SSSE method can take any value of the half angular width ($\lambda$) of the CME between 0$\arcdeg$ and 90$\arcdeg$, we use a fixed value of 30$\arcdeg$. The obtained kinematics for the tracked feature is shown in 
Figure~\ref{STAABB6Oct10} in red. The kinematics for the points at the sun-ward edge of HI FOV are not shown due to significant errors. For these points, the sum of the elongation from both observers and the separation angle between the two observers is close to 180$\arcdeg$. As pointed out by \citet{Davies2013}, in such a situation, small errors in elongation will result in significant errors in direction, and hence in distance and speed, around the aforementioned singularity. The estimated kinematics at the end of the track are used as inputs in the DBM to predict the arrival times and speeds at L1 (Table~\ref{Tab6Oct10}).

\subsection{2010 April 3 CME}
\label{Mthds3Apr10}

We used the tie-pointing method of 3D reconstruction (scc$\_$measure: \citealt{Thompson2009}) on a selected feature along the leading edge of this CME and obtained its 3D kinematics. The 3D speed, latitude, and longitude were estimated as 816 km s$^{-1}$, 25$\arcdeg$ south and 5$\arcdeg$ east of the Earth, respectively, at the outer edge of the COR2 FOV. The kinematics of this fast CME seems to be partly influenced by the presence of high-speed solar wind, as the CME experiences slight deceleration during its journey from the Sun to 1 AU; during this time, the Earth was immersed in the fast solar wind. The kinematics of this CME have been studied extensively by several authors  \citep{Mostl2010, Wood2011,Liu2011}. In contrast to the 2010 October 6 event, this allows us to assess the accuracy of various methods for the case of a fast CME moving in a fast ambient solar wind.

\begin{figure}[!htb]
\begin{center}
\includegraphics[angle=0,scale=.60]{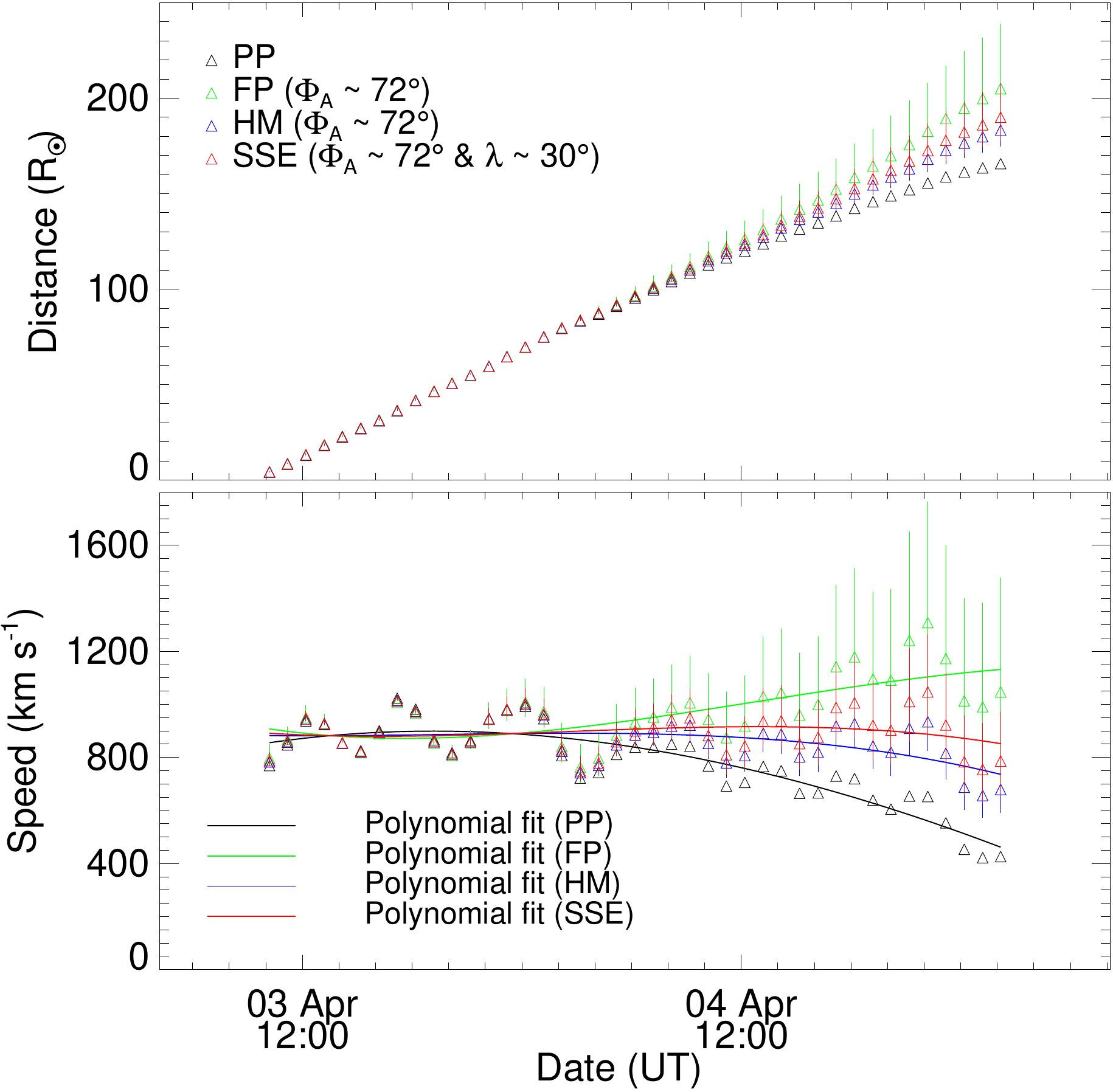} 
\caption{As Figure~\ref{STAA6Oct10}, for the 2010 April 3 CME.}
\label{STAA3Apr10}
\end{center}
\end{figure}

We constructed \textit{STEREO-A} and \textit{B} ecliptic \textit{J}-maps for this interval encompassing this CME using COR2 and HI images and extracted, from each, the time-elongation profile for the leading edge of the initial CME front. The Milky Way is visible in the HI2-B images. Therefore the CME signal is slightly difficult to track. This CME can be tracked out to 54.5$\arcdeg$ and 26.5$\arcdeg$ elongation in \textit{STEREO-A} and \textit{B} \textit{J}-maps, respectively. We implemented the seven single spacecraft methods (PP, FP, FPF, HM, HMF, SSE, and SSEF) and the three stereoscopic methods (GT, TAS, and SSSE) to derive the CME kinematics. Estimated kinematics from the PP, FP, HM, and SSE methods, applied to the time-elongation profile of the CME extracted from the \textit{STEREO-A} \textit{J}-map, is shown in Figure~\ref{STAA3Apr10}. Errors bars are calculated in the same manner as for the 2010 October 6 CME discussed in Section~\ref{ErrSing6Oct10}. We also estimated the kinematics of the CME based on the \textit{STEREO-B} \textit{J}-map. We used the kinematics from both spacecraft as inputs to the DBM to obtain the arrival time at L1 (given in Table~\ref{Tab3Apr10}). The kinematics of the tracked CME obtained using the stereoscopic methods are shown in Figure~\ref{STAABB3Apr10}. Again, results for the arrival time and speed at L1, based on the use of these kinematics as input to the DBM model, are included in Table~\ref{Tab3Apr10}. Also, results from FPF, HMF, and SSEF analysis are quoted (corrected for off-axis propagation for the HMF and SSEF cases). Results from equivalent single spacecraft fitting analyses of \textit{STEREO-B} data are also included. For this CME, we use an ambient solar wind speed of 550 km s$^{-1}$ in the DBM. Note that only the minimum value of the statistical range of the drag parameter is used because the fast ambient solar wind into which this CME was launched is characterized by a low density.

\begin{figure}[!htb]
\begin{center}
\includegraphics[angle=0,scale=.50]{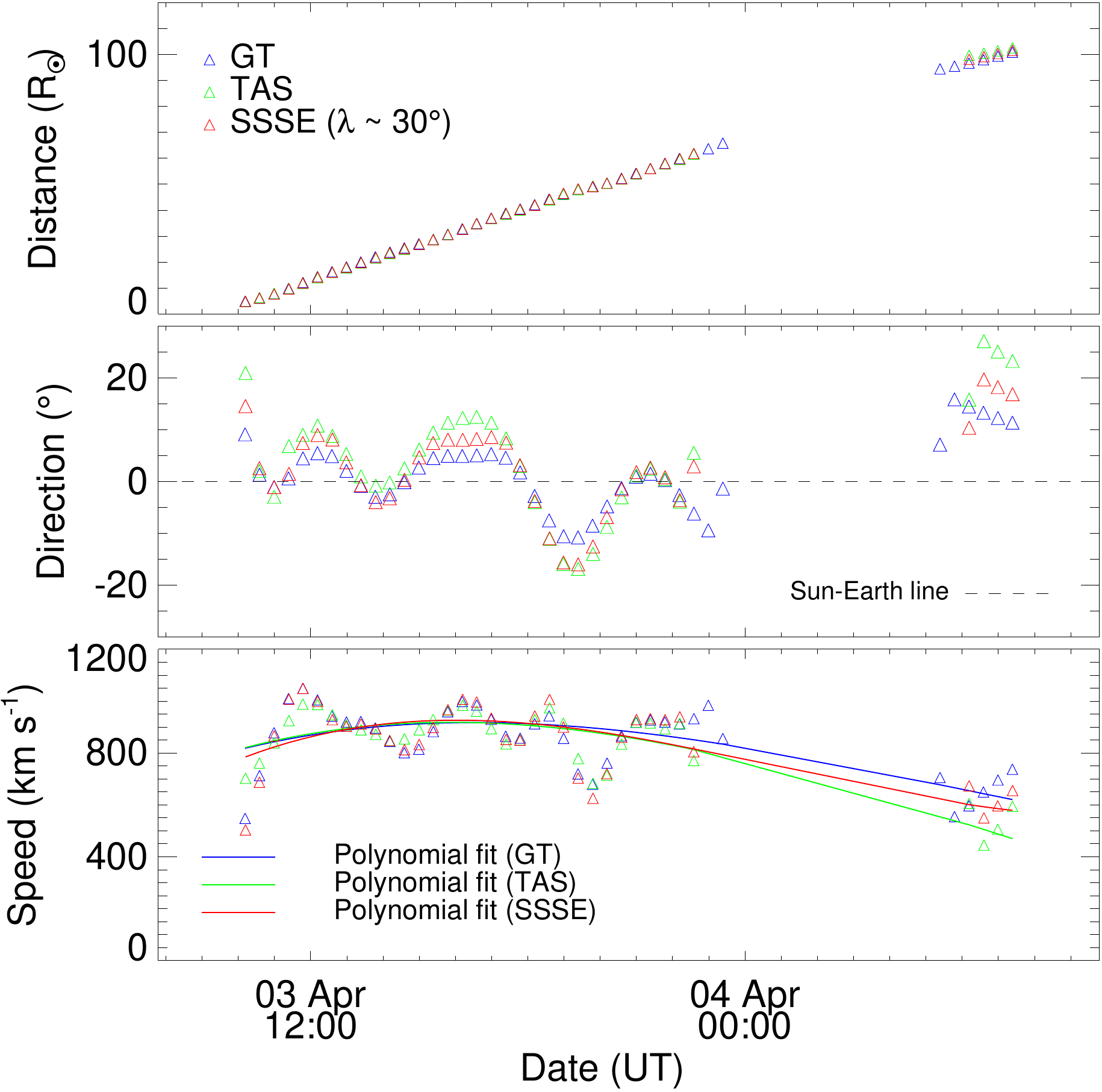} 
\caption{As Figure~\ref{STAABB6Oct10}, for the 2010 April 3 CME.}
\label{STAABB3Apr10}
\end{center}
\end{figure}

\begin{sidewaystable}
  \centering
{\scriptsize
 \begin{tabular}{ p{3.0cm}|p{2.0cm}| p{3.0cm}| p{2.5cm}|p{2.5cm}|p{2.5cm}}
    \hline
		
 Method & Kinematics as inputs in DBM [t$_{0}$, R$_{0}$ (\textit{R}$_\odot$), v$_{0}$ (km s$^{-1})$]& Predicted arrival time using kinematics + DBM (UT) [$\gamma$ = 0.2 (10$^{-7}$ km$^{-1}$)] & Predicted transit speed at L1 (km s$^{-1}$)  [$\gamma$ = 0.2 (10$^{-7}$ km$^{-1}$)] & Error in predicted arrival time (hrs) [$\gamma$ = 0.2 (10$^{-7}$ km$^{-1}$)] &  Error in predicted speed (km s$^{-1}$)  [$\gamma$ = 0.2 (10$^{-7}$ km$^{-1}$)]  \\  \hline

PP (\textit{STEREO-A}) & 05 April 02:11, 165, 426  & 05 April 23:14 & 446 	& 11.2 	& -274  \\ \hline

PP (\textit{STEREO-B}) & 04 April 05:00, 85, 866 & 05 April 12:14 &  735 &  0.2  &  15  \\  \hline

FP (\textit{STEREO-A}) &  04 April 12:11, 122, 800  & 05 April 11:14 & 727 	& -0.7 & 7  \\ \hline

FP (\textit{STEREO-B}) & 04 April 04:59, 85, 790  &  05 April 14:20 & 702  & 2.3  & -18\\ \hline

HM (\textit{STEREO-A}) & 05 April 02:11, 183, 660  & 05 April 10:53  & 653 	& -1.1 & -67   \\ \hline

HM (\textit{STEREO-B})  & 04 April 05:00, 85, 810 &  05 April 13:44 & 711  & 1.7 &  -9   \\ \hline

SSE (\textit{STEREO-A}) & 05 April 02:11, 190, 800  & 05 April 07:42  & 777 	& -4.3 & 57   \\ \hline

SSE (\textit{STEREO-B}) & 04 April 05:00, 85, 820  & 05 April 13:27  & 716 	& 1.5 & -4   \\ \hline

GT	       & 04 April 07:23, 101, 640    &  05 April 17:35	& 624	&  5.5 & -96  \\  \hline

TAS       &  04 April 07:23, 103, 580  &  05 April 19:57	 & 578 	& 7.9 & -142    \\  \hline
SSSE       & 04 April 07:23, 101, 615  &  05 April 19:51	& 574 	& 7.8  &  -146    \\  \hline		
			
 \end{tabular}

\begin{tabular}{p{3.0cm}| p{3.0cm}| p{2.5cm}|p{2.5cm}|p{2.5cm}| p{2.0cm}}
   
\multicolumn{6}{c}{Time-elongation track fitting methods} \\  \hline
   
 Methods   & Best fit parameters [t$_{(\alpha = 0)}$, $\Phi$ ($\arcdeg$), v (km s$^{-1}$)]  &   Predicted arrival time at L1 (UT)  & Error in predicted arrival time  & Error in predicted speed at L1 (km s$^{-1}$) & Longitude ($\arcdeg$) 
   \\ \hline
	
	FPF (\textit{STEREO-A})   & 03 April 08:47, 63.4, 865 & 05 April 08:18 &  -3.7 &  145      &  4  \\  \hline
	FPF  (\textit{STEREO-B})  & 03 April 09:07, 86.8, 886 & 05 April 07:30 & -4.5  &   166     &  16    \\  \hline
	 
	HMF (\textit{STEREO-A})   & 03 April 09:22, 78.5, 908  & 05 April 07:32 & -4.5  &   171      &   -11  \\  \hline
	HMF  (\textit{STEREO-B})  & 03 April 09:11, 103.5, 928 & 05 April 13:38 &  1.5  &    67    &   32 \\  \hline
	
	SSEF (\textit{STEREO-A})  & 03 April 09:11, 71, 889 & 05 April 07:39   & -4.3  &   163  &  -4  \\  \hline
	SSEF (\textit{STEREO-B})  & 03 April 09:10, 95, 907  & 05 April 17:29  & 5.5  &  12  &   24    \\  \hline

\end{tabular}
}
\caption[As Table~\ref{Tab6Oct10}, for the 2010 April 3 CME]{{The predicted arrival times and speeds (and errors therein) at L1 for the 2010 April 03 CME. Details as in the caption of Table~\ref{Tab6Oct10}.}}
\label{Tab3Apr10}

\end{sidewaystable}

\subsection{2010 February 12 CME}
\label{Mthds12Feb10}

We carried out a 3D reconstruction of a selected feature along the leading edge of this CME using the scc$\_$measure procedure \citep{Thompson2009}, from which the 3D speed, latitude, and longitude of the CME at the outer edge of COR2 FOV were estimated to be 867 km s$^{-1}$, 5$\arcdeg$ north and 10$\arcdeg$ east of the Earth, respectively. The heliospheric kinematics of this geo-effective (D$_{st}$ = -58 nT) CME (Figure~\ref{KinGT12Feb10}) shows that this fast CME continuously decelerated throughout its journey beyond the COR2 FOV to 1 AU. This CME allows us to test the efficacy of various methods to predict the arrival time of a fast CME decelerating in a slow ambient solar wind. We tracked this CME out to 48$\arcdeg$ and 53$\arcdeg$ in the \textit{STEREO-A} and B \textit{J}-maps constructed using COR2 and HI observations, respectively. We apply all the reconstruction methods to this CME as for the October 06 CME (Section~\ref{Mthds6Oct10}). Results of the PP, FP, HM, and SSE methods applied to the time-elongation profile extracted from the \textit{STEREO-A} \textit{J}-map is shown in Figure~\ref{STAA12Feb10}. Errors in distance and speed (marked with  vertical  lines) are calculated in the same manner as for the previous CMEs. Results of the stereoscopic GT, TAS, and SSSE techniques are shown in Figure~\ref{STAABB12Feb10}. Again, we do not show points near the singularity for the stereoscopic methods. The estimated kinematics used as input in the DBM, and the resultant predicted arrival times and transit speeds, are given in Table~\ref{Tab3Apr10}. Results of the single-spacecraft fitting methods (FPF, HMF, and SSEF), applied to elongation profiles from both \textit{STEREO-A} and \textit{STEREO-B}, are quoted in the bottom panel of Table~\ref{Tab12Feb10}.

\begin{figure}
\begin{center}
\includegraphics[angle=0,scale=.50]{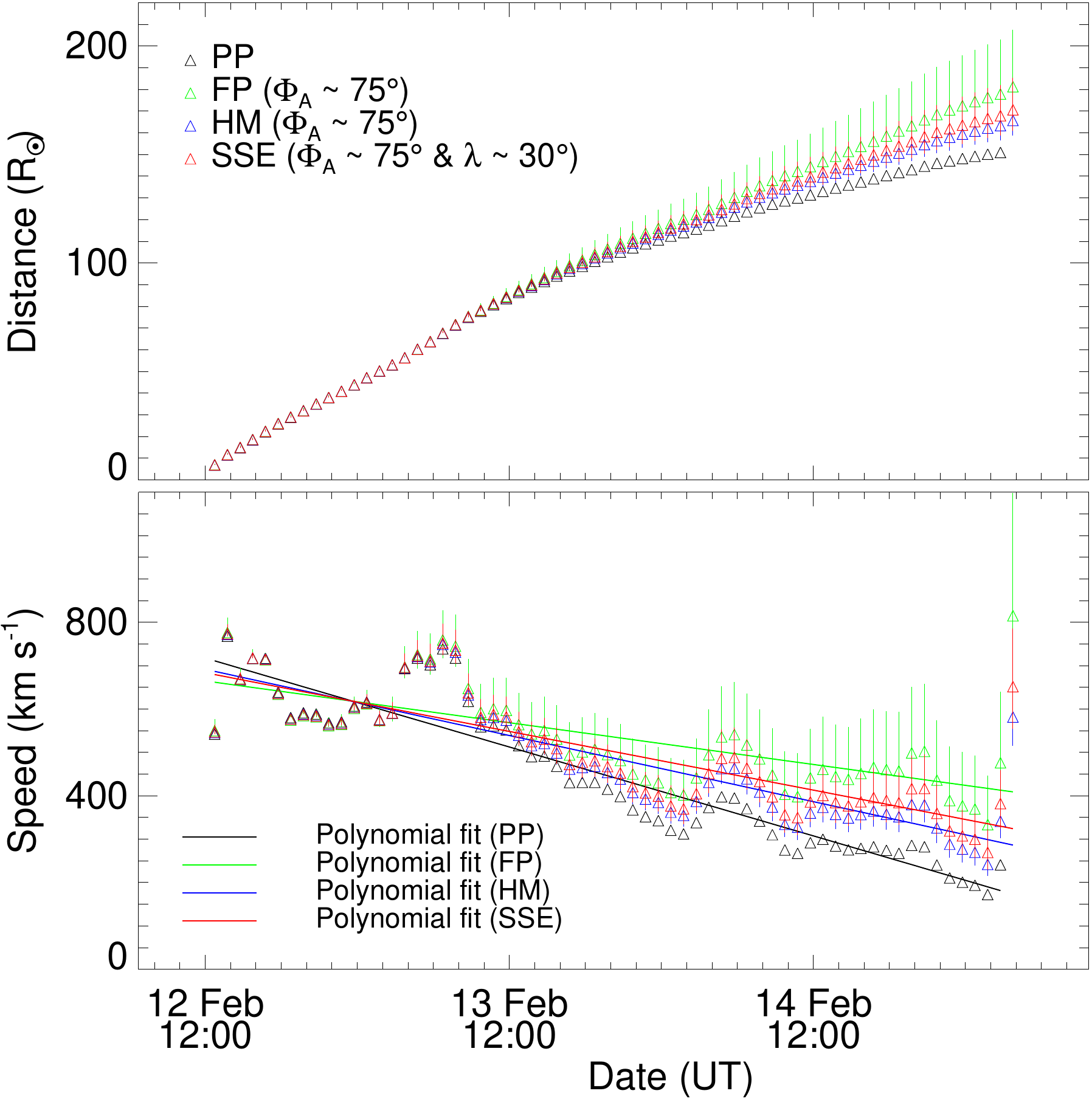}
\caption{As Figure~\ref{STAA6Oct10}, for the 2010 February 12 CME.}
\label{STAA12Feb10}
\end{center}
\end{figure}

\begin{figure}
\begin{center}
\includegraphics[angle=0,scale=.50]{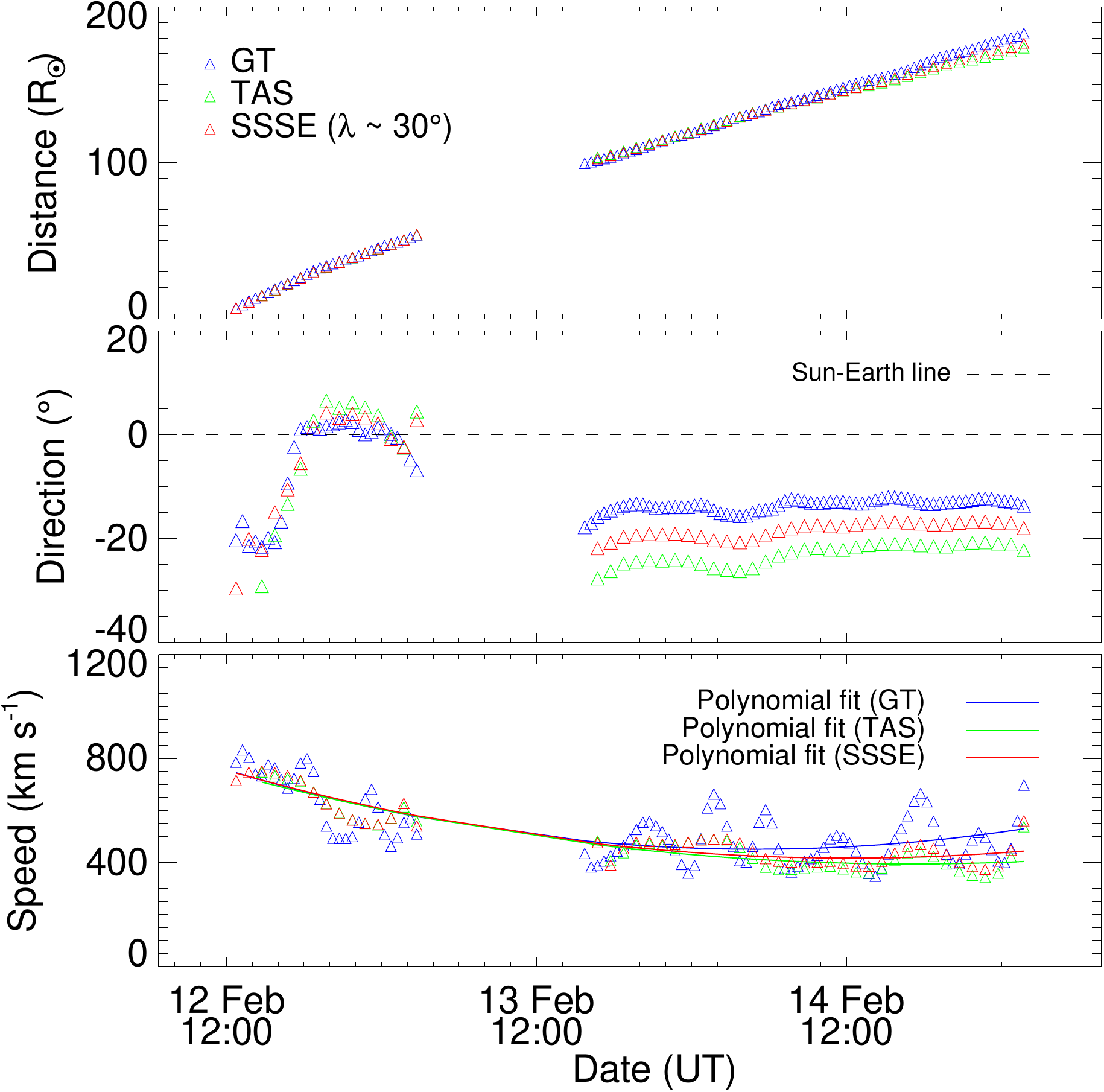}
\caption{As Figure~\ref{STAABB6Oct10}, for the 2010 February 12 CME.}
\label{STAABB12Feb10}
\end{center}
\end{figure}

\begin{sidewaystable}
  \centering
{\scriptsize
 \begin{tabular}{p{3.0cm}|p{2.0cm}| p{3.0cm}| p{2.5cm}|p{2.5cm}|p{2.5cm}}
    \hline
		
 Method & Kinematics as inputs in DBM [t$_{0}$, R$_{0}$ (\textit{R}$_\odot$), v$_{0}$ (km s$^{-1})$] &  Predicted arrival time using kinematics + DBM (UT) [$\gamma$ = 0.2 to 2.0 (10$^{-7}$ km$^{-1}$)] & Predicted transit speed at L1 (km s$^{-1}$)  [$\gamma$ = 0.2 to 2.0 (10$^{-7}$ km$^{-1}$)] & Error in predicted arrival time (hrs) [$\gamma$ = 0.2 to 2.0 (10$^{-7}$ km$^{-1}$)] &  Error in predicted transit speed (km s$^{-1}$)  [$\gamma$ = 0.2 to 2.0 (10$^{-7}$ km$^{-1}$)] \\  \hline

PP (\textit{STEREO-A})& 15 Feb 03:44, 152, 270  & 16 Feb 21:46 to 16 Feb 17:46	 &  286 to 325 & 	22.4 to 18.4 &  -34 to 5  \\ \hline

PP (\textit{STEREO-B})	& 15 Feb 01:43, 172, 330  & 16 Feb 01:26  to 16 Feb 01:16 &  331 to 335   & 	2.2 to 2	 &  11 to 15\\ \hline
  
FP (\textit{STEREO-A})& 15 Feb 03:44, 181, 400 &  15 Feb 19:02 to 15 Feb 19:23  & 	397 to 382 &	-4.3 to -4.8	& 77 to 62 \\ \hline

FP (\textit{STEREO-B})& 15 Feb 01:43, 187, 500 & 15 Feb 11:45 to 15 Feb 12:43 & 	485 to 419 &	-11.5 to -10.2	& 165 to 99 \\ \hline
  
HM (\textit{STEREO-A})	& 15 Feb 03:44, 165, 300 & 16 Feb 10:06 to 16 Feb 08:50 & 305 to 326 &	10.8 to 9.5 &  -15 to 6 \\ \hline

HM (\textit{STEREO-B})	& 15 Feb 01:43, 179, 420   &  15 Feb 17:15 to 15 Feb 17:52 & 415 to 389  	 & -6 to -5.5 &  95 to 69 \\  \hline

SSE (\textit{STEREO-A})	& 15 Feb 03:44, 171, 390 & 16 Feb 00:23 to 16 Feb 00:48 & 388 to 375 &	1.1 to 1.5 &  68 to 55 \\ \hline

SSE (\textit{STEREO-B})	& 15 Feb 01:43, 182, 450 & 15 Feb 14:58 to 15 Feb 15:48 & 441 to 400 &	-8.2 to -7.4 &  121 to 80 \\ \hline

GT         &  15 Feb 01:43, 183, 450  &  15 Feb 14:35 to 15 Feb 15:20 & 442 to 401  &  -8.7 to -7.9 & 122 to 81  \\  \hline

TAS       &  15 Feb 01:43,  174, 365  &  15 Feb 22:08 to 15 Feb 22:12 & 365 to 362 &   -1.1 to -1.0 &  45 to 42  \\  \hline
SSSE      & 15 Feb 01:43, 176, 400   &   15 Feb 19:12 to 15 Feb 19:39	& 397 to 380 	&  -4.0 to -3.6	 & 77 to 60   \\  \hline		

 \end{tabular}

\begin{tabular}{p{3.0cm}| p{3.0cm}| p{2.5cm}|p{2.5cm}|p{2.5cm}| p{2.0cm}}
\multicolumn{6}{c}{Time-elongation track fitting methods} \\  \hline
  
 Methods   & Best fit parameters [t$_{(\alpha = 0)}$, $\Phi$ ($\arcdeg$), v (km s$^{-1}$)]  &   Predicted arrival time at L1 (UT)   & Error in predicted arrival time  & Error in predicted speed at L1 (km s$^{-1}$) & Longitude ($\arcdeg$) 
   \\ \hline
	
	FPF (\textit{STEREO-A})   & 12 Feb 10:47, 93, 710 & 14 Feb 20:43 & -26.5  &  390     &  -28   \\  \hline
	FPF  (\textit{STEREO-B})  & 12 Feb 10:34, 77.7, 667 & 15 Feb 00:10  & -23  &   347   &  7 \\  \hline
	 
	HMF (\textit{STEREO-A})   & 12 Feb 11:18, 132, 926 & 17 Feb 08:16 &  33   &    41    &  -67  \\  \hline
	HMF  (\textit{STEREO-B})  & 12 Feb 11:19, 105.7, 764  & 15 Feb 04:46  & -18.5  &   305    &   35 \\  \hline
	
	SSEF (\textit{STEREO-A})  & 12 Feb 11:07, 111.7, 803 &  ---   &  ---   &   ---     &     -47   \\  \hline
	SSEF (\textit{STEREO-B})  & 12 Feb 11:04, 91.3, 714  & 15 Feb 05:37  &  -17.6  &  300   &     20 \\  \hline

\end{tabular}
}
\caption[As Table~\ref{Tab6Oct10}, for the 2010 February 12 CME]{The predicted arrival times and speeds (and errors therein) at L1 for the 2010 February 12 CME. Details as in the caption of Table~\ref{Tab6Oct10}.}
\label{Tab12Feb10}
\end{sidewaystable}

\section{Identification of Tracked CME Features Using In Situ Observations Near the Earth}

We have tracked the leading edge of the initial intensity front of the Earth-directed 2010 October 6 CME, in \textit{J}-maps derived from \textit{STEREO}/HI images and derived its kinematics based on several techniques. The in situ observations of this CME is shown in Figure~\ref{insitu6Oct10} of Chapter~\ref{Chap4:Associa}. The feature of this CME tracked in the \textit{J}-maps (Figure~\ref{Jmaps6Oct10}) corresponds to the leading edge of the initial, curved CME-associated intensity front (Figure~\ref{Evolution6Oct10}). Hence, its arrival is likely associated with LE. Thus 05:50 UT on 2010 October 11, is considered the actual arrival time of the remotely sensed feature. Furthermore, the in situ speed of the CME is approximately 355 km s$^{-1}$ at L1. Table~\ref{Tab6Oct10} summarizes the differences between the range of predicted and actual (in situ at L1) arrival times and speeds for each method.

We also identified the 2010 April 03 CME in the in situ data taken at L1. The arrival times of the shock and the CME leading and trailing boundaries, based on plasma and magnetic field signatures, are marked in Figure~\ref{insitu3Apr10} of Chapter~\ref{Chap4:Associa}. The arrival time and speed of the in situ signature, which is thought to be associated with the feature tracked in the \textit{STEREO}/HI data, are 12:00 UT on 2010 April 05 and 720 km s$^{-1}$, respectively. The in situ arrival time and speed are used to compute errors in the predicted values from each method (Table~\ref{Tab3Apr10}).

We also identified the 2010 February 12 CME in the near-Earth in situ data (see Figure~\ref{insitu12Feb10} of Chapter~\ref{Chap4:Associa}). The arrival time of this CME at L1 is considered to be 23:15 UT on 2010 February 15, and its speed is 320 km s$^{-1}$. These values are used as a reference to compute the errors in the predicted arrival times and speeds at L1 given in 
Table~\ref{Tab12Feb10}.

\section{Results and Discussion}
\label{ResDis32}

We constructed \textit{J}-maps using SECCHI/HI and COR2 images (including the latter for only two of the CMEs) to extract time-elongation profiles for three selected CMEs to analyze their kinematics. We implemented a total of ten reconstruction methods, which include four single spacecraft methods (PP, FP, HM, and SSE), three single spacecraft fitting methods (FPF, HMF, and SSEF), and three stereoscopic methods (GT, TAS, and SSSE). We examined the relative performance of these methods.

For the CMEs of October 6 and February 12, the arrival time and speed predictions are more accurate for all methods if the maximum value of the drag parameter is used in the DBM (see Tables~\ref{Tab6Oct10} and ~\ref{Tab12Feb10}). This is possibly due to the fact that these CMEs are less massive, have a large angular width, and are propagating in a dense solar wind environment \citep{Vrsnak2013}. From our study of three CMEs based on the ten aforementioned techniques, we find that there are significant errors involved in estimating their kinematic properties (up to 100 km s$^{-1}$ in speed) using HI data. Using 3D speeds determined in the COR2 FOV in the Sun-Earth direction may be preferable for reliable and advanced space weather forecasting, particularly for slow CMEs propagating in the slow solar wind.

It is worth noting that the significant contribution to arrival time errors arises due to the limitations of the methods themselves. However, implementation of the DBM may also contribute to these errors. It is also important to point out that the selected CMEs in our study propagate within $\pm$ 20$\arcdeg$ of the Sun-Earth line, so, strictly, an off-axis correction is also required for the HM, SSE, TAS, and SSSE methods before using the resultant speed as the input to the DBM. However, for these CMEs, we estimate that such a correction would decrease the speed by only a few km s$^{-1}$ and hence increase the travel time by only a few tens of minutes. We expect that if the final estimated speed of each tracked CME were taken as constant for the rest of its journey to L1, then the errors in predicted arrival time would be similar to that obtained from using the DBM with the minimum drag parameter. This is because the CMEs are tracked out to a significant fraction of 1 AU ($\approx$ 0.5 to 0.8 AU), and beyond this distance, a small drag force will have little effect on the CME dynamics.

Our assessment of the relative performance of various methods is based mainly on the difference between the CME arrival time predicted at L1 by different reconstruction methods and the ``actual'' arrival time determined in situ. We do not perform a detailed comparison of the kinematic profiles derived using each method. Of course, different kinematics profiles can lead to the same arrival time.  Moreover, it is unlikely that the same part of a CME tracked in remote imaging observations will pass through the spacecraft meant for in situ measurements. We cannot advocate, with confidence, the superiority of one method over others based only on the accuracy of arrival time predictions. It would be useful to compare the speed profile of a CME derived in the inner heliosphere using 3D  MHD modeling with that obtained using HI observations. Of course, results from MHD models also need to be considered with caution.

\subsection{Relative performance of single spacecraft methods}
\label{PerfrmSin}

In our study, we assess the performance of various reconstruction techniques based on the obtained difference between the predicted and actual arrival times of three CMEs. Of the four single spacecraft methods that do not rely on a curve-fitting approach, the PP method gives the largest range of errors (up to 25 hr) in the predicted arrival time of all CMEs. This is perhaps due to its oversimplified geometry. At large elongations (beyond $\approx$ 120 \textit{R}$_\odot$), for the CMEs of October 6 and February 12, the speed estimated using the PP technique is less than the ambient solar wind speed, which is unphysical. For the October 6 CME, the PP, FP, HM, and SSE approaches produce roughly similar errors in predicted arrival time (up to 25 hr) and transit speed (up to 100 km s$^{-1}$). For the CME of April 3, the HM and PP methods provide the most and least accurate predictions of arrival time at L1, respectively. For the CME of February 12, the SSE method provides the most precise L1 arrival time, while the PP method is the least accurate of these four methods.

The kinematics of the April 3 CME estimated using the FP method (Figure~\ref{STAA3Apr10}: green track) show a sudden unphysical late acceleration,  possibly due to its real deflection. We suggest that the FP, HM, and SSE methods, as implemented here, can give accurate results if the estimated speeds tend to a constant value far from the Sun. In the FP method, the tracked feature is assumed to correspond to the same point moving in a fixed radial direction, which is unlikely to be valid for a real CME structure \citep{Howard2011}. One major drawback of the FP method is that it does not take into account the finite cross-sectional extent of a CME. In terms of the four single viewpoint methods that enable estimation of the kinematics properties as a function of time, the HM and SSE methods provide a more accurate arrival time prediction for this CME. Of course, the assumption of a circular front in these methods may not be valid due to possible flattening of the CME front resulting from its interaction with the structured coronal magnetic field and solar wind ahead of the CME \citep{Odstrcil2005}. Also, the assumption made here in implementing the HM and SSE methods (and indeed the FP method) that the CME propagates along a fixed radial trajectory (in particular one derived close to the Sun), ignoring real or ``artificial'' heliospheric deflections, will induce errors, particularly for slow speed CMEs that are more likely to undergo deflection in the interplanetary medium \citep{Wang2004, Gui2011}. As noted previously, ``artificial deflection'' means that the observer does not detect the same feature of CME in consecutive images due to a well-known geometrical effect. We conclude that the implausible deceleration of October 6 CME to a speed less than that of the ambient solar wind speed (see Figure~\ref{STAA6Oct10}) is due to violations of the assumptions inherent in the PP, HM, and SSE methods at elongations beyond 30$\arcdeg$.

Irrespective of the event, the PP, FP, HM, and SSE methods estimate significantly different radial distance and speed profiles after approximately 100 R$_{\odot}$. This is because the assumed geometry impacts the results with increasing elongation. For the PP and FP methods, implausible acceleration or deceleration evident beyond approximately 100 R$_{\odot}$, if assumed to be real, would lead to unrealistically large errors in arrival time prediction. Therefore, among the single spacecraft methods, we suggest that methods like HM and SSE should be used to achieve reasonable arrival time predictions.

The value of the propagation direction adopted for each CME in our implementation of the FP, HM, and SSE techniques will affect the performance of each method; this is an important issue in our study.  The quoted  CME arrival times  in Tables  1, 2, and 3 based on the FP, HM, and  SSE techniques are based on a direction estimated from tie-pointing. To assess the sensitivity of our results to the exact value of the propagation direction used, we have repeated our analysis using a range of propagation directions.  As described in Section~\ref{ErrSing6Oct10}, we repeated our FP, HM, and SSE analyses using propagation directions that are +10$\arcdeg$ and -10$\arcdeg$ different from the values ($\phi$) estimated using tie-pointing.  We used these revised kinematic profiles to estimate the arrival time and transit speed of the selected CMEs in our study for $\phi$ $\pm$ 10$\arcdeg$.

For the October 6 CME analyzed using the FP method, using $\phi$+10 ($\phi$-10) resulted in predicted arrival times that are 21 hr later (earlier) for \textit{STEREO-A} and 10 hr earlier (later) for \textit{STEREO-B} than the arrival time predicted using $\phi$.  For the HM method, using $\phi$+10 ($\phi$-10) resulted in a predicted arrival time 12 hr later (earlier) for \textit{STEREO-A} and 6 hr earlier (later) for \textit{STEREO-B}. For the SSE method,  using $\phi$+10 ($\phi$-10) results in a predicted arrival time 16 hr  later (earlier) for \textit{STEREO-A} and 7 hr earlier (later) for \textit{STEREO-B}. For the 2010 April 3 CME, the deviation in arrival time from that quoted in Table 2 is less than 5 hr for \textit{STEREO-A} and less than 3 hr for \textit{STEREO-B} for all the three single spacecraft methods for  $\phi$+10. Assuming a propagation direction equal to $\phi$-10 yields the same uncertainties as for $\phi$+10, except for the FP  method applied to \textit{STEREO-A} where the uncertainty increases to 8 hr. For the 2010 February 12 CME, assuming that the propagation direction is $\phi$+10 ($\phi$-10) in FP analysis results in predicted arrival times 9 hr later (14 hr earlier) for \textit{STEREO-A} and 4 hr earlier (6 hr later) for \textit{STEREO-B} than the values quoted in Table 3 that are based on using $\phi$ directly from tie-pointing. In the case of the HM and SSE methods, using $\phi$+10 ($\phi$-10) results in predicted arrival times that are less than 9 hr later (earlier) for \textit{STEREO-A} and less than 6 hr earlier (later) for \textit{STEREO-B} than the values quoted in Table 3.

The uncertainties discussed above do not reflect the total errors involved in implementing the FP, HM and SSE methods. The uncertainties in the distance and speed due to a change in propagation direction are not significant at smaller elongations \citep{Wood2009, Howard2011}. Therefore, for a case like that of the April 3 CME as observed by \textit{STEREO-B}, in particular,  where the CME cannot be tracked out far in elongation angle, any uncertainty in propagation direction will have a minimal effect on the derived kinematic profile and also the predicted arrival time. At greater elongations, however, this effect, along with the ``well-known effect of CME geometry'', will severely limit the accuracy of these methods. Given the effects of uncertainty in propagation direction in the FP, HM, and SSE methods, we note that it may be better to combine the DBM with CME kinematics derived in the near-Sun HI1 FOV to optimize the goal of space weather prediction, at least for CMEs launched into a slow speed ambient solar wind medium.

\subsection{Relative performance of stereoscopic methods}
\label{Perfrmtwn}
For the October 6 CME, the stereoscopic GT method provides a large range of arrival time errors (up to 25 hr), while the stereoscopic TAS method predicts arrival time within 17 hr. For this CME, the errors resulting from the application of the SSSE method are intermediate between those from the GT and TAS methods. For the February 12 CME, among the three stereoscopic methods, the TAS method provides the best prediction of L1 arrival time (within 2 hr of the in situ arrival) and transit speed (within 45 km s$^{-1}$ of the in situ speed). For the April 3 CME, all of the stereoscopic methods give approximately the same arrival time errors (within 8 hr).

As in the FP (and FPF) techniques, the assumption in GT that the same point of a CME is being observed in consecutive images is likely to become increasingly invalid with increasing elongation. GT also assumes that the same point is observed simultaneously from both viewpoints. Moreover, the effect of ignoring the Thomson scattering geometry is minimized for Earth-directed events. Therefore deviations from such a configuration will also result in errors in the estimated kinematics. From our analysis, which is limited to three CMEs, we conclude that TAS technique performs most accurately, and GT performs least accurately in estimating CME arrival times and speeds among the three stereoscopic methods. The predicted SSSE arrival time is within a couple of hours of that predicted by the TAS technique. As the estimated kinematics properties from the SSSE method (here implemented with an angular half-width of 30$\arcdeg$) are intermediate between the kinematics derived from the GT and TAS methods (see Figure~\ref{STAABB6Oct10}, \ref{STAABB3Apr10}, \ref{STAABB12Feb10}), we are tempted to suggest that the SSSE method may be preferable for space weather forecasting if a reasonable estimate of a half angular width of a CME is available. In all the stereoscopic methods used, any effects due to the Thomson scattering geometry are ignored, and the assumption of self-similar expansion \citep{Xue2005} may also result in errors. However, one needs to quantify the potential errors due to ignoring real effects for each method, over different elongation ranges, and for different spacecraft separation angles, before concluding the unbiased superiority of the TAS technique.

\subsection{Relative performance of single spacecraft fitting methods}
\label{Perfrmfitng}
Results from the single spacecraft fitting techniques (FPF, HMF, and SSEF) suggest that the October 6 CME propagates eastward of the Sun-Earth line. All three methods give roughly the same launch time for the CME. Using \textit{STEREO-A} observations, the error in the predicted CME arrival time at L1 is least (within 5 hr) for HMF and largest (within 30 hr) for FPF. Arrival time errors derived from \textit{STEREO-B} observations are similar for all three methods ($\approx$ 22 hr). For the fast CME of 2010, April 3, which did not decelerate noticeably, arrival time errors are small (within 5 hr) for all three fitting methods. For this CME, SSEF predicts most accurately the arrival time at L1 while FPF is least accurate. Note that the estimated longitudes by these fitting methods are up to 40$\arcdeg$ from the Sun-Earth line for both the October 6 and April 3 CMEs. We consider these CMEs to be closely Earth-directed. For the fast, decelerating CME of 2010 February 12, arrival time errors from these single spacecraft fitting techniques are very large (18 to 33 hr), and the estimated CME longitudes (Table~\ref{Tab12Feb10}) can be more than 70$\arcdeg$ from the Sun-Earth line. Large errors in the results by applying these fitting methods to slow or decelerating CMEs are most likely due to a breakdown in their inherent assumptions of constant speed and direction. The predicted arrival times, errors therein, and errors in transit speed resulting from the application of the SSEF technique to the \textit{STEREO-A} profiles are not shown for the October 6 and April 3 CMEs. In these cases, the CME is not predicted to hit an in situ spacecraft at L1, based on the retrieved propagation direction. Note that all of the fitting methods reproduce the observed elongation track well, so we must be cautious about relying on the fitted parameters to consider one method superior.

In terms of the three fitting methods (FPF, HMF, and SSEF), we find that HMF and SSEF (applied with $\lambda$ = 30$\arcdeg$) predict arrival time and transit more accurately speed at L1 than does the FPF method. For all three CMEs, the propagation
direction derived from applying these fitting techniques separately to the \textit{STEREO-A} and \textit{STEREO-B} elongation variations show little consistency (Tables~\ref{Tab6Oct10}, ~\ref{Tab3Apr10} $\&$ ~\ref{Tab12Feb10}). However, our study suggests that CME propagation direction is best obtained using the FPF  method while it is worst from the HMF method.  \citet{Lugaz2010} has shown that  the  FPF  method can give  significant errors in propagation direction when a CME is propagating at an angle beyond 60$\arcdeg$ $\pm$ 20$\arcdeg$ from the Sun-spacecraft line; this is not in agreement with our findings. We also note that in the case of the fast 2010 February 12 CME, where the assumption of a constant CME speed is not valid, i.e.  a physical deceleration is observed, the propagation direction estimated from all fitting methods is highly erroneous. These elongation profile fitting approaches have the potential to give better results if features are tracked out to large elongations ($\approx$ 40$\arcdeg$) and the manual selection of points is done with extreme care \citep{Williams2009}. 

\section{Conclusion}
\label{Conclu}

We have studied the kinematics of eight CMEs, by exploiting the STEREO COR2 and HI observations (Section~\ref{ArrtimGT}).
3D Speeds of the selected CMEs in our study range from low ($\approx$ 335 km s$^{-1}$) to high ($\approx$ 870 km s$^{-1}$) at the exit of coronagraphic (COR2) FOV. In our study, we obtained a good agreement (within $\approx$ 100 km s$^{-1}$) between the speed
calculated using the tie-pointing method of 3D reconstruction and the speed derived by implementing the GT method using \textit{J}-maps in the COR2 FOV. The difference in the speeds using the two techniques can occur because of the estimation of speeds of different features at different latitudes. We conclude that the use of the GT method on HI data combined with DBM gives a better prediction of the CME arrival time than using only the 3D speed estimated in COR FOV. We also show that by estimating 3D speed in COR FOV and assuming that it remains constant up to L1, the arrival time cannot be predicted correctly for most CMEs. Therefore, longer tracking of CMEs using HIs observations is necessary for an improved understanding of the evolution and consequences of CMEs in the heliosphere. Combining the estimated kinematics from the GT method with DBM, errors of predicted arrival time range from 3-9 hr and that of transit speed near 1 AU range from 25-120 km s$^{-1}$.

We have applied a total of ten reconstruction methods to three Earth-directed CMEs observed by \textit{STEREO} 
(Section~\ref{Mthds3CMEs}). These three CMEs having different speeds are launched into different ambient solar wind environments. We found that stereoscopic methods are more accurate than single spacecraft methods to predict CME arrival times and speeds at L1. Irrespective of the characteristics of the CMEs, among the three stereoscopic methods, the TAS method gives the best prediction of transit speed (within a few tens of km s$^{-1}$) and arrival time (within 8 hr for fast CMEs and 17 hr for slow or fast decelerating CMEs).

We also find that the HM method (based on a propagation direction retrieved from the 3D reconstruction of COR2 data) performs best among the single spacecraft techniques. For the selected fast speed CME with no apparent deceleration, the HM method provides the best estimate of the predicted L1 arrival time (within 2 hr) and speed (within 60 km s$^{-1}$). However, for the fast but decelerating CME, this method predicts arrival time to around 10 hr, and this increases to $\approx$ 20 hr for the slow CME in our study.

Independent of the characteristics of the CMEs, our study shows that the HMF and SSEF single spacecraft fitting methods perform better than FPF. All three fitting methods provide reasonable arrival time predictions (within 5 hr of the arrival time identified in situ) for the fast speed CME that undergoes no discernible deceleration. For the slow CME and the fast but decelerating CME, the fitting methods are only accurate to 10 to 30 hr in terms of their arrival time prediction and yield relatively larger errors (up to hundreds of km s$^{-1}$) in predicted speed.

In summary, the HI imagers provide an opportunity for us to understand the association between remotely observed CME structures and in situ observations. Our study demonstrates the difficulties inherent in reliably predicting CME propagation direction and arrival time and speed at 1 AU based on such remote-sensing observations. From our research, we conclude that, although HIs provide the potential to improve the space weather forecasting for slow or decelerating CMEs, the specific assumptions in some of the currently-used 3D reconstruction methods compromise the estimates of CMEs kinematics and hence result in deviations from the actual arrival time at Earth.

\chapter{Association between Remote and In Situ Observations of CMEs}
\label{Chap4:Associa}
\rhead{Chapter~\ref{Chap4:Associa}. Association between Remote and In Situ Observations of CMEs}

\section{Introduction}
\label{introd4}

A classic CME imaged near the Sun displays the so-called three-part structure, i.e., a bright leading edge followed by
a dark cavity and finally a bright core \citep{Illing1985}. If a CME travels faster than the characteristic speed of the ambient medium, then it drives a shock \citep{Gopalswamy1998a}. The identification of CMEs in in situ spacecraft observations is carried out using various plasma properties of CMEs \citep{Zurbuchen2006}. Further, due to a lack of information regarding the evolution of a CME during its propagation between the Sun and Earth, and the process by which it manifests itself in the ambient solar wind, sometimes, its identification in in situ observations is difficult. It is believed that the leading edge, which appears bright due
to the sweeping up of coronal plasma by erupting flux ropes or the presence of pre-existing material in the overlying fields 
\citep{Riley2008}, is identified near the Earth in the in situ observations as the CME sheath region (the disturbed region in
front of the leading edge of the CME; \citealp{Forsyth2006}). The darker region is assumed to correspond to a flux rope structure
having a large magnetic field and a low plasma density and is identified as a magnetic cloud (or MC) \citep{Klein1982} in in situ observations \citep{Burlaga1991}. In a classical sense, an MC is a plasma and magnetic field structure that shows an enhanced magnetic field, a rotation in the magnetic field vector, a low plasma density, and temperature, and a plasma $\beta$ of less than
unity \citep{Burlaga1981, Lepping1990,Zurbuchen2006}. \citet{Gosling1990} have shown that 30\% of ICMEs during 1978-1982 were MCs. However, \citet{Bothmer1996}, using the in situ data of ICMEs from Helios spacecraft and the remote data of CMEs from the coronagraph onboard P78/1 spacecraft during 1979-1981, have concluded that approximately 41\% of the ejecta are MCs.  Further, analyzing the Helios data, \citet{Cane1997} has shown that approximately 60\% of ejecta are MCs. In addition,  they used the observations of Helios 1, Helios 2, and IMP 8 and showed that some ejecta observed at one spacecraft show signatures of MC but not at the other. Therefore, they concluded that MCs are a substructure of ejecta (ICME), and the signatures of ICMEs depend on where the ejecta is intercepted by in situ spacecraft. The inner-most bright feature (the CME core) has been observed in H-$\alpha$ which indicates its cooler temperature and corresponds to a solar filament \citep{Schwenn1980, Burlaga1998}. However, such cool and dense structures are rarely observed in in situ observations \citep{Lepri2010}. Hence, an association of remote observations to in situ observations of CMEs remains difficult even today. Since different structures of a CME have different characteristics and result in a different response to Earth. Therefore identification and prediction of their arrival time to Earth is a prime concern for a solar-terrestrial physicist.

\textit{STEREO} Heliospheric Imager (HI) \citep{Eyles2009} era has proved to be a boon for a solar-terrestrial physicist in improving the understanding of the propagation of the Earth-directed CME. However, identification of different features of CMEs in the heliosphere by continuous tracking using \textit{J}-maps and the prediction of their arrival time at 1 AU has been achieved with limited accuracy only \citep{Liu2010, Howard2012a,Liu2013, Mishra2013,Mishra2014a, Mishra2014b}. This is because of the challenges in extracting the faint Thomson scattered signal from the brighter background signals dominated by instrumental stray light, F-corona, and background star-field \citep{Eyles2009}. The estimation of kinematics by using derived elongations from \textit{J}-maps have been carried out using various reconstruction methods in Chapter~\ref{Chap3:ArrTim}. These methods are based on several assumptions on the geometry and evolution of CMEs, and therefore, even if tracked continuously out to larger elongations, the predicted CME arrival time is not very accurate. The errors in the predicted arrival time of CMEs which may be due to sampling of different features in remote and in in situ observations, can lead to incorrect interpretations.

It has often been challenging to associate CME features imaged near the Sun with features observed in situ. This was because of the large distance gap between the Sun and the location of in situ spacecraft and the difficulty in characterizing the true
evolution of the remotely-sensed features. However, using HI observations, it is now possible to relate the
near Sun remote observations and near-Earth in situ spacecraft observations of the CMEs. In this chapter, we describe the in situ observations of several selected CMEs, whose remote sensing observations have been described in Chapter~\ref{Chap3:ArrTim}. Among all the studied CMEs in Chapter~\ref{Chap3:ArrTim}, interestingly, a CME of 2010 October 6 shows two bright tracks in the \textit{J}-maps. In Section~\ref{LETE6Oct}, we have tracked these two bright tracks, which are probably at the front and rear edge of 2010 October 6 CME, in imaging observations. The obtained kinematics and arrival times of both tracked features have been associated with the density enhanced structures observed in in situ observations. We attempt to find an association between the three-part structures of CME as seen in the COR field of view (FOV) with features observed in HI and in in situ observations. The tracking and estimation of different CME features' kinematics can also help investigate different forces possibly acting on them.

\section{In Situ Observations of CMEs}
In this section, we describe the in situ observations of the CMEs presented in Chapter~\ref{Chap3:ArrTim}. The in situ identification criteria based on several magnetic, plasma, and compositional signatures of the CMEs are described in Section~\ref{insitu} of Chapter~\ref{Chap1:IntMot}. CME is an expanding plasma structure with enhanced magnetic field with lesser fluctuation, lower temperature, and enhanced alpha to proton ratio than ambient solar wind plasma.    

\subsection{2008 December 12 CME}
We plotted the in situ measured parameters of solar wind in Figure~\ref{insitu12Dec08} and identified 2008 December 12 CME by a combination of various signatures \citep{Zurbuchen2006}. The figure shows the predicted arrival time and transit velocity at \textit{WIND} in situ spacecraft situated at the L1 point. These predicted values are estimated with the extreme values of the average range of the drag parameter used in the Drag Based Model (DBM), as discussed in Section~\ref{arrtim12Dec08} of Chapter~\ref{Chap3:ArrTim}. In Figure~\ref{insitu12Dec08}, the first dashed vertical line (red) marks the arrival of the sheath at 11:55 UT on December 16, and the second dashed vertical line (red) marks the arrival of the leading edge of a magnetic cloud \citep{Klein1982} at 04:39 UT on December 17. The third dashed vertical line (red) marks the trailing edge of a magnetic cloud at 15:48 UT on 2008 December 17. The hatched region (blue) shows the predicted arrival time (with uncertainties due to the range of values of the drag parameter adopted in DBM) using the DBM. In the fourth panel, two red horizontal lines mark the predicted transit velocities at L1 of the tracked feature corresponding to different values of the drag parameter used in the DBM. In the third panel (from the top), the red curve shows the expected proton temperature calculated from the observed in situ proton speed \citep{Lopez1986, Lopez1987}.

\begin{figure*}[!htb]
	\centering
		\includegraphics[width=0.65\textwidth,height=0.55\textheight]{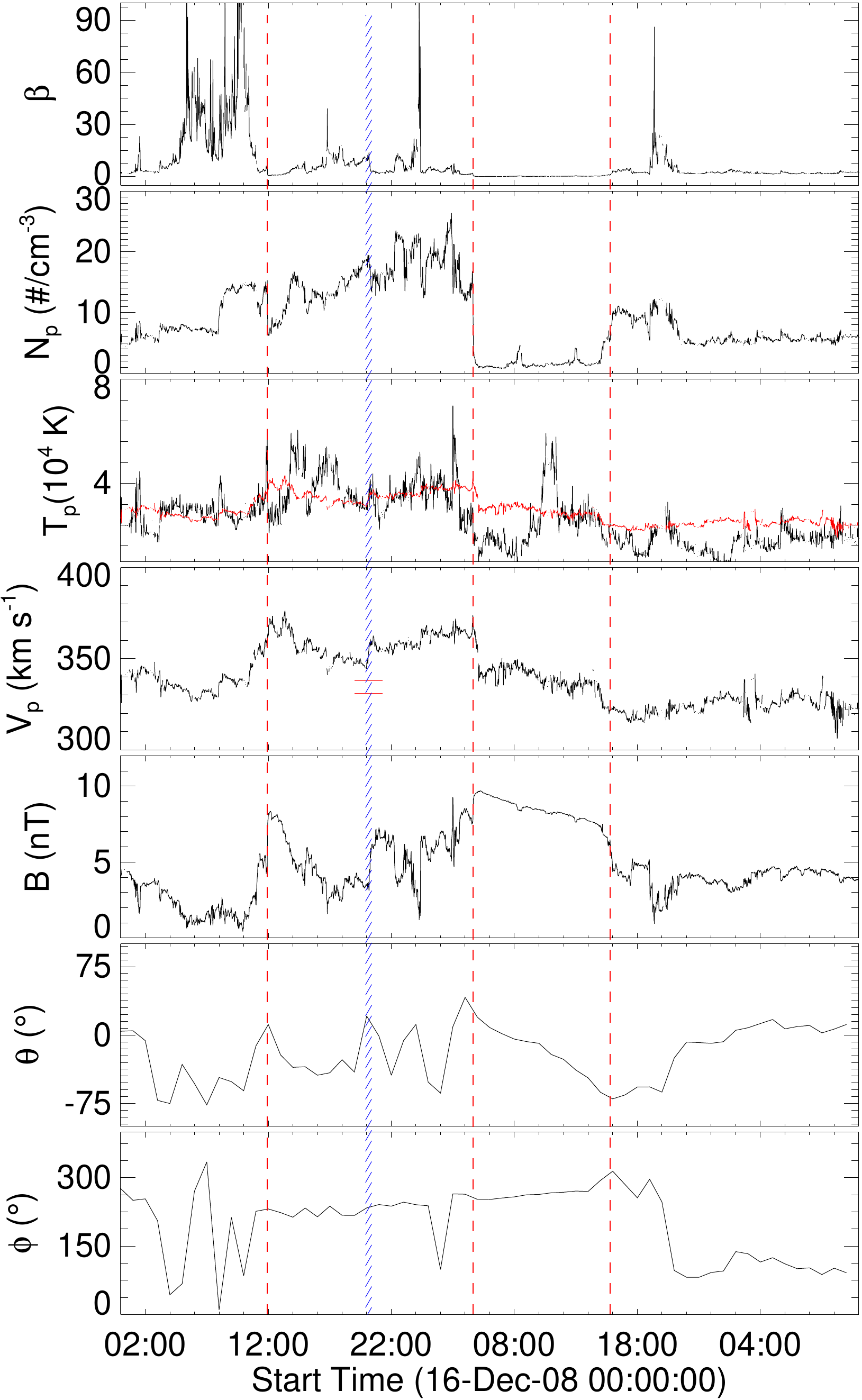}
\caption[In situ measurements of plasma and magnetic field parameters of the 2008 December 12 CME]{From top to bottom, panels show the variation of plasma beta, proton density, proton temperature, proton velocity, the magnitude of the magnetic field, and latitude and longitude of the magnetic field vector, respectively, corresponding to the CME of 2008 December 12. From the left, the first, second, and third vertical dashed lines (red) mark the arrival time of the CME sheath and the
leading and trailing edges of a magnetic cloud, respectively. The hatched region (blue) marks the interval of the predicted arrival time of the tracked feature. In the third panel from the top, the expected proton temperature is shown as a red curve, and in the fourth-panel horizontal lines (red) mark the predicted velocities of the tracked feature at L1.}
\label{insitu12Dec08}
\end{figure*}

\subsection{2010 February 7 CME}
The in situ observations of this CME is shown in Figure~\ref{insitu7Feb10}. The first vertical dashed line (red) at 01:00 UT on February 11 marks the arrival of a weak shock or CME sheath, the second vertical dashed line (red) at 12:47 UT marks the arrival of the CME leading edge, and the third vertical dashed line (red) at 23:13 UT marks the trailing edge of the CME. The hatched
region (blue) shows the predicted arrival time with uncertainties due to extreme values of the drag parameter with the same inputs
from the estimated kinematics employed in the DBM.

\begin{figure*}[!htb]
	\centering
		\includegraphics[width=0.65\textwidth,height=0.55\textheight]{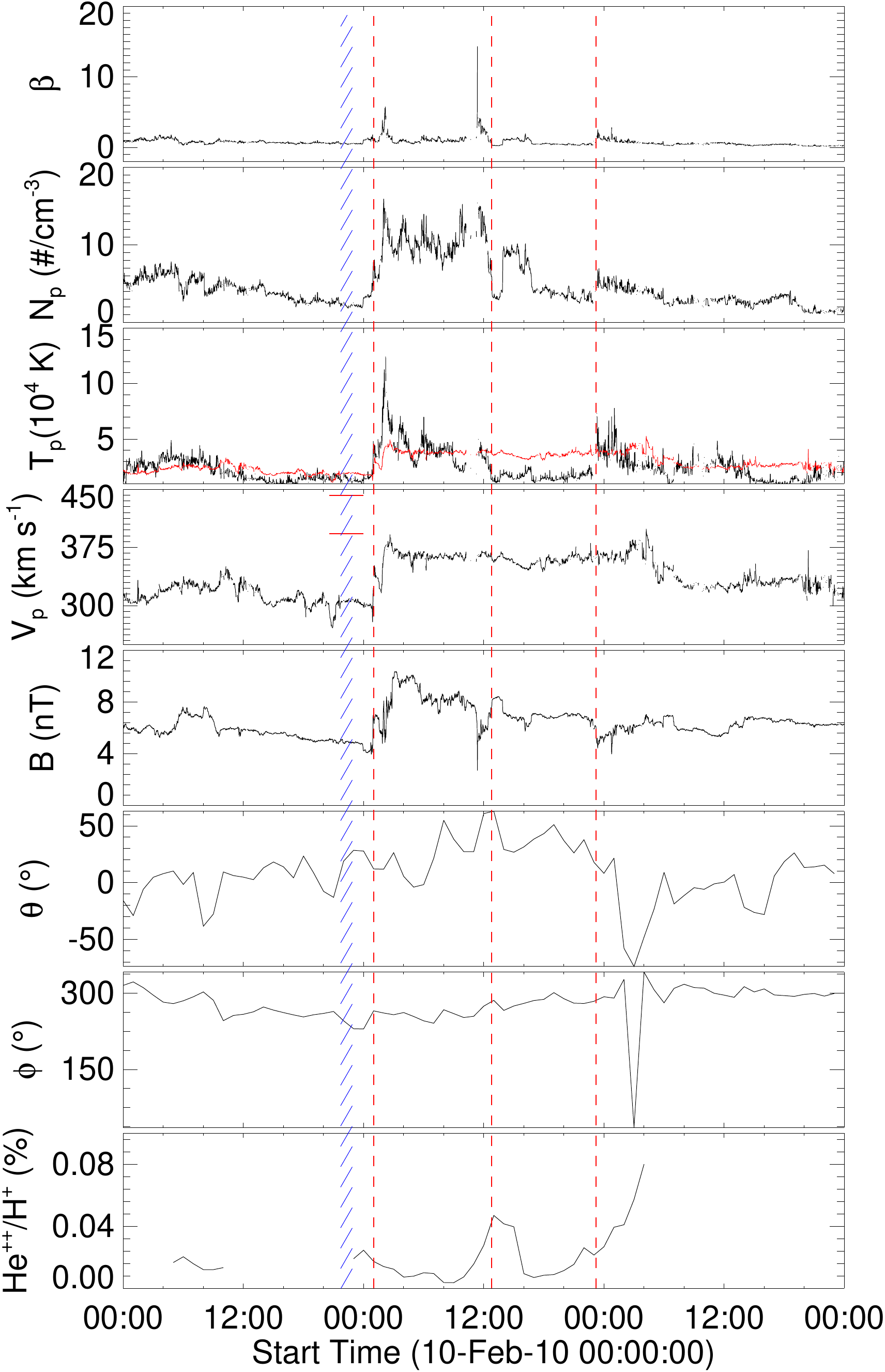}
\caption[As Figure~\ref{insitu12Dec08}, for the 2010 February 7 CME]{The panels show plasma parameters as described in Figure~\ref{insitu12Dec08} except the bottom panel shows alpha to proton ratio, corresponding to the CME of 2010 February 7.}
\label{insitu7Feb10}
\end{figure*}

\subsection{2010 February 12 CME}

The identification of the CME near the Earth is studied by analyzing the in situ data (Figure~\ref{insitu12Feb10}). In this figure, the first vertical dashed line (red) at 18:42 UT on February 15 marks the arrival of shock; the second dashed vertical line (red) at 04:32 UT on February 16 marks the arrival of the CME leading edge, and the third vertical dashed line (red) at 12:38 UT on February 16 marks the CME trailing edge. The hatched region (blue) marks the predicted arrival time (with uncertainties due to extreme values of the drag parameter used in the DBM) estimated by incorporating kinematics parameters combined with the DBM.

\begin{figure*}[!htb]
	\centering
		\includegraphics[width=0.65\textwidth,height=0.55\textheight]{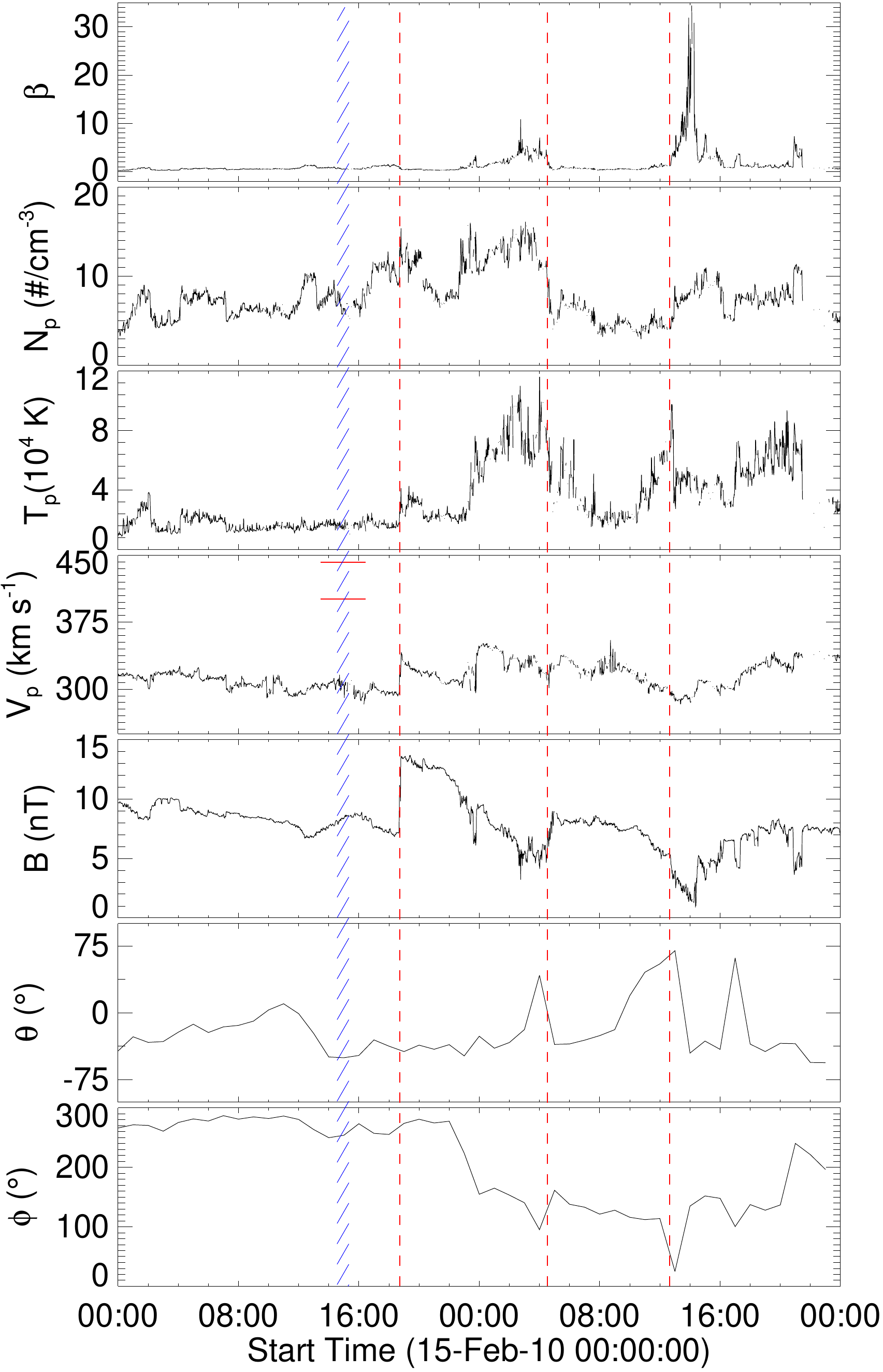}
\caption[As Figure~\ref{insitu12Dec08}, for the 2010 February 12 CME]{The panels show same plasma parameters as in Figure~\ref{insitu12Dec08}, corresponding to the CME of 2010 February 12.}
\label{insitu12Feb10}
\end{figure*}

\subsection{2010 March 14 CME}

The in situ observations for this CME is shown in Figure~\ref{insitu14Mar10}. In this figure, the first vertical dashed line (red) at 21:19 UT on March 17 marks the arrival of the CME leading edge, and the second vertical dashed line (red) at 11:26 UT on March 18
marks the arrival of the CME trailing edge. The hatched region (blue) shows the predicted arrival time of CME with the range of uncertainty due to extreme values of the range of the drag parameter employed in the DBM.

\begin{figure*}[!htb]
	\centering
		\includegraphics[width=0.65\textwidth,height=0.55\textheight]{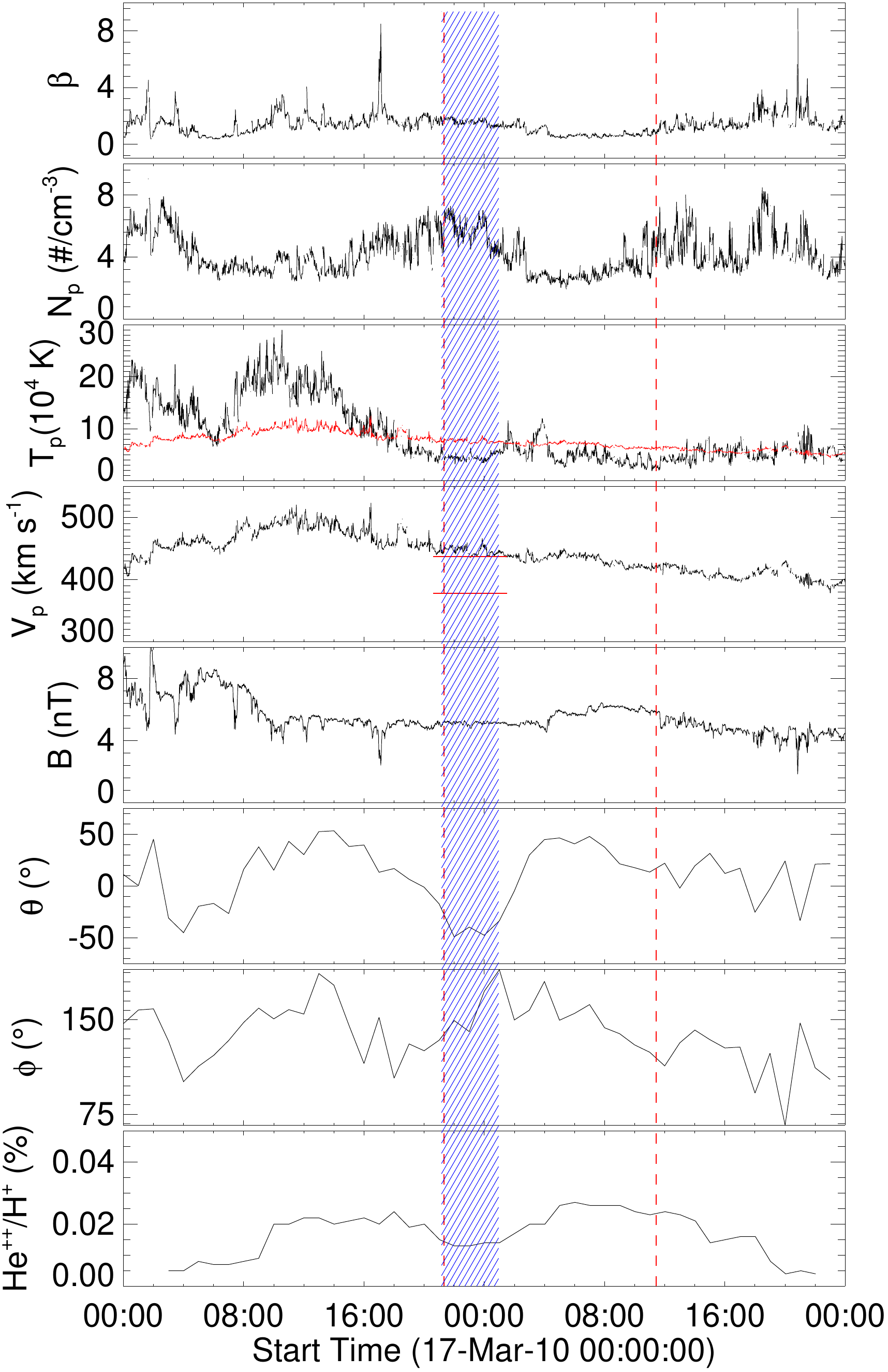}
\caption[As Figure~\ref{insitu12Dec08}, for the 2010 March 14 CME]{The panels show same plasma parameters as in Figure~\ref{insitu7Feb10}, corresponding to the CME of 2010 March 14.}
\label{insitu14Mar10}
\end{figure*}

\subsection{2010 April 3 CME}

We identified the CME in situ observations which are shown in Figure~\ref{insitu3Apr10}. Here, the first vertical dashed line (red) marks the arrival of a shock at 8:28 UT on April 5, the second vertical dashed line (red) marks the arrival of the CME leading edge at
13:43 UT and the fourth vertical dashed line (red) marks the CME trailing edge at 16:05 UT on April 6. The third vertical
dashed line (blue) marks the predicted arrival time of the CME obtained after the estimated dynamics employed in the DBM.
In the fourth panel from the top, the horizontal line (red) marks the predicted transit velocity of the CME at L1.

\begin{figure*}[!htb]
	\centering
		\includegraphics[width=0.65\textwidth,height=0.55\textheight]{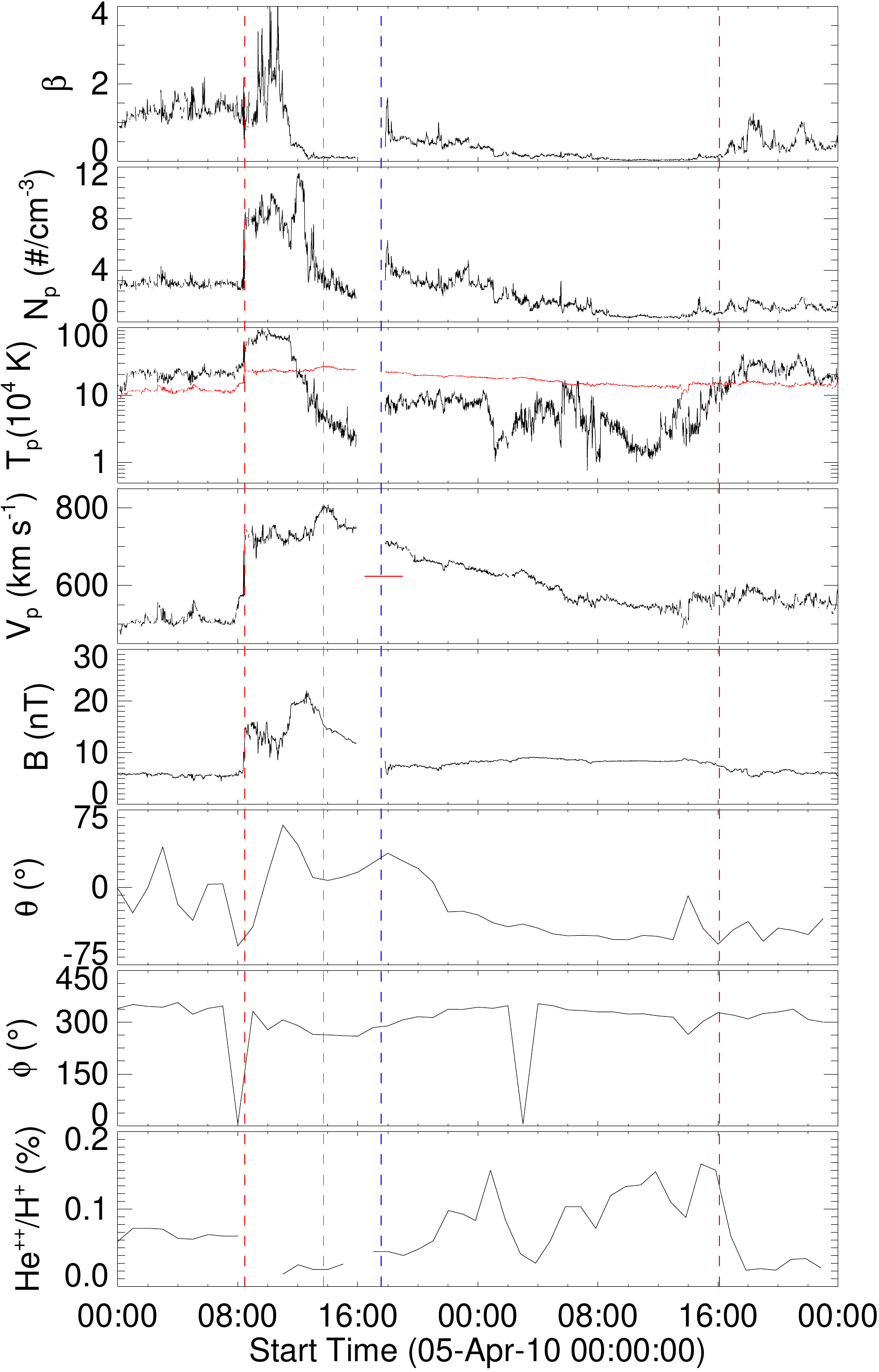}
\caption[As Figure~\ref{insitu12Dec08}, for the 2010 April 3 CME]{The panels show same plasma parameters as in Figure~\ref{insitu7Feb10}, corresponding to the CME of 2010 April 3.}
\label{insitu3Apr10}
\end{figure*}

\subsection{2010 April 8 CME}

By analyzing the in situ data taken nearly at 1AU, the identification of the CME boundary is shown in Figure~\ref{insitu8Apr10}. This figure marks a weak shock or sheath by the first dashed vertical line (red) at 12:44 UT on April 11. The leading and trailing edges of a magnetic cloud are marked by the second and third vertical dashed lines (red) at 02:10 UT and 13:52 UT on April 12, respectively. The blue hatched region shows the predicted arrival time of the CME (with uncertainties) obtained
using DBM.

\begin{figure*}[!htb]
	\centering
		\includegraphics[width=0.65\textwidth,height=0.55\textheight]{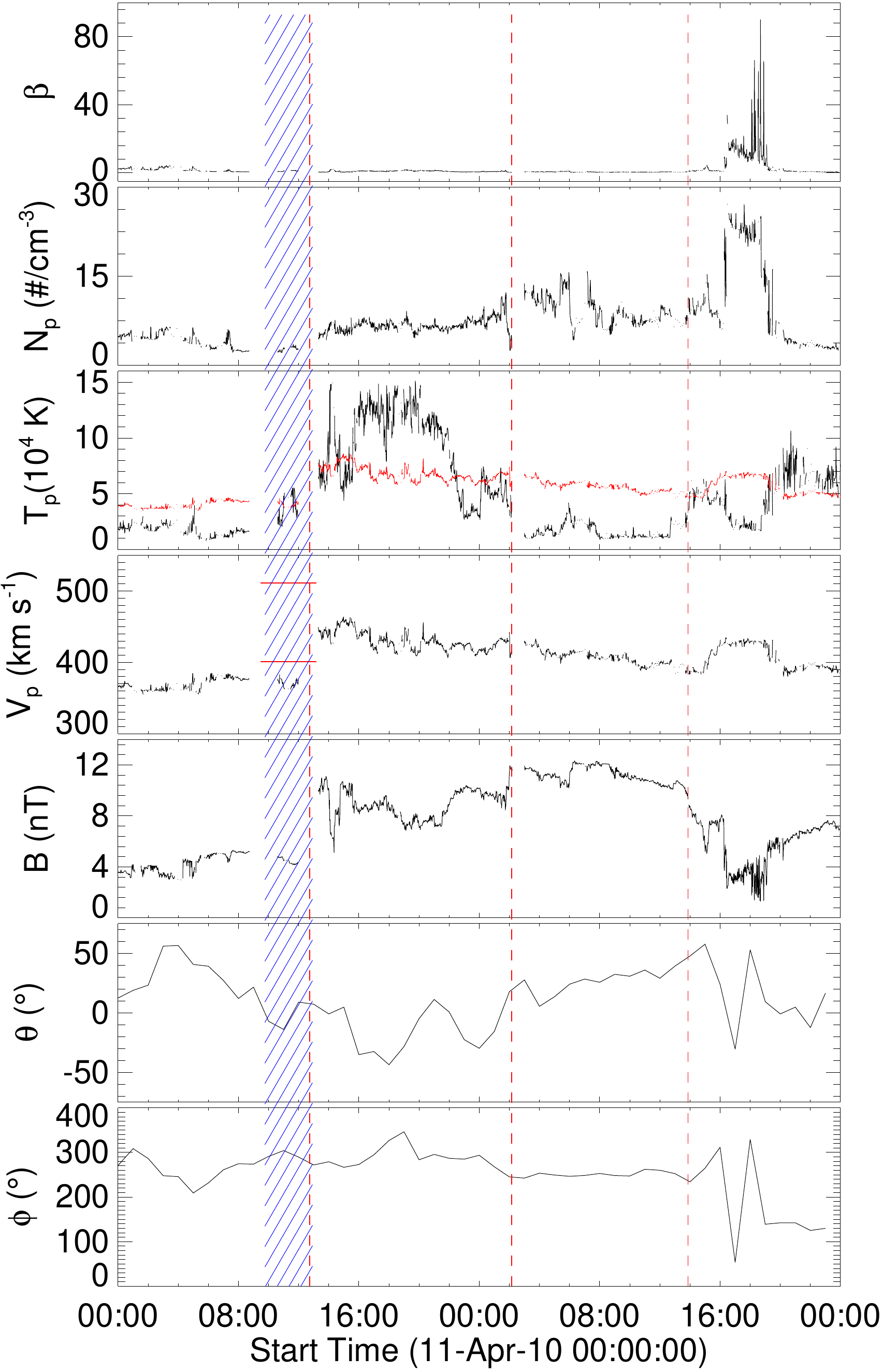}
\caption[As Figure~\ref{insitu12Dec08}, for the 2010 April 8 CME]{The panels show same plasma parameters as in Figure~\ref{insitu12Dec08},corresponding to the CME of 2010 April 8.}
\label{insitu8Apr10}
\end{figure*}

\subsection{2010 October 6 CME}

We have also identified the 2010 October 6 CME in the in situ data, based on plasma, magnetic field, and compositional signatures \citep{Zurbuchen2006}. The in situ observations from 2010 October 11-12, is shown in Figure~\ref{insitu6Oct10}. In this figure, the red curve in the third panel from the top shows the variation of the expected proton temperature \citep{Lopez1987} and the first vertical line (dotted, labeled as LE) marks the arrival of the CME leading edge at 05:50 UT on 2010 October 11 and the fourth vertical line (dashed, labeled TE) marks the trailing (rear) edge arrival at 17:16 UT. The enhanced density before the first dotted vertical line is the CME sheath region. The region bounded by the second and third vertical lines (solid), at 09:38 UT and 13:12 UT, respectively, can be classified as an MC \citep{Klein1982, Lepping1990}, as it shows an enhanced magnetic field ($>$10 nT), a decreased plasma $\beta$ ($<$ 1), and a smooth rotation in the magnetic field over a large angle ($>$ 30$\arcdeg$) \citep{Klein1982, Lepping1990}.

\begin{figure*}[!htb]
	\centering
		\includegraphics[width=0.65\textwidth,height=0.55\textheight]{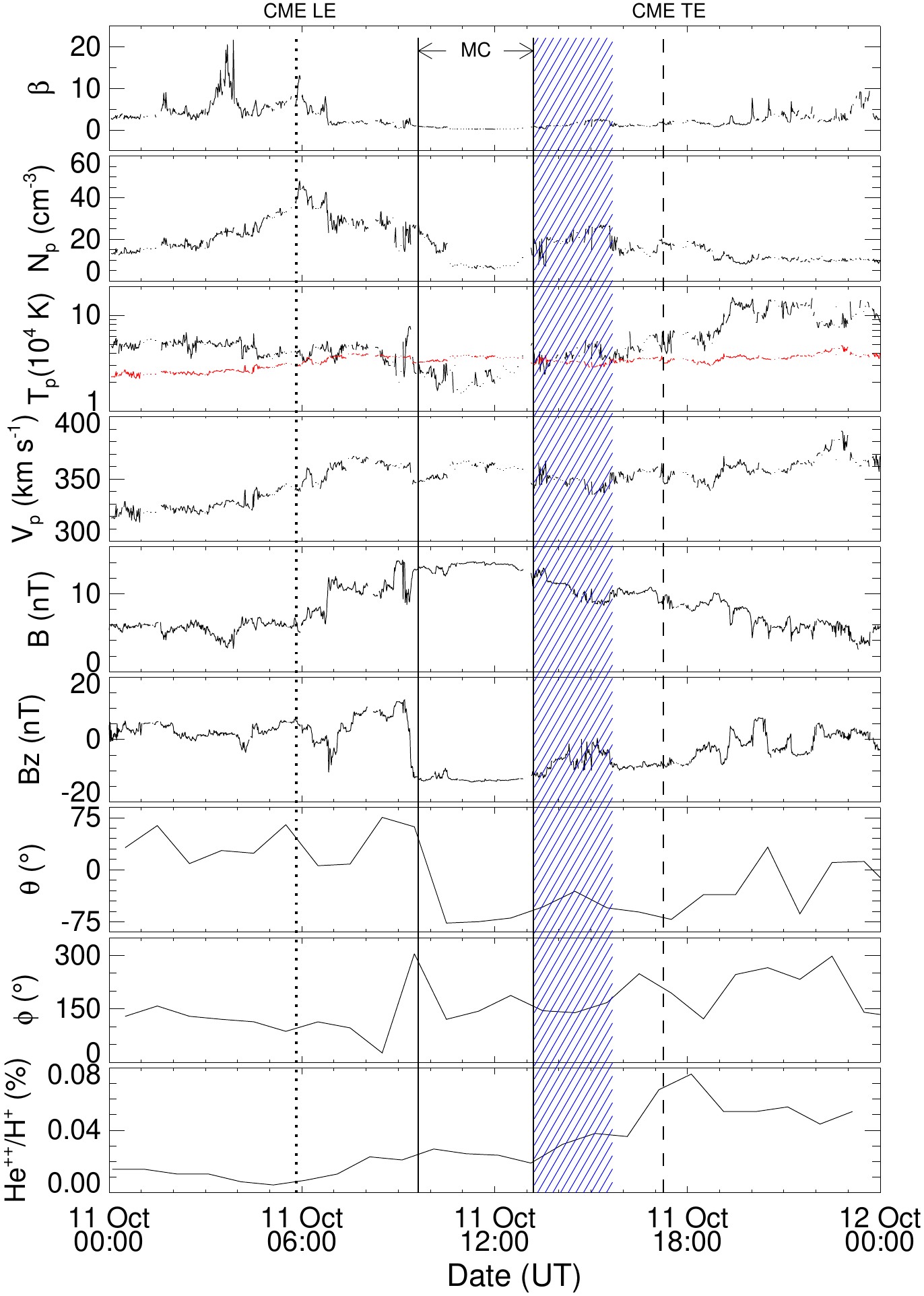}
\caption[As Figure~\ref{insitu12Dec08}, for the 2010 October 6 CME]{The panels show same plasma parameters as in Figure~\ref{insitu7Feb10}, except the z-component of magnetic field in the fourth panel from the bottom, corresponding to the CME of 2010 October 6. The hatched region with blue line marks the region associated with tracked Feature 2 shown in Figure~\ref{Jmaps} and described in Section~\ref{LETE6Oct}.}
\label{insitu6Oct10}
\end{figure*}

\subsection{2010 October 10 CME}

We analyzed the in situ data and identified the CME boundary, which is shown in Figure~\ref{insitu10Oct10}. In this figure, CME sheath arrival is marked by the first vertical dashed line (red) at 04:30 UT on 2010 October 15. The second vertical dashed line (red) marks the trailing edge of the CME sheath region at 01:38 UT on 2010 October 16. The hatched line (blue) marks the predicted arrival time (with uncertainties) obtained using DBM.

\begin{figure*}[!htb]
	\centering
		\includegraphics[width=0.65\textwidth,height=0.55\textheight]{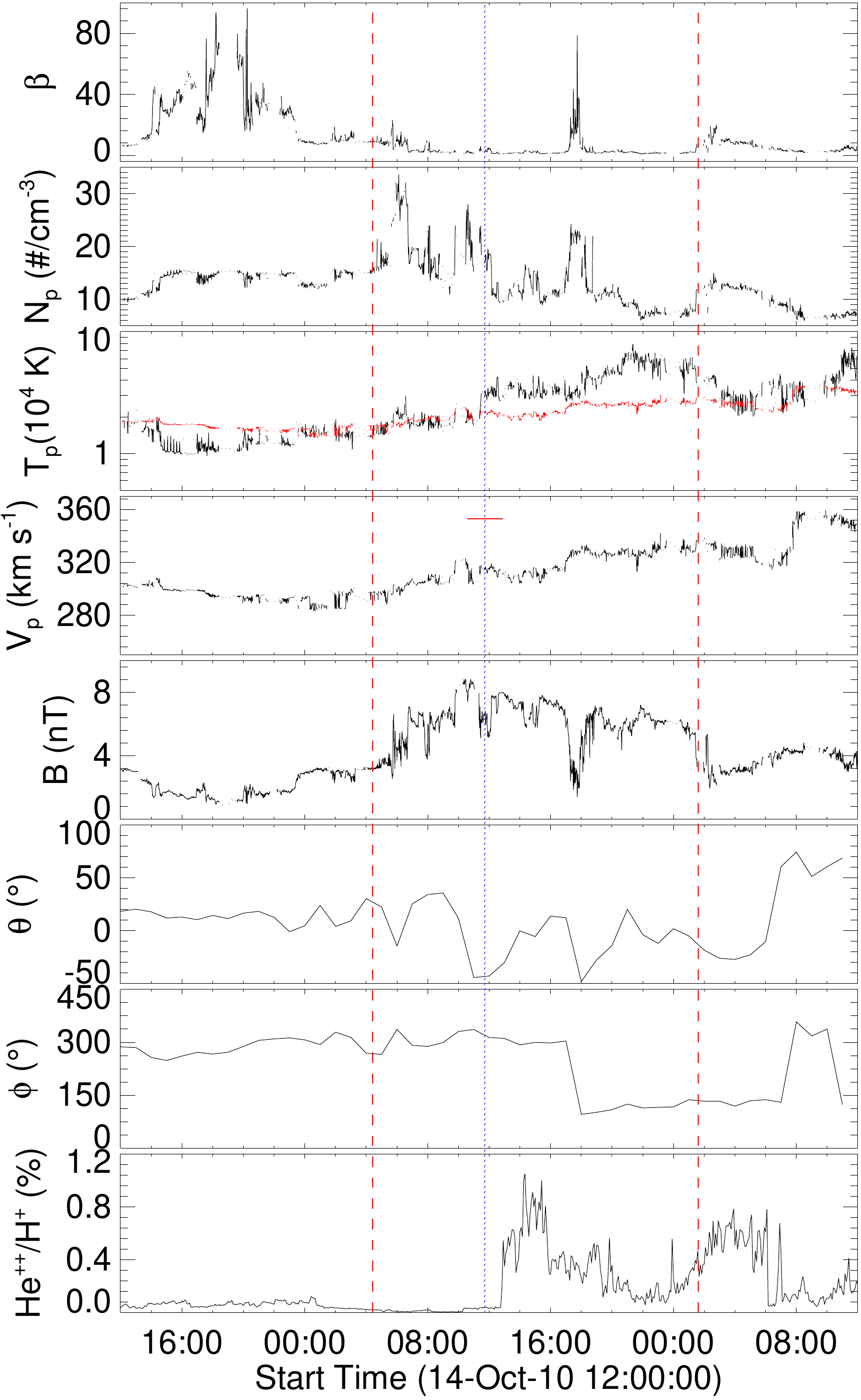}
\caption[As Figure~\ref{insitu12Dec08}, for the 2010 October 10 CME]{The panels show same plasma parameters as in Figure~\ref{insitu7Feb10}, corresponding to the CME of 2010 October 10.}
\label{insitu10Oct10}
\end{figure*}

\subsection{2010 October 26 CME}

In situ observations of solar wind taken nearly at 1 AU and the identification of 2010 October 26 CME is shown in 
Figure~\ref{insitu26Oct10}. This figure shows a sudden enhancement of density, temperature, and velocity at 10:32 UT on October 30. These changes mark the arrival of a shock, as indicated by the first dashed vertical line (red) from the left. The hatched region (blue) marks the CME's predicted arrival time (with uncertainty) using DBM. From the left, the second and fourth vertical dashed lines (red) mark the leading and trailing edges of the CME. This CME also shows MC \citep{Klein1982, Lepping1990} structure which is identified as the region between the third and fourth vertical dashed lines (red) at 01:30 and 21:35 UT on November 1, respectively. This region has a low proton beta \citep{Burlaga1981,Cane2003}, a decrease in proton density \citep{Richardson2000}, a decrease in proton temperature \citep{Gosling1973}, a monotonic decrease in proton velocity \citep{Klein1982}, and an enhanced alpha to proton ratio \citep{Borrini1982}.

\begin{figure*}[!htb]
	\centering
		\includegraphics[width=0.65\textwidth,height=0.55\textheight]{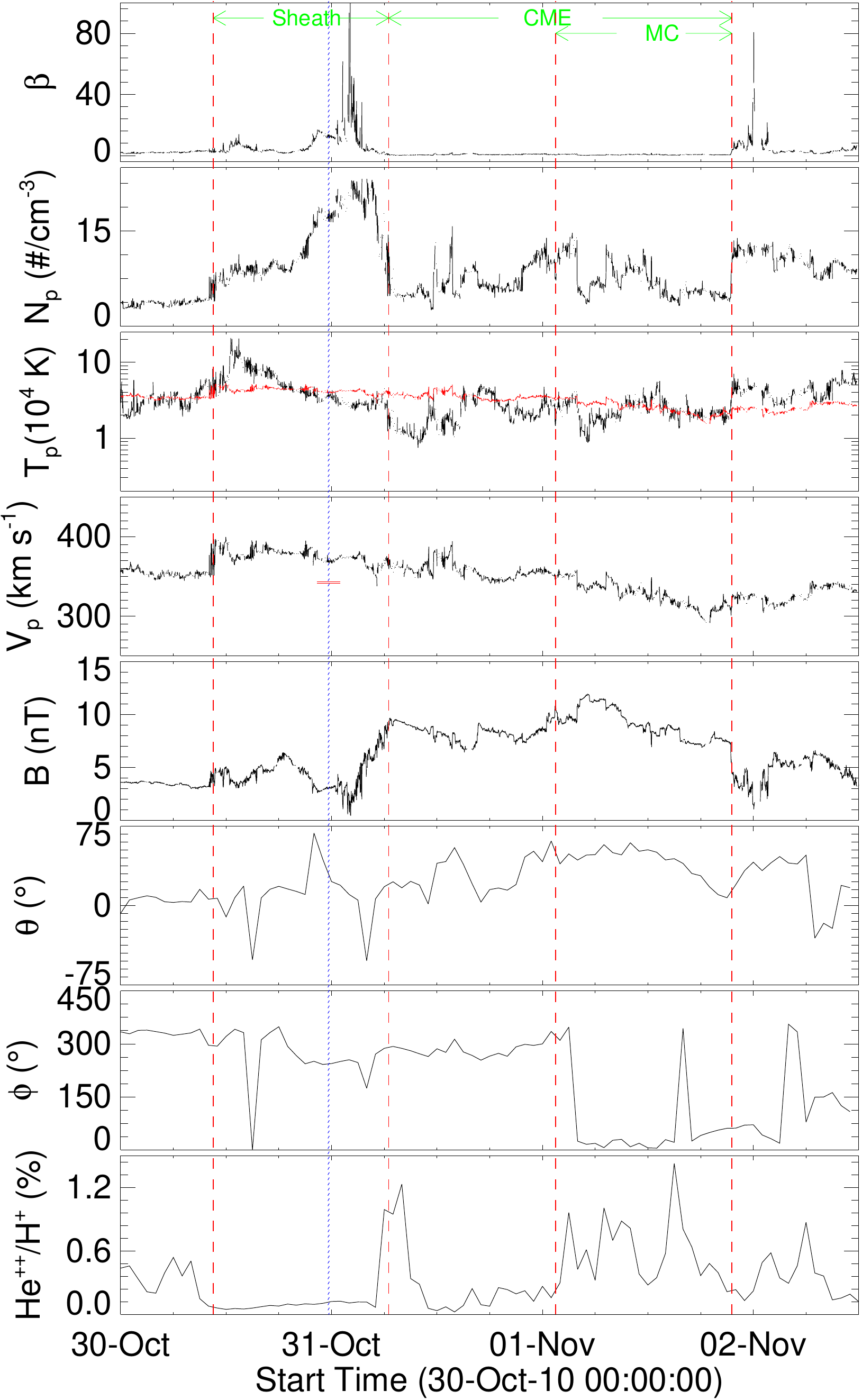}
\caption[As Figure~\ref{insitu12Dec08}, for the 2010 October 26 CME]{The panels show same plasma parameters as in Figure~\ref{insitu7Feb10}, corresponding to the CME of 2010 October 26.}
\label{insitu26Oct10}
\end{figure*}

\section{Tracking of Different Features of 2010 October 6 CME Using Remote and In Situ Observations}
\label{LETE6Oct}
The heliospheric tracking of CME features using \textit{J}-maps mainly deals with tracking the CME bright front and associating it with CME sheath observed before the leading edge in in situ data \citep{Davis2009, Liu2010,Liu2011, Mostl2011,Mishra2013}. In a rare attempt, \citet{Howard2012a} tracked a feature like a `cavity' (in coronagraph images) of a classical CME using HI images. This cavity-like feature was associated with an MC identified in in situ data near the Earth. \citet{Deforest2011} have also attempted to identify different CME structures in \textit{STEREO}/HI2 observations and compared them with in situ features detected near 1 AU.

As a CME shows large-scale inhomogeneous structures in terms of density and magnetic field, it is likely that these structures will be acted upon by unequal forces resulting in different kinematics. Occasionally, suppose a feature of a CME is missed by an in situ spacecraft. In that case, the sequential tracking of other structures might be helpful in relating the remote observations and in situ. Such a study is of two-fold importance. Scientifically, understanding the physical nature of various features (leading edge, cavity, and core) of a CME can help in the theoretical modeling of a CME and understanding its heliospheric evolution. On the other hand, different features of a CME may lead to different perturbations in the Earth's magnetosphere because of their dissimilar plasma and magnetic field properties, which also need to be understood.

In this section, we study the evolution of the front and the rear edge of a filament associated geoeffective CME of 2010 October 6. The filament was located in north-east (NE) quadrant of the solar disc as seen from the Earth's perspective. For this study, we used the white light observations of CME from twin \textit{STEREO} and in situ plasma and magnetic field parameters of the solar wind from the \textit{ACE} and \textit{WIND} spacecraft.

\subsection{Remote sensing observations of 2010 October 6 CME}
The evolution of 2010 October 6 CME in COR and HI FOV are shown in Figure~\ref{Evolution6Oct10} of Chapter~\ref{Chap3:ArrTim}. The front/leading edge (hereafter termed as F1) and core/associated filament material (hereafter termed as F2) of the CME could be easily identified in COR1 and COR2 images. As the CME reached near the edge of COR2 FOV, \textit{STEREO-A} had a data gap, and when the CME appeared in HI1-A FOV, it was diffused, and the leading edge and core could not be distinguished as in COR FOV. In the \textit{J}-maps for this CME (Figure~\ref{Jmaps}), two bright tracks having positive inclination are noticed where they are termed as Feature 1 and Feature 2 as we are not certain that these features are the same as F1 and F2 noticed in COR FOV.

\begin{figure}[!htb]
\begin{center}
\includegraphics[height=8cm, width=7cm]{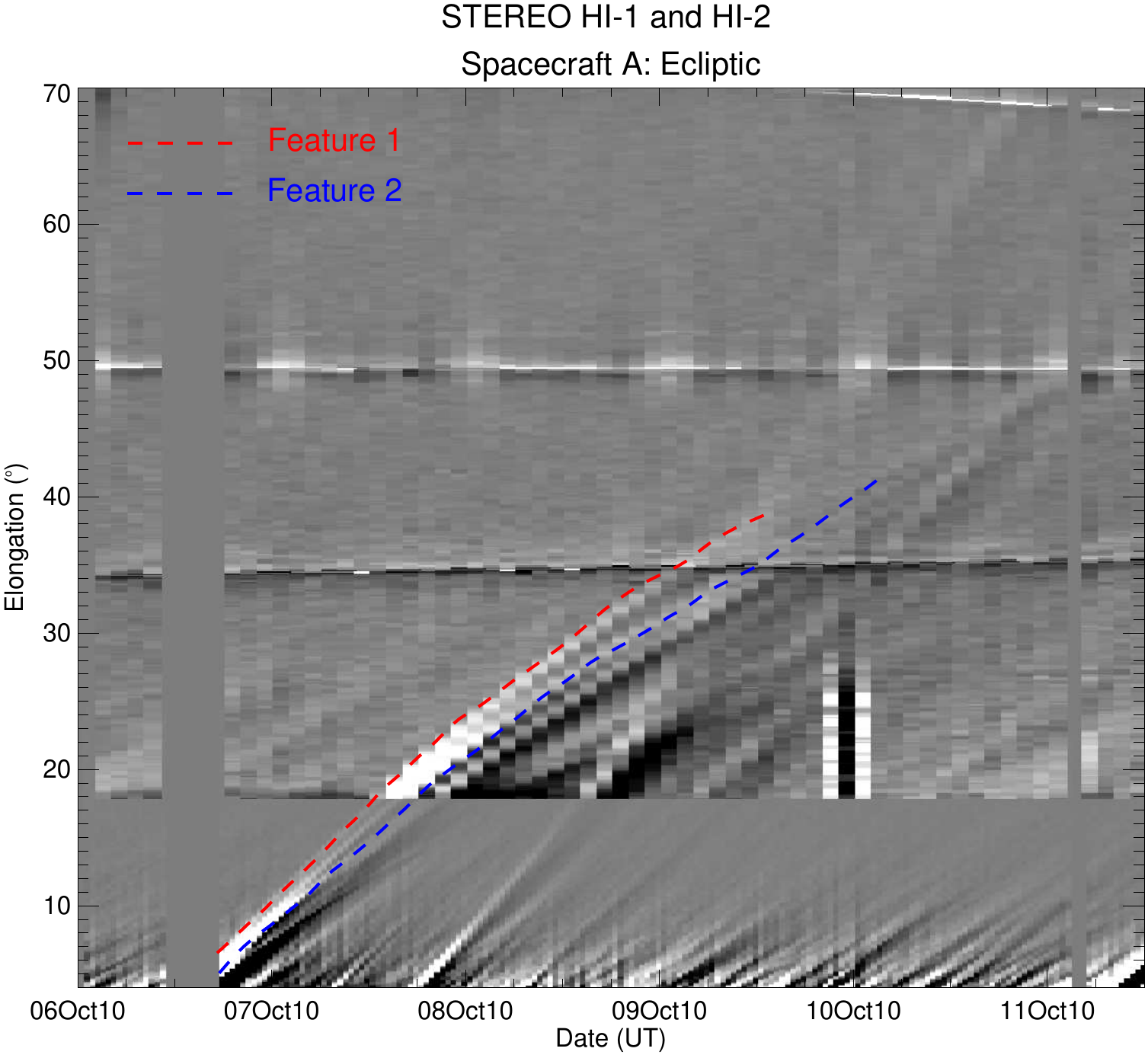}
\hspace{0.15cm}
\includegraphics[height=8cm, width=7cm]{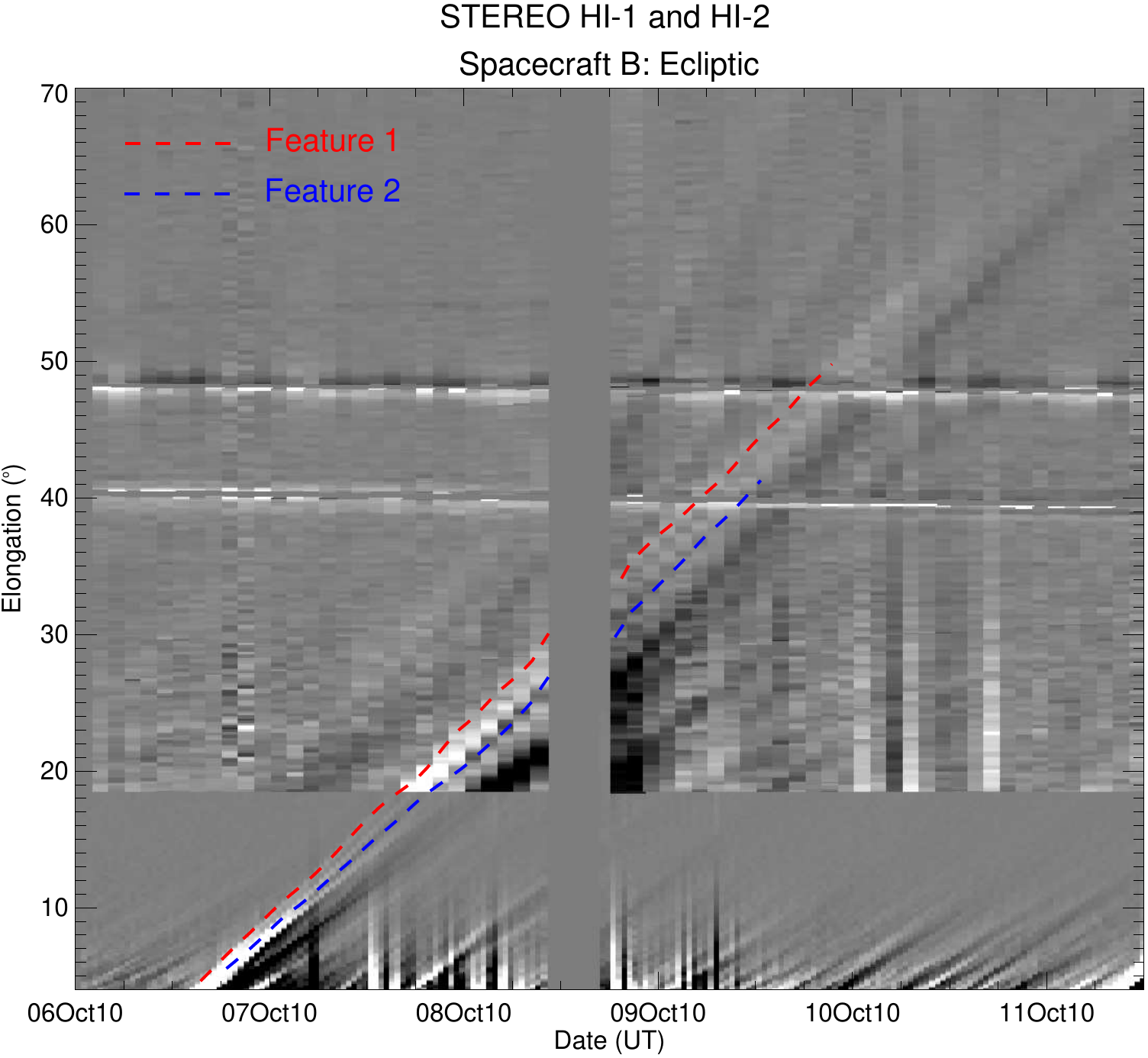}
\caption[\textit{J}-maps for the 2010 October 6 CME]{Time-elongation maps (\textit{J}-maps) for \textit{STEREO-A} (left) and \textit{STEREO-B} (right) constructed from running differences images of the HI1 and HI2, for the time interval from 2010 October 6, 00:00 UT  to 2010 October 11, 12:00 UT. Two features (marked as Feature 1 and Feature 2 with red and blue line, respectively) are tracked corresponding to density enhancement features.}
\label{Jmaps}
\end{center}
\end{figure}

It must be noted that kinematics of tracked Feature 1, using ten reconstruction methods (Point-P (PP), Fixed-Phi (FP), Fixed-Phi Fitting (FPF), Harmonic Mean (HM), Harmonic Mean Fitting (HMF), Self-Similar Expansion (SSE), Self-Similar Expansion Fitting (SSEF), Geometric Triangulation (GT), Tangent to A Sphere (TAS), Stereoscopic Self-Similar Expansion (SSSE)), has been estimated and described in Section~\ref{Mthds6Oct10} of Chapter~\ref{Chap3:ArrTim}. To consider the maximum uncertainties in our analysis, various reconstruction methods for the estimation of kinematics are used. The obtained kinematics are used as inputs in the DBM \citep{Vrsnak2013} to estimate the arrival time (for details see, Figure~\ref{STAA6Oct10}, ~\ref{FF_HMF6Oct10}, \ref{STAABB6Oct10} and Table~\ref{Tab6Oct10} in Chapter~\ref{Chap3:ArrTim}). For identification of the second tracked feature in in situ observations, we followed the same analysis of Feature 2 as described for Feature 1. 

\subsection{Reconstruction of 2010 October 6 CME}
\label{recons}

We applied the tie-pointing method of 3D reconstruction (scc\_measure: Thompson 2009) for a feature on the leading edge (F1) and filament/core (F2) part of the CME observed in COR1 and COR2 images to estimate their 3D dynamics. Figure~\ref plots the estimated 3D coordinates for tracked features (F1 \& F2) are plotted in Figure~\ref{3DCOR}. 

\begin{figure}[!htb]
\begin{center}
\includegraphics[angle=0,scale=1.0]{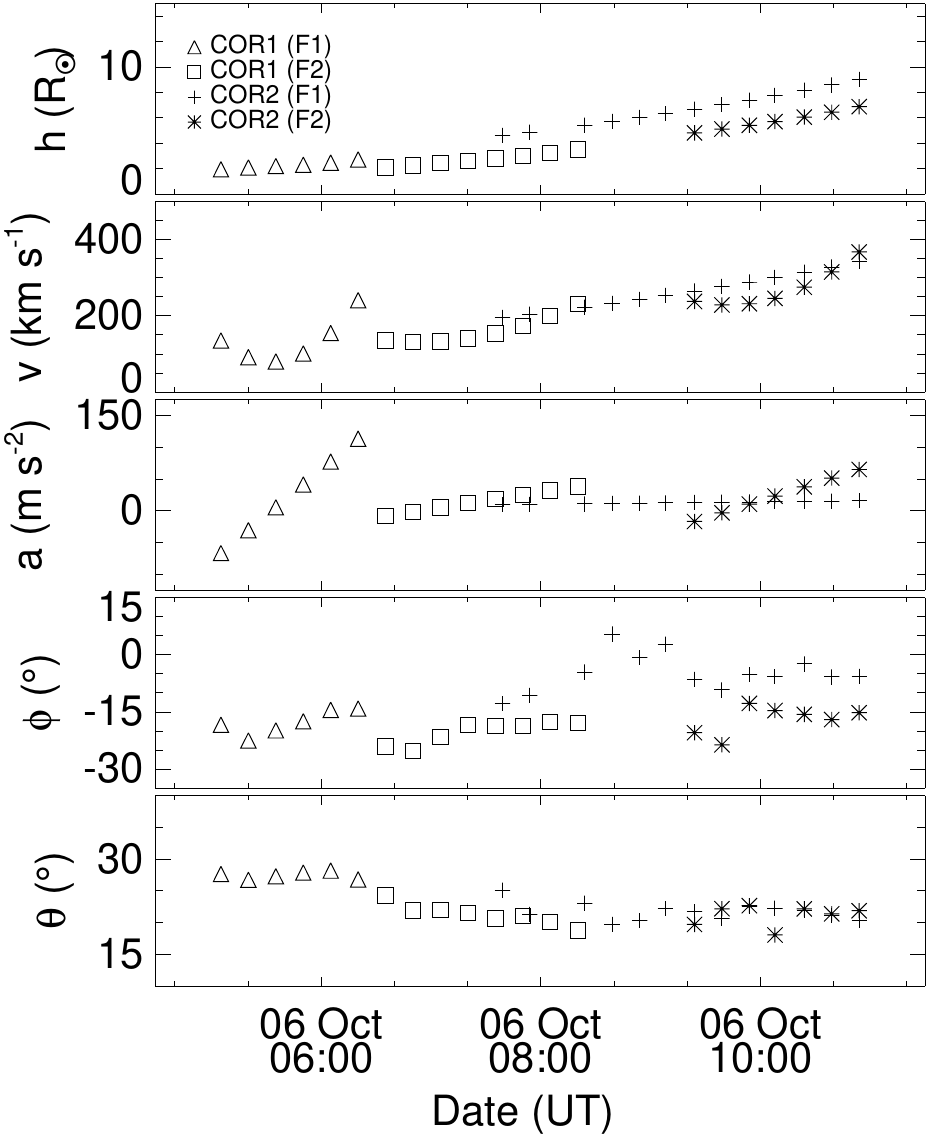}
\caption[Estimated height, speed, acceleration, longitude and latitude of the leading edge and core of the 2010 October 6 CME]{Top to bottom panels show the evolution of height, speed, acceleration, longitude and latitude, respectively. Here F1 and 
F2 correspond to features on the leading edge and core/filament of the 2010 October 6 CME.}
\label{3DCOR}
\end{center}
\end{figure}

From the Figure~\ref{3DCOR}, it is clear that both F1 \& F2 move at around 30$^\circ$ to 20$^\circ$ north in coronagraphic FOV 
and are eastward to the Sun-Earth line. The estimated kinematics show that they follow approximately the same trajectory in 3D space. Leading-edge (F1) shows acceleration in COR1 FOV while filament (F2) accelerates more in COR2 FOV than COR1 FOV. Leading-edge (F1) becomes too diffuse to be tracked at the time when filament (F2) enters the COR1 FOV. From Figure~\ref{3DCOR}, it is clear that the separation between the features F1 and F2 is approximately 1.0 \textit{R}$_\odot$ in COR1 FOV. F1 and F2 could also be tracked in COR2 FOV up to 10:54 UT on 2010 October 6. Because of a data gap after 10:54 UT until 17:39 UT in science images from \textit{STEREO-A}, these features could not be tracked, and 3D reconstruction was not possible. At the outermost tracked points in COR2 FOV, the separation between F1 and F2 increased to 3 \textit{R}$_\odot$.

To follow the CME features in HI FOV, we carefully examined the \textit{J}-maps (Figure~\ref{Jmaps}) where two positively inclined bright tracks (marked with red and blue dashed lines) which correspond to outward motion of two density structures are noticed. Such bright tracks may be due to features of either two different CMEs or different features (front and rear edge) of the same CME passing along the ecliptic. As no other Earth-directed CMEs have been reported close to the occurrence of 2010 October 6 CME in COR1 CME catalog (\url{http://cor1.gsfc.nasa.gov/catalog/}) or LASCO CME catalog (\url{http://cdaw.gsfc.nasa.gov/CME_list/}), we assume that these tracked features are features of Earth-directed geoeffective CME of 2010 October 6.

We derived the elongation angles of both moving features (Feature 1 and Feature 2) in the \textit{J}-maps by tracking them manually. These elongation angles are converted to distance using various reconstruction methods based on the assumptions involved regarding geometry, Thomson scattering effects and nature of propagation, and observation from single or multiple viewpoints. Using \textit{J}-maps, Feature 1 and Feature 2 could be tracked out to 39$^\circ$ and 41$^\circ$ elongation, respectively, in \textit{STEREO-A} \textit{J}-maps. In \textit{STEREO-B} \textit{J}-maps, Features 1 and 2 could be tracked out to 50$^\circ$ and 41$^\circ$, respectively.

We used the single spacecraft PP, FP, HM, and SSE methods to derive the kinematics of the tracked Feature 2. The FP, HM, and SSE methods require the direction of propagation (longitude) of tracked feature, which is obtained from the longitude estimates 
(Figure~\ref{3DCOR}) by using the tie-pointing reconstruction method in COR2 FOV. We fixed the propagation direction of this Feature 2  as 10$^\circ$ east from the Sun-Earth line, which corresponds to a longitude difference of 93$^\circ$ and 68$^\circ$ from the \textit{STEREO-A} and \textit{B}, respectively. In the SSE method, we need to fix the angular half-width \citep{Davies2012} of CME, which is taken as 30$^\circ$ in our case. The obtained kinematics profile of tracked Feature 2 from \textit{STEREO-A} viewpoints is estimated and is shown in Figure~\ref{KinAA}. We also tracked Feature 2 in \textit{STEREO-B} \textit{J}-maps and estimated the kinematics using all the single spacecraft methods. The estimated kinematics at outermost tracked points has been used in the DBM   corresponding to the extreme range of drag parameter \citep{Vrsnak2013} to obtain the arrival time of Feature 2 at L1 (Table~\ref{KinarrL1}).

We also applied the three fitting methods (FPF, HMF, and SSEF) to obtain a set of speed, propagation direction, and launch time parameters that best reproduce the observed elongation-time profiles of Feature 2. The obtained best-fit parameters for elongation-time profiles from \textit{STEREO-A} and \textit{B} \textit{J}-maps, the estimated arrival time derived using the same is shown in Table~\ref{KinarrL1}.

\begin{figure}[!htb]
\begin{center}
\includegraphics[angle=0,scale=0.6]{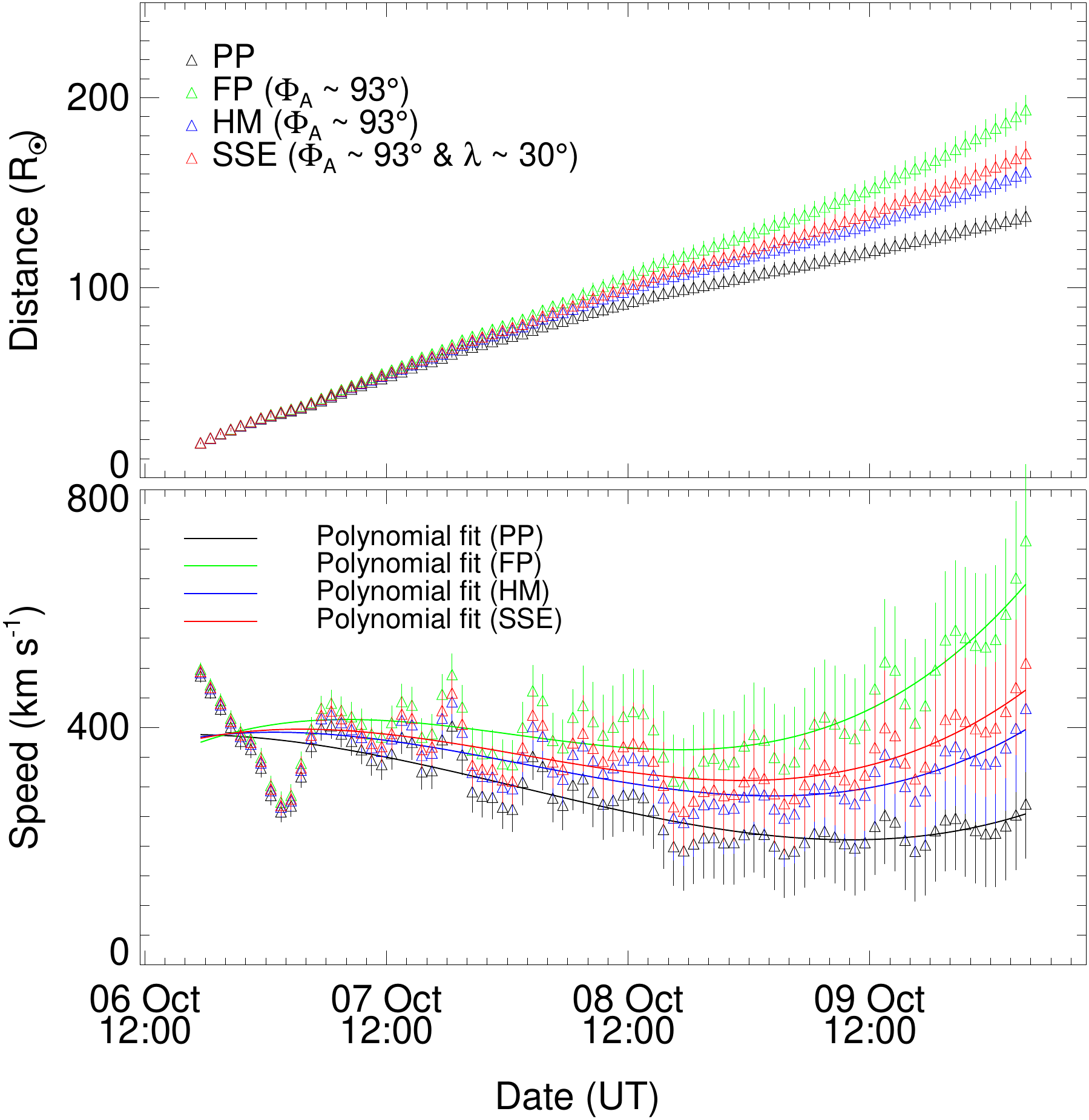}
\caption[Distance and speed profiles using PP, FP, HM, and SSE methods for the tracked Feature 2 of the 2010 October 6 CME]{Top panel shows the estimated distance profiles of Feature 2, based on the application of the single spacecraft (PP, FP, HM, and SSE)  methods.  The bottom panel shows speed profiles derived from the adjacent distances using three-point Lagrange interpolation (solid line shows the polynomial fit). Vertical lines show the errors bars derived taking the uncertainties of 3\% and 4\% in the estimated distance in HI 1 and HI2 FOV, respectively.}
\label{KinAA}
\end{center}
\end{figure}

\begin{sidewaystable}
  \centering
{\scriptsize
 \begin{tabular}{p{3.0cm}|p{2.0cm}| p{3.0cm}| p{2.5cm}|p{2.5cm}|p{2.5cm}}
    \hline
		
 Method & Kinematics as inputs in DBM [t$_{0}$, R$_{0}$ (\textit{R}$_\odot$), v$_{0}$ (km s$^{-1})$] &  Predicted arrival time using kinematics + DBM (UT) [$\gamma$ = 0.2 to 2.0 (10$^{-7}$ km$^{-1}$)] & Predicted transit speed at L1 (km s$^{-1}$)  [$\gamma$ = 0.2 to 2.0 (10$^{-7}$ km$^{-1}$)] & Error in predicted arrival time (hrs) [$\gamma$ = 0.2 to 2.0 (10$^{-7}$ km$^{-1}$)] &  Error in predicted transit speed (km s$^{-1}$)  [$\gamma$ = 0.2 to 2.0 (10$^{-7}$ km$^{-1}$)] \\  \hline

PP (\textit{STEREO-A})	& 10 Oct 03:30, 137, 230  & 12 Oct. 12:51 to 12 Oct. 03:22	 &  270 to 327   & 	23.7 to 14.2	 &  -85 to -28 \\ \hline

PP (\textit{STEREO-B})	& 09 Oct 12:40, 137, 360  & 11 Oct 05:14 to 11 Oct 05:21	 &  360 to 358   & 	-8 to -7.8	 &  5 to 3 \\ \hline
  
FP (\textit{STEREO-A})	& 09 Oct 10:30, 149, 380  &  10 Oct 18:54 to 10 Oct 19:27  & 	378 to 368  &	-18.2 to -17.6	& 23 to 13  \\ \hline

FP (\textit{STEREO-B})	& 08 Oct 23:39, 130, 464    & 10 Oct 11:05 to 10 Oct 15:28 & 	438 to 377 &	-26.3 to -21.7	& 83 to 22  \\ \hline
  
HM (\textit{STEREO-A})	& 09 Oct 11:30, 132, 285  & 11 Oct. 16:48 to 11 Oct. 12:43	 & 298 to 330  &	3.6 to -0.4	 &  -57 to -25  \\ \hline

HM (\textit{STEREO-B})	& 09 Oct 12:40, 158, 430  & 10 Oct 13:29 to 10 Oct 15:07  & 420 to 382 	 & -23.7 to -22.1	 &  65 to 27 \\  \hline
GT         & 09 Oct 12:40, 159, 450   & 10 Oct 12:03 to 10 Oct 14:06 & 436 to 385  &  -25.2 to -23.2  & 81 to 30   \\  \hline

TAS       &  09 Oct 12:40, 131, 280    & 11 Oct 19:21 to 11 Oct 14:41  & 295 to 330 &   6.1 to 1.4  & -60 to -25   \\  \hline
SSSE &  09 Oct 12:40, 154, 385 &  10 Oct 18:09 to 10 Oct 18:45	& 383 to 370 	& -19.1 to -18.5	 & 28 to 15    \\  \hline		

 \end{tabular}

\begin{tabular}{p{3.0cm}| p{3.0cm}| p{2.5cm}|p{2.5cm}|p{2.5cm}| p{2.0cm}}
\multicolumn{6}{c}{Time-elongation track fitting methods} \\  \hline
  
 Methods   & Best fit parameters [t$_{(\alpha = 0)}$, $\Phi$ ($^\circ$), v (km s$^{-1}$)]  &   Predicted arrival time at L1 (UT)  & Error in predicted arrival time  & Error in predicted speed at L1 (km s$^{-1}$) & Longitude ($^\circ$) 
   \\ \hline
	
	FPF (\textit{STEREO-A})   & 06 Oct 07:44, 90, 378 & 10 Oct 20:29 & -16.9  &  23      &  -7   \\  \hline
	FPF  (\textit{STEREO-B})  & 06 Oct 04:14, 56.2, 369 & 10 Oct 19:25  & -17.7  &    14   &  -21.5 \\  \hline
	 
	HMF (\textit{STEREO-A})   & 06 Oct 08:54, 125, 467 & 11 Oct 07:48 & -5.4  &    -18    &  -42   \\  \hline
	HMF  (\textit{STEREO-B})  & 06 Oct 05:51, 58, 378 & 11 Oct 01:35 &  -11.6  &    -10   &   -19.8 \\  \hline
	
	SSEF (\textit{STEREO-A})  & 06 Oct 08:40, 114, 436  & 11 Oct 09:02   & -4.2  &   -13  &  -31 \\  \hline
	SSEF (\textit{STEREO-B})  & 06 Oct 05:37, 57.4, 377 & 11 Oct 05:43  & -7.5  &  -12   &   -20.4   \\  \hline

\end{tabular}
}
\caption[Results of different reconstruction techniques applied to trailing feature of the 2010 October 6 CME]{The estimated kinematics (used as input in the DBM) and predicted arrival time and speed (and errors therein) of Feature 2 at L1 corresponding to the extreme range of drag parameter used in the DBM is shown. The bottom panel gives best fit parameters, predicted arrival time, and speed (and errors therein) estimated from time-elongation track fitting methods. The \textit{STEREO-A} and \textit{B} shown in parentheses for each method denote the spacecraft from which the images are obtained to derive elongation. Negative (positive) errors in predicted arrival time correspond to a predicted arrival time before (after) the actual CME arrival time determined from in situ measurements. Negative (positive) errors in predicted  speed correspond to a predicted speed that is less (more) than the actual speed of CME  at L1.}
\label{KinarrL1}
\end{sidewaystable}

For associating remotely observed features with in situ observations, we must correctly ascertain their direction of propagation in the heliosphere. The heliospheric direction of propagation of tracked Feature 1 has been compared using various methods and was found to be directed towards the Earth (see, Section~\ref{Mthds6Oct10} of Chapter~\ref{Chap3:ArrTim}). If Feature 2 is not directed towards the Earth, then when both features are at an equal heliocentric distance from the Sun, they will have different elongation angles. Further, if their direction of propagation is different at the instant when they have the same elongation angles, they will be at a different heliocentric distances from the Sun. Therefore, we must check the continuous evolution of Feature 2, particularly its propagation direction, which can not be determined using single spacecraft methods (e.g., PP, FP, HM, SSE, FPF, HMF, SSEF). 

We also implemented the stereoscopic reconstruction methods, which require elongation measurements of a moving feature from two view directions. We used the GT, TAS, and SSSE methods to estimate the kinematics of tracked Feature 2 in the heliosphere (Figure~\ref{KinAABB}). In this figure, the kinematics at the sun-ward edge of HI1 FOV is not shown due to the occurrence of a singularity, where small uncertainties in elongation measurements lead to larger errors in kinematics (for details, see, Section~\ref{Rmt7Feb10} Chapter~\ref{Chap3:ArrTim}). Using the kinematics at the outermost points in the DBM, the obtained arrival time, transit speed at L1, and errors therein are given in Table~\ref{KinarrL1}. 

\begin{figure}[!htb]
\begin{center}
\includegraphics[angle=0,scale=0.5]{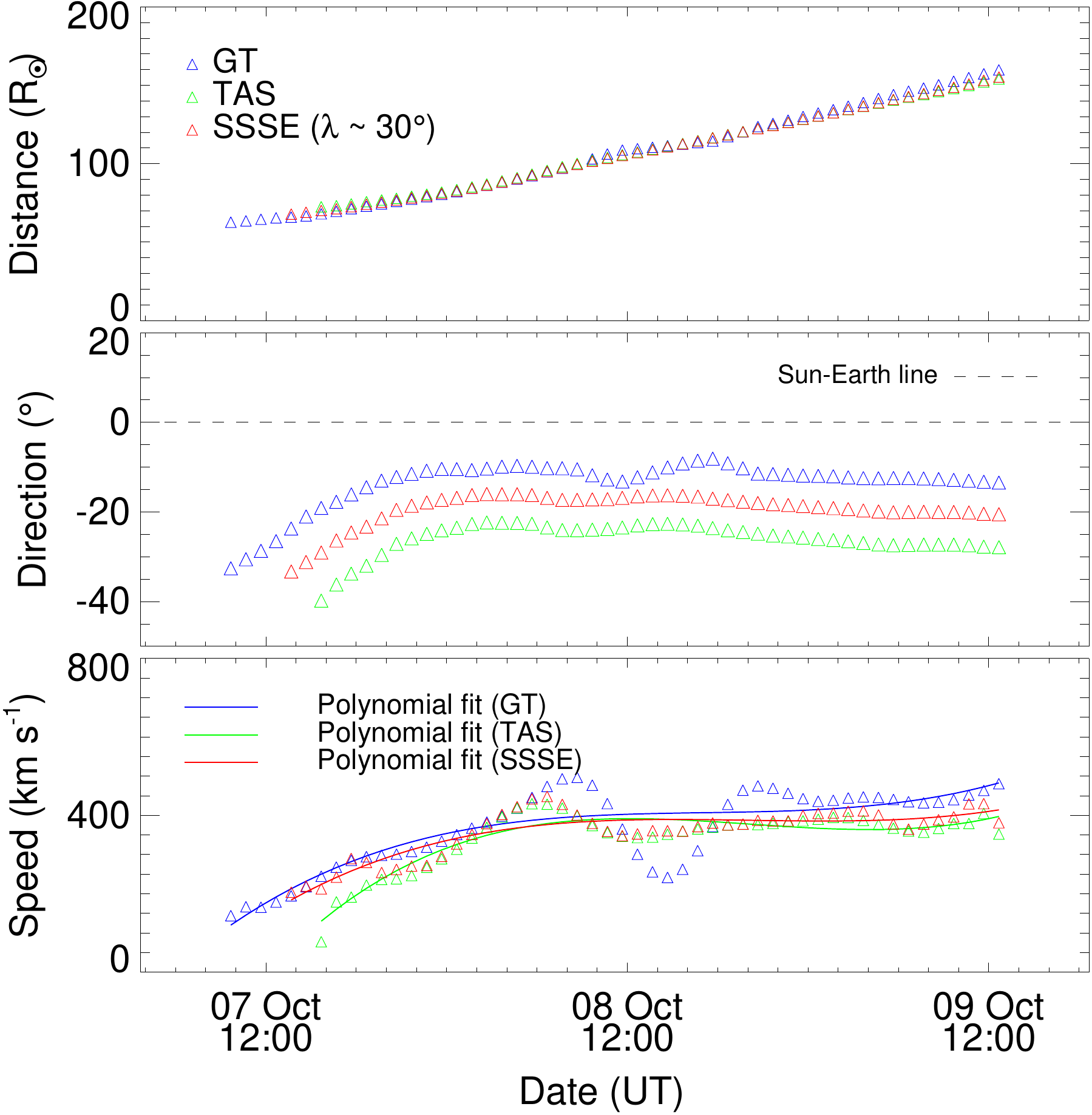}
\caption[Distance, propagation direction and speed using stereoscopic (GT, TAS and SSSE) methods for the tracked Feature 2 of the 2010 October 6 CME]{From top to bottom, panels show the estimated distance, propagation direction and speed of the tracked Feature 2 using GT, TAS and SSSE methods. In the middle panel, dashed horizontal line marks the Sun-Earth line.}
\label{KinAABB}
\end{center}
\end{figure}

From Figure~\ref{KinAA} (top panel), we see that the PP method gives the lowest estimate of the distance while the FP method gives the highest estimate. HM, and SSE methods give an intermediate distance estimate between PP and FP methods. The observed unphysical late acceleration for Feature 2 (Figure~\ref{KinAA}, lower panel) is attributed to possible deflection of the tracked feature far from the Sun. Although real deflection of features far from the Sun is rarely observed, however, it refers to the infeasibility of tracking the same feature sequentially at larger elongation due to the overall expansion of the CME. In our analysis, we have taken care that if the estimated kinematics becomes unreliable at any instant, the value of speeds prior to those points are considered inputs in the DBM. Also, speed obtained by implementing fitting methods (FPF, HMF, and SSEF) is corrected for off-axis correction before using it for estimation of the arrival time at L1 \citep{Mostl2011, Mostl2013}. From Table~\ref{KinarrL1}, we also notice that different fitting methods give different estimates of the propagation direction of Feature 2. In light of our earlier study in \citet{Mishra2014}, we rely on the estimates of direction from stereoscopic methods such as GT, TAS, and SSSE.

From Figure~\ref{KinAABB} for Feature 2 and Figure~\ref{STAABB6Oct10} for Feature 1 in Chapter~\ref{Chap3:ArrTim}, we can see that both tracked Features are moving approximately in the same direction in the heliosphere. Therefore, both these Earth-directed features are likely to be detected by in situ spacecraft located at L1. Since we have no information on the plasma and magnetic field properties associated with these tracked features from remote sensing observations, we can only rely on their estimated arrival time at L1 to relate features in situ data with remotely tracked Feature 1 and Feature 2.

\subsection{Comparison of kinematics of tracked features}
\label{SepF1F2}
We tracked the front (leading) and the rear (trailing) edge of the intensity front in the \textit{STEREO}/HI \textit{J}-maps and derived the kinematics of these two CME features. Figure~\ref shows the separation between Feature 1 and Feature 2 based on their height derived from different reconstruction methods is shown in Figure~\ref{CompF1F2}. We found that the separation of the two features increased with time (Figure~\ref{3DCOR} \& ~\ref{CompF1F2}) during their evolution from the COR to HI FOV. The estimated large separation may be due to unequal driving forces on different features, an overall CME expansion, or a combination of both. From Figure~\ref{CompF1F2}, it is evident that the separation increased from 4.4 \textit{R}$_\odot$ to 11.3 \textit{R}$_\odot$ (as found by using the PP method) between the first and last common tracked points, respectively. The FP method showed a continuously increasing separation, ranging from 4.6 \textit{R}$_\odot$ to 21.7 \textit{R}$_\odot$. Using the HM method, the separation was found to increase from 4.5 \textit{R}$_\odot$ to 15.4 \textit{R}$_\odot$. For the GT, TAS, and SSSE techniques, the separation between tracked features (Feature 1 and Feature 2) increased up to 16.0, 12.5, and 13.5 \textit{R}$_\odot$, respectively. It is noted that the separation between the two features (estimated from GT, TAS, and SSSE) became approximately constant at large distances of $\approx$ 120 \textit{R}$_\odot$.

\begin{figure}[!htb]
\begin{center}
\includegraphics[angle=0,scale=0.8]{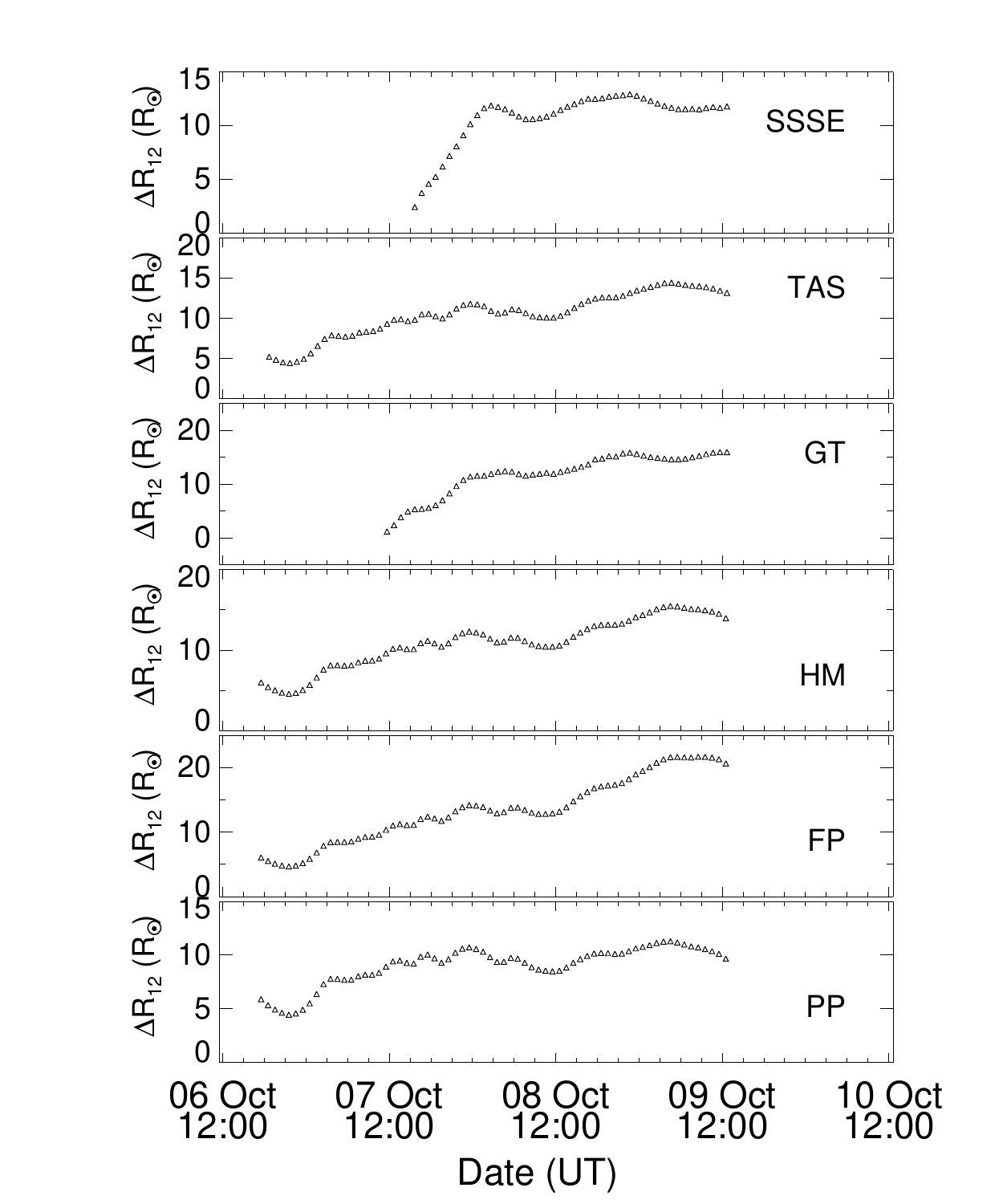}
\caption[Separation between tracked features of 2010 October 6 CME]{Variation in separation between both tracked features of the 2010 October 6 CME with time. Different panels show computed separation of the two features using different methods for the estimation of distance of tracked Feature 1 and Feature 2. From top to bottom, panels corresponds to SSSE, TAS, GT (twin spacecraft reconstruction), HM, SSE, FP, and PP (single spacecraft reconstruction) methods.}
\label{CompF1F2}
\end{center}
\end{figure}

Due to the occurrence of singularity at the sunward edge of HI1 FOV in GT, TAS, and SSSE methods, the separation between both tracked features (Feature 1 and Feature 2) is estimated to be slightly lesser than the observed separation in COR FOV. However, these initial erroneous estimates of separation can be discarded by considering the results obtained from GT, TAS, and SSSE methods at higher elongation and estimates from other methods. Based on the results of all the reconstruction methods applied on SECCHI/HI observations, the estimated separation between Feature 1 and 2 ranges between 11.2 \textit{R}$_\odot$ to 21.7 \textit{R}$_\odot$.

Several authors have shown that, at a large distance from the Sun, the kinematic evolution of CMEs is mostly attributed to the drag force between the CMEs and the ambient solar wind \citep{Cargill2004, Manoharan2006,Vrsnak2010}. In the present case, both features are tracked at sufficiently larger distances beyond the HI1 FOV. They attain speed close to ambient solar wind speed, so driving forces on Feature 1 and Feature 2 will become equal. The expansion speed of CME at larger distances is more likely due to the high internal thermal pressure of CME than the ambient solar wind pressure \citep{Schwenn2005, Gopalswamy2009,Michalek2009, Poomvises2010}. In our case, we notice that the expansion speed of CME is $\approx$ 165 km s$^{-1}$ at the entrance of HI FOV, which becomes negligibly small compared to the radial speed of the CME at a distance of 115 \textit{R}$_\odot$. Therefore, under the aforementioned constraints regarding expansion and equal drag on both features of the CME, it is expected that the final estimate of the separation between Feature 1 and Feature 2 will be approximately maintained out to the L1 point. If a constant speed of 350 km s$^{-1}$ is assumed for both features beyond the last observation points in HI FOV, then the difference in their arrival time at L1 will range from 6.1 to 11.9 hr.

\subsection{Identification of filament plasma using near-Earth in situ observations}
\label{insituobs}
The identification of the 2010 October 6 CME using in situ observations is shown in Figure~\ref{insitu6Oct10}. The tracked Feature 1 is the leading edge of the first bright positively inclined track in the \textit{J}-maps (Figure~\ref{Jmaps}). Hence, the arrival of enhanced density associated with this Feature 1 can be associated with the in situ measured density enhancement (LE) at 05:50 UT on 2010 October 11 \citep{Mishra2014}. In Figure~\ref{insitu6Oct10}, we do not notice a monotonic decrease in the speed of CME, i.e., there is no expansion. This is consistent with the finding that separation between Feature 1 and Feature 2 has become constant well before reaching the L1. As discussed in the previous Section~\ref{SepF1F2}, we expect the Feature 2 to arrive at L1 approximately 6 to 12 hr after the arrival of Feature 1. Therefore, we associate the arrival of Feature 2 with a second peak in proton density between 13:14 and 15:40 UT on 2010 October 11, associated with a low proton temperature ($\approx$ 10$^{4}$ K). Since the Feature 2 follows the MC and is associated with dense (maximum N$_{p}$ = 27 cm$^{-3}$), cold material (minimum T$_{p}$ = 2.4 $\times$ 10$^{4}$ K) observed in situ, it possibly corresponds to the core of a classical three-part structure CME. We need to examine the plasma composition and charge state and the root mean square deviations of bulk velocity in high-resolution data \citep{Burlaga1998, Gopalswamy1998,Lepri2010, Sharma2012,Sharma2013} to confirm the presence of filament material in the associated CME at L1. 

Hourly-resolution \textit{ACE}/SWICS \citep{Gloeckler1998} data show that during the density enhancement, the C$^{+6}$/C$^{+5}$ ratio reduces to ~0.5; its average value is $\approx$ 1.0 after that. The O$^{+7}$/O$^{+6}$ ratio is roughly constant ($\approx$ 0.2) before and after the density enhancement between 13:14 to 15:40 UT. Moreover, the average charges state of Carbon (+5), Oxygen (+6), and Iron (+10) remain roughly constant over this period. The Fe/O ratio is enhanced up to 0.24, which is three times larger than its ambient solar wind value.

The alpha to proton ratio (Figure~\ref{insitu6Oct10}, lowest panel) is elevated over the entire CME interval peaking at 0.07 at 18:30 UT 2010, October 11, which is seven times the ambient solar wind value. During the time interval of second density enhancement alpha to proton ratio is approximately three times greater than its value in the ambient solar wind (before the leading edge of CME). Between 13:14 and 15:40 UT, the thermal speeds of He$^{+2}$, C$^{+5}$, O$^{+7}$, Fe$^{+10}$ are depressed, with minimum value of 18, 13, 12, and 13 km s$^{-1}$, respectively. These signatures are indicative of filament material from 13:14 to 15:40 UT.

Previously, using very restrictive criteria, attempts have been made to identify filament plasma near 1 AU \citep{Lepri2010}, and in a few cases filament signatures have been identified \citep{Burlaga1998, Gopalswamy1998}. In these studies, a selected threshold value for the thermal and compositional characteristics of the plasma was used to identify cold filament. However, these values will likely vary for CMEs associated with eruptive prominences at different phases of the solar cycle, and hence for different solar wind conditions. \citet{Sharma2012} adopted less stringent criteria for the identification of filaments in in situ data, based on the thermal speed signatures of various ions, charge states of carbon, oxygen, and iron, and relative abundances of He$^{+2}$/H$^{+}$, Fe/O, C$^{+6}$/C$^{+5}$, O$^{+7}$/O$^{+6}$ for the events of 2003 November 18 and 2010 August 1. Based on the criteria of \citet{Sharma2012}, many of the signatures observed between 13:00 and 16:00 UT on 11 October 2010 suggest that second density enhancement in in situ data associated with Feature 2 can be identified as filament material. Further, \citet{Sharma2012} observed a depression in the average charge state of carbon (+4 to +5), oxygen (+ 5 to +6), and iron (+8); however, for the present case, we do not notice such a depression. In addition, we find a drop in the C$^{+6}$/C$^{+5}$ ratio to 0.5 in the filament region, while the minimum value of the C$^{+6}$/C$^{+5}$ ratio reported by \citet{Sharma2012} was nearly 0.15 in the cold plasma region. Therefore,   although the CME of 2010 October 6 was associated with a filament eruption, all the criteria required to definitively identify filament plasma are not fulfilled. This may be due to in situ observations taken at a single point or may occur due to the heating of the cold filament material during its journey to L1 \citep{Skoug1999, Sharma2012}.

\section{Results and Discussion}
\label{Resdis}
For nine selected Earth-directed CMEs, the association between remote and in situ observations was carried out. In this study, we tracked the first density enhancement of CMEs in the \textit{J}-maps. The density enhancement in the shock-sheath region of the CME identified in situ observations is considered as a reference for the actual arrival time of the remotely observed tracked feature. Based on our analysis, we found that it is difficult to track a CME-driven shock in the \textit{J}-maps constructed from HI observations. We also realize that using difference images for tracking cavity (observed as MC in in situ) may be extremely difficult. It is because of the extremely low density associated with a cavity (flux rope), which on taking the running difference becomes indistinguishable from the brighter background. However, our analysis shows that the arrival time of the first density peak of the CME at L1 can be predicted reasonably well using \textit{J}-maps constructed from SECCHI/HI observations. Therefore, we conclude that an association between remote and in situ observations of CMEs can be made if they are tracked by estimating their 3D kinematics out to larger elongation using HI observations.

In several studies carried out previously without the construction of \textit{J}-maps, different authors have adopted different signatures of CMEs near 1 AU as references in correlating remote sensing observations with in situ observations. For instance, in the derivation of the ECA time prediction model, \citep{Gopalswamy2000} adopted the start time of the magnetic cloud and low proton beta ($\beta$ $<$ 1) as a reference for the actual arrival time of the CME. In another study, \citep{Schwenn2005}, an IP shock was taken as a reliable in situ signature for the arrival of a CME. \citet{Zhang2003} considered the minimum Dst index of the associated geomagnetic storm as the ICME arrival time. \citet{Kilpua2012} considered the arrival time of the CME leading edge at in situ spacecraft as the arrival of a CME. In our selected CMEs, the CMEs of 2010 February 7, and 2010 April 8 drive extremely weak shocks, which seem to arrival of the sheath structure; however, CME of 2010 February 12, 2010 April 3, and 2010 October 6 drive a clear IP shock identified in in situ observations at L1. Furthermore, four events, i.e., 2008 December 12, 2010 April 8, 2010 October 6, and 2010 October 26, could be categorized as magnetic clouds in in situ measurements. Also, for all the CMEs, except 2010 October 10, leading and trailing boundaries of the CMEs could be well-identified using in situ data. Hence, we can infer that specific signatures (i.e. for shock, leading edge, and MC) in situ observations cannot always be strictly found and used to pinpoint the actual (reference) arrival time of a remotely tracked features of the CME. The absence of MC for a few selected CMEs may be because either it was not intersected by the in situ spacecraft or it was veritably absent or less structured \citep{Gopalswamy2006, Webb2012} in those CMEs.

For 2010 October 6 CME, two bright features were tracked, which are associated with two enhanced density structures in in situ observations at L1. Feature 1 is associated with the sheath region of CME, while Feature 2 is associated with density enhancement at the rear edge of the magnetic cloud. The plasma parameters in the second density structure show signatures of filament material. Association between remotely tracked Feature 2 and second density enhancement measured in situ relies on the kinematics of both features. Since filament follows the cavity (flux rope) in imaging (COR) observations, therefore, based on the in situ ( plasma and compositional) data, we expect that the second peak in density corresponding to Feature 2 at the rear edge of MC is due to the arrival of the filament structure. Based on our approach of continuously tracking density structures at the front and rear edge of CME and comparing them with in situ data, we highlight that it is possible to track a filament in the heliosphere using HI images. As mentioned in Section~\ref{recons}, F1 and F2 have been tracked as leading edge and filament in COR2 FOV as such, while Feature 1 and Feature 2 tracked as brightness enhancement (in HI FOV) could be associated as sheath and filament material, respectively, in in situ data. Therefore, one can infer that Feature 1 and Feature 2 represent the continuous tracking of F1 and F2 further out in the heliosphere.

Although we believe that remotely observed features have been successfully associated with in situ structures for the CMEs studied here, it is important to remember the fundamental difference between both sets of observations. While the imaging observations record the density information at all depths (azimuth) of the heliosphere along the line of sight, the in situ observations measure density at a particular depth and azimuth in the heliosphere at a time. Moreover, due to a fixed single point location of in situ spacecraft, it is difficult to claim that remotely tracked features are certainly intersected by the in situ spacecraft unless we track them up to L1, which could not be done in the present case.

\chapter{Interplanetary Consequences of CMEs}
\label{Chap5:Conseque}
\rhead{Chapter~\ref{Chap5:Conseque}. Interplanetary Consequences of CMEs}

CMEs are frequent expulsions of massive magnetized plasma from the solar corona into the heliosphere. They are known to cause several interplanetary consequences and space weather effects. Such consequences are described in Section~\ref{HelioConse} of 
Chapter~\ref{Chap1:IntMot}. If the CMEs are directed toward the Earth and have enhanced southward magnetic field, they can result in severe geomagnetic storms \citep{Dungey1961, Gosling1990,Echer2008}. This chapter focuses on understanding the heliospheric and geomagnetic consequences of interacting CMEs. The CMEs in the heliosphere can affect the other solar wind structures (e.g., CIRs/CMEs) if they collide or interact with them. Here, we also investigate the morphological and kinematic evolution of CMEs launched in quick succession from the Sun. 

\section{Interaction of CMEs} 
\label{IntCMEsIntro}
The typical transit time of CMEs from the Sun to the Earth is between 1 to 4 days, and the number of CMEs launched from the Sun is about 5 per day around solar maximum \citep{StCyr2000, Webb2012}. Sometimes, CMEs are observed to erupt in quick succession, which, under certain favorable initial conditions, can interact or merge during their propagation in the heliosphere. Therefore, the interaction of CMEs in the heliosphere is expected to be more frequent near the solar maximum. In such a scenario, space weather prediction schemes may not be successful without considering the post-interaction CME characteristics.  The possibility of CME-CME interaction has been reported much earlier by analyzing in situ observations of CMEs by Pioneer 9 spacecraft \citep{Intriligator1976}. The compound streams (interaction of CME-CIR or CME-CME) were first inferred by \citet{Burlaga1987} using observations from \textit{Helios} and \textit{ISEE-3} spacecraft. They showed that such compound streams formed due to interactions have amplified parameters responsible for producing major geomagnetic storms. Using a wide field of view 
(FOV) coronagraphic observations of LASCO \citep{Brueckner1995} on-board SOHO and long-wavelength radio observations, \citet{Gopalswamy2001apj} first time provided evidence for CME-CME interaction. \citet{Burlaga2002} identified a set of successive halo CMEs directed toward the Earth and found that they appeared as complex ejecta near 1 AU \citep{Burlaga2001}. They inferred that these CMEs launched successively, merged en route from the Sun to the Earth, and formed complex ejecta in which the identity of individual CMEs was lost. Thus, these interactions are of great importance from a space weather point of view.

\citet{Farrugia2004, Farrugia2006} showed that CME-CME interactions are important as they can result in an extended period of enhanced southward magnetic field, which can cause intense geomagnetic storms. Such interactions help to understand the collisions between large-scale magnetized plasmoids and hence various plasma processes. Also, suppose a shock from a following CME penetrates a preceding CME. In that case, it provides a unique opportunity to study the evolution of the shock strength and structure and its effect on preceding CME plasma parameters \citep{Lugaz2005, Mostl2012,Liu2012}. Since estimating the arrival time of CMEs at the Earth is crucial for predicting space weather effects near the Earth, and CME-CME interactions are responsible for changing the dynamics of interacting CMEs in the heliosphere. Therefore, such interactions need to be examined in detail. Further, reconnection between magnetic flux tubes of CMEs can be explored by studying cases of CME-CME interactions \citep{Gopalswamy2001apj, Wang2003} which are also known to lead to solar energetic particles (SEPs) events \citep{Gopalswamy2002}.

Before the era of wide-angle imaging far from the Sun, the understanding of involved physical mechanisms in CME-CME or CME-shock interaction was achieved mainly from magnetohydrodynamic (MHD) numerical simulations of the interaction of a shock wave with a magnetic cloud (MC)  \citep{Vandas1997, Vandas2004,Xiong2006}, the interaction of two ejecta \citep{Gonzalez-Esparza2004,Lugaz2005, Wang2005}, and the interaction of two MCs \citep{Xiong2007, Xiong2009}. Also, limited efforts regarding CME-CME interaction studies were made by analyzing imaging observations near the Sun and in situ observations near the Earth. \citet{Wang2003} have shown that a forward shock can cause an intense southward magnetic field of long duration in the preceding MC. Such modifications in the preceding cloud are important for space weather prediction. Therefore, events involving CME-CME and CME-shock interactions are important candidates to investigate their kinematics and arrival time near theEarth.

After the launch of twin \textit{STEREO} \citep{Kaiser2008} in 2006, it is possible to continuously image the CMEs from its lift-off in the corona up to the Earth and beyond. Such twin spacecraft observations also enable to determination the 3D locations of CMEs features in the heliosphere and hence provide direct evidence of CME-CME interaction using images from Heliospheric Imager (HI) on the SECCHI package \citep{Howard2008}. However, immediately after the launch of \textit{STEREO}, during deep extended solar minimum, not many interacting CMEs were observed. As the current solar cycle progressed, CME interaction appears to be a fairly common phenomenon, particularly around solar maximum. The interacting CMEs of 2010 August 1 have been studied extensively by using primarily the \textit{STEREO}/HI (white light imaging), near-Earth in situ and, \textit{STEREO}/Waves radio observations \citep{Harrison2012,Liu2012,Mostl2012,Temmer2012,Martinez-Oliveros2012,Webb2013}. Also, \citet{Lugaz2012} have reported a clear deflection of 2010 May 23-24 CMEs after their interaction. Therefore, only a few studies on CME interaction have been reported as mentioned above. Several key questions regarding CME interaction need to be addressed, which are not well understood quantitatively, \textit{viz}. 

\begin{enumerate}
\item{How do the dynamics of CMEs change after the interaction? What is the regime of interaction, \textit{i.e.} elastic, inelastic, or super-elastic? \citep{Lugaz2012,Shen2012,Mishra2014a,Mishra2014b}.} 
\item{What are the consequences of the interaction of CME-shock structure? How does the overtaking shock change the plasma and magnetic field properties in the preceding magnetic cloud? \citep{Lugaz2005,Lugaz2012,Liu2012}.}
\item{What are the favorable conditions for the merging of CMEs and the role of magnetic reconnection in it? \citep{Gopalswamy2001apj}.} 
\item{What is the possibility for the production of a reverse shock at the CME-CME interaction site? \citep{Lugaz2005}.}
\item{Do these interacted structures produce different geomagnetic consequences than individual CMEs, on their arrival to the magnetosphere? \citep{Farrugia2006}.} 
\item{What are the favorable conditions for the deflection of CMEs and enhanced radio emission during CME-CME interaction? \citep{Lugaz2012,Martinez-Oliveros2012}.} 

\end{enumerate}

In light of the aforementioned questions and only a few studies reported, it is clear that the prediction of the arrival time of interacting CMEs and association of remote observations of such CMEs with in situ observations are challenging. Our study selected two cases of interacting CMEs of 2011 February 13-15 and 2012 November 9-10. We exploited the remote observations of these CMEs from \textit{SOHO}/LASCO, SECCHI/COR, and SECCHI/HI as well as near-Earth observations from \textit{ACE} and \textit{WIND} spacecraft to achieve our goals.

\section{Interacting CMEs of 2011 February 13-15}
\label{IntFeb}
We had selected three Earth-directed interacting CMEs launched during 2011, February 13 - 15, from NOAA AR 11158, when the twin \textit{STEREO} spacecraft were separated by approximately 180$^\circ$. The interaction of these CMEs has been studied based on imaging and in situ observations from \textit{STEREO} and \textit{WIND} spacecraft, respectively. These interacting CMEs have also been studied by \citet{Maricic2014} and \citet{Temmer2014}.

\citet{Maricic2014} used the plane of sky approximation and Harmonic Mean (HM) method \citep{Lugaz2009} to convert the derived elongation-time profiles from single \textit{STEREO} spacecraft to distance-time profiles of the leading edge of CMEs and then estimated their arrival time at L1. Their approach seems to be less reliable at larger elongation, where the direction of propagation and structure of CME plays a crucial role. Further, \citet{Maricic2014} did not construct the \textit{J}-maps and could track a CME feature only up to small elongations ($\approx$ 25$\arcdeg$). However, in our analysis, we have constructed \textit{J}-maps \citep{Sheeley1999, Davies2009} which allow us to track the CMEs to significantly greater elongations ($\approx$ 45$\arcdeg$). Recent studies have shown that tracking of CMEs to larger elongation using \textit{J}-maps and subsequent stereoscopic reconstruction gives more precise kinematics and estimates of the arrival time of CMEs than using single spacecraft reconstruction methods \citep{Williams2009, Liu2010,Lugaz2010, Mishra2013,Colaninno2013, Mishra2014}. In our study, we have applied the stereoscopic methods (Stereoscopic Self-Similar Expansion (SSSE); \citealt{Davies2013}, Tangent to A Sphere (TAS); \citealt{Lugaz2010.apj} and Geometric Triangulation (GT); \citealt{Liu2010}) which apart from the dynamics also yield the time-variations of the direction of propagation of CMEs. We have estimated the kinematics of overall CME structure using Graduated Cylindrical Shell (GCS) model \citep{Thernisien2009} in COR2 FOV while \citet{Maricic2014} estimated the kinematics of a single tracked feature. The kinematics of the overall CME structure in COR2 FOV helps determine the probability of collision of CMEs beyond COR2 FOV. \citet{Temmer2014} studied the interaction of February 14-15 CMEs corresponding to different position angles measured over the entire latitudinal extent of these CMEs. In this context, the present study is important as it also focuses on understanding the nature of collision by calculating momentum and energy exchange during the collision phase of the CMEs.

At the time of 2011 February interacting CMEs, the unique positioning of twin \textit{STEREO} spacecraft with $\approx$ 180$^\circ$ separation between them, the first time since its launch, enticed us to do an additional study on the geometrical evolution of these Earth-directed CMEs in COR2 FOV from identical multiple viewpoints. The location of the active region 
(S20E04 to S20W12 during February 13-15) for these CMEs allowed its SECCHI/COR coronagraph to observe these Earth-directed CMEs at the limb (i.e., plane orthogonal to the Sun-\textit{STEREO} line) contrary to \textit{SOHO}/LASCO observations, which always record such CMEs as halos. In this scenario, the CME observations are least affected by the projection effects in both SECCHI/COR-A and B FOV, and hence crucial parameters, i.e., widths, speeds, etc. that define the geoeffectiveness of CMEs can be determined with reasonable accuracy. Morphological studies have been carried out earlier, assuming either a cone or ice cream cone model for CMEs for estimating the true angular width, central position angle, radial speed, and acceleration of halo CMEs \citep{Zhao2002, Michalek2003,Xie2004, Xue2005}. Also, \citet{Howard1982, Fisher1984} suggested that the geometrical properties of CMEs can be described by a  cone model, which can be used to estimate their mass. All the cone models assume that the angular width of CMEs remains constant beyond a few solar radii as they propagate through the solar corona. \citet{Vrsnak2010, Vrsnak2013} also assumed a constant cone angular width of a CME for developing the drag-based model (DBM) of propagation of CMEs. Our study of the morphological evolution of the selected CMEs is expected to provide results that can help to refine the cone model by incorporating the possible variation in angular width of CMEs corresponding to their different speed, i.e., slower, comparable, and faster than the ambient solar wind speed.

In Section~\ref{IntFebMorphoCOR}, we present the morphological evolution of CMEs. In Section~\ref{IntFebEvolution}, the kinematics and interaction of CMEs in the heliosphere are discussed. In Section~\ref{IntFebCompWidth}, the angular widths of CMEs determined from 2D images are compared with the angular widths derived from the GCS model. In Section~\ref{IntFebCollision}, we focus on the nature of the collision and estimate the energy and momentum transfer during the collision of CMEs. In Section~\ref{IntFebInsitu}, in situ observations of CMEs are described. Section~\ref{IntFebArrtime} \& ~\ref{IntFebGeomag} describes the arrival of CMEs at L1 and their geomagnetic response, respectively.  The main results of the present study are discussed in Section~\ref{IntFebResDis}.

\subsection{Morphological evolution of interacting CMEs in the COR FOV}
\label{IntFebMorphoCOR}

CME of February 13 (hereinafter, CME1) was observed by \textit{SOHO}/LASCO-C2 at 18:36 UT on 2011 February 13 as a faint partial halo CME with an angular width of 276$^{\circ}$ and a linear speed of 370 km s$^{-1}$. In SECCHI/COR1-A and B, this CME appeared at 17:45 UT in SE and SW quadrants, respectively. The CME of February 14 (hereinafter, CME2) was first recorded by \textit{SOHO}/LASCO-C2 at 18:24 UT on 2011 February 14 as a halo with a linear speed of $\approx$ 325 km s$^{-1}$. The CME2 appeared in SECCHI/COR1-A and B at 17:45 UT at the east and west limb, respectively. In \textit{SOHO}/LASCO-C2 FOV, CME of February 15 (hereinafter, CME3) was first observed at 02:24 UT on 2011 February 15 as a halo with a linear speed of $\approx$ 670 km s$^{-1}$. In SECCHI/COR1-A and B images, the CME3 was first observed at 02:05 UT at the east and west limbs respectively. At the time of observation of CMEs presented in this study, \textit{STEREO-A} was $\approx$ 87$^{\circ}$ west, and \textit{STEREO-B} was $\approx$ 94$^{\circ}$ east of the Earth. They were approximately in an ecliptic plane with 0.96 AU and 1.0 AU distance from the Sun.

We measured the geometrical properties (e.g., cone angle) of the selected CMEs by analyzing the SECCHI/COR2 images to study the deviation of CMEs from the ideal cone model. We based our analysis on the concept that slow and fast speed CMEs interact with solar wind differently, and hence, the deviation of each from the cone model might be different. We did not use COR1 images, as near the Sun, magnetic forces are dominant within a few solar radii, and CMEs are not fully developed. We excluded the CME1 for morphological study as it was very faint in COR2 FOV. We selected CME2 and CME3, which had different speeds in COR2 FOV for the morphological analysis. 

We consider the ice-cream model of \citet{Xue2005} wherein CMEs are assumed to have a symmetrical cone shape combined with a sphere. The apex of the cone and center of the globe both are located at the center of the Sun, and CME is assumed to move radially outward, having constant cone angular width beyond a few solar radii from the Sun. We measured the cone angular width of CMEs using COR2 images and estimated the cone area,\\
i.e., $A$ = $\pi$$r^{2}$$\Theta$/360, where $\Theta$ is the cone angular width in degrees, and $r$ is the radius of the sphere ,which is equal to the distance between the front edge of CME and the center of the Sun. Hence, the estimated area using the above equation, is the area of the CME as it appears from sideways (perpendicular to its direction of propagation).

To calculate the cone area, we processed the SECCHI/COR2 images of CME2 and CME3 and then subtracted a background image from them. Further, we enclosed the CME area by manual clicking and outlining the CME boundary. We used a few initial points on each side of the CME flank close to the coronagraph occulter to fit a cone model. These points are used to estimate the position angle at both flanks (near the apex of cone) of CMEs. The difference in position angle at both flanks is the 2D angular width of CME. In the top panel of the Figure~\ref{CME2_contour}, the evolution of slow speed CME2 as observed in COR2-A images is shown with overlaid contour enclosing the entire CME, and over-plotted lines denoting limiting position angle at both the flanks of CME. We repeated this analysis for the fast speed CME3, and its appearance in COR2-A FOV is shown in the bottom panel of Figure~\ref{CME2_contour}.

\begin{figure}[!htb]
 \centering
  \includegraphics[scale=0.35]{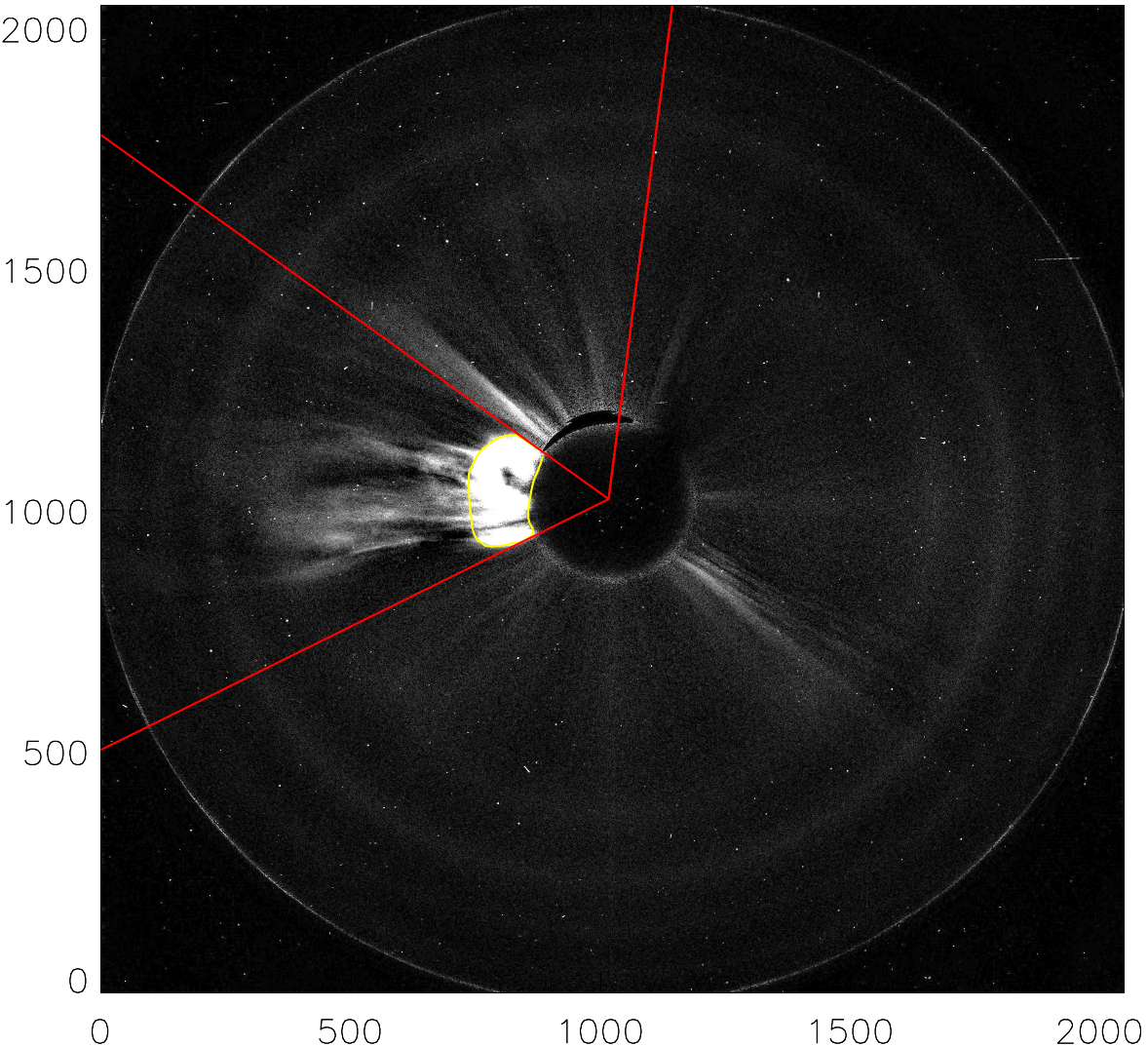}
  \includegraphics[scale=0.35]{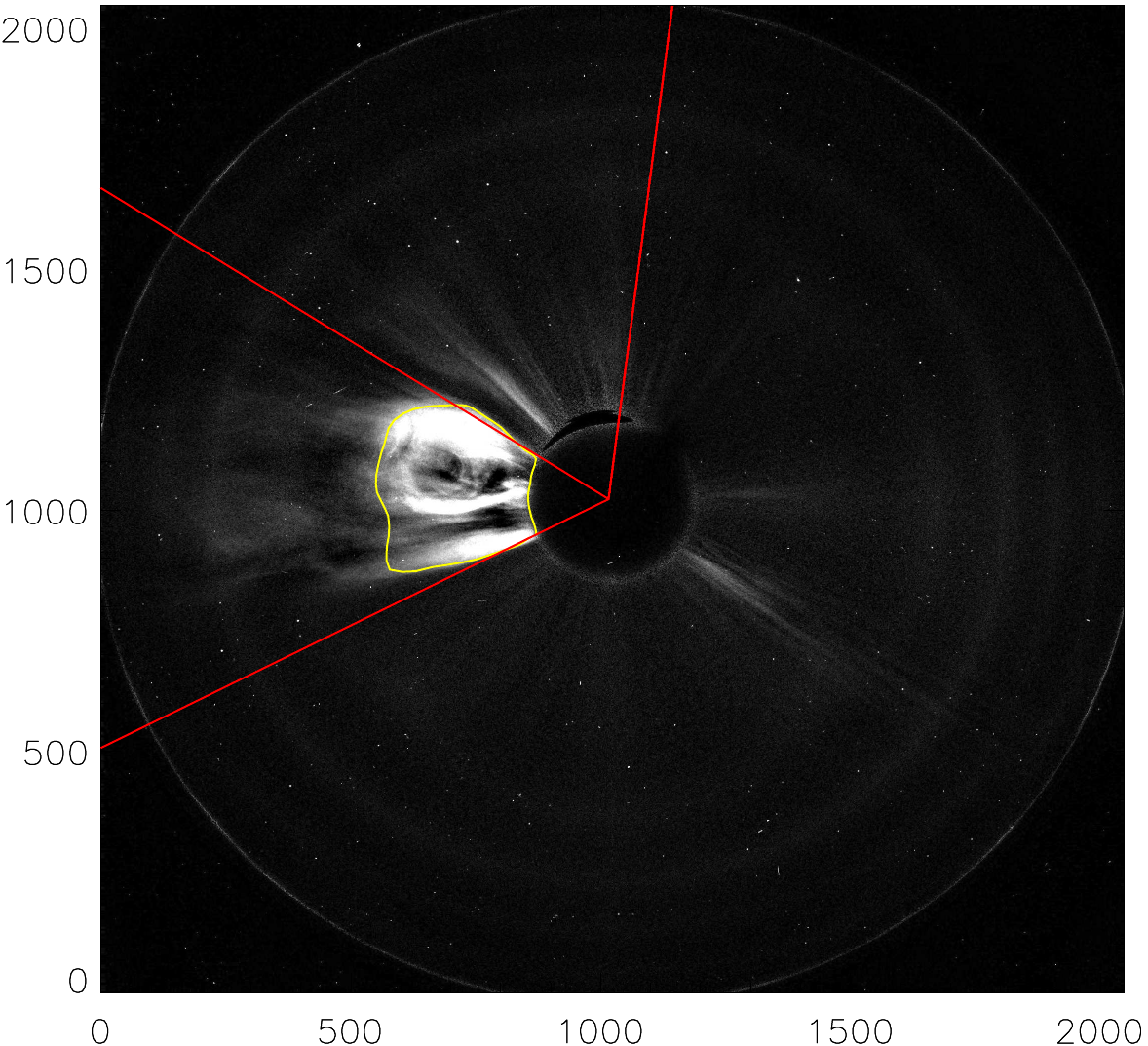}
	\includegraphics[scale=0.35]{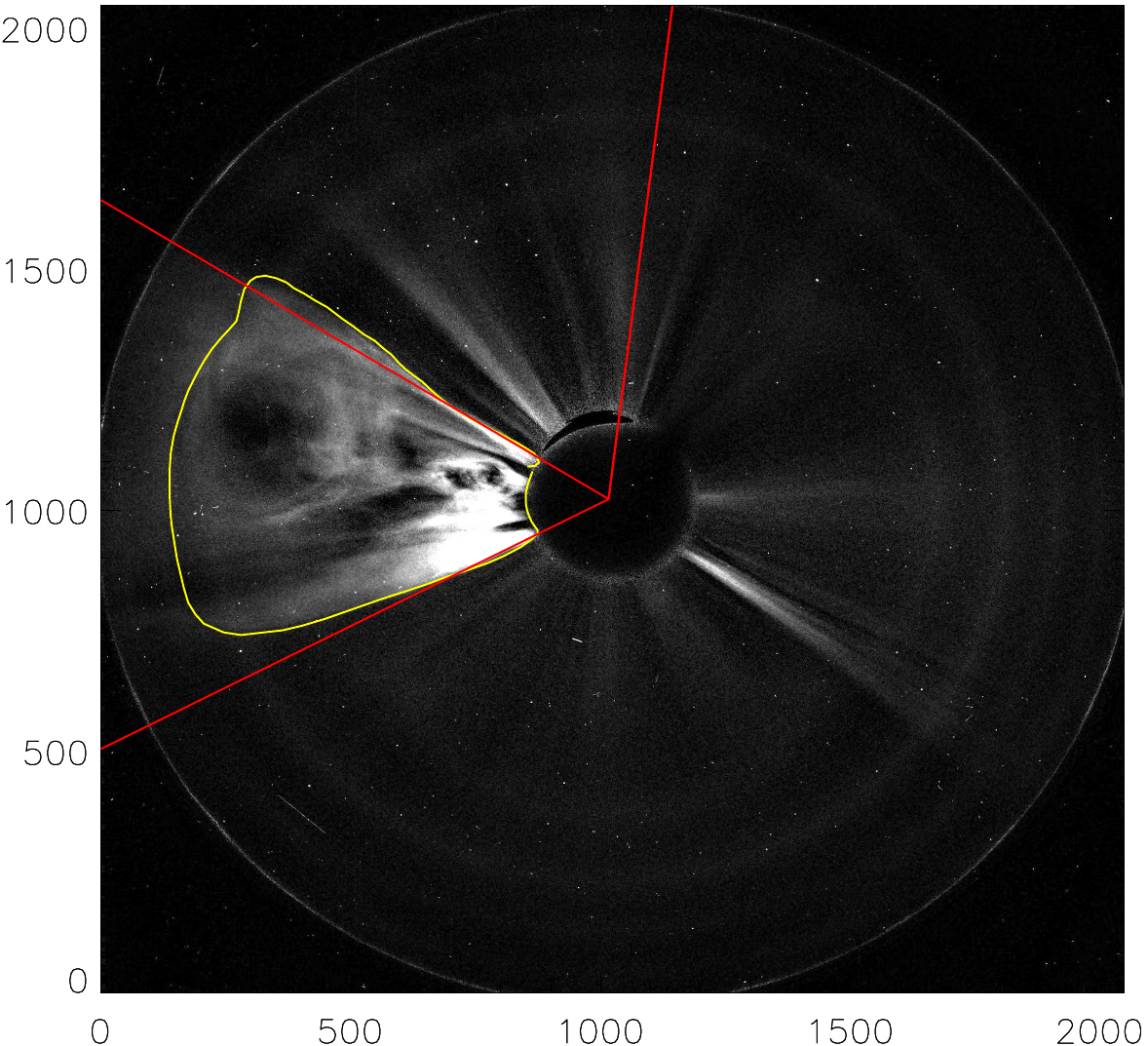}
 
  \includegraphics[scale=0.35]{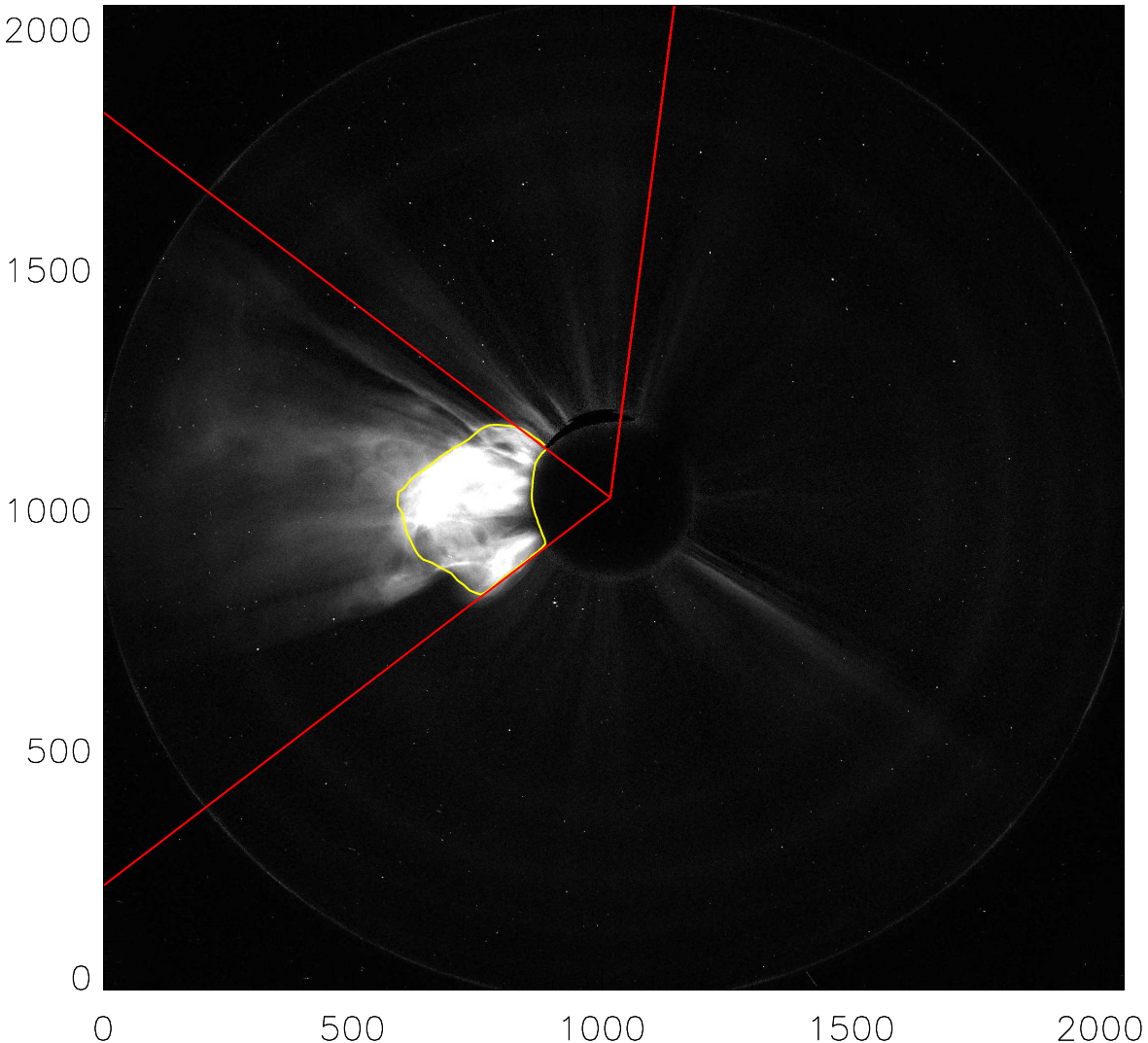}
  \includegraphics[scale=0.35]{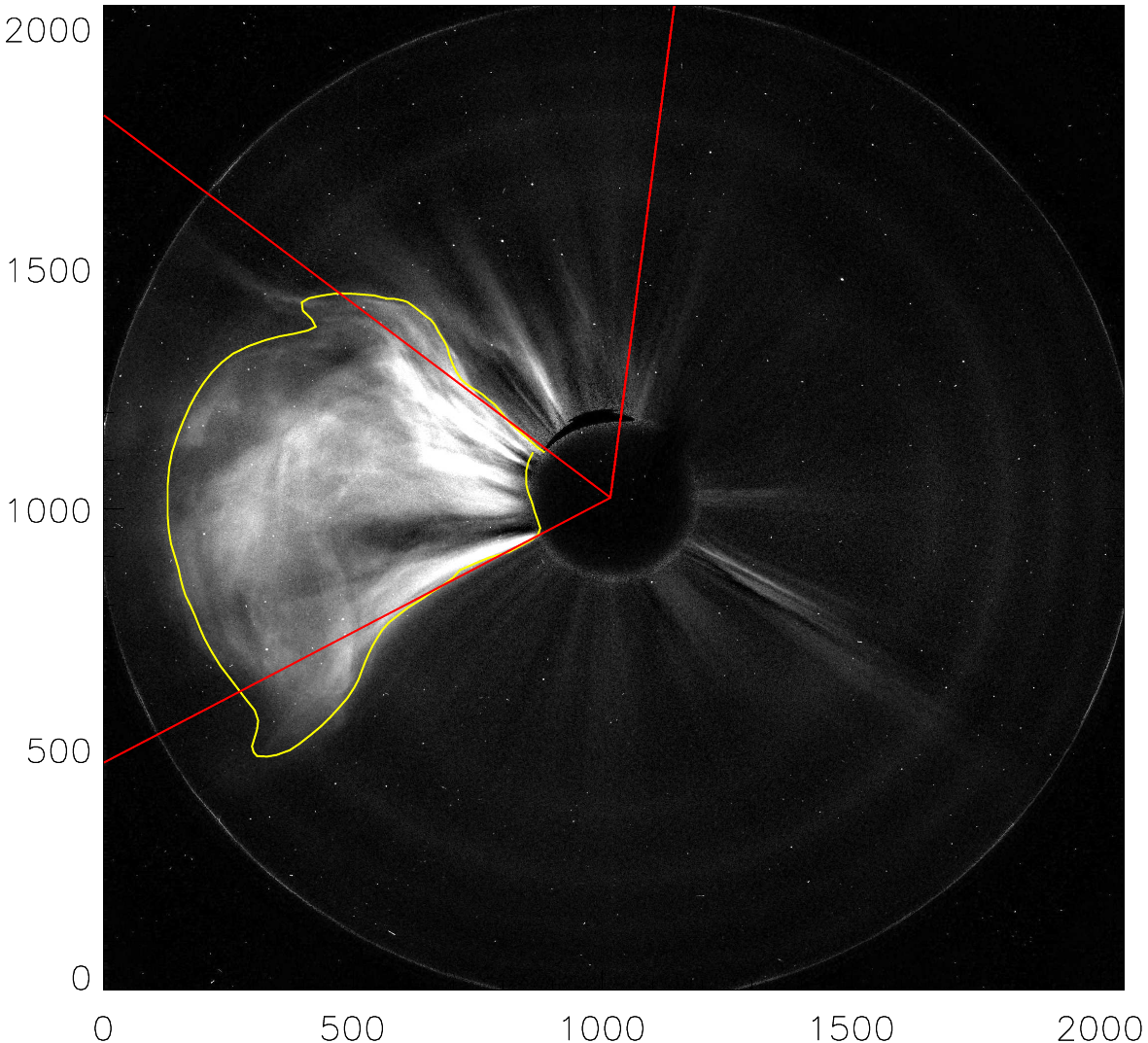}
	\includegraphics[scale=0.35]{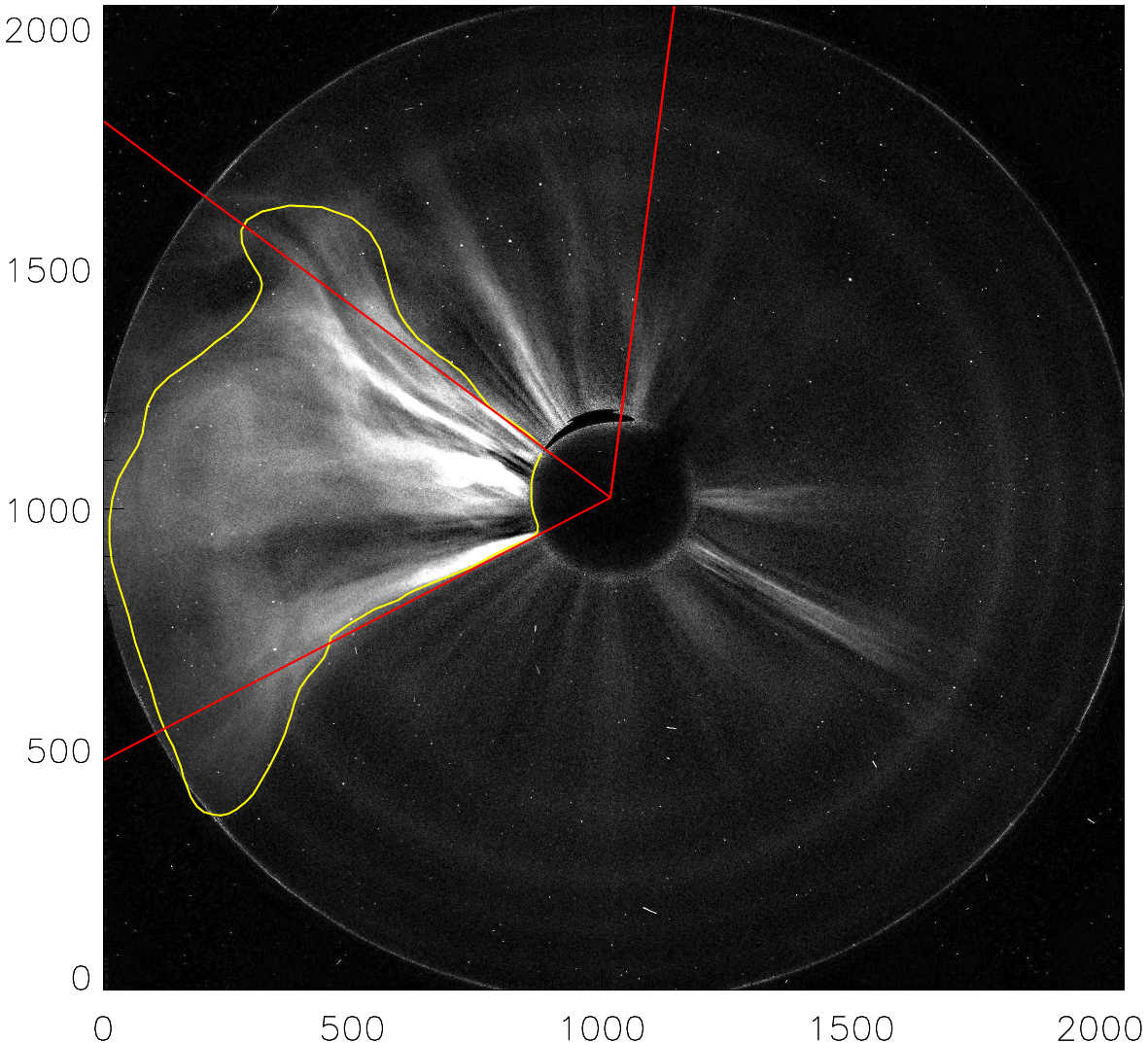}
 \caption[CMEs of the 2011 February 14 and 15 in COR2-A FOV]{The left, middle and right images in the top panel of the figure show CME2 in COR2-A FOV at 18:24 UT, 20:24 UT, and 23:54 UT on 14 February, respectively. Similarly, the bottom panel's left, middle and right images show CME3 in COR2-A FOV at 02:39 UT, 04:24 UT, and 05:54 UT on 15 February, respectively. The vertical red lines mark the zero degree position angle in a helio-projective radial coordinate system. The other two red lines forming the edges of the CME cone are marked at the position angle of the CME flanks. The contour with a yellow curve encloses the CME area completely.}
\label{CME2_contour}
\end{figure}

The estimated 2D cone angular width for CME2 and CME3 in COR2 FOV is shown in Figure~\ref{theta_CMEs}. From the figure, it is evident that slow speed CME2 has a nearly constant (between 60$^{\circ}$ to 57$^{\circ}$) 2D angular width in the COR2 FOV. For the fast speed CME3, the 2D angular width was $\approx$ 80$^{\circ}$ in the beginning, which then decreased to 62$^{\circ}$ as it crossed the outer edge of COR2 FOV. From the contour in Figure~\ref{CME2_contour} (top panel), it can be seen that CME2 followed the cone model, and a slight spill of CME2 on the upper edge is compensated by a void on the lower edge. For the fast speed CME3 (bottom panel of Figure~\ref{CME2_contour}), we noticed a significant spill on both sides (upper and lower edge) which increased with time in COR2-A FOV. The appearance and the variations of angular width of CME2 and CME3 in COR2-B images are the same as in COR2-A.

\begin{figure}[!htb]
 \centering
  \includegraphics[scale=0.9]{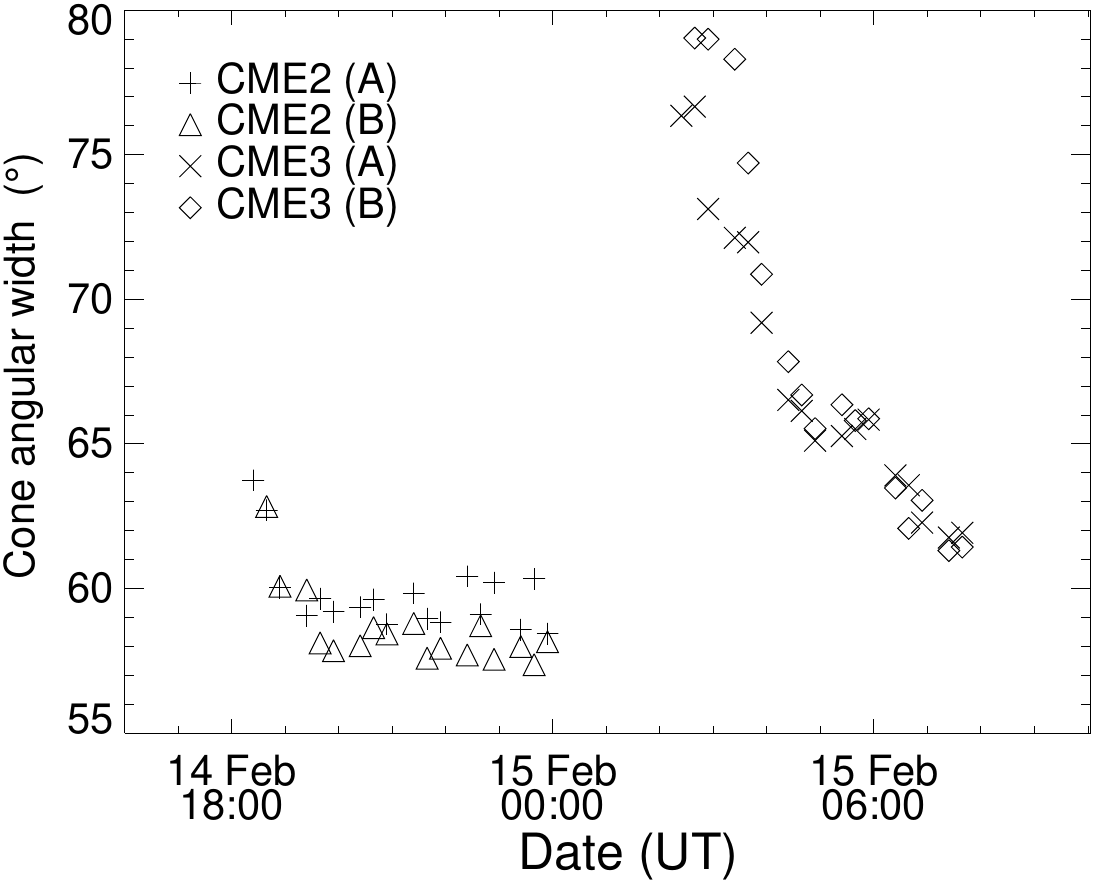}
 \caption[Time-variation of the estimated 2D cone angular width]{Time variation of estimated 2D cone angular width of CME2 and CME3 from both COR-A and COR-B images.}
\label{theta_CMEs}
 \end{figure}

To calculate the ice cream cone model area (cone area) for both CMEs, we marked a point along the CME leading edge on each image, the distance from the center of the Sun gave the radius of the sphere located on the cone-like CME. Since some part of the CME is blocked by the occulter in all the images. Therefore the area of CME blocked by the occulter has been subtracted from the sector (cone) area to compare it with the actual contour area of the CME. We also calculated the actual area enclosed by the CME contour (contour area). For convenience, both cone and contour areas have been measured in units of pixel$^{2}$, as it is the difference in the actual and the sector area that we are interested in. In the top panels of Figure~\ref{CME2_area} and ~\ref{CME3_area}, the blue curve represents the cone area (i.e., area obtained by approximating the CME with a cone model), and red curve represents the actual contour area. From these figures, we note that for both CME2 and CME3, time-variation of cone and contour area show a parabolic pattern in COR2-A and B FOV, which implies that area, $A$ is proportional to $r^{2}$. Therefore, we consider that both fast and slow CMEs follow the cone model to a certain extent. 

\begin{figure}[!htb]
 \centering
  \includegraphics[scale=0.7]{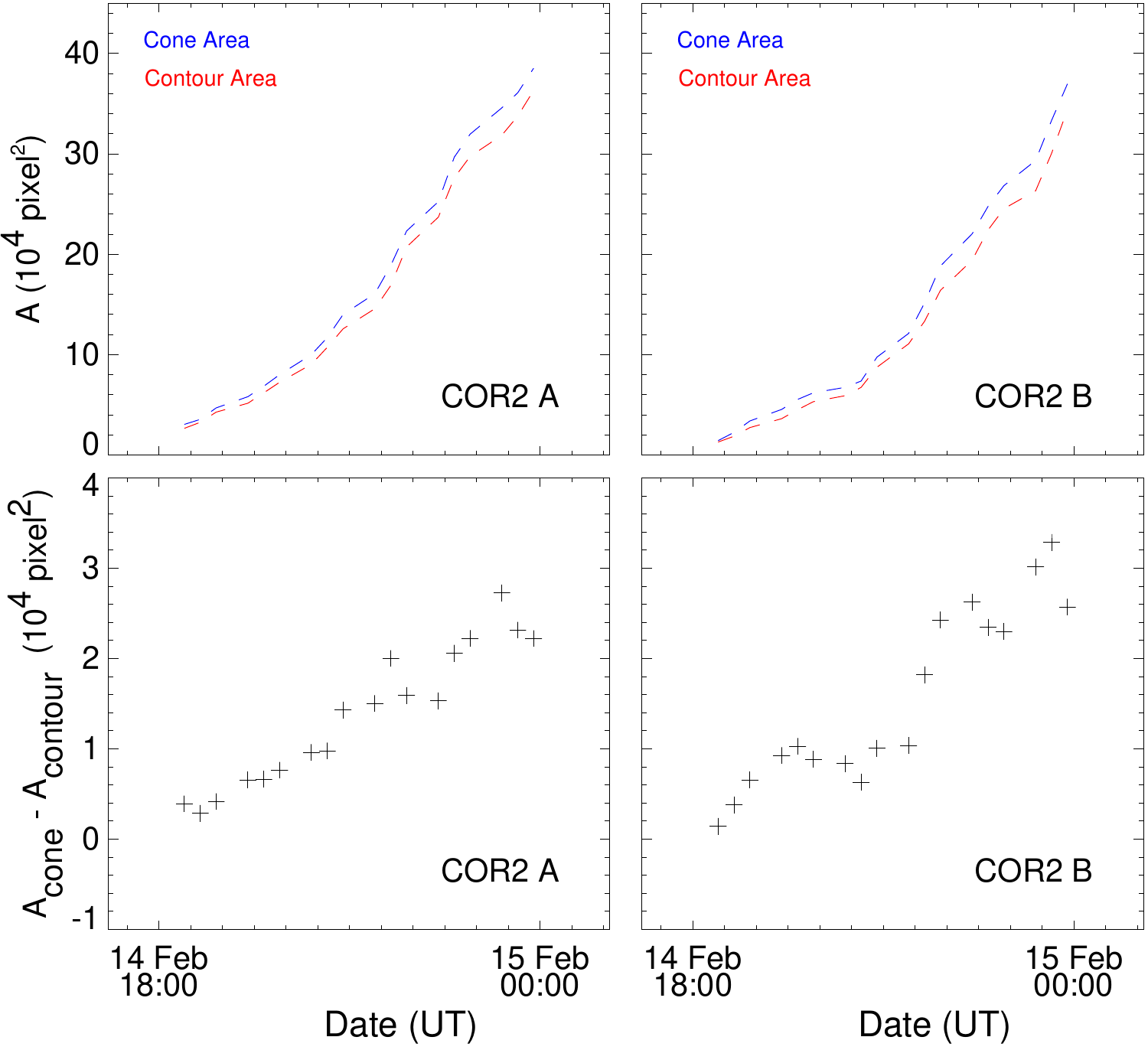}
 \caption[Variations in cone and contour area of the 2011 February 14 CME]{In the top left panel, time variations in cone and contour area of CME2 estimated in COR2-A FOV are shown with blue and red, respectively. Its variations in COR2-B FOV are shown in the top right panel. In the bottom left and right panels, the cone and contour area difference, estimated for COR2-A and B FOV, respectively, are shown.}
\label{CME2_area}
 \end{figure}

\begin{figure}[!htb] 
 \centering
  \includegraphics[scale=0.7]{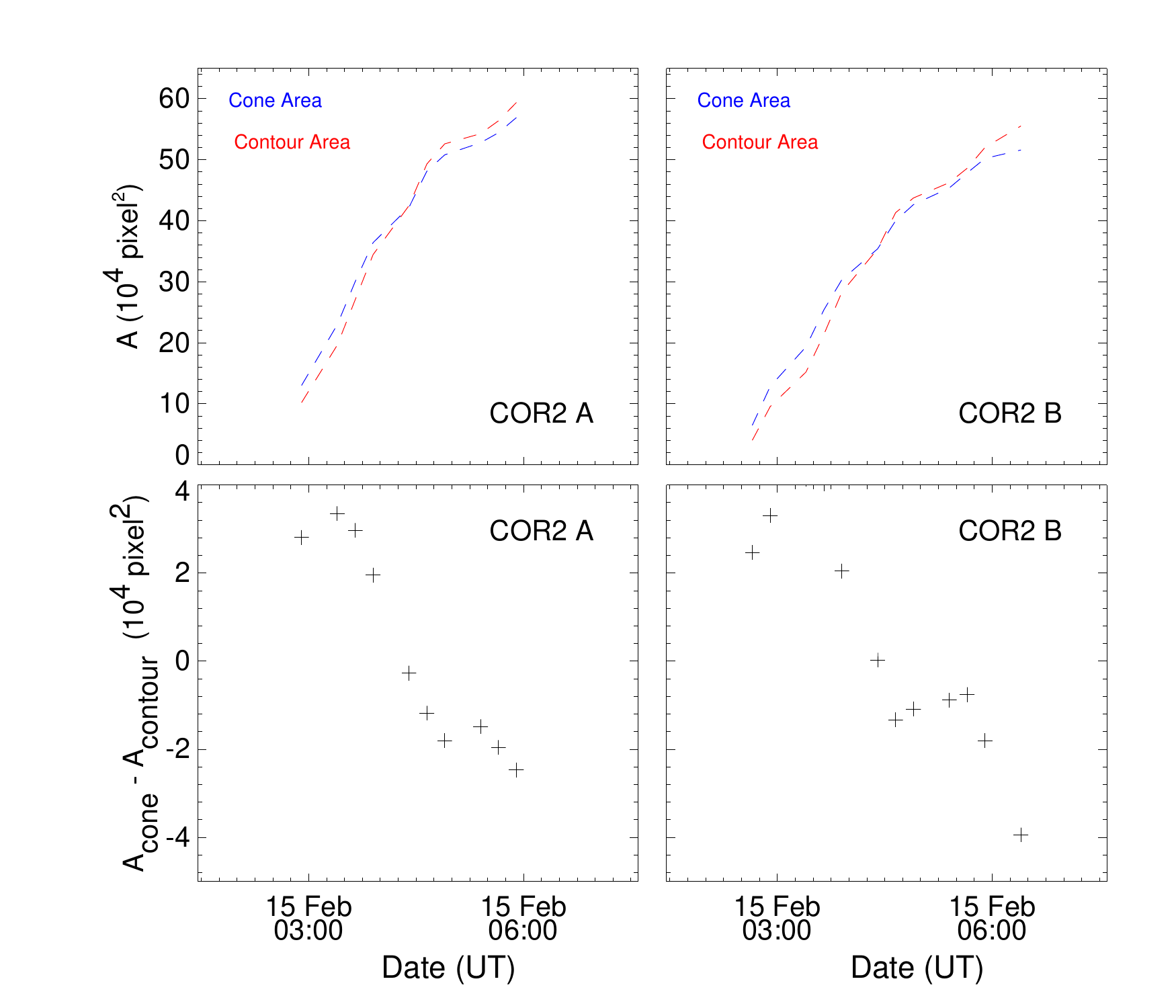}
 \caption{As Figure~\ref{CME2_area}, for the 2011 February 15 CME.}
\label{CME3_area}
 \end{figure}

From the top panel of Figure~\ref{CME2_area}, we find that the cone area is larger than the contour area, and both increased with time. The bottom panels show that the difference in cone area and contour area is positive and increased as the CME2 propagated in the outer corona. For CME3, in the top panel of Figure~\ref{CME3_area}, we find that the line representing the cone and contour area intersect one another in both COR2-A and B FOV. In the bottom panels, the difference in cone and contour area decreased from positive (2.8 $\times$ 10$^{4}$ pixel$^{2}$ in COR2-A and 2.4 $\times$ 10$^{4}$ pixel$^{2}$ in COR2-B) to negative values and remained so (-2.4 $\times$ 10$^{4}$ pixel$^{2}$ in COR2-A and -3.9 $\times$ 10$^{4}$ pixel$^{2}$ in COR2-B) as the CME3 propagated through the outer corona. This is suggestive that at a certain height during its propagation in COR2 FOV, the contour area became larger than the estimated cone area. These findings indicate dissimilar morphological evolution for slow and fast speed CMEs in the corona.

\subsection{Kinematic evolution and interaction of CMEs in the heliosphere}
\label{IntFebEvolution}

\subsubsection{3D reconstruction in COR FOV}
\label{IntFebReconsCOR}

The launch of CME1, CME2, and CME3 from the same active region in quick succession indicates a possibility of their interaction as they move out from the Sun into the heliosphere. To estimate the 3D kinematics of these CMEs, we have carried out the 3D reconstruction of CMEs using the Graduated Cylindrical Shell (GCS) model developed by \citet{Thernisien2009}. Before reconstruction, the total brightness images were processed, and then a pre-event image was subtracted from a sequence of contemporaneous images from SECCHI/COR2-B, SOHO/LASCO, and SECCHI/COR2-A to which the GCS model was applied. The images of CME1, CME2, and CME3 overlaid with the fitted GCS wireframed contour (hollow croissant) are shown in Figure~\ref{CME123_FM}.

\begin{figure}[!htb]
 \centering
 \includegraphics[scale=0.38]{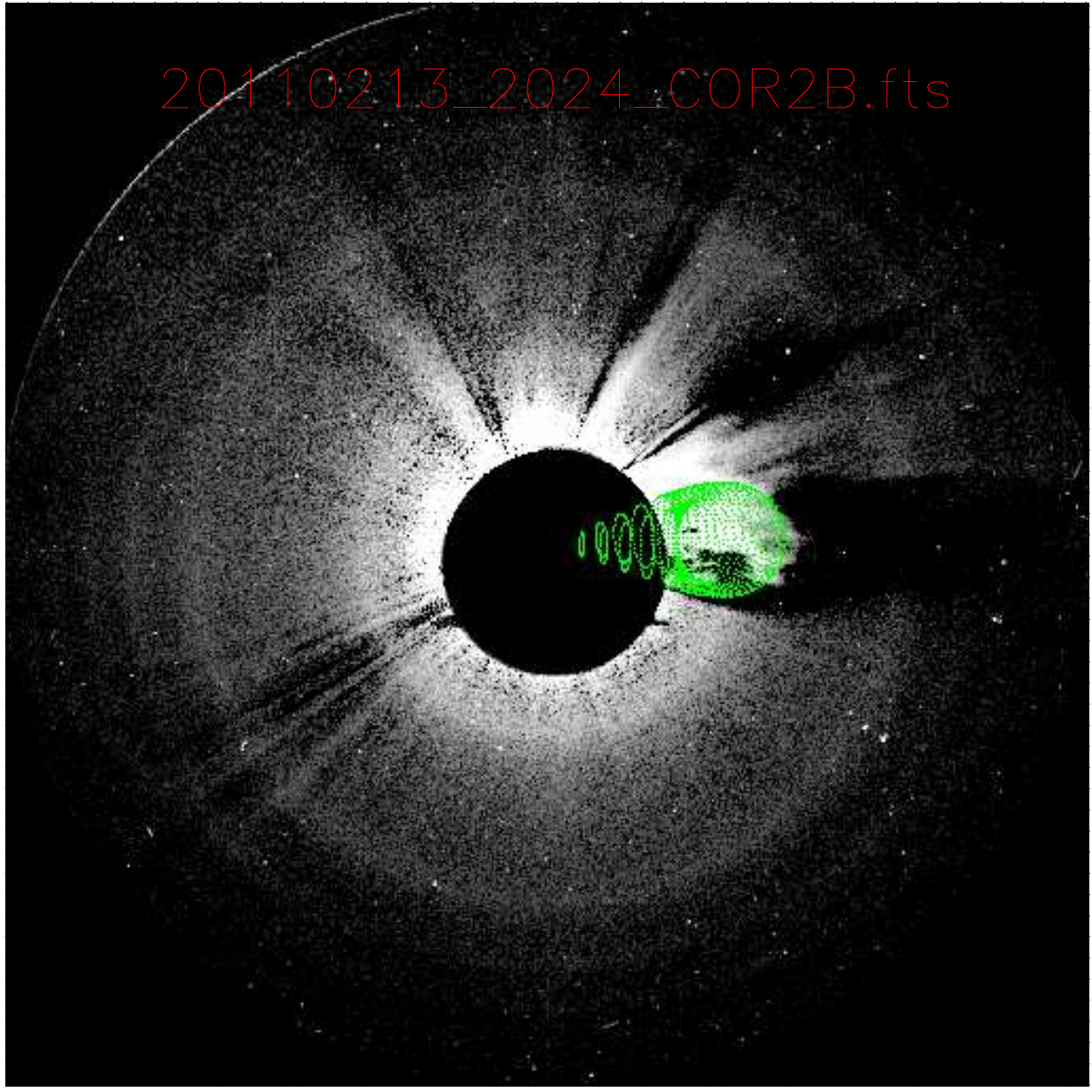}
 \includegraphics[scale=0.38]{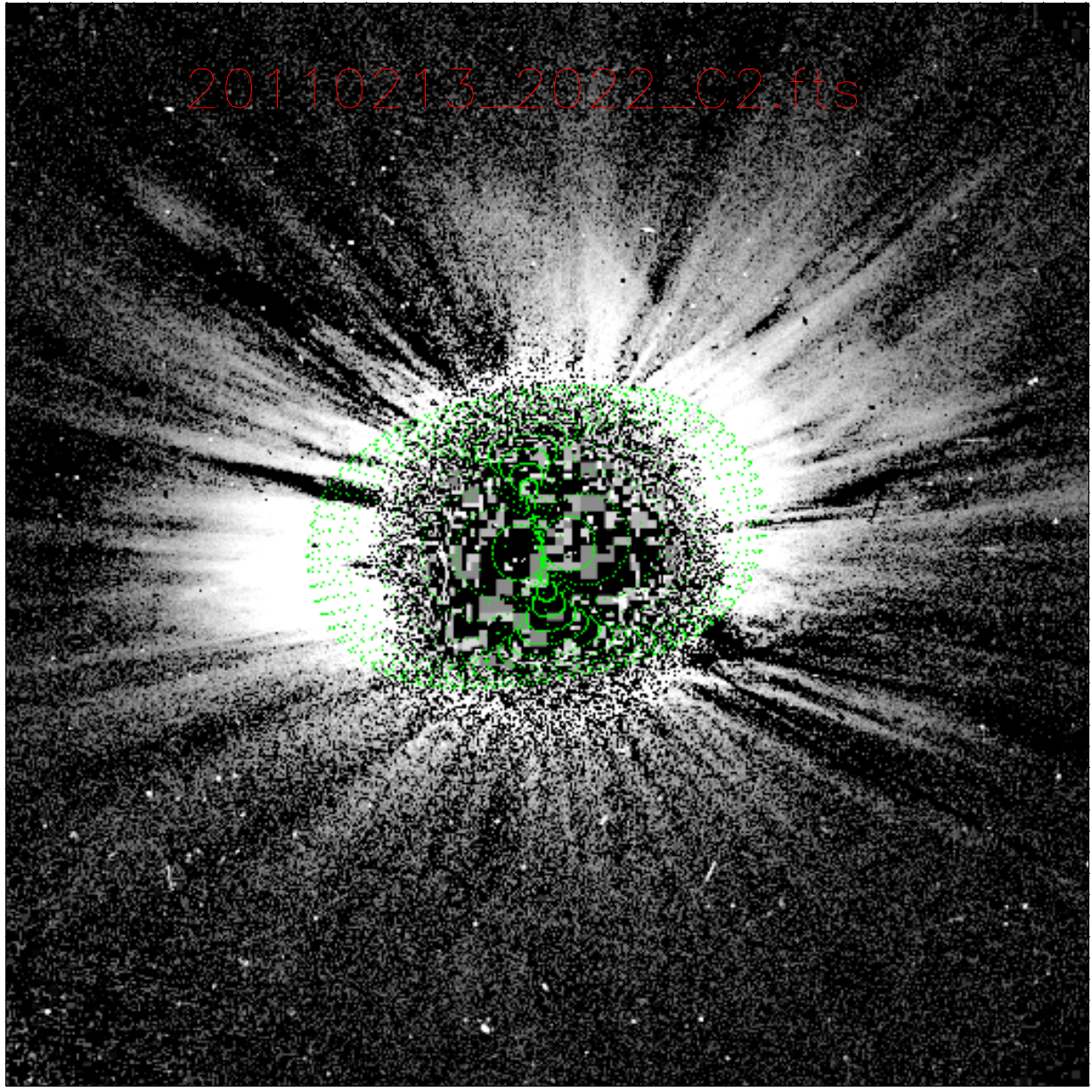}
 \includegraphics[scale=0.38]{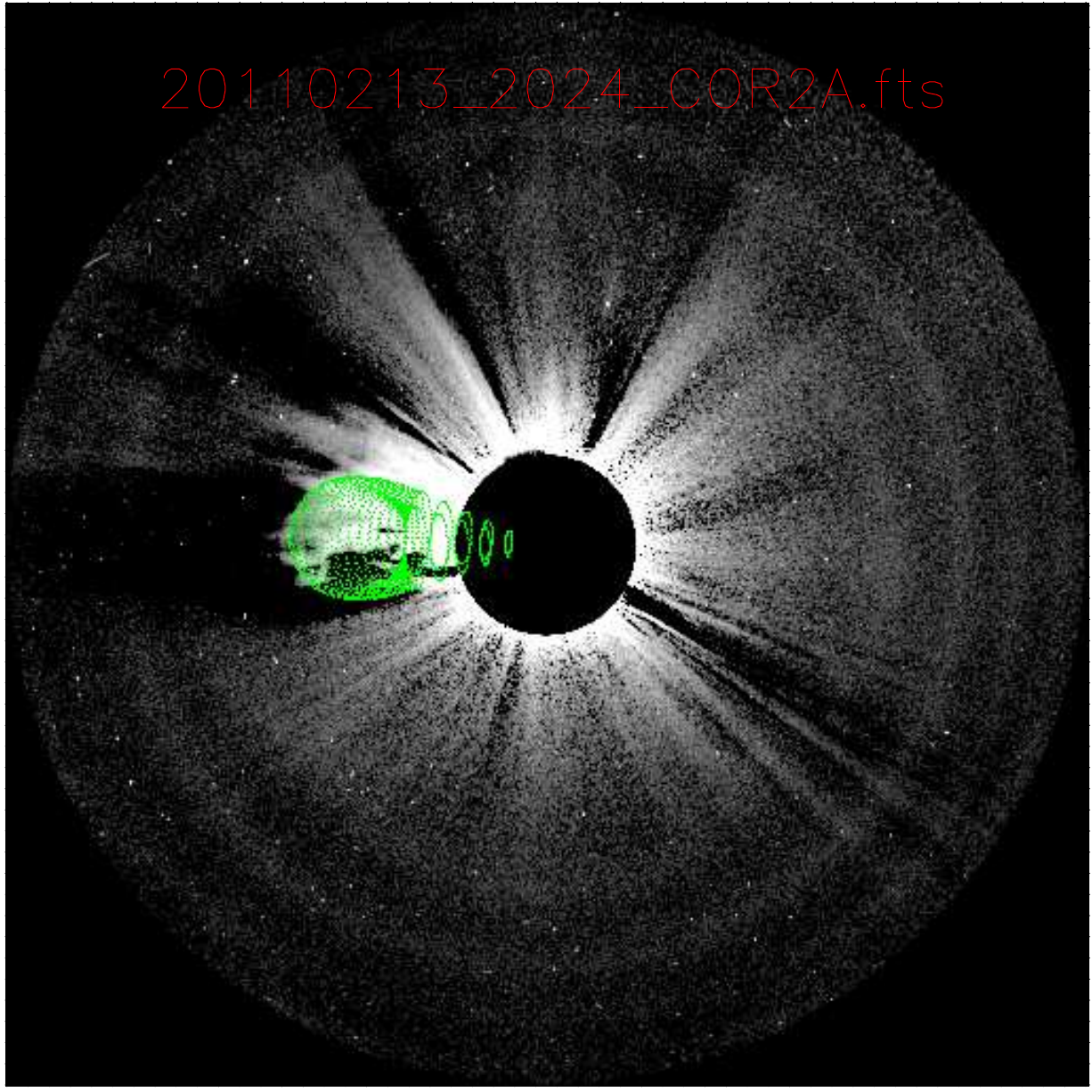}

  \includegraphics[scale=0.38]{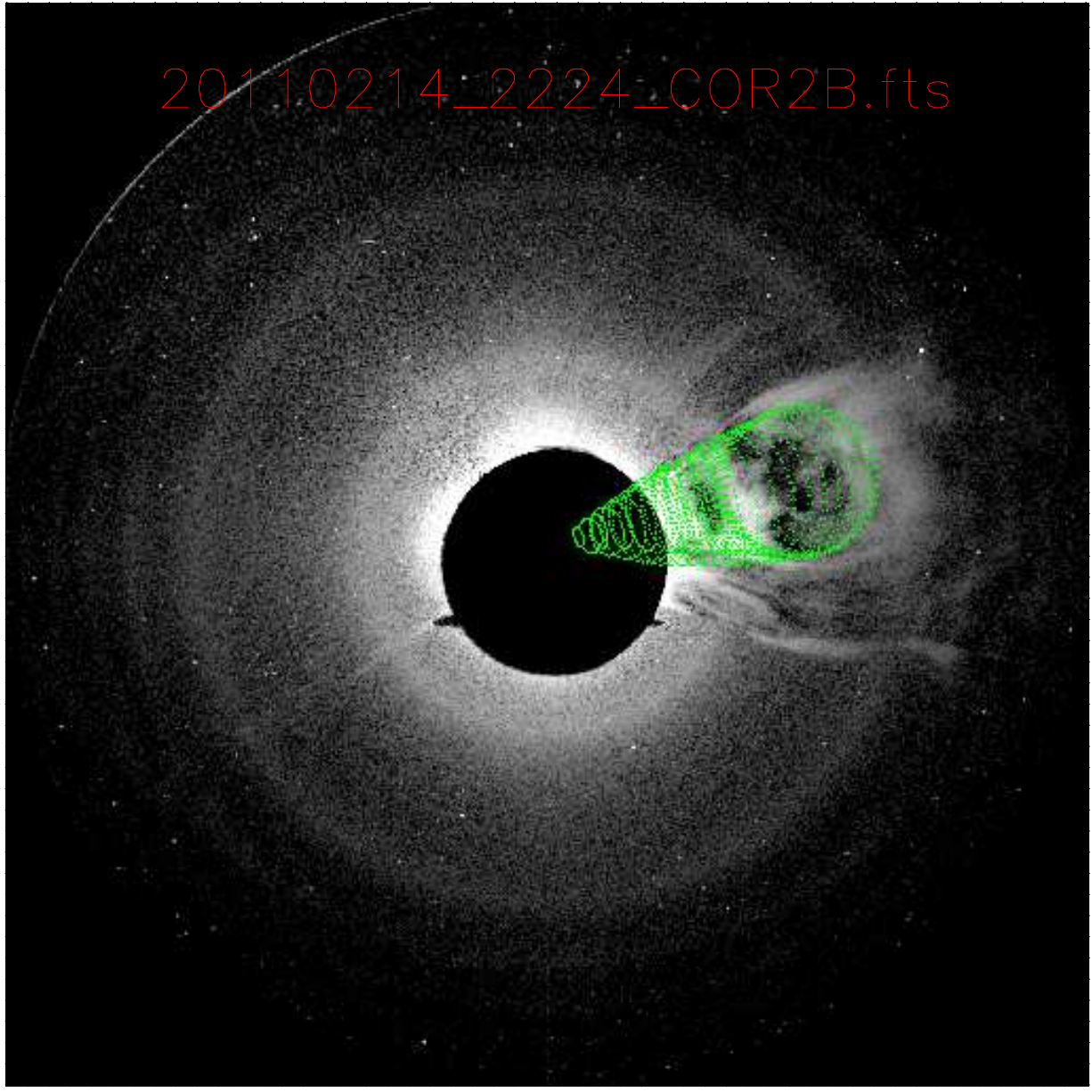}
	\includegraphics[scale=0.38]{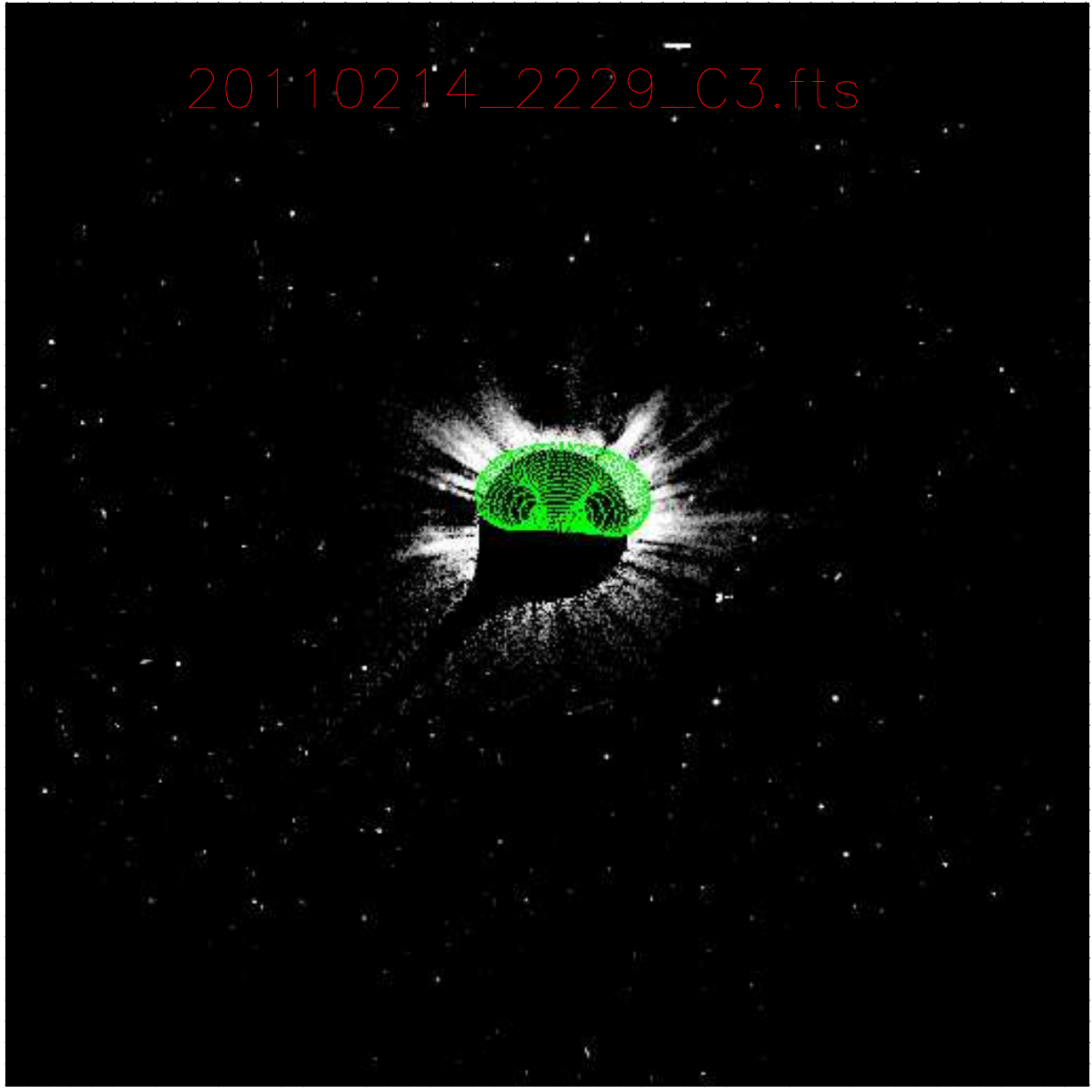}
	\includegraphics[scale=0.38]{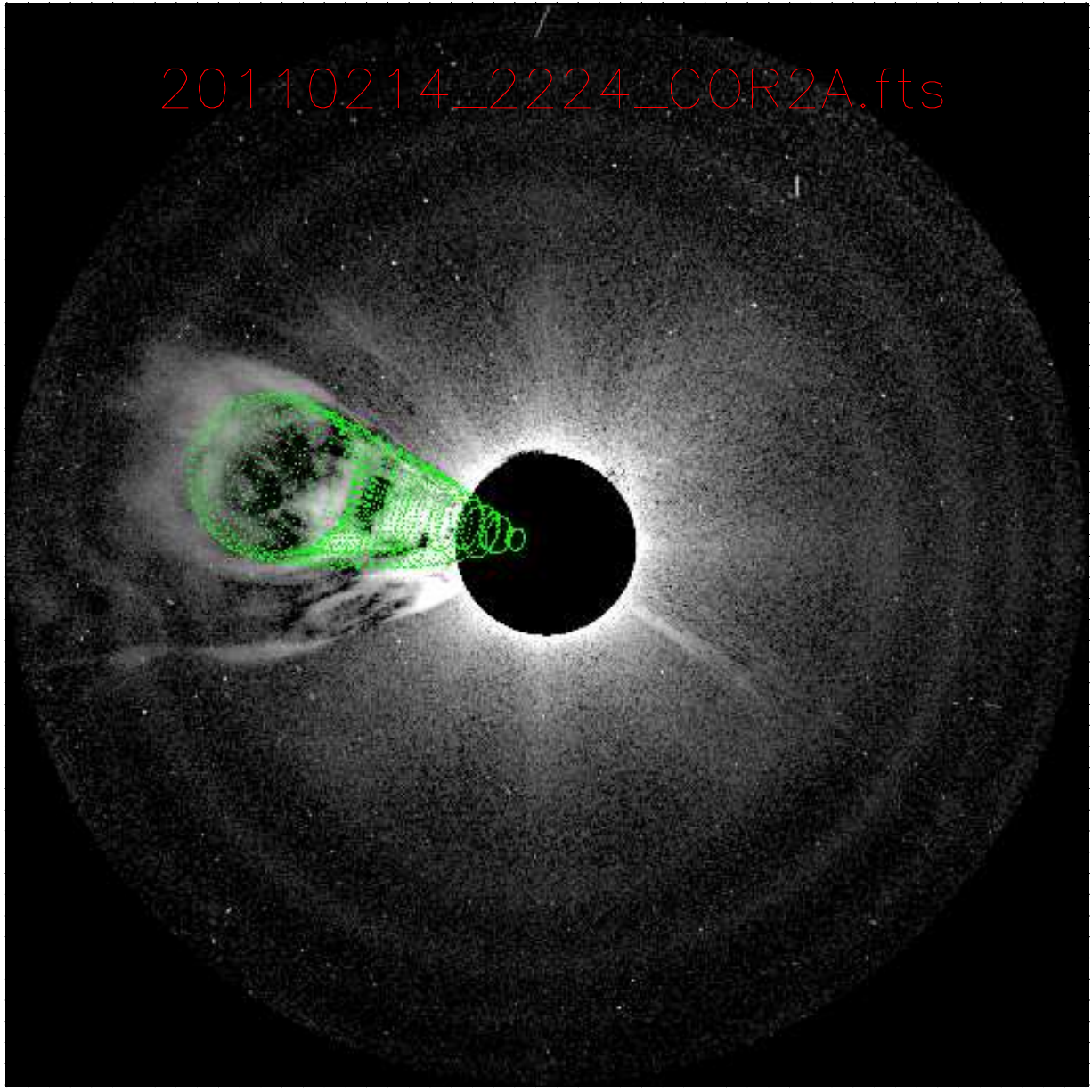}

	\includegraphics[scale=0.38]{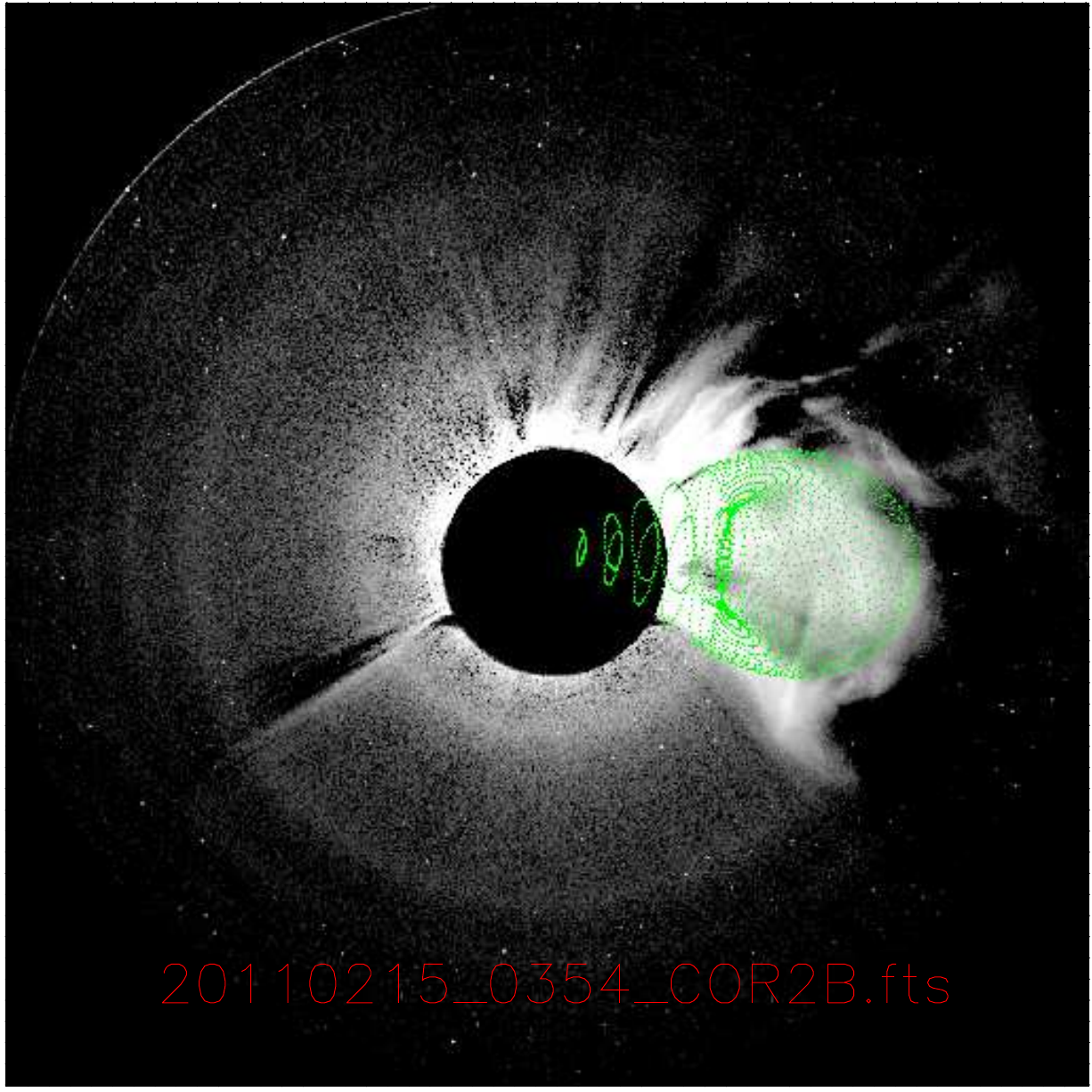}
	 \includegraphics[scale=0.38]{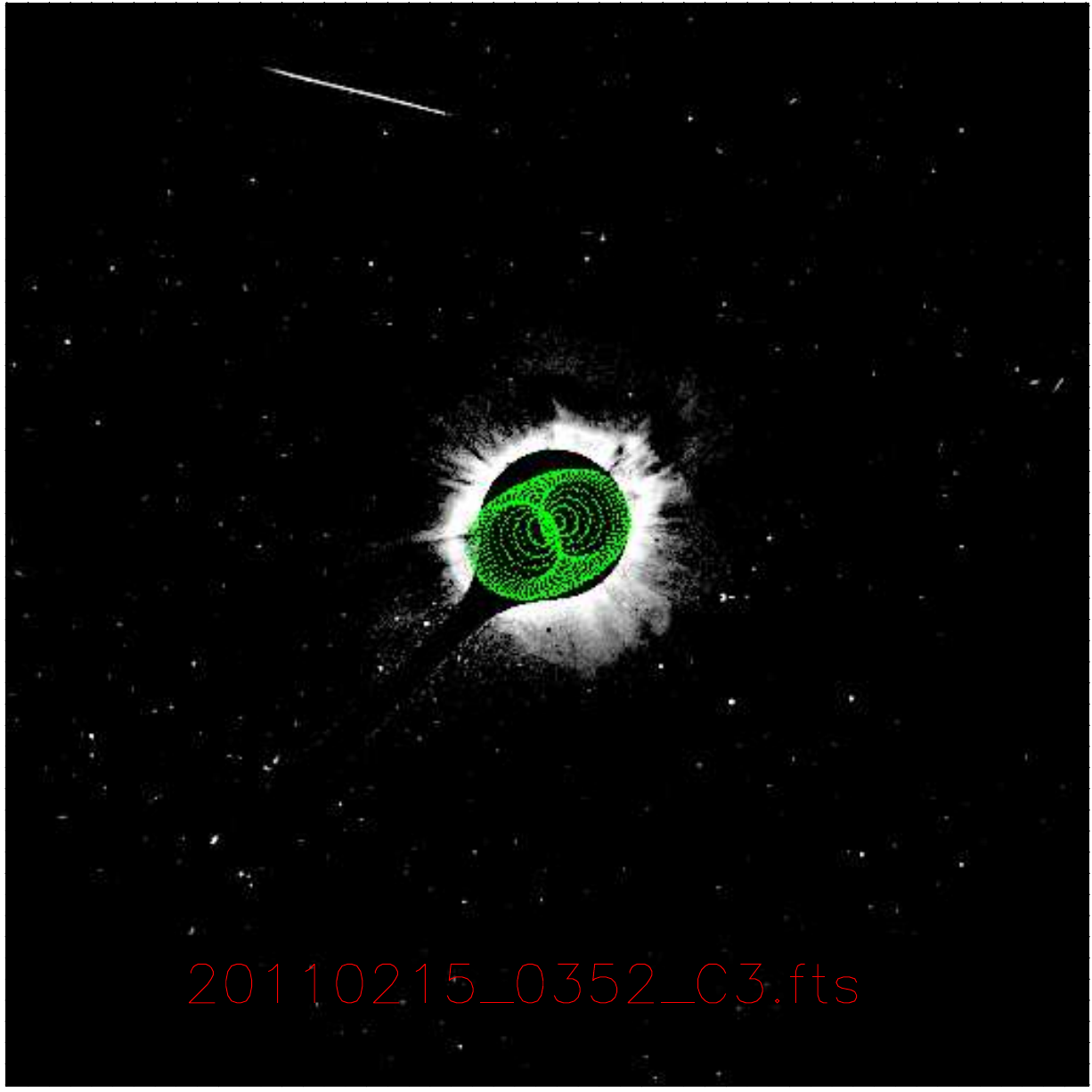}
	 \includegraphics[scale=0.38]{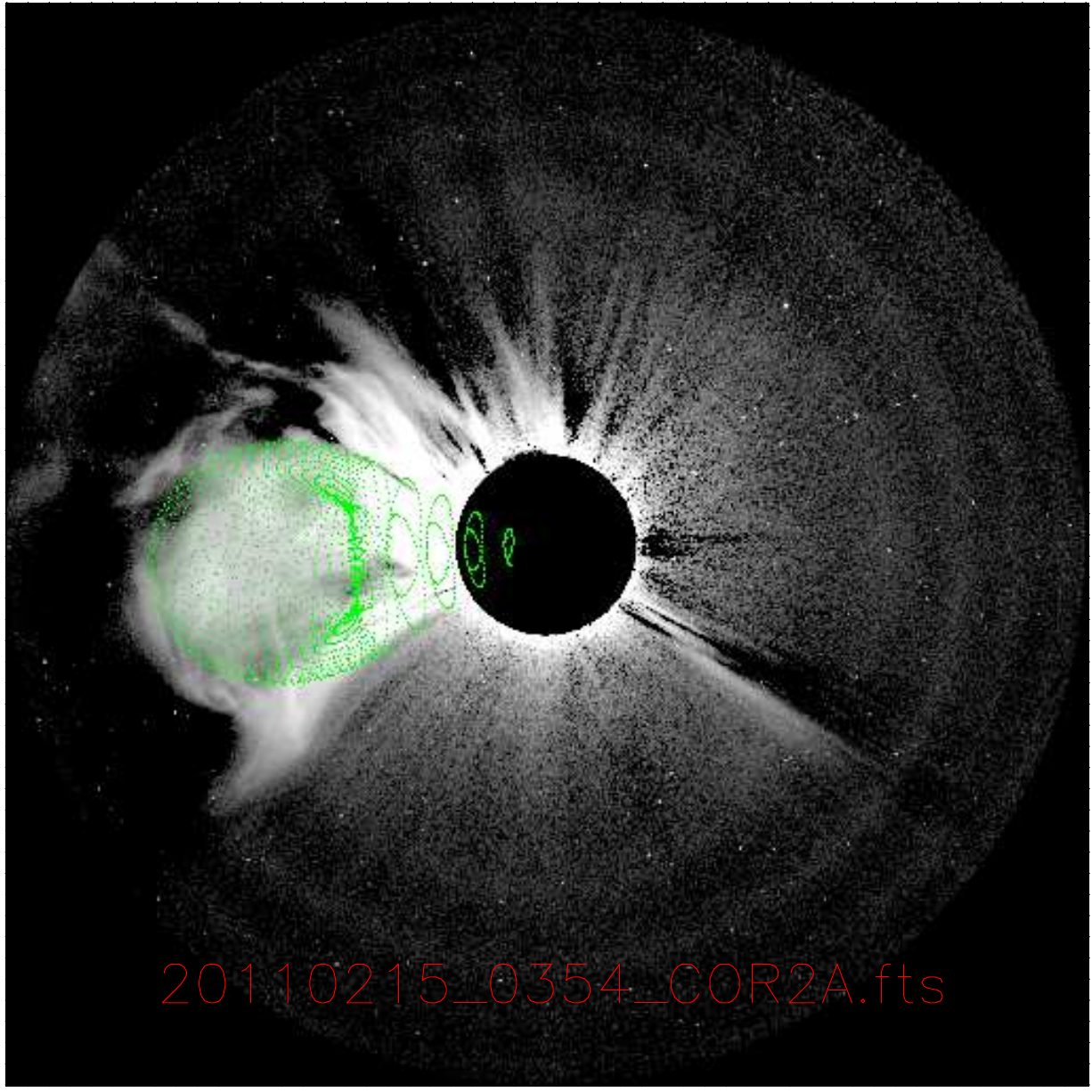}
	
 \caption[The contemporaneous image triplets for CMEs from SECCHI/COR2-B (left), SOHO/LASCO (C2 or C3) (middle), and SECCHI/COR2-A (right) with GCS wireframe (with green) overlaid on it]{The contemporaneous image triplets for CMEs from SECCHI/COR2-B (left), SOHO/LASCO (C2 or C3) (middle) and SECCHI/COR2-A (right) are shown with GCS wireframe (with green) overlaid on it. The top, middle and bottom panels show the images of CME1 around 20:24 UT on February 13, CME2 around 22:24 UT on February 14, and CME3 around 03:54 UT on February 15, respectively.}
\label{CME123_FM}
 \end{figure}

The 3D kinematics estimated for these CMEs in COR2 FOV is shown in Figure~\ref{CME123_3DFM}. As the CME1 was faint and non-structured, GCS model fitting could be done only for three consecutive images in COR2 FOV. The estimated longitudes ($\phi$) for CME1, CME2 and CME3 at their last estimated height of 8.2 \textit{R}$_{\odot}$, 10.1 \textit{R}$_{\odot}$ and 11.1 \textit{R}$_{\odot}$ are -2$^{\circ}$, 6$^{\circ}$ and -3$^{\circ}$, respectively. The estimated latitudes ($\theta$) at these heights are -6$^{\circ}$, 4$^{\circ}$, -11$^{\circ}$ for CME1, CME2 and CME3, respectively. The estimated 3D speed at their last estimated heights for CME1 (February 13, 20:54 UT), CME2 (February 14, 22:24 UT) and CME3 (February 15, 03:54 UT) is found to be 618 km s$^{-1}$, 418 km s$^{-1}$ and 581 km s$^{-1}$, respectively. From the kinematics plot (Figure~\ref{CME123_3DFM}), it is clear that CME3 was faster than the preceding CME2 and headed approximately in the same direction towards the Earth. Moreover, the launch of CME3 preceded that of CME2 by $\approx$ 9 hr; therefore, it is expected that these CMEs would interact at a certain distance in the heliosphere. Since the direction of propagation of CME1 and CME2 was also the same, there is a possibility of interaction between them if CME1 decelerates, and CME2 accelerates beyond the estimated height in COR2 FOV. From the 3D reconstruction in COR2 FOV, we found that the speed of CME3 decreased very rapidly from 1100 km s$^{-1}$ at 6 \textit{R}$_{\odot}$ to 580 km s$^{-1}$ at 11 \textit{R}$_{\odot}$ during 02:39 UT to 03:54 UT on 2011 February 15. A quick deceleration of fast CME3 within 1.5 hr is most likely due to the interaction between CME2 and CME3. The terminology `interaction' and `collision' have been used for two specific scenarios. By `interaction,' we mean that one CME causes deceleration or acceleration of another. However, no obvious signature of merging of propagation tracks of features corresponding to the two CMEs is noticed in \textit{J}-maps.  The `collision' is referred to as the phase during which the tracked features of two CMEs moving at different speeds come in close contact with each other until they achieve an approximately equal speed or their trend of acceleration is reversed, or they get separated from each other. The fast deceleration of CME3 from the beginning of the COR2 FOV may occur due to various possibilities. It may be either due to the presence of dense material of the preceding CME2 or due to the decrease of magnetic driving forces of CME3, or due to the overlying curved magnetic field lines of the preceding CME2, which can act as a magnetic barrier for CME3 \citep{Temmer2008, Temmer2010,Temmer2012}.

\begin{figure}[!htb]
 \centering
  \includegraphics[scale=0.95]{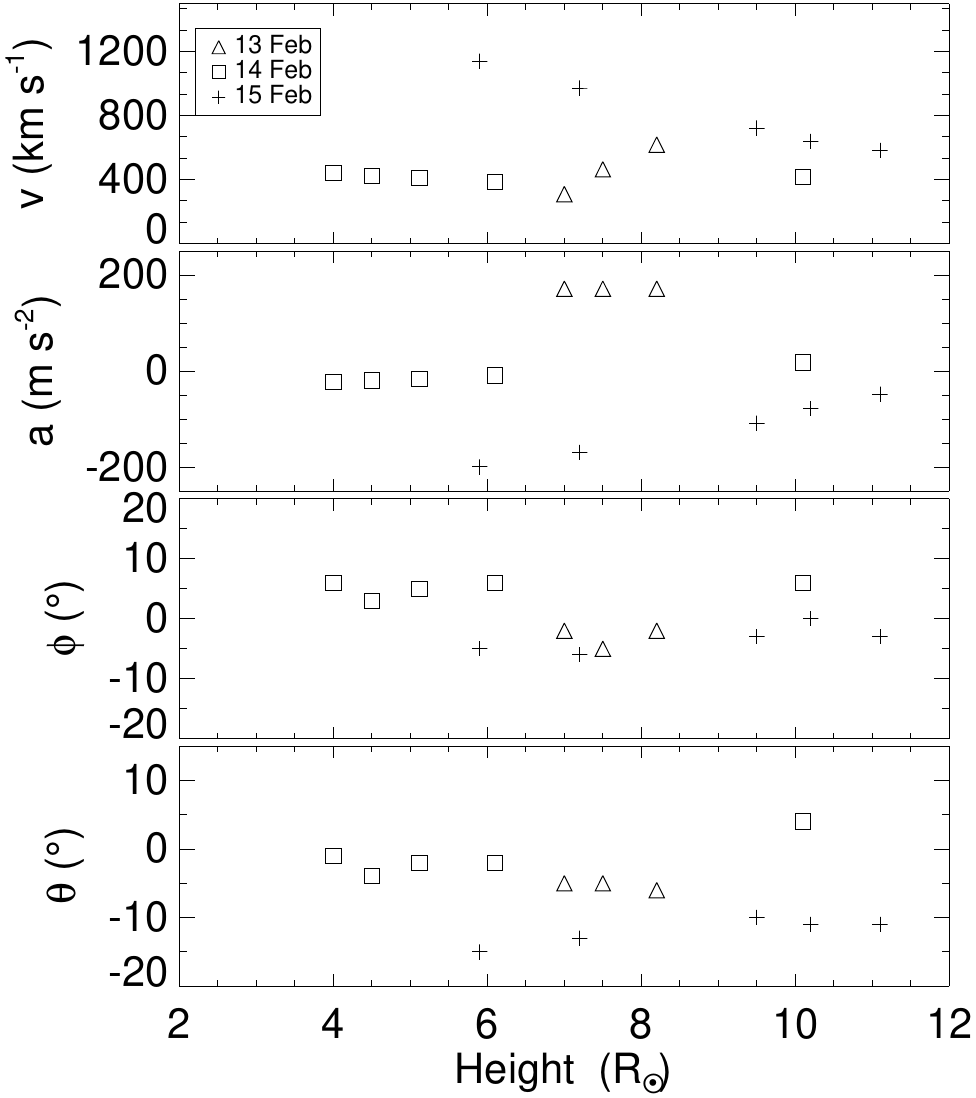}
 \caption[Estimated radial velocity, acceleration, longitude and latitude of 2011 February 13-15 CMEs using GCS model]{Top to bottom panels show the variations of radial velocity, acceleration, longitude and latitude of CME1, CME2, and CME3 with radial height from the Sun.}
\label{CME123_3DFM}
 \end{figure}

\subsubsection{Comparison of angular widths of CMEs derived from GCS model}
\label{IntFebCompWidth}

As discussed in Section~\ref{IntFebMorphoCOR}, we have estimated the cone angular width of CMEs using 2D COR2 images; it appears relevant to compare this to angular width determined from the GCS model of 3D reconstruction. Using the GCS 3D reconstruction technique, apart from the kinematics of CMEs (explained in Section~\ref{IntFebReconsCOR}), we also obtained the aspect ratio ($\kappa$) of the GCS model for CME1, CME2, and CME3 as 0.25, 0.28 and 0.37, respectively, at the last point of estimated distance in COR2 FOV. The aspect ratio is a parameter relative to the spatial extent of the CME. Also, we found the tilt angle ($\gamma$) around the axis of symmetry of the GCS model as 7$\arcdeg$, -8$\arcdeg$ and 25$\arcdeg$ for CME1, CME2, and CME3, respectively. The positive (negative) value of the tilt shows that the rotation is anticlockwise (clockwise) out of the ecliptic plane. The angular width (2$\alpha$) between both legs of a CME (in a GCS representation) is 34$\arcdeg$, 64$\arcdeg$ and 36$\arcdeg$ for CME1, CME2 and CME3, respectively. These values are in agreement (within $\pm$ 10\%) with the values obtained by \citet{Temmer2014}. It is to be noted that the measured 2D angular width of CME depends on the orientation of the GCS flux ropes. For ecliptic orientation of the flux ropes, i.e., $\gamma$ = 0$\arcdeg$, the angular width of CME seen in 2D images is equal to the 3D edge-on angular width ($\omega$$_{EO}$ = 2$\delta$) of GCS model, where $\delta$ = $\arcsin$($\kappa$). For $\gamma$ = 90$\arcdeg$, the measured 2D width is equal to 3D face-on angular width ($\omega$$_{FO}$ = 2$\alpha$+2$\delta$) of the GCS modeled CME.

We converted the GCS modeled 3D width to 2D angular width for CME2 using the expression, $\omega$$_{2D}$ = $\omega$$_{EO}$$\cos$($\gamma$)+$\omega$$_{FO}$$\sin$($\gamma$), and find that CME2 has approximately constant 2D angular width in COR2 FOV. We find that the fast speed CME3 has $\gamma$ = 21$\arcdeg$, $\kappa$=0.40 and $\alpha$ = 16$\arcdeg$ in the beginning of the COR2 FOV while $\gamma$ = 21$\arcdeg$, $\kappa$=0.31 and $\alpha$ = 18$\arcdeg$ at the last measured point in COR2 FOV. Hence, as  CME3 propagates further in COR2 FOV, its 2D angular width (derived using GCS modeled 3D width) decreases from 77$\arcdeg$ to 63$\arcdeg$. These findings are in accordance with the observed 2D angular width of CME2 and CME3 (Figure~\ref{theta_CMEs}). The possibility of rotation of CMEs has been discussed theoretically (\citealt{Lynch2009}, and references therein) and reported in low corona observations \citep{Lynch2010}. Such changes in the 2D measured angular width are also possible due to rotation (change in CME orientation, i.e., tilt, angle) of fast speed CME towards or away from the equator as shown by \citet{Yurchyshyn2009} who suggested a  higher rotation rate for a faster CME \citep{Lynch2010, Poomvises2010}. However, we must emphasize that based on the GCS modeling, we could not infer any noticeable rotation (change in $\gamma$) or deflection (change in $\phi$ in Figure~\ref{CME123_3DFM}) in the COR2 FOV for the selected CMEs. The uncertainties involved in the estimation of 3D and observed 2D angular widths are discussed in Section~\ref{IntFebResDis}.

\citet{Vourlidas2011} reported that despite the rapid rotation of CMEs, there are no significant projection effects (change in angular width) in single coronagraphic observations. They showed, for a CME launched on 2010 June 16, the projected (2D) angular width of CME altered by only 10$\arcdeg$ between 2 to 15 \textit{R}$_{\odot}$ while CME rotated by 60$\arcdeg$ over the same height range. It must be noted that a rotation of $\approx$ 40$\arcdeg$ within 6 hr is required for the observed large variations in the 2D angular width of CME3, which is indeed not found in our analysis. Therefore, we consider that the observed decrease in the angular width of CME3 is not because of its rotation but maybe due to its interaction with the solar wind or dense material of the preceding CME2.

\subsubsection{Reconstruction of CMEs in HI FOV}
\label{IntFebRecnsHI}

Based on the kinematics observed close to the Sun, i.e., using COR observations, we consider the possibility that these Earth-directed CMEs have a chance of interaction, and therefore we estimated their kinematics in HI FOV. We examined the base difference images in HI FOV to notice any density depletion or enhancement due to the passage of a CME. We noticed that CME3 and CME2 meet in the HI1 FOV (Figure~\ref{CME23_caught}). In this collision, the leading edge of CME3 flattened significantly. This observation motivated us to investigate the pre and post-collision kinematics of CMEs. Therefore, we tracked the CMEs in the heliosphere by constructing the time-elongation map (\textit{J}-map) \citep{Davies2009}, as described in detail in Section~\ref{Jmapsmthd} of  Chapter~\ref{Chap2:DataMthd}.

\begin{figure}[!htb]
 \centering
  \includegraphics[scale=0.55]{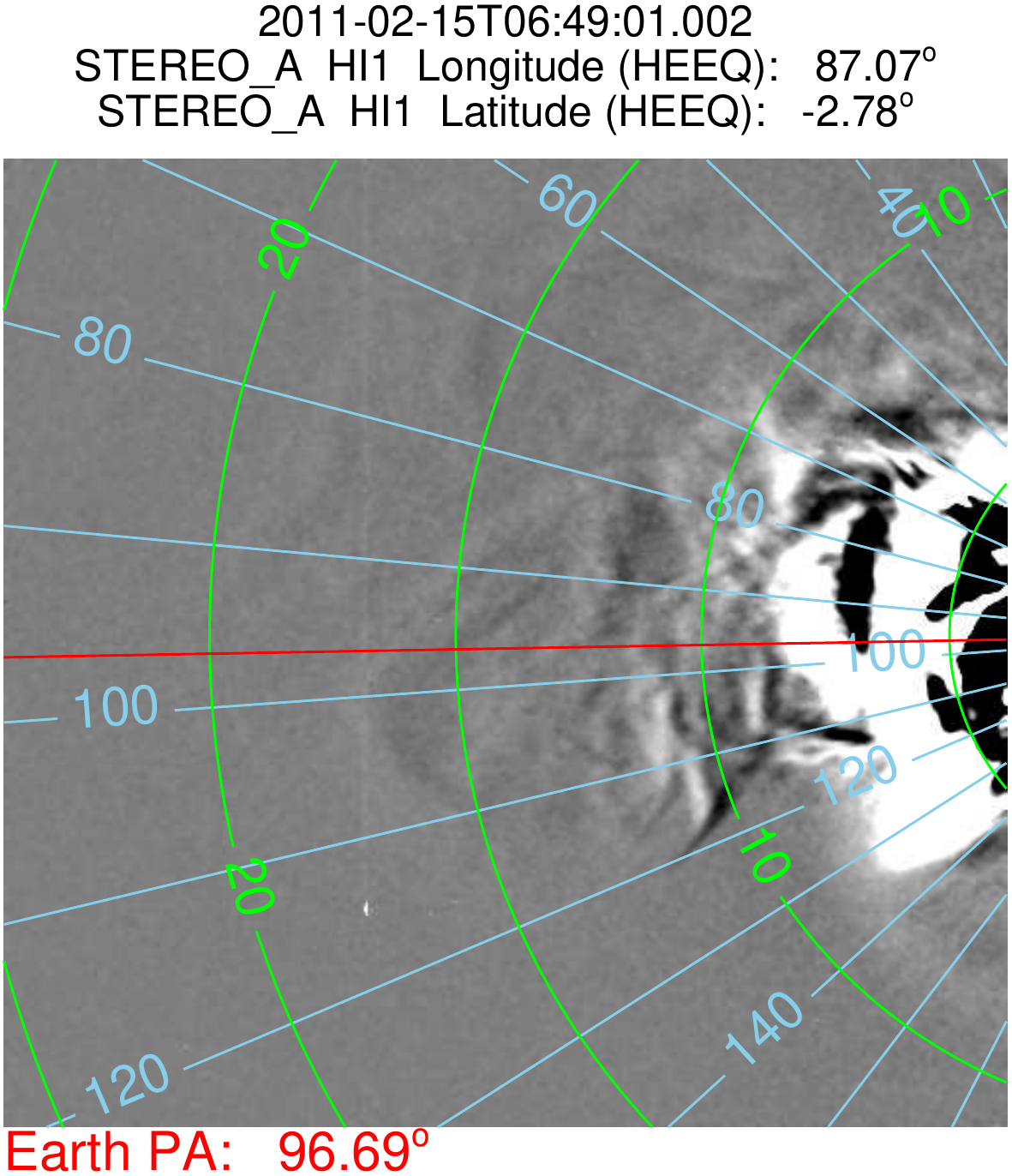}
	\hspace{5pt}
	\includegraphics[scale=0.55]{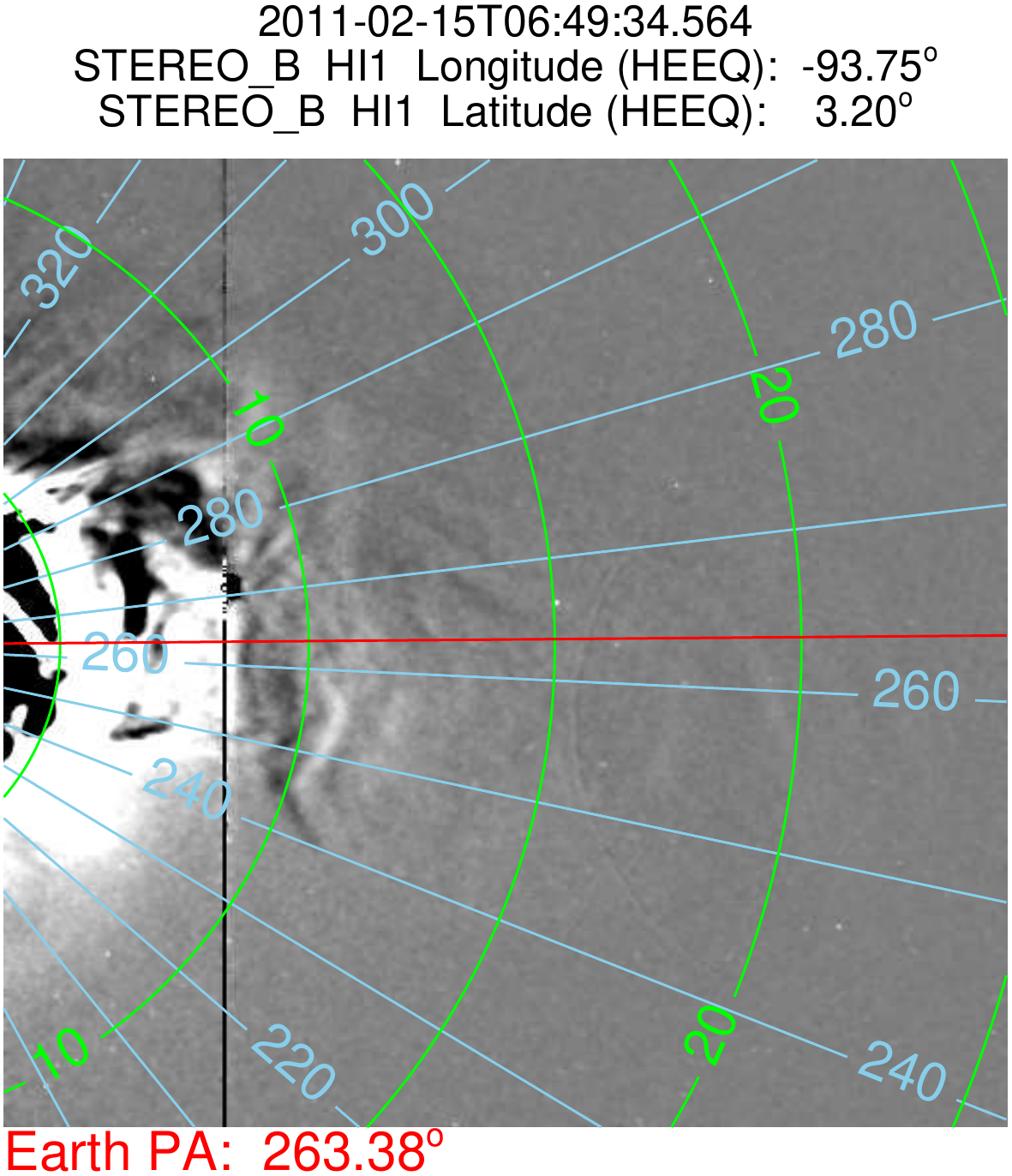}
 \caption[The base difference HI1-A and HI1-B images show the collision of the CMEs launched on 2011 February 14 and 15]{The base difference HI1-A (left) and HI1-B (right) images show the collision of CME2 and CME3. The leading edge of CME3 got flattened during collision with the trailing edge of CME2. The over drawn contours on the images with green and blue represent the elongation and position angle, respectively. The red line is along the ecliptic at the position angle of the Earth.}
\label{CME23_caught}
 \end{figure}

The constructed \textit{J}-map in the ecliptic plane for these CMEs in HI-A and B FOV are shown in Figure~\ref{IntFebJ-maps}. By manually tracking the bright leading fronts, we derived the elongation-time profiles for all three CMEs. In this figure, derived elongations for the interacting CMEs are overplotted with dotted colored lines. The CME1 is very faint and could be tracked out to $\approx$ 13$^{\circ}$ in the \textit{STEREO-A} and \textit{STEREO-B} \textit{J}-maps. However, CME2 and CME3 could be tracked out to  44$^{\circ}$ and 46$^{\circ}$ in \textit{STEREO-A} \textit{J}-maps, respectively, and out to $\approx$ 42$^{\circ}$ in \textit{STEREO-B} \textit{J}-maps. The \textit{J}-maps also show that the bright tracks of CME2 and CME3 approach close to each other, suggesting their possible collision in HI1 FOV.

\begin{figure}[!htb]
 \centering
  \includegraphics[scale=0.42]{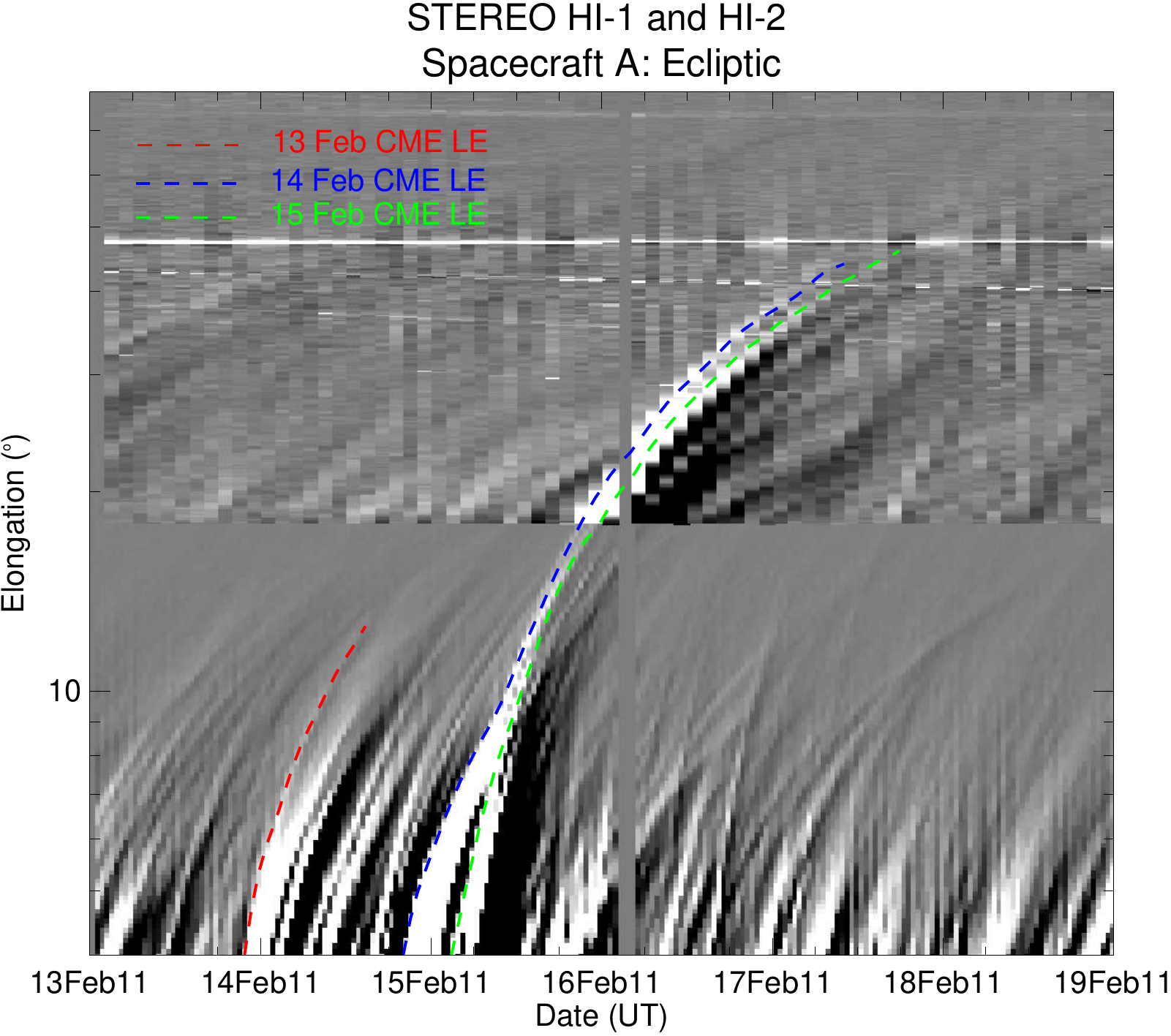}
	\includegraphics[scale=0.42]{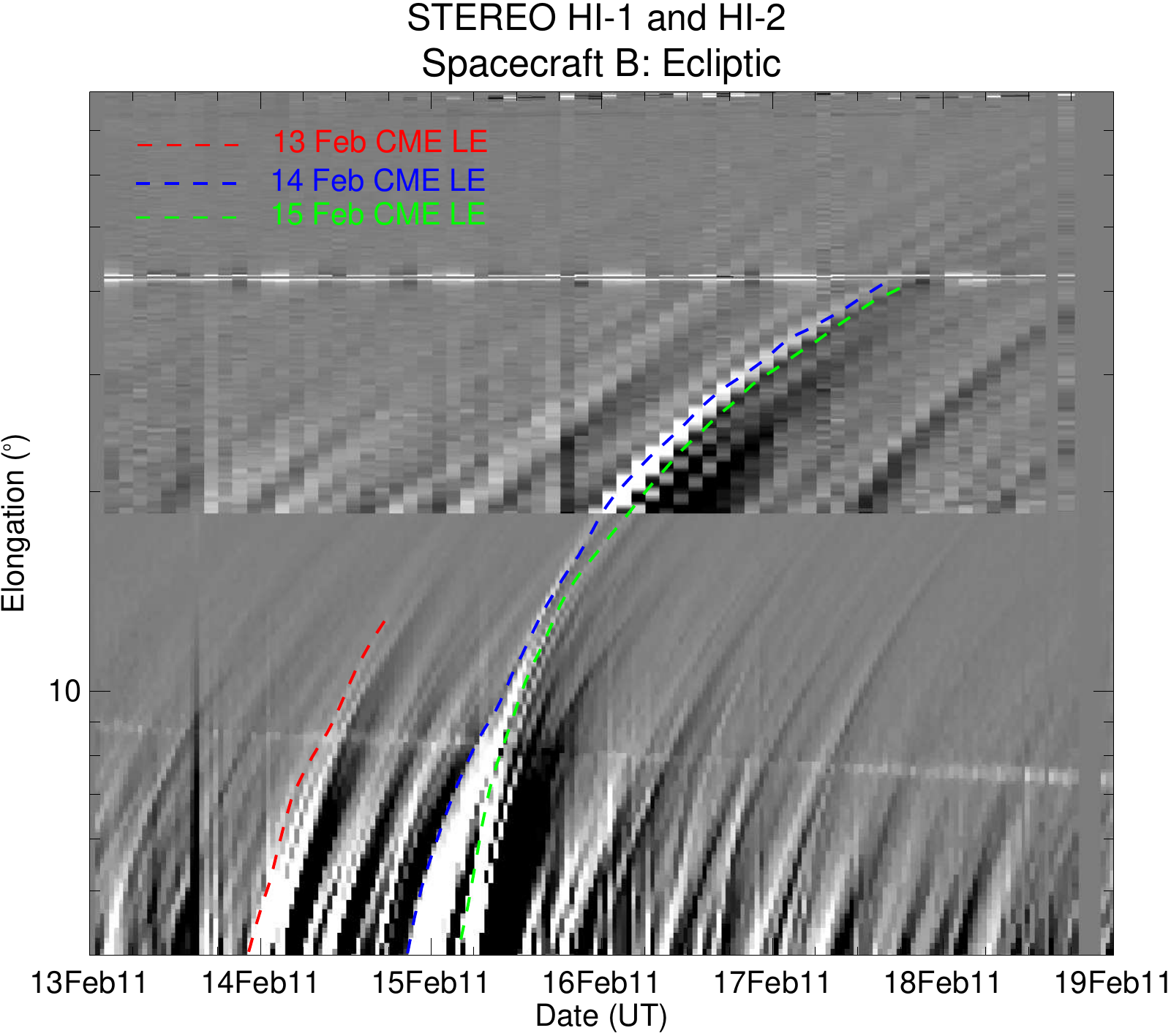}
 \caption[\textit{J}-maps for the 2011 February 13-15 CMEs]{Time-elongation maps (\textit{J}-maps) for \textit{STEREO-A} (left) and \textit{STEREO-B} (right) using running differences of images HI1 and HI2 are shown for the interval of 2011 February 13 to 19. The tracks of CME1, CME2 and CME3 are shown with red, blue, and green, respectively.}
\label{IntFebJ-maps}
 \end{figure}

Various stereoscopic reconstruction methods have been developed to estimate the kinematics of CMEs using SECCHI/HI images \citep{Liu2010, Lugaz2010.apj,Davies2013}. The selected CMEs in our study have a cone angular width of $\approx$ 60$^\circ$. Therefore, it is preferable to use those reconstruction methods which take into account the geometry of CMEs with similar angular widths. Keeping these points in mind, we implemented the SSSE \citep{Davies2013} method on the derived time-elongation profiles for all the three CMEs to estimate their kinematics. While applying this method, we fixed the cross-sectional angular half-width of the CMEs subtended at the Sun equal to 30$^{\circ}$. Using the SSSE method, for all the three CMEs, the kinematics, i.e., estimated height, direction, and speed, were obtained (Figure~\ref{CME123_SSSE}). The speed was derived from the adjacent distance points using numerical differentiation with three-point Lagrange interpolation and therefore have systematic fluctuations. Estimating the speed in this way can provide short time variations in CME speed during its interaction with solar wind or other plasma density structures in the solar wind.
On the other hand, the smoothed speed can also be derived if the estimated distance is fitted with a polynomial, but the information about variations in speed will be lost. Also, by fitting a polynomial for the derived fluctuating speeds, the speed can be shown with minimal fluctuations. Therefore, we have made a compromise and fitted the estimated distance during each 5 hr interval into a first-order polynomial and derived the speed, which is shown with horizontal solid lines in the bottom panel of Figure~\ref{CME123_SSSE}. The error bars for the estimated parameters are also shown in this figure with vertical solid lines at each data point. Section~\ref describes the detailed procedure of error estimation of errors is described in Section~\ref{IntFebCompKinem}. CME2 and CME3 seem to follow the same trajectory and are approximately Earth-directed, as noted from the direction of propagation. However, unexpected variations in the direction of propagation of both CMEs were noticed, which are discussed in Section~\ref{IntFebResDis}. In Figure~\ref{CME123_SSSE}, we noticed a jump in the speed of CME2 and CME3 at 08:25 UT on 2011 February 15. Within 18 hr, after an increase in speed is observed, the speed of the CME2 increased from about 300 km s$^{-1}$ to 600 km s$^{-1}$. During this time, the speed of CME3 decreased from about 525 km s$^{-1}$ to 400 km s$^{-1}$. Later, both CMEs achieved a similar speed of $\approx$ 500 km s$^{-1}$. Such a finding of acceleration of one and deceleration of another CME supports a possible collision between CME2 and CME3. The collision phase is shown in the top and bottom panels of Figure~\ref{CME123_SSSE} (the region between the two dashed vertical lines, from the left). After the collision, we find that both the CMEs move in close contact with each other.

\begin{figure}[!htb]
 \centering
  \includegraphics[scale=0.70]{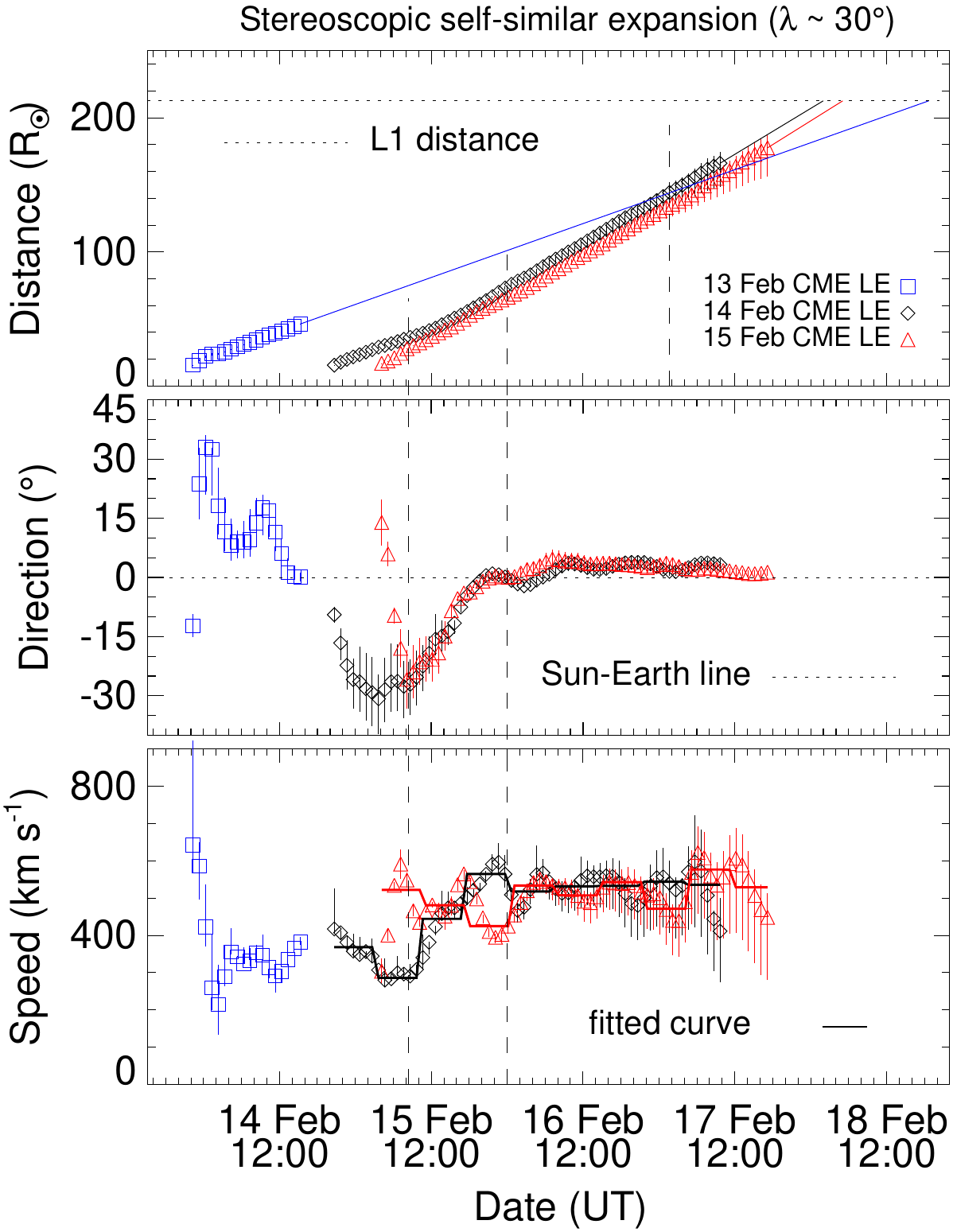}
 \caption[Estimates of distance, propagation direction, and speed of the 2011 February 13-15 CMEs using SSSE method]{From top to bottom, distance, propagation direction, and speed (as obtained using SSSE method) of CME1 (blue), CME2 (black) and CME3 (red) are shown. In the top panel, the horizontal dashed line marks the heliocentric distance of the L1 point. In the middle panel, the dashed horizontal line marks the Sun-Earth line. In the bottom panel, the speed shown with symbols is estimated from the differentiation of adjacent distance points using three-point Lagrange interpolation. The speed shown with a solid line is determined by differentiating the fitted first order polynomial for the estimated distance for each 5 hr interval. From the left, the first and second vertical dashed lines mark the start and end of the collision phase of CME3 and CME2. The top panel's rightmost vertical dashed line marks the inferred interaction between CME2 and CME1. The vertical solid lines at each data point show the error bars, explained in Section~\ref{IntFebCompKinem}.}
\label{CME123_SSSE}
\end{figure}

The strong deceleration of CME3 observed before the merging of the bright tracks (enhanced density front of CMEs) in \textit{J}-maps suggests possible interaction of CME3 with CME2. This is possible as we track the leading front of the CMEs using \textit{J}-maps, the trailing edge of CME2 can cause an obstacle for the CME3 leading front much earlier depending on its spatial scale. From the observed timings, it is clear that interaction of CME3 with CME2 had started $\approx$ 5 hr before their collision in HI FOV. Our analysis also shows that the leading front of CME3 reflects the effect of interaction (i.e. strong deceleration) at 6 \textit{R}$_{\odot}$ while leading edge of CME2 shows this effect (i.e. acceleration) at 28 \textit{R}$_{\odot}$. Therefore, the force acting on the trailing edge of CME2 takes approximately $\approx$ 5.7 hr to reach the leading front of this CME. Based on these values, the propagation speed of disturbance responsible for the acceleration of the leading front of CME2 should be $\approx$ 750 km s$^{-1}$. From the Radio and Plasma Wave Experiment (WAVES) \citep{Bougeret1995} on board \textit{WIND} spacecraft, we noticed a type II burst during 02:10-07:00 UT in the 16000-400 kHz range. Such radio bursts provide information on CME-driven shock \citep{Gopalswamy2000a}. This shock is associated with the fast-speed CME3. The average shock cone angle ($\approx$ 100$^\circ$) as seen from the Sun is significantly greater than the average angular size ($\approx$ 45$^\circ$) of any CME \citep{Schwenn2006}. This shock likely traveled across CME2. Therefore, the acceleration of CME2, observed in HI1 FOV, may be due to the combined effect of the shock and the leading front of CME3. As previously mentioned, CME1 was very faint, and its kinematics could be estimated up to 46 \textit{R}$_{\odot}$ only. Based on the linear extrapolation of the height-time curves of CME1 and CME2, we infer that they should meet each other at 144 \textit{R}$_{\odot}$ at 01:40 UT on 2011 February 17.

\subsubsection{Comparison of kinematics derived from other stereoscopic methods}
\label{IntFebCompKinem}
To examine the range of uncertainties in the estimated kinematics of the CMEs of 2011 February 13-15, by implementing SSSE method, we applied other stereoscopic methods, viz. TAS \citep{Lugaz2010.apj} and GT \citep{Liu2010} method to all the three CMEs. Using the kinematics estimated from TAS and GT methods, we found that the leading edge of CME3 caught the leading edge of CME2 at the approximately same distance (within a few solar radii) as obtained from the SSSE method. Based on linear extrapolation of the height-time profile of CME2 and CME1 estimated from TAS and GT methods, we inferred that CME2 would have reached CME1 at 157 \textit{R}$_{\odot}$ at 03:35 UT on 2011 February 17 and at 138 \textit{R}$_{\odot}$ on 20:24 UT on February 17, respectively.

The kinematics derived from the SSSE method is shown in Figure~\ref{CME123_SSSE}, and cognizance of the involved uncertainties is important. However, the actual uncertainties in the derived kinematics owe to several factors (geometry, elongation measurements, Thomson scattering, line of sight integration effect, breakdown of assumptions in the method itself), and its quantification is challenging. \citet{Davies2013} have shown that GT and TAS methods are special cases of SSSE method corresponding to two extreme cross-sectional extent (geometry) of a CME, i.e., corresponding to angular half-width of $\lambda$ = 0$\arcdeg$ and $\lambda$ = 90$\arcdeg$, respectively. We estimated the uncertainties by considering different geometry in the three implemented stereoscopic techniques (GT, TAS, and SSSE). Such uncertainties are shown with error bars with vertical solid lines in Figure~\ref{CME123_SSSE}. We estimated the absolute difference between kinematics values derived from SSSE and GT method and display it as a vertical lower error (lower segment of error bars). Similarly, the absolute difference between the SSSE method and TAS method kinematics values is displayed as a vertical upper error. From Figure~\ref{CME123_SSSE}, we notice that the results from all three methods are in reasonable agreement.

Further, we attempted to examine the contribution of errors in the kinematics due to errors in tracking (i.e., elongation measurements) of a selected feature. Following the error analysis approach of \citet{Liu2010a}, we consider an uncertainty of 10 pixels in elongation measurements from both \textit{STEREO} viewpoints which correspond to elongation uncertainty of 0.04$\arcdeg$, 0.2$\arcdeg$ and 0.7$\arcdeg$ in COR2, HI1 and HI2 FOV, respectively. This leads to an uncertainly of 0.20-0.35 \textit{R}$_{\odot}$, 0.21-0.75 \textit{R}$_{\odot}$, and 0.19-0.74 \textit{R}$_{\odot}$ in the estimated distance for CME1, CME2 and CME3, respectively. Such small uncertainties in the distance are expected to result in an error of less than $\approx$ 100 km s$^{-1}$ in speed. However, similar elongation uncertainty leads to crucially larger uncertainty in the estimated direction of propagation of CMEs when they are close to the entrance of HI1 FOV, where singularity occurs (described in Section~\ref{Rmt7Feb10} of Chapter~\ref{Chap3:ArrTim}). The estimated propagation direction of CMEs from the GT method is shown in Figure~\ref{CME123_GTErr} in which vertical lines at each data point show the uncertainty in the direction.

\begin{figure}[htb]
 \centering
  \includegraphics[scale=0.90]{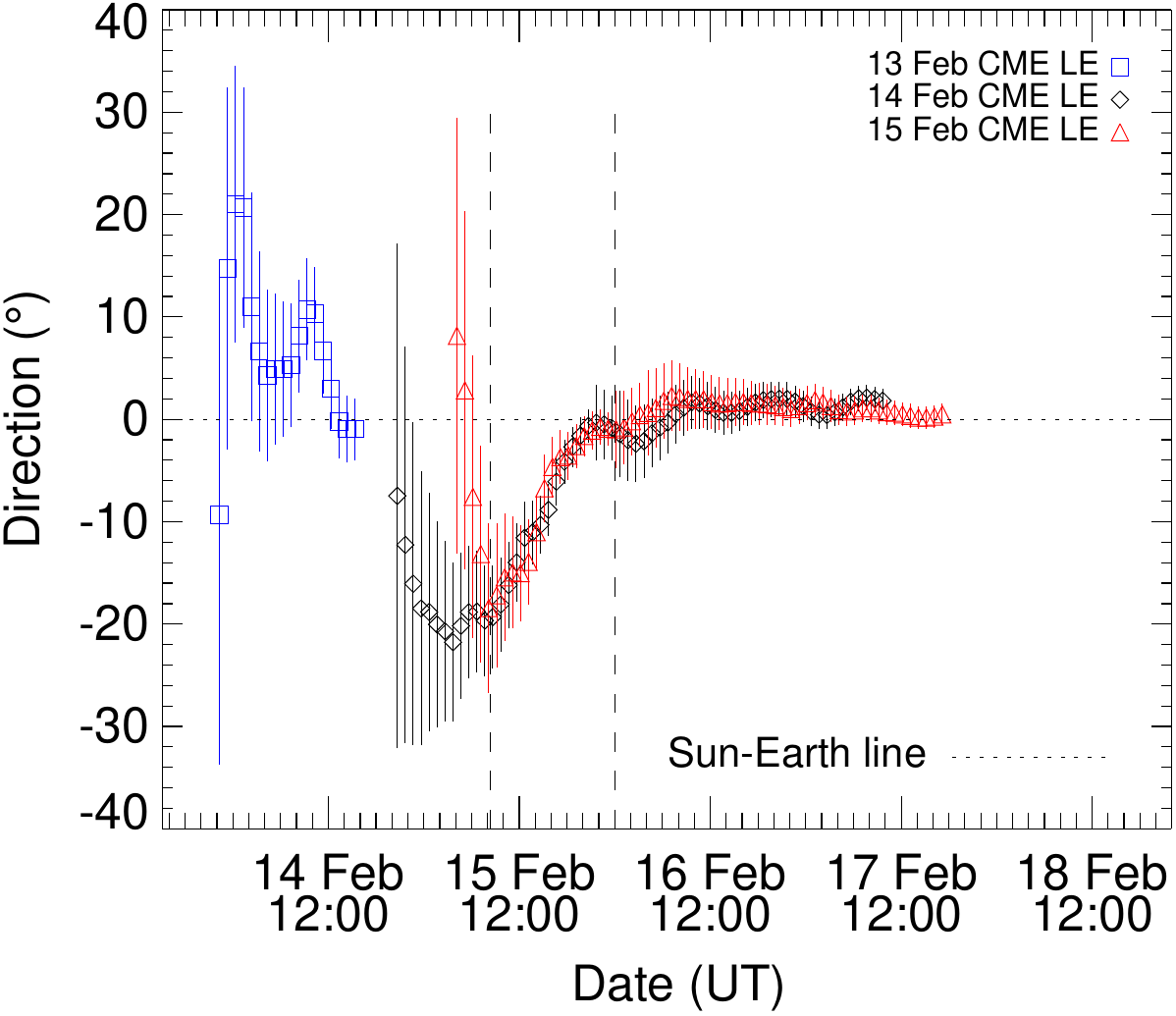}
 \caption[Estimated direction of propagation of CMEs using GT method]{The estimates of direction for CME1 (blue), CME2 (black), and CME3 (red) from the GT method are shown. The vertical lines at each data point show the error bars, which are calculated mathematically considering uncertainty of 10 pixels in elongation measurements. The two vertical dashed lines mark the start and end of the collision phase of CME2 and 
CME3. The horizontal dashed line denotes the Sun-Earth line. The negative (positive) direction angle stands for east (west).}\label{CME123_GTErr}
 \end{figure}

For the selected CMEs in our study, we find that the uncertainties in the estimated kinematics from stereoscopic methods owe mostly due to errors in elongation measurements rather than geometry. Due to the large separation between the two \textit{STEREO} viewpoints and occurrence of singularity, small observational errors in the elongation measurements yield significantly larger errors in the kinematics (especially in the direction), irrespective of the geometry considered for the CMEs \citep{Davies2013, Mishra2014}.  

\subsection{Energy, momentum exchange and nature of collision between CMEs of 2011 February 14 and 15}
\label{IntFebCollision}
The dynamics and structure of CMEs are likely to change when they collide with one another. Therefore, estimation  of 
post-collision kinematics is crucial for predicting the space weather consequences. As the CMEs are large-scale magnetized plasmoids that interact with each other, it is worth investigating the nature of collision for CMEs which is expected to be different than the collision of gaseous bubbles with no internal magnetic field. In collision dynamics, the total momentum of colliding bodies is conserved irrespective of the nature of the collision provided that external forces are absent.

We attempt to investigate the nature of collision for CME2 and CME3. As the CME3 follows the trajectory of CME2 before and after the collision, we simply use the velocity derived from the SSSE method to deal with the collision dynamics. We did not take into account the 3D velocity components and intricate mathematics for determining the motion of centroid of colliding CMEs, as used in \citet{Shen2012}. We studied one-dimensional collision dynamics similar to the head-on collision case for the interacting CMEs. We note that the start of the collision phase (marked by the dashed vertical line) occurred at the instant when the speed of CME2 starts to increase while the speed of CME3 starts to decrease (Figure~\ref{CME123_SSSE}, bottom panel). This trend of speeds is maintained up to 18 hr, where the collision phase ends. Later, CME2 and CME3 show a trend of deceleration and acceleration, respectively, reaching a constant speed of 500 km s$^{-1}$. From the obtained velocity profiles (Figure~\ref{CME123_SSSE}), we notice that velocity of CME2 and CME3 before the collision are u$_{1}$ = 300 and u$_{2}$ = 525 km s$^{-1}$, respectively. After the collision and exchange of velocity, the velocity of CME2 and CME3 is found to be v$_{1}$ = 600 and v$_{2}$ = 400 km s$^{-1}$, respectively.  If the true mass of CME2 and CME3 be m$_{1}$ and m$_{2}$, respectively, then the conservation of momentum requires m$_{1}$u$_{1}$ + m$_{2}$u$_{2}$ = m$_{1}$v$_{1}$ + m$_{2}$v$_{2}$. To examine the momentum conservation for the case of colliding  CMEs, we need to calculate the true mass of both CMEs, which is discussed in the following section.

\subsubsection{Estimation of true mass of CMEs}
\label{IntFebTrumas}
Since the appearance of a CME is due to Thomson scattered photospheric light from the electrons in the CME \citep{Minnaert1930, Billings1966,Howard2009}, the recorded scattered intensity can be converted into the number of electrons, and hence the mass of a CME can be estimated, if the composition of the CME is known. But an observer from different vantage points receives a different amount of scattered light from the electrons; therefore, the true location of electrons in the CME, \textit{i.e.} the propagation direction of the CME must be known to estimate the actual mass of the CME. However, historically, the mass of a CME has been calculated using the plane of sky approximation, which resulted in an underestimated value \citep{Munro1979, Poland1981,Vourlidas2000}. Although the propagation direction of the CME was calculated in our study using the forward modeling \citep{Thernisien2009} method and described in Section~\ref{IntFebReconsCOR}, to avoid any bias in the reconstruction methods, we use a different method based on the Thomson scattering theory to estimate 3D propagation direction and then the true mass of CMEs in COR2 FOV \citep{Colaninno2009}. Before applying this approach, base difference images were obtained following the procedure described in \citet{Vourlidas2000, Vourlidas2010,Bein2013}. To estimate the projected mass of CMEs in base difference COR2-A and B images, we selected a region of interest (ROI) that enclosed the full extent of a CME. The intensity at each pixel was then converted to the number of electrons at each pixel and then the mass per pixel was obtained. The total mass of CME was calculated by summing the mass at each pixel inside this ROI. In this way, we estimated the projected mass of CME, M$_{A}$ and M$_{B}$ from two viewpoints of \textit{STEREO-A} and \textit{STEREO-B} in COR2 FOV.

According to \citet{Colaninno2009}, CME masses  M$_{A}$ and M$_{B}$ are expected to be equal as the same CME volume is observed from two different angles. Any difference between these two masses, must be due to incorrect use of the  propagation angle in the Thomson scattering calculation. Based on this assumption, they derived an equation for true mass (M$_{T}$) as a function of projected mass and 3D direction of propagation of CME (see their equations 7 and 8). We used a slightly different approach to solve these equations, viz.

\begin{equation}
M_{A}/M_{B} = B_{e}(\theta_{A})/B_{e}(\theta_{A} + \Delta)
\end{equation}

where $\theta$$_{A}$ is the angle of direction of propagation of CME measured from the plane of sky of \textit{STEREO-A}, B$_{e}$($\theta$$_{A}$) is the brightness of a single electron at an angular distance of $\theta$$_{A}$ from the plane of sky and $\Delta$ is the summation of longitude of both \textit{STEREO-A} and \textit{STEREO-B} from the Sun-Earth line. Once we obtained the measured values of M$_{A}$ and M$_{B}$, we derived its ratio and calculate $\theta$$_{A}$. In this way, we obtained multiple values of $\theta$$_{A}$ which result in the same value of the ratio of M$_{A}$ and M$_{B}$. The correct and unique value of $\theta$$_{A}$ was confirmed by visual inspection of CME images in COR FOV. The true mass of CME was then estimated using the equation (4) of \citet{Colaninno2009}. Here we must emphasize that estimation of 3D propagation direction of CMEs ($\theta$$_{A}$) using aforementioned approach has large errors if $\Delta$ approaches 180$^\circ$. This is a severe limitation of the method of true mass estimation and arises because in such a scenario a CME from the Sun, despite its propagation in any direction (not only towards the Earth), will be measured at an equal propagation angle from the plane of the sky of both spacecraft. Therefore, in principle, both the estimated M$_{A}$ and M$_{B}$ should be exactly equal, and any deviation (which is likely) will result in the highly erroneous value of $\theta$$_{A}$, and consequently in the true mass of CME. Such a limitation has also been reported by \citet{Colaninno2009} for a very small spacecraft separation angles. This implies that an accurate propagation direction cannot be derived with this method unless we adjust the separation angle between \textit{STEREO} spacecraft slightly. Hence, we use a slightly different value of $\Delta$ $\approx$ 160$^\circ$ for our case. By repeating our analysis several times for these CMEs, we noted that a change in $\Delta$ by 20$^\circ$ has a negligibly small effect on the mass of a CME. 

We also estimated the accurate mass using the 3D propagation direction obtained from another method (GCS forward fitting model) and found that these results are within $\approx$ 15\% of estimates from the method of \citet{Colaninno2009}, therefore can be used for further analysis.

\begin{table}
  \centering

 \begin{tabular}{|p{2.0cm}| p{5.0cm}| p{5.0cm}|}
    \hline
Parameters &  February 14 CME & February 15 CME  \\ \hline
M$_{A}$  &  5.30 $\times$ 10$^{12}$ kg at $\approx$ 10 \textit{R}$_{\odot}$ & 4.56 $\times$ 10$^{12}$ kg at $\approx$ 12 \textit{R}$_{\odot}$ \\ \hline

M$_{B}$  & 4.38 $\times$ 10$^{12}$ kg at $\approx$ 10 \textit{R}$_{\odot}$ & 4.77 $\times$ 10$^{12}$ kg at $\approx$ 12 \textit{R}$_{\odot}$ \\ \hline

Direction & 24$^{\circ}$ east from the Sun-Earth line & 30$^{\circ}$ east from the Sun-Earth line  \\ \hline

True mass & $m_{1}$ = 5.40 $\times$ 10$^{12}$ kg  &  $m_{2}$ = 4.78 $\times$ 10$^{12}$ kg  \\ \hline

 \end{tabular}

\caption[The estimates of mass and direction for the 2011 February 14 and 15 CMEs]{The estimates of mass and direction for 2011 February 14 and 15 CMEs. M$_{A}$ and M$_{B}$ are the estimated mass from two viewpoints of \textit{STEREO-A} and \textit{STEREO-B}, respectively.}
\label{massFeb}
\end{table}

The estimated mass and direction of propagation for CME2 and CME3 are given in Table~\ref{massFeb}. We also noticed that mass of CMEs increased with distance from the Sun. We interpret this increase in mass as an observational artifact due to the emergence of CME material from behind the occulter of the coronagraphs. However the possibility of a small real increase in CME mass can not be ignored completely.

\subsubsection{Estimation of coefficient of restitution}
\label{IntFebCoefResti}

The coefficient of restitution measures the bounciness (efficiency to rebound) of a pair of objects in a collision and is defined as the ratio of their relative velocity of separation to their relative velocity of approach. Hence, for $e$ $<$ 1, $e$ = 1, and $e$ $>$ 1, the collision is termed inelastic, elastic, and super-elastic, and consequently, the kinetic energy of the system after the collision is found to decrease, stay equal, and increase than before the collision, respectively. Our analysis shows that the masses of CME become constant after $\approx$ 10 \textit{R}$_{\odot}$ therefore, we assume that they remain constant before and after their collision in HI FOV. Combining the equation of conservation of momentum with a coefficient of restitution, the velocities of CME2 and CME3 after the collision can be estimated theoretically ($v_{1th},v_{2th}$).

\begin{equation}  \label{v12th}
  v_{1th} = \frac{m_{1}u_{1} + m_{2}u_{2} + m_{2} e (u_{2} - u_{1})}{(m_{1} + {m_{2}})};
	v_{2th} = \frac{m_{1}u_{1} + m_{2}u_{2} + m_{1} e (u_{1} - u_{2})}{(m_{1} + {m_{2}})}
\end{equation}
 
where $e$ is the coefficient of restitution, $e$ = $v_{2}$ - $v_{1}$/$u_{1}$ - $u_{2}$ and signifies the nature of collision.

Using the velocity ($u_{1}$,$u_{2}$) = (300,525) km s$^{-1}$ and true mass values ($m_{1}$,$m_{2}$) = (5.40 $\times$ 10$^{12}$, 4.78 $\times$ 10$^{12}$) kg, we calculate a set of theoretical values of final velocity ($v_{1th}$,$v_{2th}$) after the collision of CMEs from equation~(\ref{v12th}) corresponding to a set of different values of coefficient of restitution ($e$). We define a parameter called variance, $\sigma = \sqrt{(v_{1th} - v_{1})^{2} + (v_{2th} - v_{2})^{2}}$. Considering the theoretically estimated final velocity from the equation~(\ref{v12th}) and variance ($\sigma$) values, one can obtain the most suitable value of $e$, corresponding to which the theoretically estimated final velocity ($v_{1th}$,$v_{2th}$) is found to be the closest to the observed final velocity ($v_{1}$,$v_{2}$) of CMEs. This implies that the computed variance is minimum at this $e$ value.

We have estimated the total kinetic energy of the system before the collision as  9.01 $\times$ 10$^{23}$ J. The individual kinetic energy of CME2 and CME3 is 2.43 $\times$ 10$^{23}$ J and 6.58 $\times$ 10$^{23}$ J, respectively. We note that momentum of CME2 and CME3 is  1.6 $\times$ 10$^{18}$ N s and 2.5 $\times$ 10$^{18}$ N s, respectively, just before their observed collision. Hence, the total momentum of the system is equal to 4.13 $\times$ 10$^{18}$ N s. We consider ($v_{1}$,$v_{2}$) the estimated final velocity (from SSSE method) as (600,400) km s$^{-1}$  (Figure~\ref{CME123_SSSE}). We found that ($v_{1th}$,$v_{2th}$) = (495,304) km s$^{-1}$  and the minimum value of $\sigma$ is 142 corresponding to $e$ = 0.85. For this value of $e$, the momentum is found to be conserved, and nature of collision is in an inelastic regime. Such a collision resulted in a decrease in total kinetic energy of the system by 2\% of its value before the collision. If the measured velocity values are directly used, then $e$ is estimated as 0.89, which is approximately equal to that obtained from the aforementioned theoretical approach.

To account for uncertainties in the results, we repeated our computation by taking an uncertainty of $\pm$ 100 km s$^{-1}$ in the estimated final velocity after the collision of CMEs. For example, if we use ($v_{1}$,$v_{2}$) = (700,500), then minimum value of $\sigma$ = 288 is found corresponding to $e$ = 0.80. The estimate for $\sigma$ is found to be minimum and is equal to 2.0 corresponding to $e$ = 0.90, when ($v_{1}$,$v_{2}$) = (500,300) km s$^{-1}$ is used and in this case ($v_{1th}$,$v_{2th}$) = (501,298) is obtained. This implies that keeping the conservation of momentum as a necessary condition, the combination of 
($u_{1}$,$u_{2}$) = (300,525) km s$^{-1}$ and ($v_{1}$,$v_{2}$) = (500,300) km s$^{-1}$ with $e$ = 0.90 best suits the observed case of collision of CME2 and CME3. In this case, the total kinetic energy after the collision decreased by only 1.3\%, the kinetic energy of CME2 increased by 177\%, and the kinetic energy of CME3 decreased by 67\% of its value before the collision. It implies that the observed collision is in the inelastic regime but closer to the elastic regime. For this case, the momentum of CME2 increased by 68\%, and momentum of CME3 decreased by 35\% of its value before their collision. Our analysis, therefore, shows that there is a huge transfer of momentum and kinetic energy during the collision phase of CMEs.    

\begin{figure}[htb]
 \centering
  \includegraphics[scale=0.70]{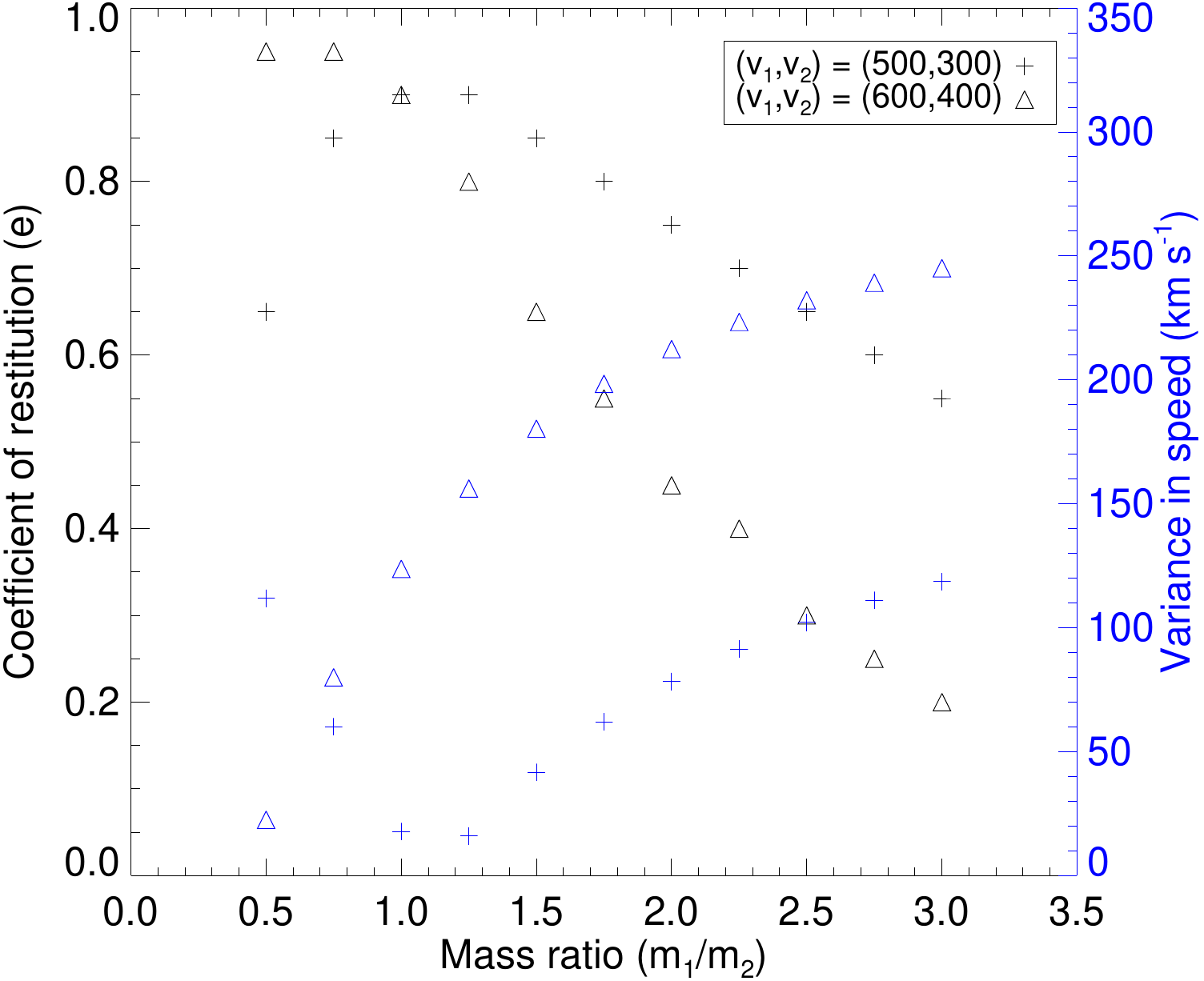}
 \caption[Coefficient of restitution values corresponding to different mass ratios of interacting CMEs of the 2011 February 14 and 15]{The best suited coefficient of restitution values corresponding to different mass ratios of CME2 and CME3 are shown for  their observed velocity in post-collision phase (in black). We also show the variance in speed corresponding to these coefficients of restitution (in blue).}
\label{CME12_mass}
 \end{figure}

It is worth examining the effect of uncertainty in mass in the estimation of the value of $e$ or the nature of the collision. We have estimated the true mass, which is also uncertain and difficult to quantify \citep{Colaninno2009}. However, a straightforward uncertainty arises from the assumption that CME structure lies in the plane of the 3D propagation direction of CME. \citet{Vourlidas2000} have shown that such a simplified assumption can cause the underestimation of CME mass by up to 15\%. Applying this uncertainty to the estimated true mass of CME2 and CME3, their mass ratio ($m_{1}$/$m_{2}$ = 1.12) can range between 0.97 to 1.28. To examine the effect of larger uncertainties in the mass, we arbitrarily change the mass ratio between 0.5 to 3.0 in the step of 0.25 and repeat the analysis above (using Equation~(\ref{v12th}) and calculating $\sigma$ value) to estimate the value of $e$ corresponding to each mass ratio. The variation of $e$ with mass ratio is shown in Figure~\ref{CME12_mass} corresponding to the observed final velocity ($v_{1}$,$v_{2}$) = (600,400) km s$^{-1}$ after the collision of CMEs in our case. We have shown earlier that the best suited final velocity of CMEs for our observed case of CME collision is ($v_{1}$,$v_{2}$) = (500,300) km s$^{-1}$.  Therefore, corresponding to this velocity, the variation of $e$ with the mass ratio is shown (Figure~\ref{CME12_mass}). We have also plotted the estimated minimum variance corresponding to each obtained value of $e$. It is evident that even if a large arbitrary mass ratio is considered, the collision remains in the inelastic regime. It never reaches a completely inelastic ($e$ = 0), elastic ($e$ = 1) or super-elastic ($e$ $>$ 1) regime. 

\subsection{In situ observations, arrival time and geomagnetic response of interacting CMEs of 2011 February 13-15}
\label{IntFebInsitu}

\subsubsection{In situ observations}
We analyzed the \textit{WIND} spacecraft plasma and magnetic field observations taken from CDAWeb $(http://cdaweb.gsfc.nasa.gov/)$. We attempted to identify the CMEs based on the criterion of \citet{Zurbuchen2006}. The variations in plasma and magnetic field parameters from 2011 February 17 at 20:00 UT to February 20 at 04:00 UT, are shown in Figure~\ref{CME123_insitu}. The regions marked as R1, R2, and R3 are associated with CME1, CME2, and CME3, respectively, and S marks the arrival of shock driven by CME3. In the region R3, the latitude and longitude of the magnetic field vector (from top, 6th and 7th panel of Figure~\ref{CME123_insitu}) rotated, and plasma beta ($\beta$) was found to be less than 1. Therefore, this region (R3) may be termed a MC.

\begin{figure}[!htb]
 \centering
  \includegraphics[height=0.6\textheight, width = 0.7\textwidth]{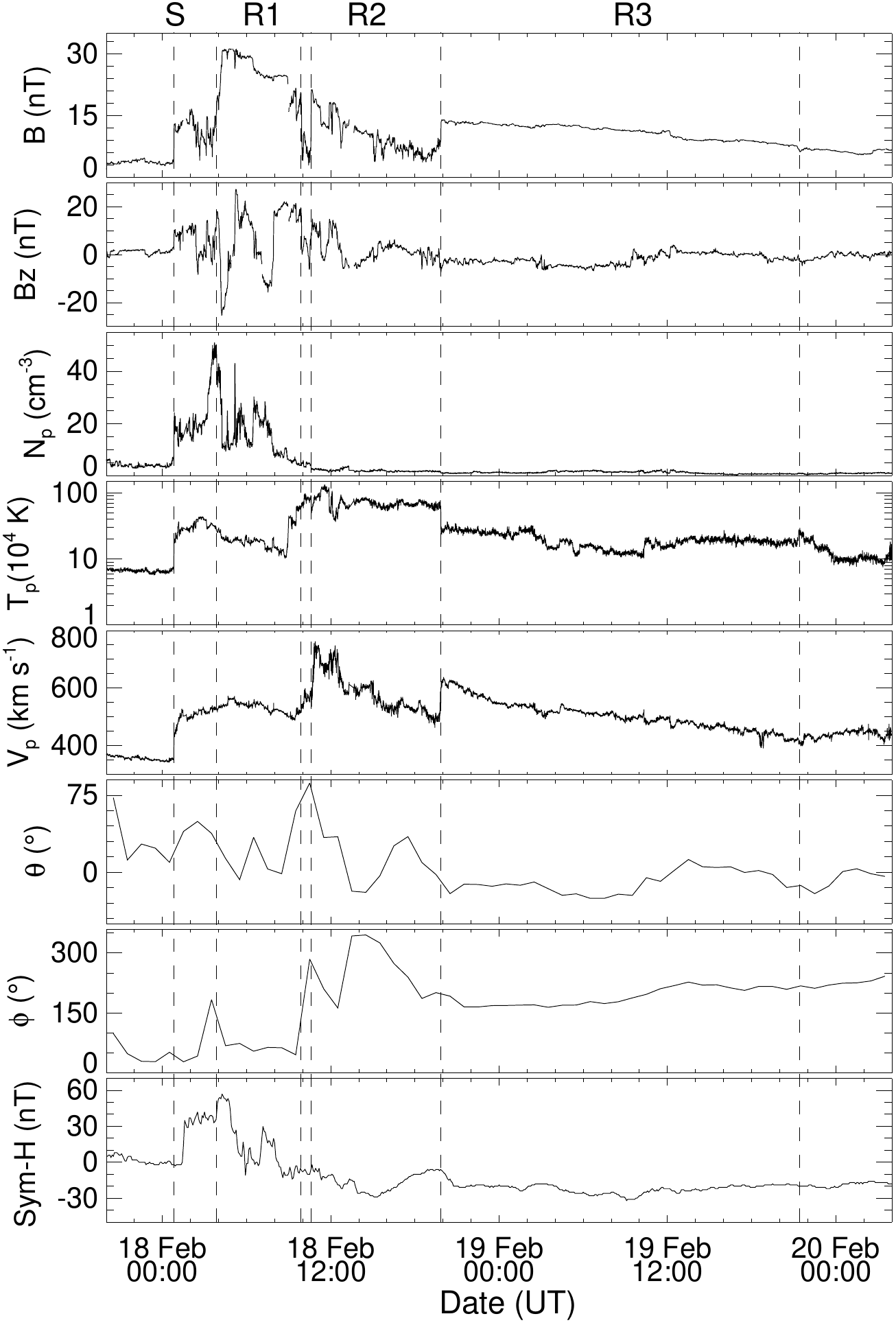}
 \caption[In situ measurements of the 2011 February 13-15 CMEs]{From top to bottom, panels show the variations of magnetic field strength, a southward component of the magnetic field, proton density, proton temperature, proton flow speed, latitude, the longitude of magnetic field vector and longitudinally symmetric disturbance index for horizontal (dipole) direction, respectively. From the left, first, second, third, fourth, fifth, and sixth vertical lines mark the arrival of shock, leading edge of CME1, trailing edge of CME1, leading edge of CME2, trailing edge of CME2, and trailing edge of CME3. S, R1, R2, and R3 stand for the arrival of shock, the bounded interval for CME1, CME2, and CME3 structures, respectively.}
\label{CME123_insitu}
\end{figure}

The region bounded between 09:52 UT and 10:37 UT on February 18 with third and fourth dashed lines, from the left, shows sharp decrease in magnetic field strength, enhanced temperature, and flow speed, and sudden change in longitude of magnetic field vector. This region lasted for less than an hour but represents a separate structure between R1 and R2 which could be a magnetic reconnection signature between field lines of region R1 and R2 \citep{Wang2003, Gosling2005}, however an in-depth analysis is required to confirm this. In situ observations also reveal that region R2 is overheated $\approx$ 10$^{6}$ K, perhaps because it is squeezed between the regions R1 and R3. Region R2 shows a high speed of 750 km s$^{-1}$ at the front and a low speed of 450 km s$^{-1}$ at its trailing edge. Such observations may indicate an extremely fast expansion of R2 due to magnetic reconnection at its front edge, as also suggested by \citet{Maricic2014}. From an overall inspection of in situ data, it is clear that the plasma is heated ($\approx$ 10$^{5}$ K for regions R1 and R3 and $\approx$ 10$^{6}$ K for region R2 ) than what is observed ($\approx$ 10$^{4}$ K) in general, in CMEs. Such signatures of compression and heating due to CME-CME interaction and passage of CME driven shock through the preceding CME have also been reported in earlier studies \citep{Lugaz2005, Liu2012,Temmer2012}. In situ data also shows that the spatial scale of CME1 (R1) and CME2 (R2) is smaller than CME3 (R3) and which may happen due to their compression by the following CME or shock for each.

\subsubsection{Estimation of arrival time of CMEs}
\label{IntFebArrtime}

If the measured 3D speeds (Figure~\ref{CME123_3DFM}) of CME2 and CME3 at the final height are assumed to be constant for the distance beyond COR2 FOV, then CME3 should have caught the CME2 at 39 \textit{R}$_{\odot}$ on 2011 February 15 at 17:00 UT. However, our analysis of HI observations (using \textit{J}-maps) shows that these two CMEs collided $\approx$ 7 hr earlier (at $\approx$ 28 \textit{R}$_{\odot}$). This can happen due to several reasons firstly, if two different features are tracked in COR and HI observations. Secondly, a deceleration of CME2 beyond COR2 FOV may also be responsible for this. Taking 3D speed estimated in COR2 FOV as a constant up to L1, the arrival times of CME1, CME2, and CME3 at L1 will be at 13:00 UT on February 16, at 20:10 UT on February 18 and 23:20 UT on February 17, respectively. However, as discussed in section~\ref{IntFebRecnsHI}, after the collision between CME2 and CME3, the dynamics of CMEs changed. Therefore, we extrapolated linearly the height-time plot up to L1 by taking a few last points in the post-collision phase of these CMEs and obtained their arrival time. Such extrapolation may contribute to uncertainties in arrival times of CMEs \citep{Colaninno2013}. From these extrapolations (shown in the top panel of Figure~\ref{CME123_SSSE}), the obtained arrival time of CME2 and CME3 at L1 is on 2011 February 18 at 02:00 UT and 05:00 UT, respectively. These extrapolated arrival time for CME2 and CME3 is 12 hr earlier and 6 hr later, respectively than that estimated from measurements made in COR2 FOV. Based on these results, we infer that after the collision of CME2 and CME3, CME2 gained kinetic energy and momentum at the cost of the kinetic energy and momentum of CME3. The arrival time of CME3 is also estimated (within an error of 0.8 to 8.6 hr) by \citet{Colaninno2013} by applying various fitting approaches to the deprojected height-time plots derived by applying the GCS model to CMEs observed SECCHI images.

We associate the starting times of in situ structures marked as R1, R2, and R3 (in Figure~\ref{CME123_insitu}) with the actual arrival of CME1, CME2, and CME3, respectively. We find that the marked leading edge of CME1 at L1 is $\approx$ 14 hr earlier than that estimated by extrapolation. The extrapolated arrival time for CME1 is 18:40 UT on February 18. This difference can be explained by assuming a possible acceleration of CME1 beyond the tracked points in HI FOV. We have extrapolated CME1 height-time tracks from its pre-interaction phase because CME1 could not be tracked in \textit{J}-maps up to longer elongations where the interaction is inferred. This highlights the possibility that after its interaction with CME2 or CME3 driven shock (discussed in section~\ref{IntFebRecnsHI}), the CME1 has accelerated.

The actual arrival times of CME2 and CME3 leading edge (shown in Figure~\ref{CME123_insitu}) are $\approx$ 8.5 and 15 hr later, respectively, than obtained by direct linear extrapolation of the height-time curve (Figure~\ref{CME123_SSSE}). From the aforementioned arrival time estimates, we notice an improvement in arrival time estimation of CME2 and CME3 by a few (up to 10) hr when the post-collision speeds are used instead of using the pre-collision speeds. The average measured (actual) transit speed of CME2 and CME3 at L1 is approximately 100 km s$^{-1}$ larger than its speed in remote observations in the post-collision phase. Such an inconsistency of delayed arrival is possible only if it is assumed that CME2 and CME3 over-expand before reaching L1 or in situ spacecraft is not hit by the nose of these CMEs \citep{Maricic2014}. The short duration of CME2 in in situ data with a lack of magnetic cloud signature favors a flank encounter of CME2 at the spacecraft. The late arrival of CME3 may also be due to its higher deceleration than estimated in HI FOV. If the remotely tracked feature is incorrectly identified in the in situ data, such inconsistency may also arise. 

\subsubsection{Geomagnetic response of interacting CMEs}
\label{IntFebGeomag}
In the bottom panel of Figure~\ref{CME123_insitu}, longitudinally symmetric disturbance (Sym-H) \citep{Iyemori1990} index is plotted. This index is similar to the hourly disturbance storm time (Dst) \citep{Sugiura1964} index. Still, it uses 1 minute values recorded from a different set of stations and slightly different coordinate systems and method to determine base values. The effect of solar wind dynamic pressure is more clearly seen in Sym-H index than in the hourly Dst index. We observed a sudden increase in Sym-H index up to 30 nT around 01:30 UT on February 18 which is within an hour of the arrival of interplanetary shock. The Sym-H index continued to rise, and around 04:15 UT reached 57 nT. The first steep rise in this index marked by the shock is represented by the enhanced magnetic field, speed, and density. The second peak in Sym-H is primarily due to a corresponding peak in magnetic field strength and density. However no peak in speed was observed at this time. During the passage of region R1, the z- component of the interplanetary magnetic field (Bz) began to turn negative at 04:07 UT and remained so up to one hour. During this period, it reached -25 nT at 04:15 UT and then turned to positive values around 05:00 UT. We noticed that  Bz turned negative a second time at 07:07 UT and remained so for 47 minutes, reaching a value of -15 nT at 07:31 UT on February 18. From the Sym-H plot, it is clear that the negative turning of Bz twice caused a fast decrease in elevated sym-H values. \citet{Dungey1961} has shown that the negative Bz values and process of magnetic reconnection at the magnetosphere enables magnetized plasma to transfer its energy into the magnetosphere and form a ring current.

Succinctly, we infer that the arrival of magnetized plasma can be attributed to the strong storm's sudden commencement (SSC) (Dst = 57 nT) and short duration (47 minutes) negative Bz field therein resulting in a minor geomagnetic storm (Dst = -32 nT). The intensity of SSC is independent of the peak value of depression in the horizontal component of the magnetic field during the main phase of a geomagnetic storm. Our analysis supports the idea of collision (or interaction) of multiple CMEs, which can enhance the magnetic field strength, density, and temperature within CMEs \citep{Liu2012, Mostl2012}. Such enhanced parameters can increase the conductivity of CME plasma and result in intense induced electric current in CME when it propagates towards the Earth's magnetic field. This induced electric current within CME plasma causes its intense shielding from Earth's field and increases the magnetic field intensity around the Earth, which is manifested as SSC \citep{Chapman1931}.  

\subsection{Results and Discussion on 2011 February 13-15 CMEs}
\label{IntFebResDis}
We summarize our results on the analysis of the interaction of three Earth-directed CMEs launched in succession during 2011, February 13-15, on three main aspects. These include the morphological \& kinematic study of interacting CMEs and near-Earth manifestations.

\subsubsection{Morphological evolution of CMEs}
\label{IntFebResDis1}
We have studied the morphological properties of Earth-directed CMEs (CME2 and CME3) when the separation angle between \textit{STEREO-A} and \textit{STEREO-B} was 180$^\circ$. Comparing the morphological evolution of CMEs with the cone model, we found that slow speed CME2 maintained a constant angular width in the corona. Still, the angular width of fast speed CME3 decreased monotonically as it propagated further in the corona. The possible explanation for this is that when CME3 was launched from the Sun, its leading edge suddenly experienced the ambient solar wind pressure, and resulted in its flattening \citep{Odstrcil2005}, causing a large angular width. However, as the CME3 propagated further in the corona, there was a decrease in interaction between the solar wind and the part of CME (i.e., near apex of the cone), which decides the angular width. Therefore, a decrease in angular width is noticed away from the Sun.

The difference in the cone and contour area of CME2 increases linearly with the radial height of the CME leading edge (bottom panel of Figure~\ref{CME2_area}). This can be explained by the fact that the CME2 interacted with the solar wind such that its leading edge (especially the nose) stretched out thereby increasing the value of $r$ (distance between Sun-center and nose of CME) and also the ice-cream cone area. For the fast speed CME3, we find that the contour area is less than the cone area close to the Sun, but as the CME propagated further in the corona, its contour area became larger than the cone area (Figure~\ref{CME3_area}). This can be possibly explained by the concept that, contrary to the behavior of CME2, as the CME3 propagated further in the corona, its front flattened due to drag force, leading to spilling some CME mass outside the cone, i.e., at the flanks of CME. This flattening resulted in a lower estimated value of $r$ and hence a decrease in the estimated cone area. Therefore, a negative value is obtained for the difference between cone area and contour area (bottom panel of Figure~\ref{CME3_area}). From Figure~\ref{CME2_area} and~\ref{CME3_area} (bottom panels), we can say that slow and fast speed CMEs deviate from the cone model differently.

Our analysis shows that the estimated 2D angular width (converted from 3D) follows the same trend as observed in 2D images (Figure~\ref{theta_CMEs}) but has slightly different (within 5\% for CME3 and 15\% for CME2) value at a certain height. We also emphasize that the GCS model parameters ($\gamma$, $\kappa$ and $\alpha$) are very sensitive \citep{Thernisien2009} and can only be fitted with limited accuracy, especially for fast CME whose front gets distorted (see, Figure~\ref{CME3_area}) due to possible interaction with the solar wind. Also, the estimation of these parameters depends on the visual agreement between GCS modeled CME and the observed CME and is user dependent. It is to be noted that the minor error in these sensitive parameters can lead to significant errors in the 3D edge-on and face-on width of GCS modeled CME. This is the reason, despite a reasonably good agreement between GCS model parameters derived in our study with those derived in \citet{Temmer2014}, the 3D values of angular widths for fast CME of February 15 (CME3) do not match well with their results.  Although, we acknowledge that measurements of observed 2D width (Figure~\ref{theta_CMEs}) also have some error (within 5$\arcdeg$), which is relatively small than the involved uncertainties in the 3D or 2D angular width estimated from the GCS model. In light of the aforementioned uncertainties and results, further work needs to be carried out to investigate the change in angular width of fast CMEs.

\subsubsection{Kinematic evolution of interacting CMEs}
\label{IntFebResDis2}
The 3D speed and direction estimated for three selected CMEs in COR2 FOV suggest their possible interaction in the interplanetary medium. We have found that CME3 is the fastest among all the three CMEs and shows strong deceleration in the COR2 FOV because of the preceding CME2, which acts as a barrier for it. From the analysis of the kinematics of CMEs in the heliosphere using stereoscopic methods, we have noted that a collision between CME3 and CME2 took place around 24 \textit{R}$_\odot$-28 \textit{R}$_\odot$. As the CME1 was faint and could not be tracked up to HI2 FOV in \textit{J}-maps, we inferred based on the extrapolation of distances that CME2 caught up with CME1 between 138 \textit{R}$_\odot$ to 157 \textit{R}$_\odot$.

It may be noted that using three stereoscopic methods in our study, the estimates of velocity and location of collision for three selected CMEs are approximately the same (within a reasonable error of a few tens of km s$^{-1}$ and within a few solar radii) to those obtained by \citet{Maricic2014} using single spacecraft method.

We have identified collision signatures between CMEs in the kinematics profiles as an exchange in their speed. We analyzed momentum and energy exchange during the collision phase of CME2 and CME3 and found that the nature of collision was in an inelastic regime, reaching close to elastic. Our study found that even after considering reasonable uncertainties in derived mass and velocity parameters, the coefficient of restitution ($e$) lies between 0.78 to 0.90 for the interacting CME2 and CME3. This implies that the total kinetic energy of the system of CMEs after the collision is less than its value before the collision.

\subsubsection{Interacting CMEs near the Earth}
\label{IntFebResDis3}

We have examined the interaction and the collision signatures of CMEs in the in situ (\textit{WIND}) observations. The interacting CMEs could be identified as a separate entity in in situ observations, therefore, could not be termed as complex ejecta as defined by \citet{Burlaga2002}. The in situ observations suggest that a shock launched by the fastest CME of February 15 (CME3) passed through the CME of February 14 (CME2) and CME of February 13 (CME1) and caused compression, heating, and acceleration, in particular for the CME of February 14 which is sandwiched between preceding CME (CME1) and the following CME (CME3). Our analysis shows that the interacting CMEs resulted in a minor geomagnetic storm with a strong long, duration SSC. This is in contrast to the results of \citet{Farrugia2004, Farrugia2006,Mishra2014a}, which suggested that interaction of CMEs leads to long-duration southward component of magnetic field and, therefore to intense geomagnetic storms.

\section{Interacting CMEs of 2012 November 9-10}
\label{IntNov}
This section focuses on the identification, evolution, and propagation of two CMEs launched on 2012 November 9-10 as they traveled from the corona to the inner heliosphere. This study, in which the launch time of the two selected CMEs on November 9 and 10 are separated by about 14 hr, provides an opportunity to understand the CME-CME interaction unambiguously.  For these CMEs, we have followed a similar approach to obtain kinematics in COR and HI FOV and understand the energy and momentum exchange, geomagnetic response, and arrival time as for the previous case of 2011 February 13-15 CMEs. The CME of November 9 and 10 are referred to as CME1 and CME2 throughout this Section~\ref{IntNov}.

\subsection{3D reconstruction in COR2 FOV}
\label{IntNovRecnsCOR}

A partial halo CME1 (angular width of $276^{\circ}$) was observed in SOHO/LASCO-C2 FOV at 15:12 UT on 2012 November 9 (\url{http://cdaw.gsfc.nasa.gov/CME_list/}). In the SECCHI/COR1-B and A, this CME was observed at 14:25 UT in the SW and SE quadrants, respectively. Another partial halo CME2 (angular width $\cong 210^{\circ}$) was observed in LASCO/C2 FOV at 05:12 UT on November 10. This CME was detected by SECCHI/COR1-B and A in the SW and SE quadrants, respectively, at 05:05 UT on 2012 November 10. 
At the time of observations around   November 9, the \textit{STEREO-A} and \textit{STEREO-B} were $127^{\circ}$ westward and $123^{\circ}$ eastward from the Sun-Earth line at a distance of 0.96 AU and 1.08 AU from the Sun, respectively.

Figure~\ref{IntNovscc}(a) and (b) show 3D kinematics of CME1 and CME2, respectively, from the 3D reconstruction of selected features along the leading edges of corresponding CMEs using a tie-pointing procedure (scc\_measure: \citealt{Thompson2009}) on SECCHI/COR2 data.  

\begin{figure}[!htb]
\begin{center}
\includegraphics[scale=0.70]{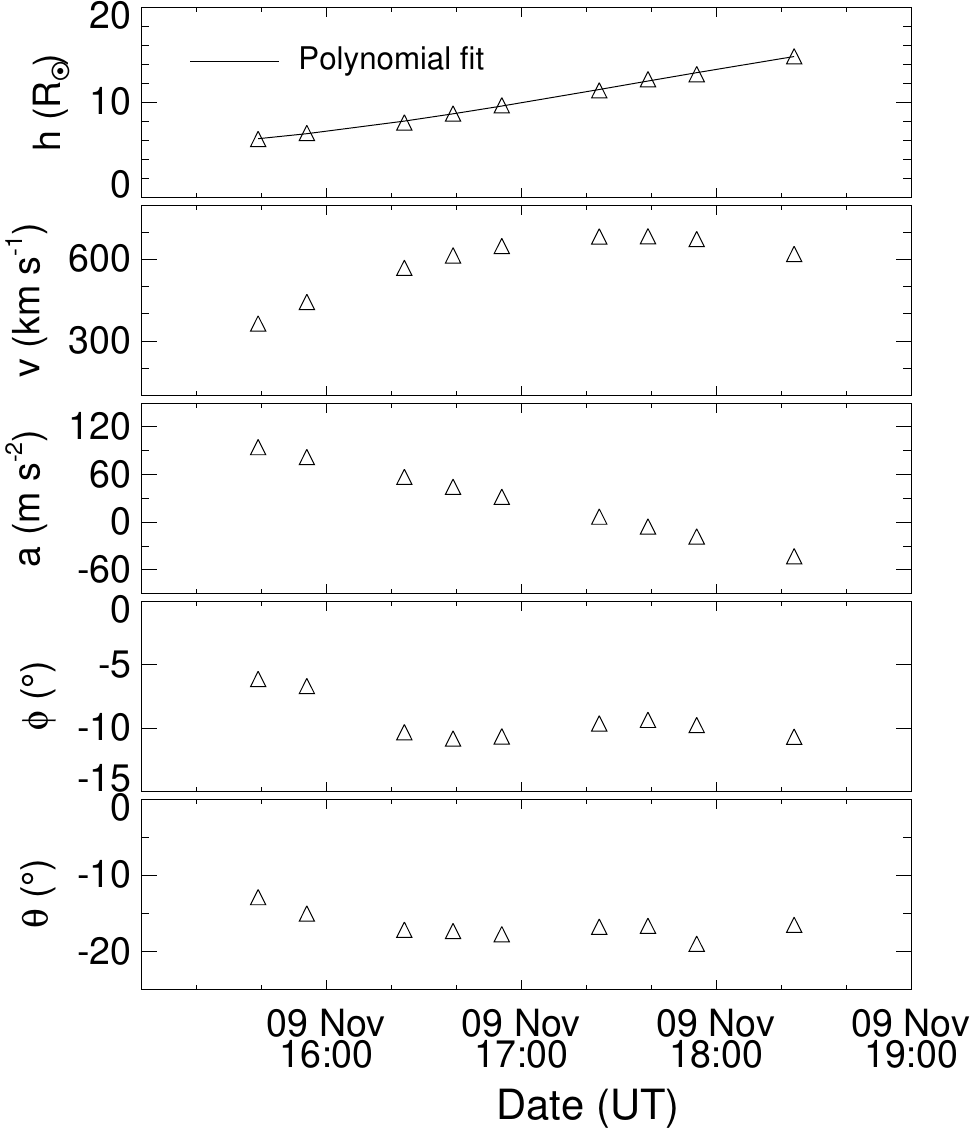}
\includegraphics[scale=0.70]{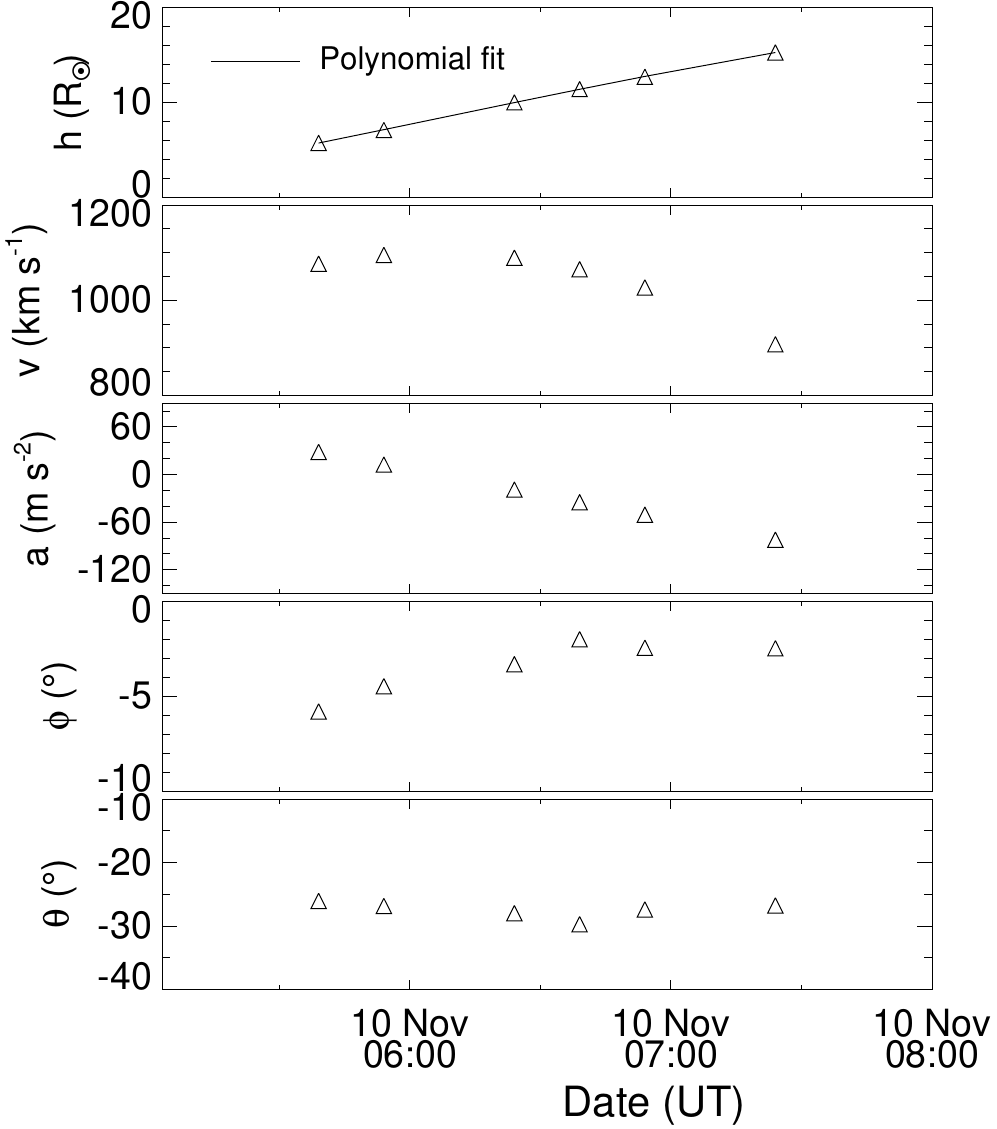}			
\caption[Estimated 3D kinematics of the 2012 November 9 and 10 CMEs using tie-poining method]{From top to bottom panels, 3D height, velocity, acceleration, longitude, and latitude of the selected feature along the leading edge as derived from the tie-pointing method have been plotted as a function of time for CME1 (left) and CME2 (right)}
\label{IntNovscc}
\end{center}
\end{figure}

As the separation angle of the \textit{STEREO} spacecraft was large, it is better to use two independent methods to confirm the results of 3D reconstruction. Therefore, we have visually fitted both CMEs in the SECCHI/COR2 FOV using the graduated cylindrical shell (GCS) model \citep{Thernisien2009, Thernisien2011} for the 3D reconstruction of these CMEs. We used contemporaneous image triplets of CME1 around 17:39 UT on November 9 from \textit{STEREO-A}/COR2, \textit{STEREO-B}/COR2, and \textit{SOHO}/LASCO-C3. The best visual fit for CME1 is found in the direction of W02S14, with a half angle of 19$^{\circ}$, tilt angle of 9$^{\circ}$ around the axis of symmetry of the model (\textit{i.e.} rotated 9$^{\circ}$ anticlockwise out of the ecliptic plane), and an aspect ratio of 0.52. At this time, CME1 was at a distance of about 9.6 \textit{R}$_{\odot}$ from the Sun. Using the obtained fitted values of half-angle and aspect ratio of CME1, its 3D face-on and edge-on angular width are estimated as 100$^\circ$ and 62$^\circ$, respectively. We also carried out the visual fitting using the GCS model for CME2 around 06:39 UT on November 10 by exploiting the concurrent image triplets of \textit{STEREO-A}/COR2, \textit{STEREO-B}/COR2, and \textit{SOHO}/LASCO-C3.  The best fit for CME2 is obtained in the direction of W06S25, at a tilt angle of $9^{\circ}$ around the axis of symmetry of the model, with a half angle of $12^{\circ}$, and an aspect ratio of 0.19. At this time, CME2 was at a distance of 8.2 \textit{R$_{\odot}$} from the Sun. Using the GCS model fitted values, the 3D face-on and edge-on angular width of CME2 is estimated as 46$^\circ$ and 22$^\circ$, respectively. The GCS model fit for CME1 and CME2 is shown in Figures~\ref{IntNovFM}. We find that the estimated latitude and longitude of both CMEs from the GCS model and the tie-pointing procedure agree within few degrees. This shows that the results of both tie-pointing and the GCS model are reliable and can be used for estimating the kinematics of CMEs in the coronagraphic FOV.

\begin{figure}[!htb]
\begin{center}
\includegraphics[scale=0.32]{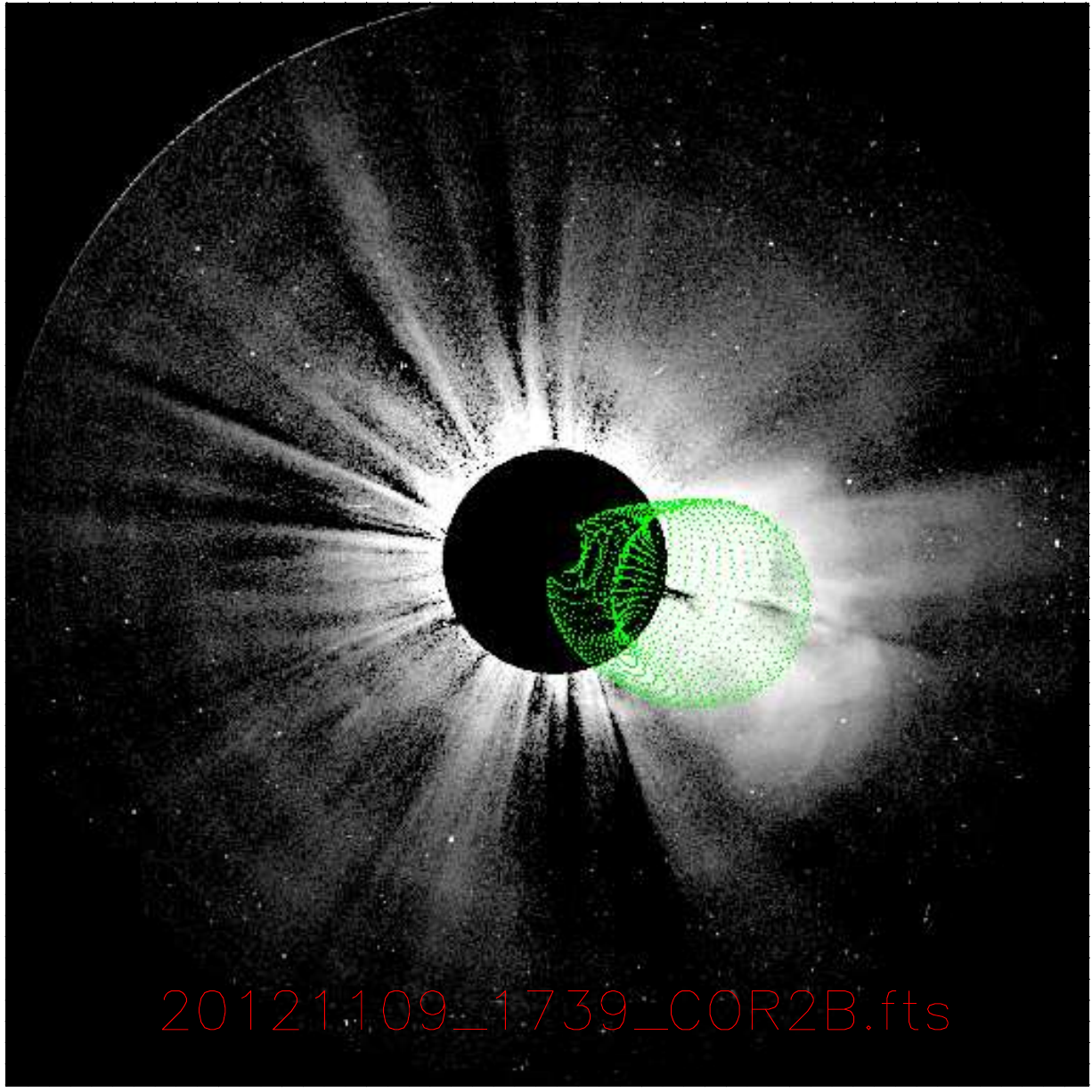}
\includegraphics[scale=0.32]{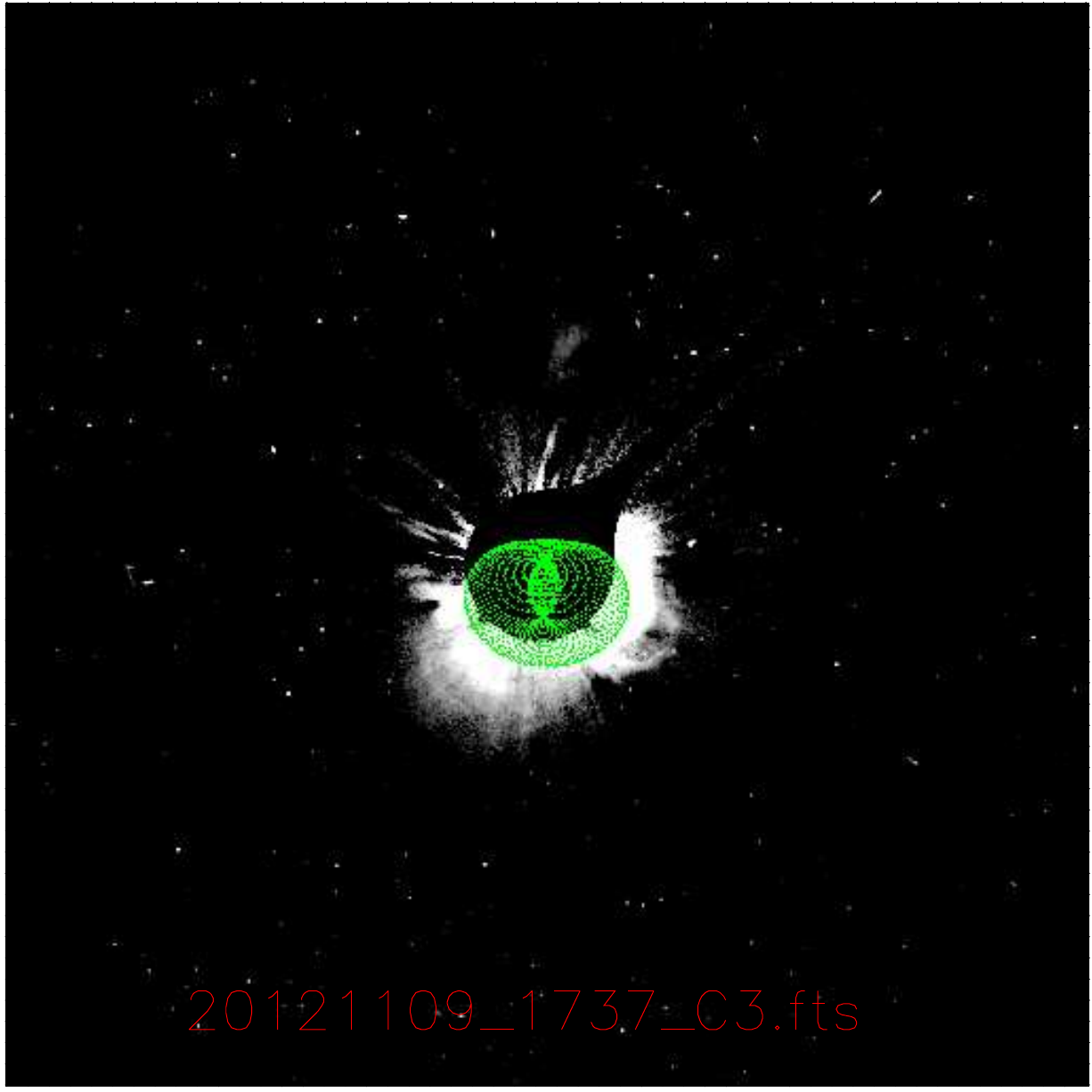}
\includegraphics[scale=0.32]{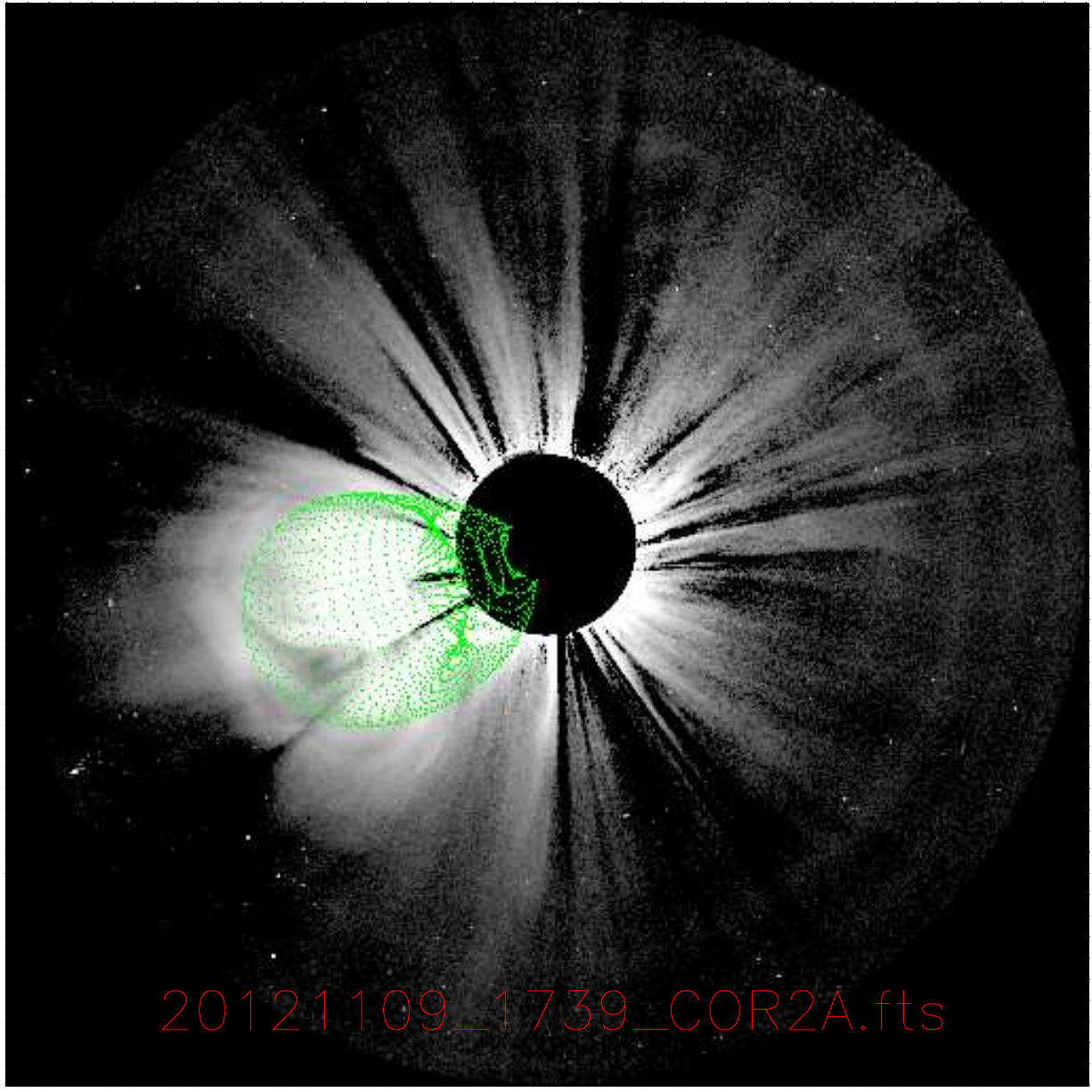}
\includegraphics[scale=0.32]{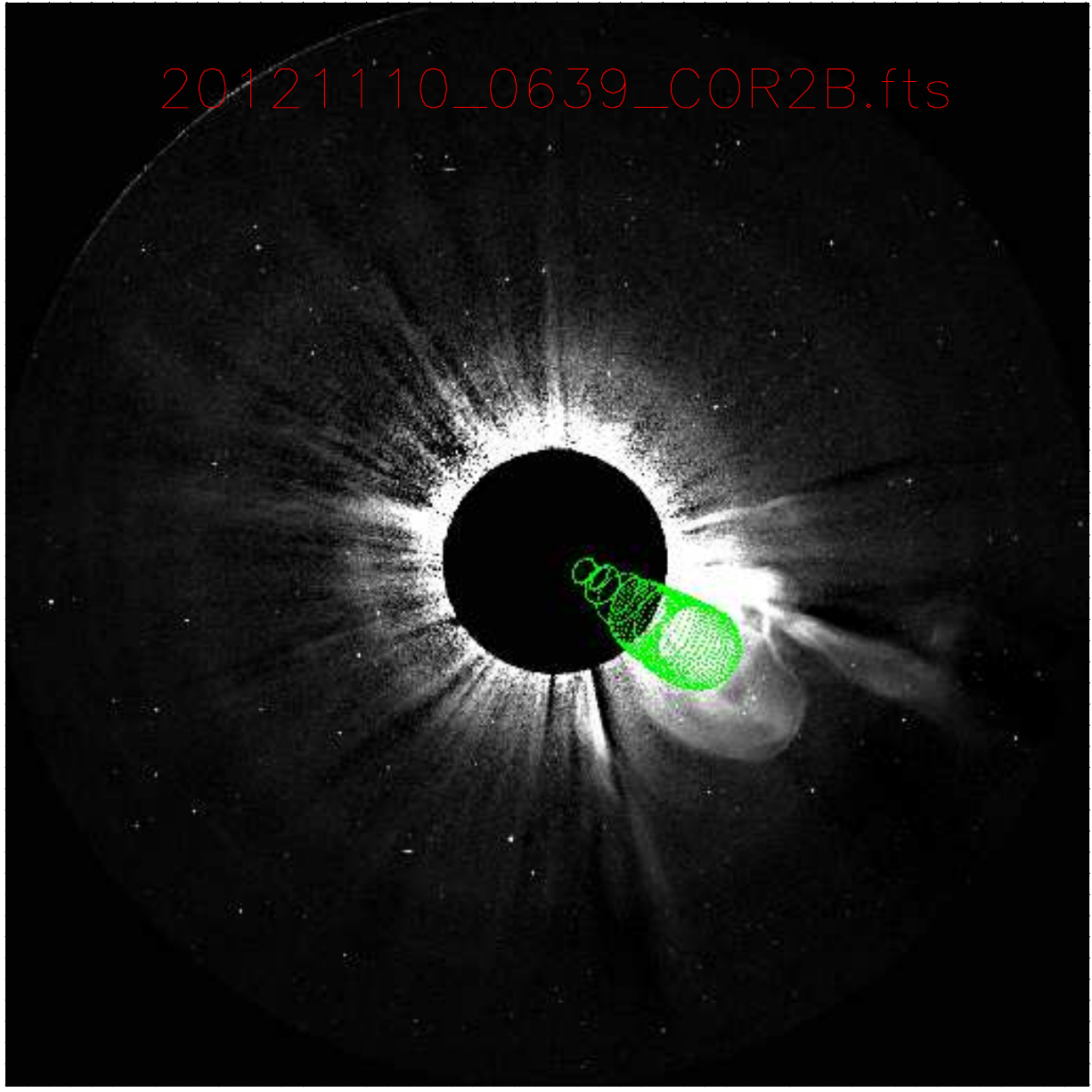}
\includegraphics[scale=0.32]{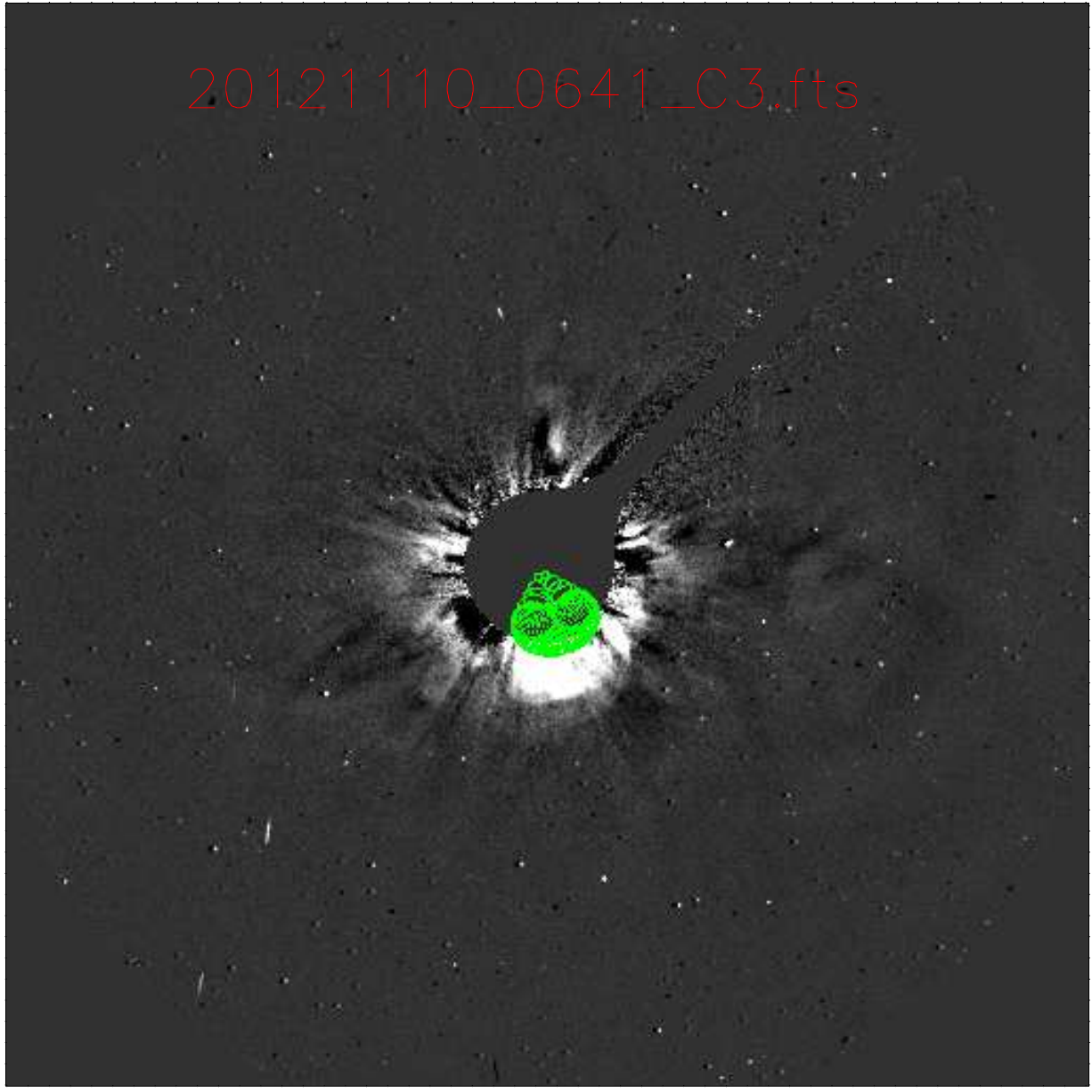}
\includegraphics[scale=0.32]{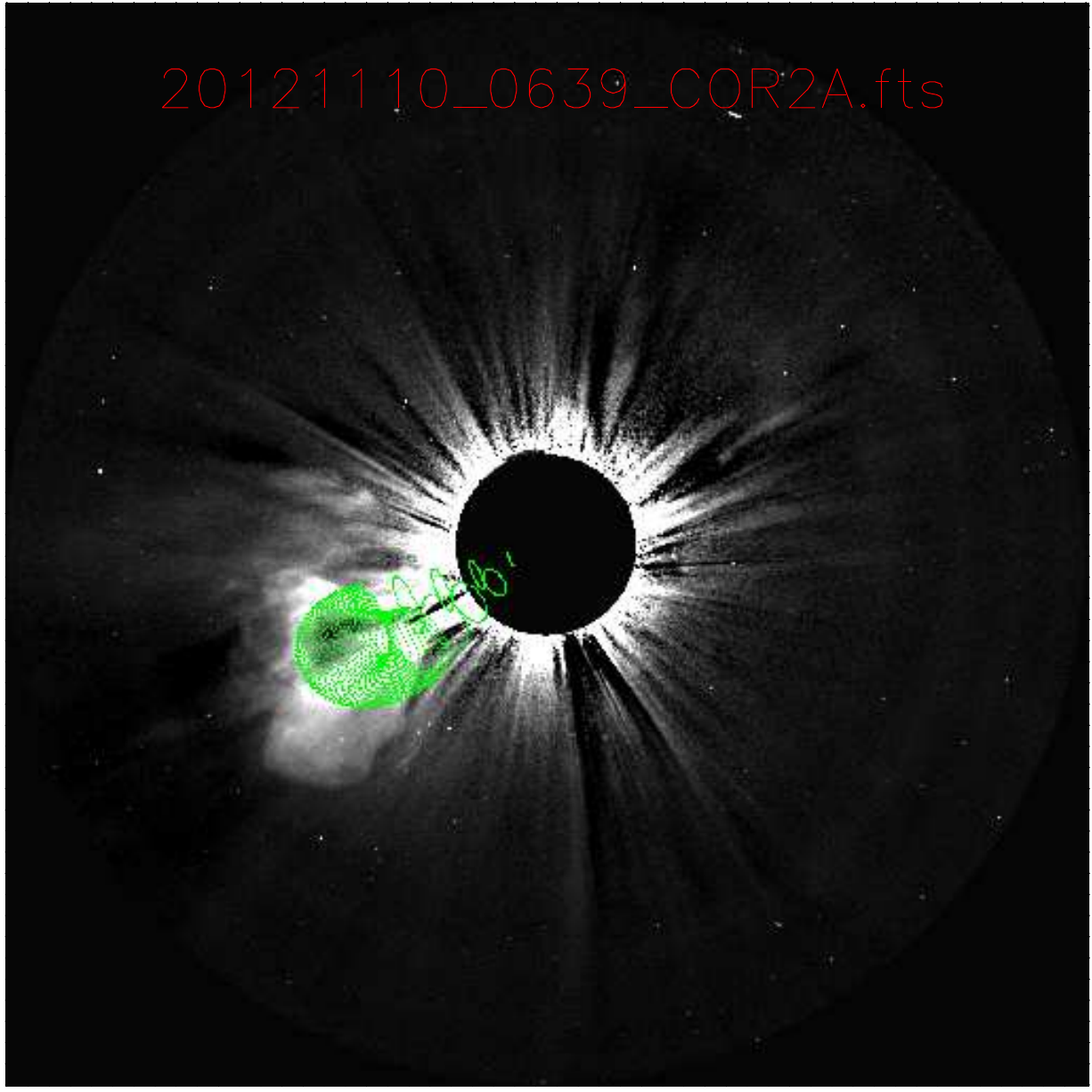}

\caption[GCS model applied on the 2012 November 9 and 10 CMEs images taken from \textit{STEREO}/COR2-B, \textit{SOHO}/LASCO-C3, and \textit{STEREO}/COR2-A]{The GCS model wireframe overlaid on the CME1 (top) and CME2 (bottom) images, respectively. The triplet of concurrent images around 17:39 UT on November 9 and around 06:39 UT on November 10 corresponding to CME1 and CME2, respectively, are from \textit{STEREO}/COR2-B (left), \textit{SOHO}/LASCO-C3 (middle), and \textit{STEREO}/COR2-A (right).}
\label{IntNovFM}
\end{center}
\end{figure}

From a study of 3D kinematics of the two CMEs using the tie-pointing approach, we found that CME1 was slow (620 km s$^{-1}$) around 15 \textit{R$_{\odot}$} while CME2 was faster (910 km s$^{-1}$) at a distance of approximately 15 \textit{R$_{\odot}$} (Figure~\ref{IntNovscc}). The latitude ($\approx$ -15$^{\circ}$ to -25$^{\circ}$) and longitude ($\approx$ -3$^{\circ}$ to -10$^{\circ}$) of the reconstructed features for both CMEs suggest that these were Earth-directed and could interact in the interplanetary medium. If the speed of the CMEs is assumed to be constant beyond the outer edge of the COR2 FOV, then these CMEs should have collided at approximately 130 \textit{R$_{\odot}$} on  November 11 around 02:30 UT. However, earlier studies have shown that the speed of a CME can significantly change after the coronagraphic FOV \citep{Lindsay1999, Gopalswamy2000,Cargill2004, Manoharan2006,Vrsnak2010}. Therefore, further tracking of CME features in the heliosphere is required to determine the exact location and time of interaction of these CMEs.

\subsection{Tracking of CMEs in HI FOV}
\label{IntNovRecnsHI}
The evolution of CME1 and CME2 in the running difference images of COR2, HI1, and HI2 FOV is shown in Figure~\ref{IntNovevolution}. To track and estimate the arrival times of the CMEs in the heliosphere using HI1 and HI2 images, we constructed the so-called \textit{J}-maps. We derived the variation in the elongation of selected features with time. The positively inclined bright features in the \textit{J}-maps (Figure~\ref{IntNovJ-maps}) correspond to the enhanced density structure of the CMEs. We tracked the leading and trailing edges (marked in green and red) of a bright feature corresponding to the slow CME1 and the top edge (blue) of a bright feature corresponding to the fast CME2. Prior to interpreting the tracked features of CME1 in the \textit{J}-map as its leading and trailing edges, respectively, the derived elongations corresponding to all the three tracked features were overplotted on the base difference HI1 images. On careful inspection of the sequence of these images, we noticed that the two tracked features of CME1 correspond to the first density enhancement at the front and the second density enhancement at the rear edge of CME1, respectively. The base difference HI1-A images with overplotted contours of elongation corresponding to the tracked features of CME1 and CME2 are shown in Figure~\ref{IntNovbasediff}. It is to be noted that the term `trailing edge' of CME1 does not correspond to the rear-most portion of CME1 but refers to the part behind the front and is not an image artifact in the \textit{J}-map constructed from running difference images.  

\begin{figure}[!htb]
\begin{center}
\includegraphics[scale=0.37]{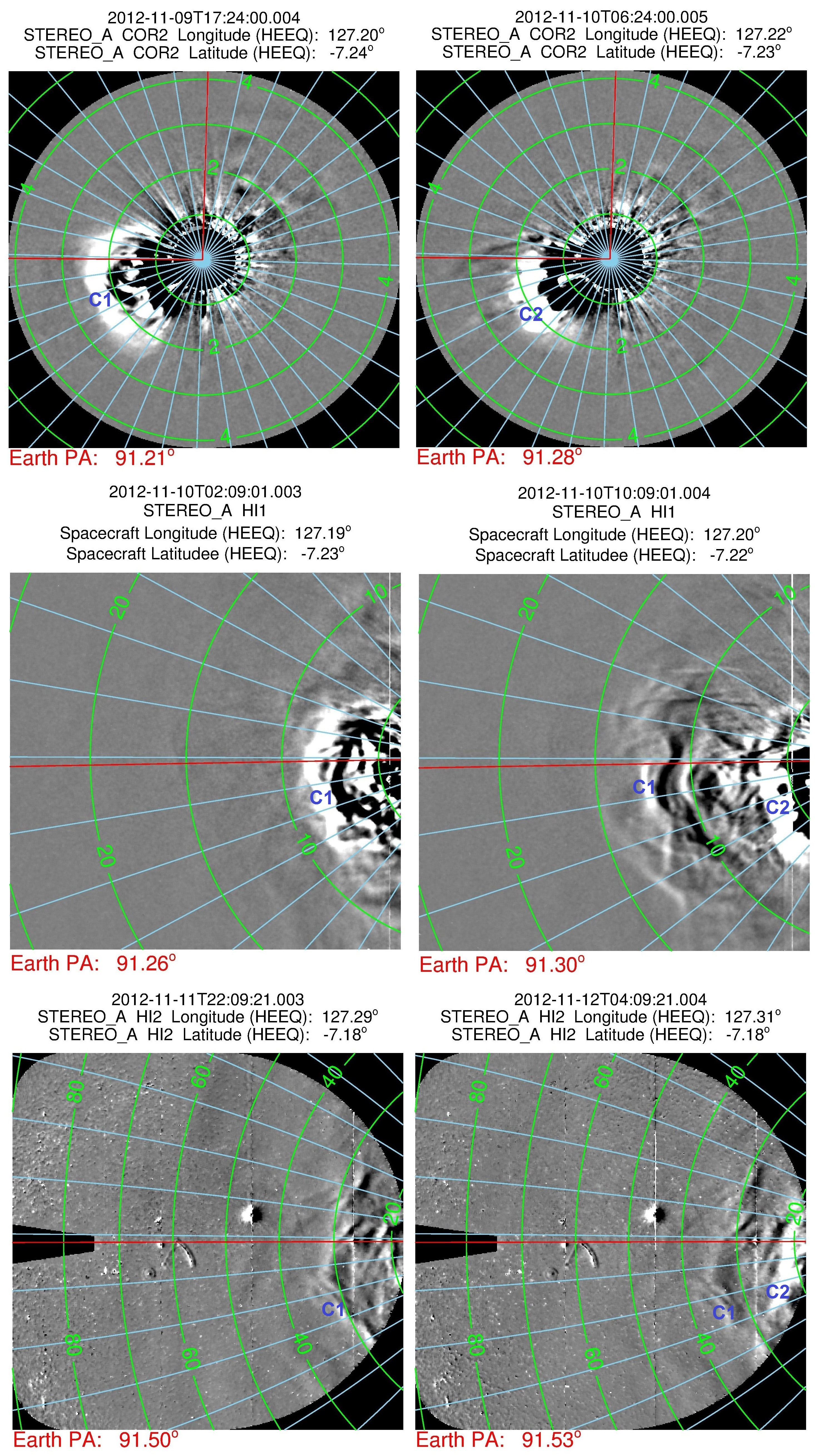}
\caption[Evolution of the 2012 November 9 and 10 CMEs in COR2, HI1, and HI2 images]{Evolution of the 2012 November 9 and 10 CMEs in running difference images of COR2 (top), HI1 (middle), and HI2 (bottom) FOV. The Left and right panels show observations from \textit{STEREO}/SECCHI-A spacecraft at two different times. The images are overlaid by contours of elongation (green) and position angle (blue). In each image, the position angle is overlaid in an interval of $10^{\circ}$ and the horizontal red line is along the ecliptic at the position angle of Earth. In the upper panel (both left and right), the vertical red line marks the $0^{\circ}$ position angle. In each panel, C1 and C2 correspond to CME1 and CME2.}
\label{IntNovevolution}
\end{center}
\end{figure}

Since the tracked features of CME2 were not observed well in the \textit{STEREO-B} ecliptic \textit{J}-map, we could not implement the stereoscopic reconstruction technique to estimate the CME kinematics. Instead, we used the harmonic mean (HM) method \citep{Lugaz2009}. The choice of this method is based on our earlier study on a comparison of various reconstruction techniques involving both stereoscopic and single spacecraft observations \citep{Mishra2014}. It was found that the HM method is the most suitable for arrival time prediction amongst the single spacecraft techniques. In Figure~\ref{IntNovscc}, we have shown that the estimated longitude of CME1 and CME2 are approximately 10$^{\circ}$ and 2$^{\circ}$ east of the Sun-Earth line. We assume that beyond the COR2 FOV, CMEs continued to propagate along the same direction \textit{i.e.} we ignore any heliospheric longitudinal deflection of CMEs. Therefore, the aforesaid estimated values of longitude were used in the HM approximation to convert the derived elongation from the \textit{J}-maps to radial distance from the Sun. 

\begin{figure}[!htb]
\begin{center}
\includegraphics[scale=0.6]{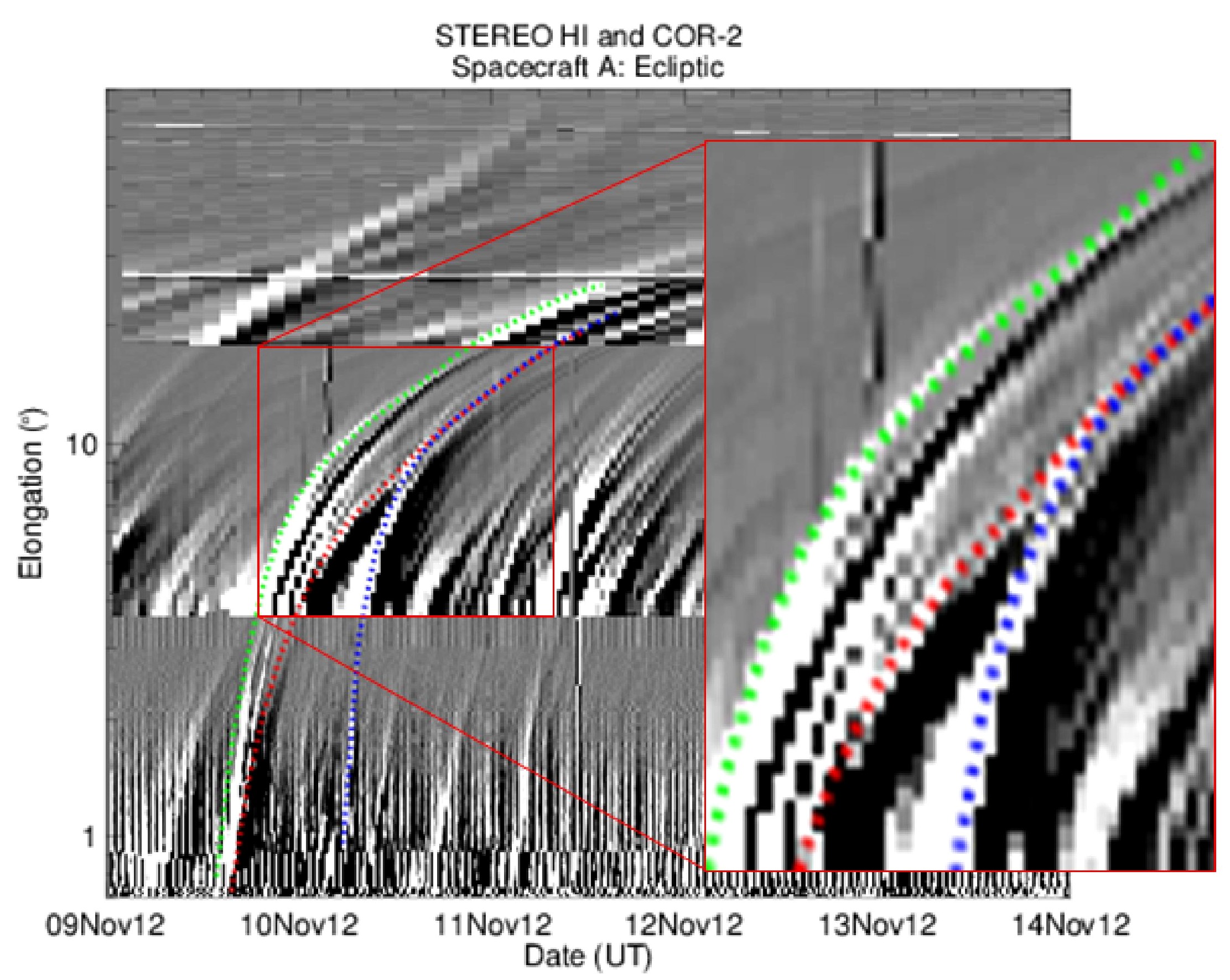}
\caption[\textit{J}-map for the 2012 November 9 and 10 CMEs]{Time-elongation map (\textit{J}-map) using the COR2 and HI observations of \textit{STEREO}/SECCHI spacecraft during the interval of 2012 November 9-14 is shown. The features corresponding to the CME1 leading edge (LE), CME1 trailing edge (TE), and CME2 leading edge are (LE) tracked and over plotted on the \textit{J}-map with green, red, and blue, respectively. The red rectangle (rightmost) is an enlarged plot of the red rectangle (on the left) which clearly shows that the red and blue tracks meet in the HI1 FOV.}
\label{IntNovJ-maps}
\end{center}
\end{figure}

Figure~\ref{IntNovkinematics} shows the distance and speed estimated for different tracked features of the two CMEs. The leading edge (LE) of CME1 has a higher speed ($\approx$ 500 km s$^{-1}$) than its trailing edge (TE) speed ($\approx$ 350 km s$^{-1}$), averaged over a few data points at the entrance of HI1 FOV. Also, the LE and TE of CME1 have lower speeds than the LE of CME2. LE of CME2 shows a large radial speed of approximately 1100 km s$^{-1}$ (ecliptic speed = 950 km s$^{-1}$) in the COR2 FOV (\textit{i.e.} 2.5-15 \textit{R$_{\odot}$}) (Figure~\ref{IntNovscc}). Beyond 10 \textit{R$_{\odot}$} distance, the LE of CME2 continuously decelerated for $\approx$ 10 hr up to 46 \textit{R$_{\odot}$} where its speed reduced to 430 km s$^{-1}$. The fast deceleration of LE of the CME2, starting from COR2 FOV, seems to be due to possible interaction with the preceding CME1. The CME1 likely had large spatial scale due to which the trailing plasma and magnetic fields from CME1 created a sufficiently dense ambient medium acting as a huge drag force for CME2 which resulted in its observed deceleration. The extremely fast deceleration of CME2 LE can also be due to the closed magnetic structure of CME1, which may act like a magnetic obstacle for the CME2 \citep{Temmer2012}, also reported for our previous case of 2011 February events.

\begin{figure}[!htb]
\begin{center}
\includegraphics[height=6cm, width=3.9cm]{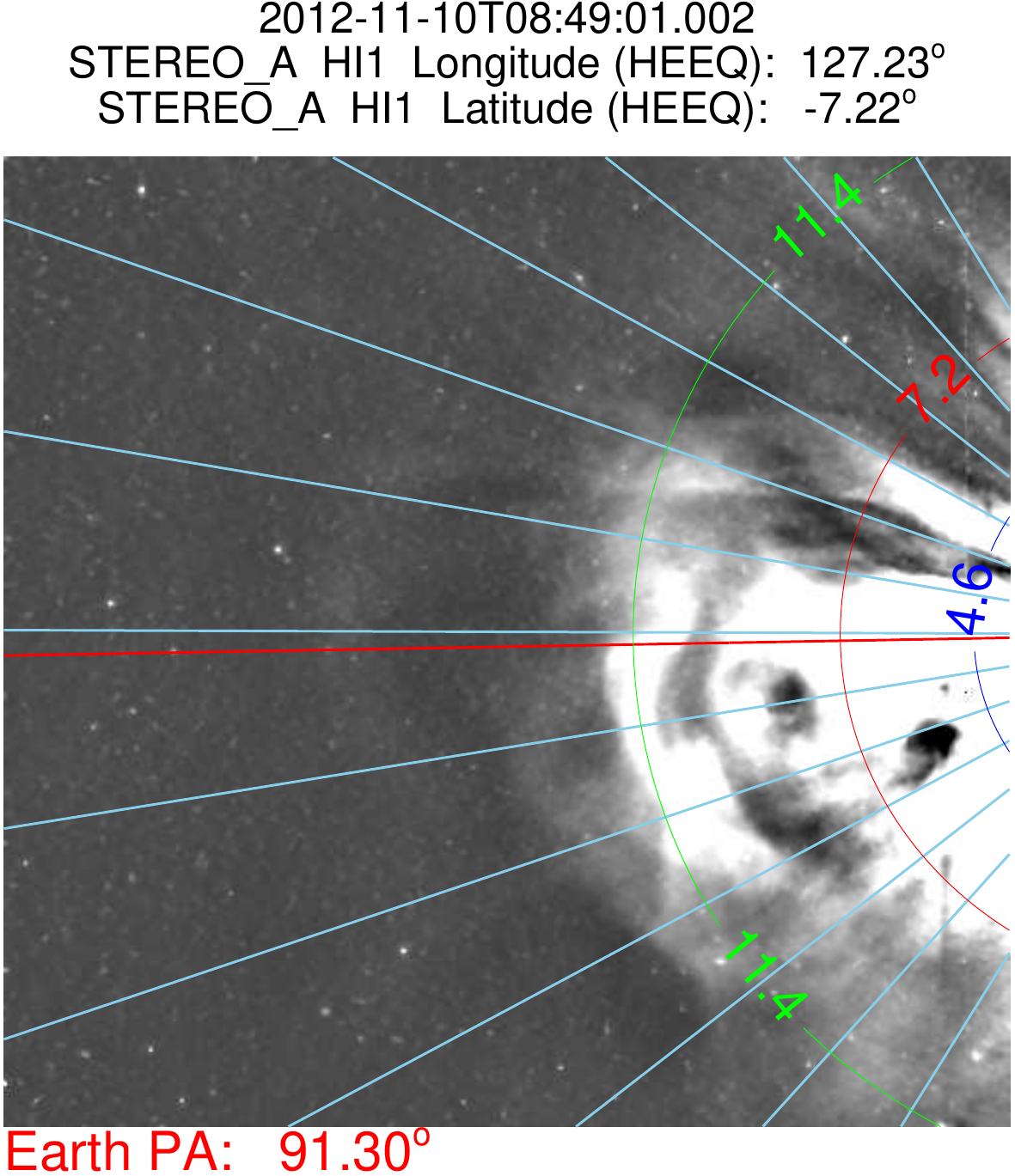}
\includegraphics[height=6cm, width=3.9cm]{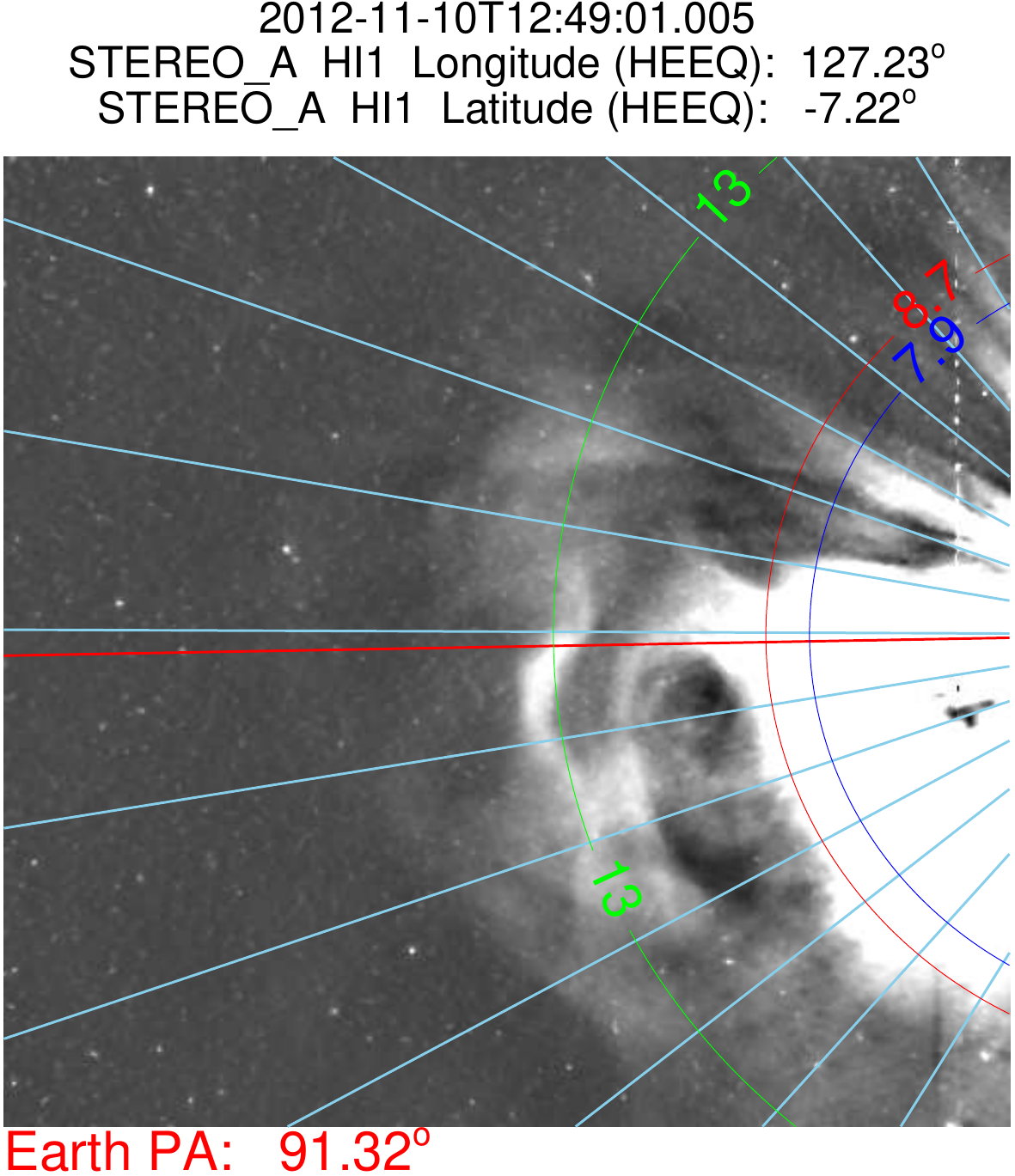}
\includegraphics[height=6cm, width=3.9cm]{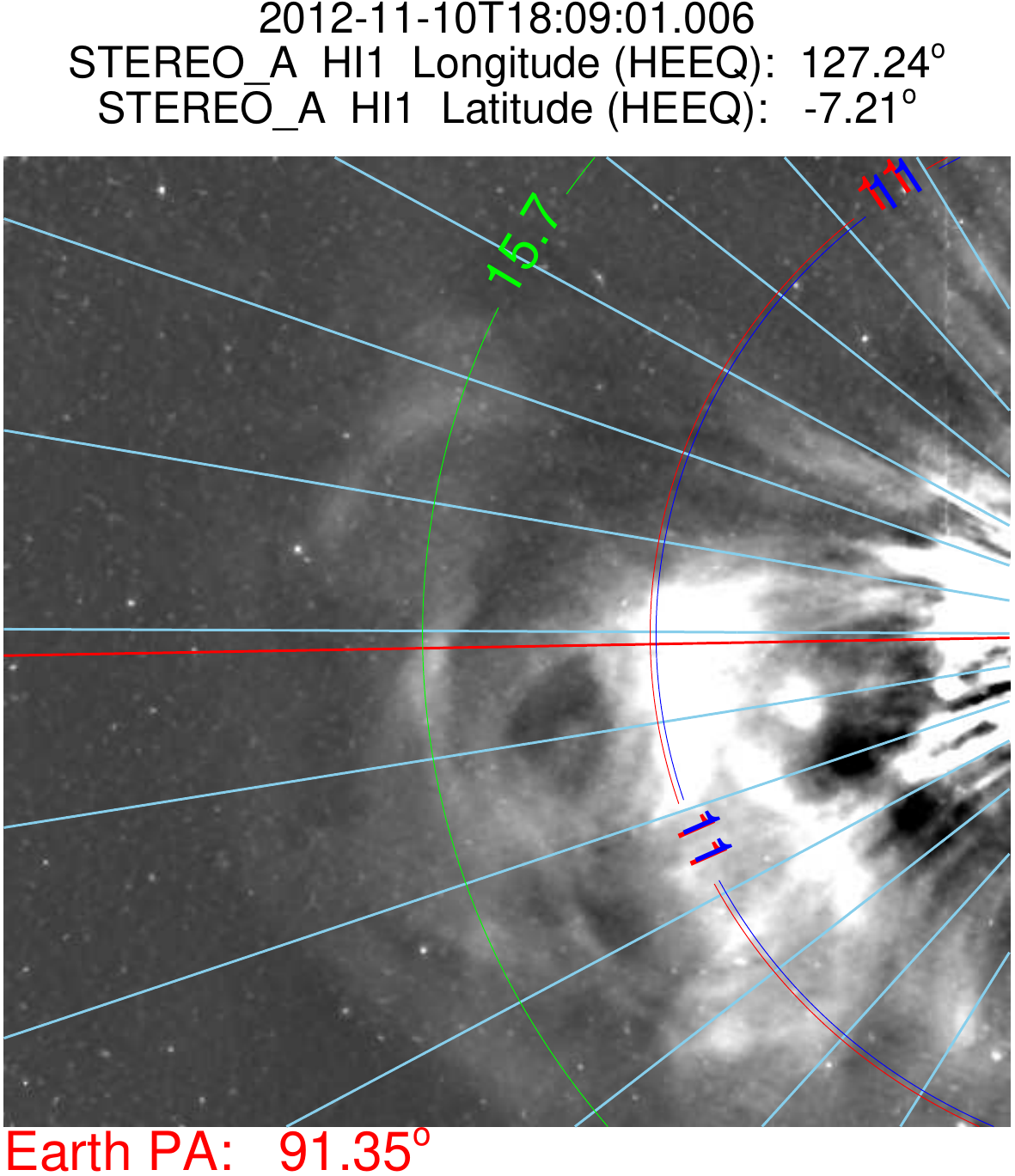}
\caption[Contours of derived elongation of tracked features overplotted on the base difference HI1-A images]{Contours of derived elongation of tracked CME1 LE (green), CME1 TE (red), and CME2 LE (blue) features from the \textit{J}-map at three epochs (08:49, 12:49, and 18:09 UT, 2012 November 10) are overplotted on the base difference HI1-A images. In each image, the position angle (sky blue) is overlaid in interval of $10^{\circ}$ and the horizontal red line is along the ecliptic at the position angle of Earth.}
\label{IntNovbasediff}
\end{center}
\end{figure}

From the estimated kinematics of tracked features in the heliosphere (Figure~\ref{IntNovkinematics}), it is clear that around November 10 at 11:30 UT, the speed of CME1 TE started to increase from 365 km s$^{-1}$ with a simultaneous decrease in speed of CME2 LE from 625 km s$^{-1}$. Such an observation of acceleration of CME1 and deceleration of CME2 provides evidence for the commencement of collision \citep{Temmer2012,Lugaz2012,Shen2012,Maricic2014,Temmer2014,Mishra2014}.

\begin{figure}[!htb]
\begin{center}
\includegraphics[scale=0.75]{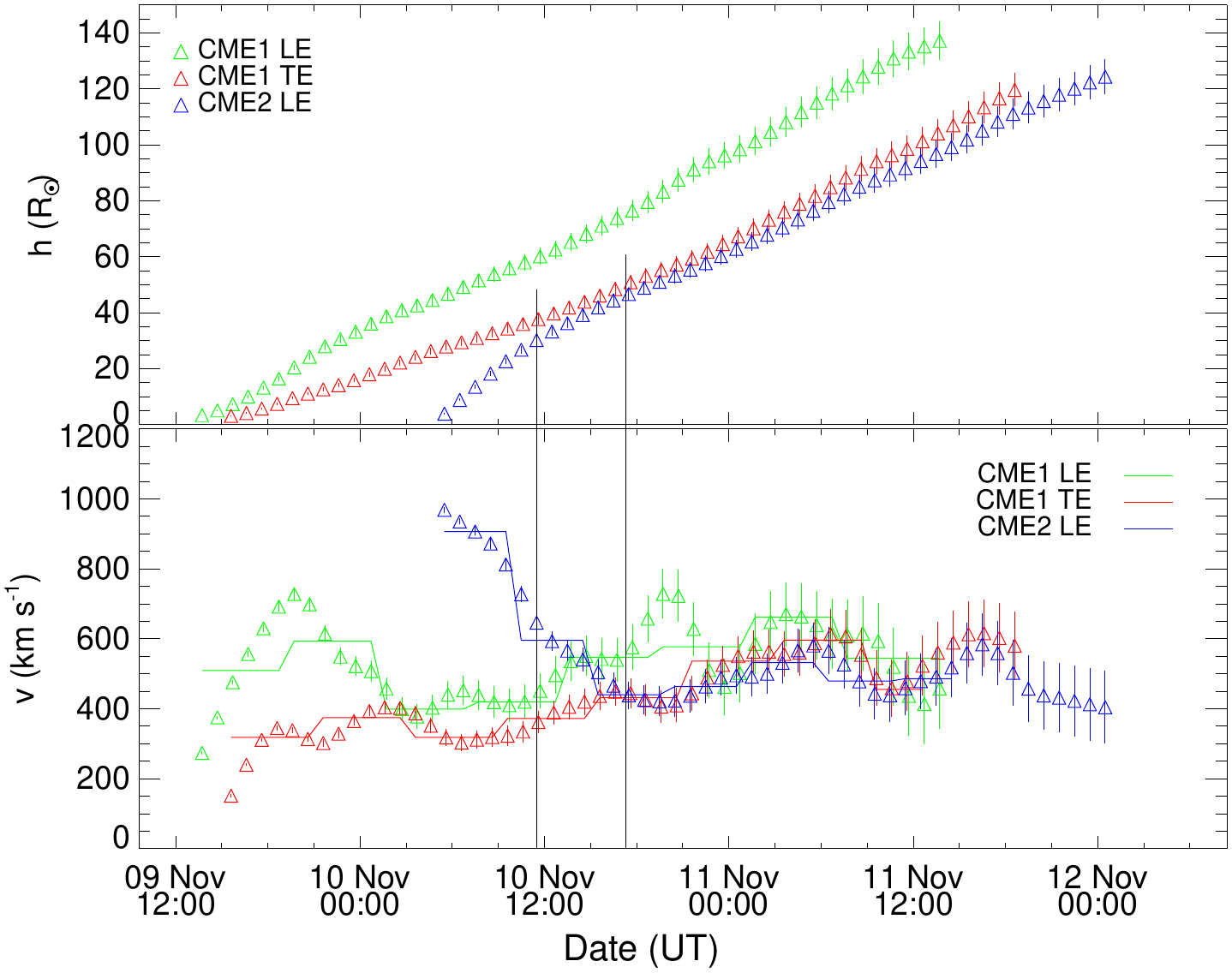}
\caption[Estimated distance and speed of the tracked features using HM method for the 2012 November 9 and 10 CMEs]{Distance (top) and speed (bottom) of the tracked features CME1 LE, CME1 TE, and CME2 LE as marked in Figure~\ref{IntNovJ-maps} with green, red, and blue, respectively. Triangles in panel (bottom) are calculated from the differentiation of adjacent distance points using the three-point Lagrange interpolation. Line segments in panel (bottom) are the speeds estimated by differentiating a linear fit to the estimated distance points for an interval of approximately 5 hr. In both panels, vertical lines show the error bars. We have assumed a fractional error of 5\% in the estimated distance, which is used to determine the uncertainties in the speeds. The two vertical lines (black) mark the start and end of the collision phase.}
\label{IntNovkinematics}
\end{center}
\end{figure}

Carefully observing the variations of speed of tracked CME1 TE and CME2 LE, the start and end boundaries of the collision phase are drawn as vertical lines in Figure~\ref{IntNovkinematics} (bottom). We noticed that at the end of the collision phase around 10 November at 17:15 UT, the speed of CME1 TE is $\approx$ 450 km s$^{-1}$ and the speed of CME2 LE is 430 km s$^{-1}$. At the beginning of the marked collision phase, CME1 TE was at a distance of 37 \textit{R$_\odot$}, and CME2 was at 30 \textit{R$_\odot$}. At the end of the marked collision phase, they were at a distance of 50 \textit{R$_\odot$} and 46 \textit{R$_\odot$}, respectively. Here, we must highlight that the estimated heliocentric distance of CME1 TE and CME2 LE was never found to be equal throughout the collision phase. This may be because of the large-scale structure of CMEs, and therefore we suggest that the tracked TE of CME1 and LE of CME2 using \textit{J}-maps are not strictly the rear-most trail of CME1 and outermost front of CME2, respectively.

\subsection{Momentum, energy exchange, and nature of collision between CMEs of 2012 November 9 and 10}
\label{IntNovMomentum}

To understand the momentum exchange during the collision of CMEs, we followed a similar approach as described in Section~\ref{IntFebTrumas}. The estimated mass and propagation direction of November 9 (CME1) and November 10 (CME2) CMEs is given in Table~\ref{massNov}.

\begin{table}
  \centering

 \begin{tabular}{|p{2.0cm}| p{5.0cm}| p{5.0cm}|}
    \hline
Parameters &  November 9 CME & November 10 CME  \\ \hline
M$_{A}$  &  4.60 $\times$ 10$^{12}$ kg at $\approx$ 15 \textit{R$_{\odot}$}  & 2.25 $\times$ 10$^{12}$ kg at $\approx$ 15 \textit{R}$_{\odot}$ \\ \hline

M$_{B}$  & 2.81 $\times$ 10$^{12}$ kg at $\approx$ 15 \textit{R}$_{\odot}$ & 1.31 $\times$ 10$^{12}$ kg at $\approx$ 12 \textit{R}$_{\odot}$ \\ \hline

Direction & 19$^{\circ}$ west from the Sun-Earth line & 21$^{\circ}$ east from the Sun-Earth line  \\ \hline

True mass & $M_{1}$ = 4.66 $\times$ 10$^{12}$ kg &  $M_{2}$ = 2.27 $\times$ 10$^{12}$ kg  \\ \hline

 \end{tabular}

\caption[The estimates of mass and direction for the 2012 November 9 and 10 CMEs]{The estimates of mass and direction for 2012 November 9 and 10 CMEs. M$_{A}$ and M$_{B}$ are the estimated mass from two viewpoints of \textit{STEREO-A} and \textit{STEREO-B}, respectively.}
\label{massNov}
\end{table}

We assume that after crossing the COR2 FOV and during the collision of CME1 and CME2, their estimated true masses ($M_{1}$ and $M_{2}$) remain constant. From Figure~\ref{IntNovkinematics} (bottom), the observed velocity of CME1 and CME2 before the collision is estimated as ($u_{1},u_{2}$) = (365,625) km s$^{-1}$ and observed velocity of CME1 and CME2 after the collision is ($v_{1},v_{2}$) = (450,430) km s$^{-1}$. We restrict ourselves not to estimating the value of the coefficient of restitution ($e$) directly by using the pre and post-collision velocity of CMEs. This is because the velocities estimated from the reconstruction method have some errors and do not guarantee the conservation of momentum, a necessary condition for collision, and therefore can lead to erroneous estimation of the $e$ value.

We used the approach described in Section~\ref{IntFebCoefResti}. The value of $e$ = 0.1 is found with ($v_{\rm 1th},v_{\rm 2th}$) = (458,432) km s$^{-1}$ and $\sigma$ = 9. With this theoretically estimated $e$ value, the total kinetic energy of CMEs is found to decrease by 6.7\% of its value before the collision. Using the estimated velocities of the CMEs before and after the collision from Figure~\ref{IntNovkinematics} (bottom), the value for $e$ is found to be 0.08, which is approximately the same as obtained from iterations described above. Therefore, our analysis suggests that the observed collision between the CMEs is close to perfectly inelastic. The kinetic energy of CME1 and CME2 before the collision was 3.1 $\times$ 10$^{23}$ J and 4.4 $\times$ 10$^{23}$ J, respectively. After the collision, based on the observed speeds, we found that the kinetic energy of CME1 increased by 51\%, and that of CME2 decreased by 54.5\% of their respective values before the collision. We also noticed that after the collision, the momentum of CME1 increased by 23\%, and the momentum of CME2 decreased by 31\% of their values before the collision. Such calculations support the claim that a significant exchange of kinetic energy and momentum takes place during the CME-CME collision \citep{Temmer2012, Lugaz2012,Shen2012, Maricic2014} as also observed for the 2011 February events (Section~\ref{IntFebCoefResti}). 

\begin{figure}[!htb]
\begin{center}
\includegraphics[scale=0.75]{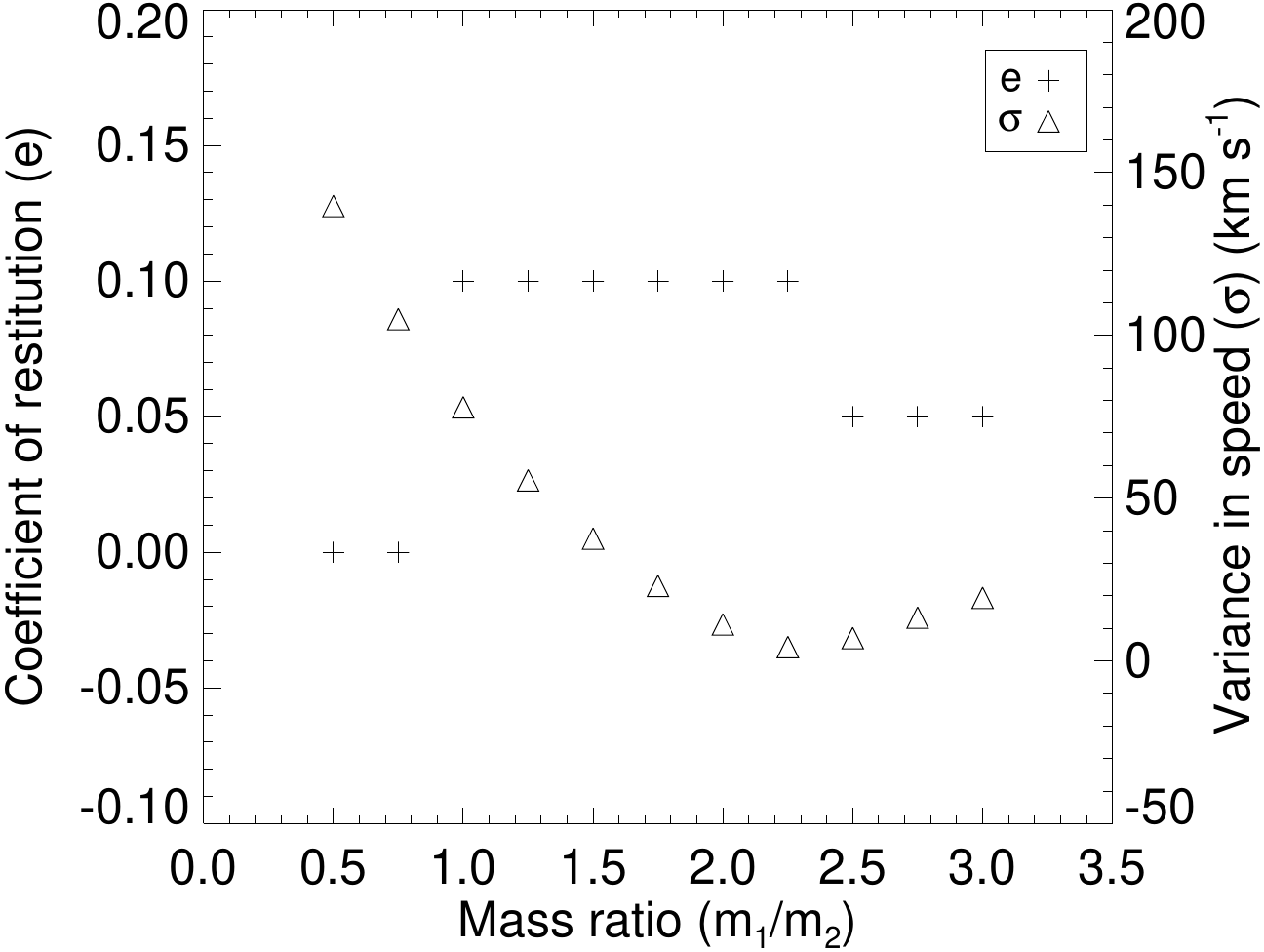}
\caption[Coefficient of restitution corresponding to different mass ratios of interacting CMEs of the 2012 November 9 and 10]{The best suited coefficient of restitution ($e$) corresponding to different mass ratios of CME1 and CME2 are shown with pluses symbol and corresponding variance ($\sigma$) in velocity is shown with triangles symbol.}
\label{IntNovmass}
\end{center}
\end{figure}

Considering uncertainty of $\pm$ 100 km s$^{-1}$ in the speed, we repeated the above analysis. As there are many sources of errors in estimating the true mass of CMEs, the effect of uncertainty in mass is also examined. In our case, the mass ratio ($M_{1}/M_{2}$) is equal to 2.05 and Figure~\ref{IntNovmass} shows the variation of $e$ and $\sigma$ with varying mass ratio. It is clear that despite taking large uncertainties in the mass of CME1 and CME2, the nature of the observed collision remains close to perfectly inelastic for the case of interacting CMEs of 2012 November 9-10.

\subsection{In situ observations, arrival time, and geomagnetic response of interacting CMEs of 2012 November 9 and 10}
\subsubsection{In situ identification of tracked CME features}
\label{IntNovInsitu}
\label{Insitu}
We analyzed the in situ data taken from the \textit{WIND} spacecraft located at L1 to identify the tracked density enhanced features of CMEs. Figure~\ref{IntNovinsitu} shows the magnetic field and plasma measurements from 12:00 UT, November 12 to 12:00 UT, November 15. The arrival of a forward shock (labeled as S) marked by a sudden enhancement in speed, temperature, and density is noticed at 22:20 UT on November 12. The region between the first and second vertical lines represents the turbulent sheath region. Based on the CME identification criteria of \citet{Zurbuchen2006}, the region bounded between the second vertical lines at 08:52 UT on November 13 and the third vertical line at 02:25 UT on November 14 is identified as the CME structure. We observed an expansion of the CME in this region characterized by a monotonic decrease in proton speed and temperature. Based on the predicted arrival times, which were derived using the estimated kinematics of the remotely observed tracked features of CME1 and CME2 as inputs in the DBM (explained in Section~\ref{IntNovarrtime}), this region was associated with Earth-directed CME1 launched on November 9.

During the passage of CME1, the magnetic field was observed to be high ($\approx$ 20 nT), and plasma beta was less than unity ($\beta$ $<$ 1) with smooth rotation in the magnetic field vector. Also, the latitude ($\theta$) value of the magnetic field vector decreased from 43$^\circ$ to -43$^\circ$ and its longitude ($\phi$) decreased from 203$^\circ$ to 74$^\circ$. Therefore, this region can be classified as an MC and based on the observed arrival time; it was associated with CME1. Due to the interaction of the CMEs, the region associated with CME1 is found to be at a higher temperature than found generally for a normal isolated CME.

The magnetic field strength (after the time marked by the third vertical line) decreased, reaching a minimum value of 6 nT around 4:00 UT on November 14. This interval of a sudden drop in the magnetic field was associated with a sudden rise in density, temperature, and plasma $\beta$, suggestive of a possible magnetic hole (MH) \citep{Tsurutani2006} which is considered as a signature of magnetic reconnection \citep{Burlaga1978}. Another region of magnetic field depression from 08:05 UT to 10:15 UT on November 14, reaching a minimum value of 3 nT, is also noticed. Corresponding to this minimum magnetic field value, the plasma $\beta$ and temperature were found to increase. The region from 03:45 UT-08:05 UT on November 14, bounded between two distinct MH like structures, has an enhanced magnetic field ($\approx$ 15 nT) and plasma $\beta$ less than unity. This region, between two MH, seems to be a magnetic field remnant of reconnecting CME structures.

\begin{figure}[!htb]
\begin{center}
\includegraphics[scale=0.65]{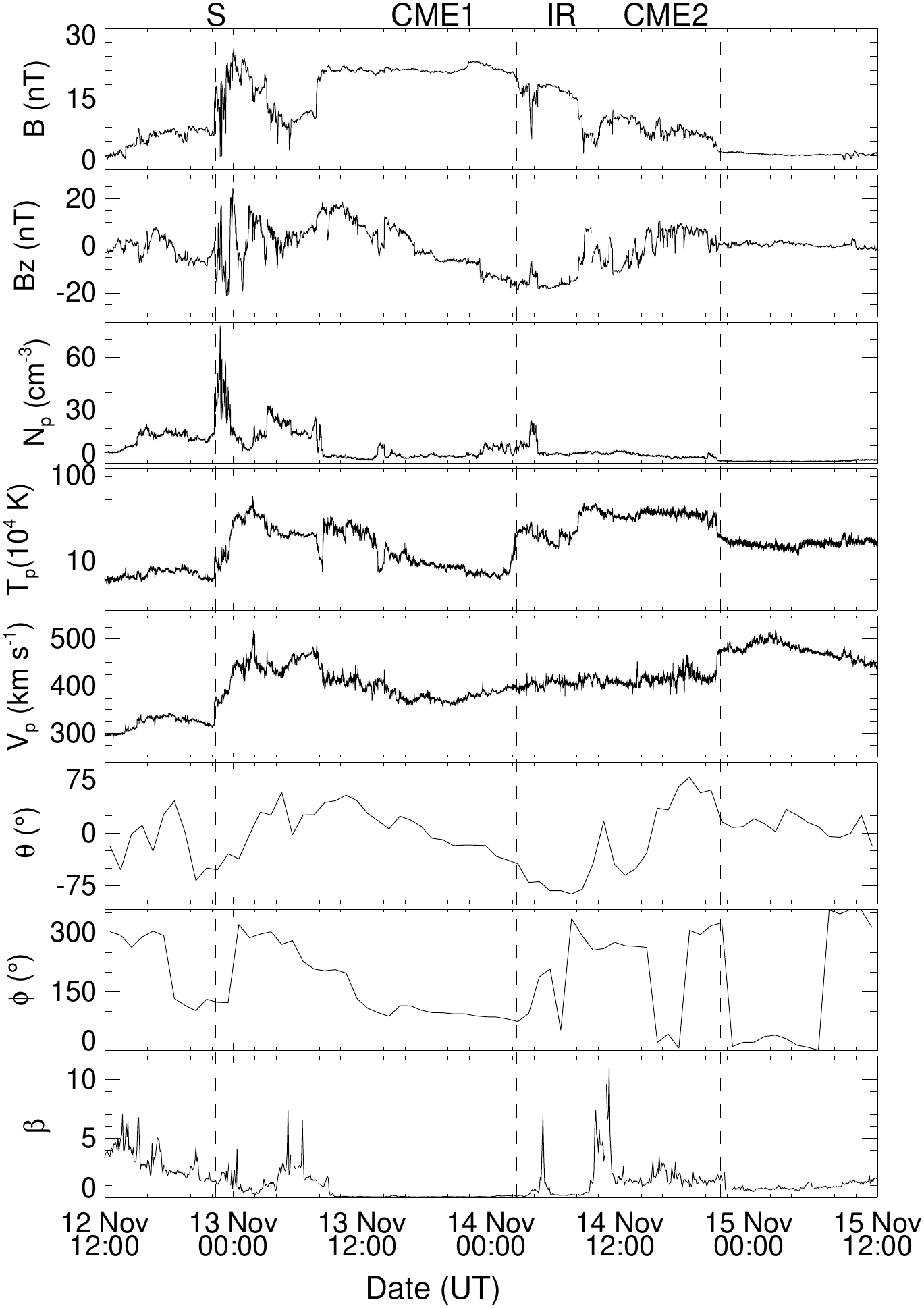}
\caption[In situ measurements of interacting CMEs of the 2012 November 9 and 10]{From top to bottom, total magnetic field strength, the $z$-component of the magnetic field, proton density, proton temperature, proton speed, latitude and longitude of magnetic field vector, and plasma beta ($\beta$) are shown for the time interval of 12:00 UT on November 12 to 12:00 UT on November 15. The vertical dashed lines mark, from left to right, the arrival of shock, CME1 leading edge (LE), CME1 trailing edge (TE), CME2 LE, and CME2 TE, respectively. The Interaction region (IR) is between the third and fourth vertical lines.}
\label{IntNovinsitu}
\end{center}
\end{figure}

Based on the extremely elevated interval of plasma beta, temperature, observation of MH, and probably sudden fast rotation in the magnetic field vector in the region bounded between the third and fourth vertical lines in Figure~\ref{IntNovinsitu}, this region was identified as the interaction region (labeled as IR) of CME1 TE with CME2 LE.

Another structure was identified based on the elevated fluctuating magnetic field and temperature during 12:00 UT-21:21 UT on November 14, bounded between the fourth and fifth vertical lines in Figure~\ref{IntNovinsitu}. During this interval, we noticed a high magnetic field (9 nT) with no monotonic decrease in temperature and speed profile, as well as plasma $\beta$ $>$ 1. From these observations, \textit{i.e.} lack of MC signatures and the short duration (9.5 hr) of the structure associated with CME2, we infer that the \textit{WIND} spacecraft perhaps intersected the flank of CME2 \citep{Mostl2010}. This is also confirmed by the estimated latitude (-25$^\circ$) of CME2 using the GCS model (described in Section~\ref{IntNovRecnsCOR}). On examining the evolution of CME2 in the HI1-A movie, we noticed that CME2 was directed towards the southern hemisphere, which was a favorable condition for its flank to encounter with the \textit{WIND} spacecraft.

We observed that the magnitude of the magnetic field in the MC region was constant around $\approx$ 20 nT, which may be due to the passage of the shock from CME1. We also noticed that the average temperature in the first half of CME1 was higher ($\approx$ 10$^{5}$ K) than found in general ($\approx$ 10$^{4}$ K). The high temperature of the CME1 may occur due to its collision with CME2, thereby resulting in its compression. Another possibility of high temperature of CME1 is due to the passage of the forward shock driven by CME2 as reported in earlier studies \citep{Lugaz2005, Liu2012,Maricic2014, Mishra2014a}. However, in the present case, the following CME2 was also observed with a high temperature ($\approx$ 5 $\times$ 10$^{5}$ K), which has not been reported in earlier studies of interacting CMEs. We noticed significant high density at the front of CME1, possibly due to overall compression by sweeping the plasma of CME1 at its leading edge by CME2 driven shock.

\subsubsection{Arrival time of tracked features}
\label{IntNovarrtime}
We used the estimated speed at the last point of measurement (up to where CMEs could be tracked unambiguously in HI) and used it as input to the DBM developed by \citet{Vrsnak2013} to estimate the arrival time of tracked features at L1. The estimated speed, time, and distance ($v_{0}, t_{0}$, and \textit{$R_\odot$}) of LE of CME1 (a green track in the \textit{J}-map) are used as inputs to the DBM, corresponding to an extreme range of the drag parameter, to predict its arrival time and transit speed at L1. In situ observations show a peak in density, $\approx$ 0.5 hr after the shock arrival around 23:00 UT on November 12 with a transit speed of 375 km s$^{-1}$, which is expected to be the actual arrival of the tracked LE feature corresponding to CME1. The predicted values of the arrival time and transit speed of the features and errors from the actual values are shown in Table~\ref{IntNovTabarr}.

\begin{sidewaystable}
  \centering
 \begin{tabular}{|p{1.5cm}|p{2.7cm} |p{4.5cm}| p{3.0cm}|p{3.0cm}|p{3.2cm}|}
    \hline	
 Tracked features & Kinematics as inputs in DBM [$t_{0}$, $R_{0}$ ($R_\odot$), v$_{0}$ (km s$^{-1})$]& Predicted arrival time (UT) using kinematics + DBM [$\gamma$ = 0.2 - 2.0 (10$^{-7}$ km$^{-1}$)] & Predicted transit speed (km s$^{-1}$) at L1   [$\gamma$ = 0.2 - 2.0 (10$^{-7}$ km$^{-1}$)] & Error in predicted arrival time (hr) [$\gamma$ = 0.2 - 2.0 (10$^{-7}$ km$^{-1}$)] &  Error in predicted speed (km s$^{-1}$)  [$\gamma$ = 0.2 - 2.0 (10$^{-7}$ km$^{-1}$)]  \\  \hline

CME1 LE & Nov 11 13:42, 545, 137  & Nov 12 18:10 - Nov 13 00:30  & 490 to 380 	& -5 to 1.5 	& 115 to 5  \\ \hline

CME1 TE & Nov 11 18:35, 120, 470  & Nov 13 15:25                  & 375        &  -8         &  -15  \\  \hline

CME2 LE & Nov 12 00:30, 124, 455  & Nov 13 19:40                 	& 375   	   &  -16        &  -35  \\ \hline
\end{tabular}
\caption[Estimated kinematics, predicted arrival time and transit speed at L1 for the tracked features of the 2012 November 9 and 10]{the First column lists the tracked features of CMEs, and second column shows their estimated kinematics (by the HM technique) which is used as input to the DBM. The predicted arrival time  and transit speed of the tracked features at  L1, corresponding to the extreme range of the drag parameter (used in the DBM), are shown in the third and fourth columns. Errors in predicted arrival time and speed, based on a comparison with the in situ arrival time and speeds, are shown in the fifth and sixth columns. The errors in arrival time with a negative (positive) signs indicate that the predicted arrival is earlier (later) than the actual arrival time. The errors in transit speed with a negative (positive) signs indicate that the predicted transit speed is slower (faster) than the transit speed measured in situ.}
\label{IntNovTabarr}
\end{sidewaystable}

For the TE feature of CME1 and LE of CME2, we assume they encountered the dense ambient solar wind medium created by the preceding CME1 LE. Therefore, their kinematics with the maximum value of the statistical range of the drag parameter (2.0 $\times$ $10^{-7}$ km$^{-1}$) are used as inputs to the DBM. We consider that the TE of CME1 corresponds to the density enhancement at the trailing front of CME1. At the rear edge of CME1, a density enhancement (12 particles cm$^{-3}$) was observed around 23:30 UT in \textit{WIND} observations on November 13, which is considered the actual arrival of TE of CME1. The predicted arrival time and transit speed of this feature and errors therein are given in Table~\ref{IntNovTabarr}. We further notice that in in situ data, an enhancement in density corresponding to LE of CME2 around 12:00 UT on 14 November (marked as the arrival of CME2 with the fourth vertical line in Figure~\ref{IntNovinsitu}) was observed, which can be considered as the actual arrival time of LE of CME2. The predicted values of CME2 LE, using its kinematics with DBM and errors therein, are also listed in Table~\ref{IntNovTabarr}. The error for CME2 LE is large, but several factors can lead to such large errors, which are discussed in Section~\ref{IntNovResults}. It must be highlighted that if the 3D speed of CMEs estimated at the final height in COR2 FOV is assumed to be constant up to L1, then the predicted arrival time of CME1 and CME2 will be $\approx$ 10-16 hr and 44 hr earlier, respectively, than the predicted arrival times using post-collision speeds combined with DBM. This emphasizes HI observations and post-collision speeds of CMEs as inputs to the DBM for an improved arrival time prediction of interacting CMEs. Similarly, using HI observations, \citet{Colaninno2013} have shown that a linear fit to the deprojected height-time data above 50 \textit{$R_\odot$} gives a 12 hr improvement over the CME arrival time estimated using LASCO data.

\subsubsection{Geomagnetic consequences of interacting CMEs}
\label{IntNovgeomagnetic}

As mentioned in Section~\ref{IntCMEsIntro}, very few studies have been dedicated to the study of the geomagnetic consequences of interacting CMEs. The CMEs of 2012 November 9-10 resulted in a single strong geomagnetic storm with Dst index $\approx$ -108 nT at 8:00 UT on November 14, therefore it is important to investigate the impact of the interaction of the CME1 of November 9  with the CME2 of November 10 on the terrestrial magnetosphere-ionosphere system in details.

Figures~\ref{IntNovgeomag}(a-e) reveal the variations in the solar wind parameters in the geocentric solar ecliptic (GSE) coordinate system during 2012 November 12-15. These parameters include solar wind proton density (in cm$^{-3}$), velocity (in km s$^{-1}$, negative $X$-direction), ram pressure (in nPa), the $Z$-component of the interplanetary magnetic field (IMF B$_{z}$, in nT) and the $Y$-component of the interplanetary electric field (IEF$_{y}$, in mV m$^{-1}$) respectively. These data with a cadence of 1 min are taken from the NASA GSFC CDAWeb (\url{www.cdaweb.gsfc.nasa.gov/istp_ public/}). It is also important to note that the solar wind parameters presented in Figures~\ref{IntNovgeomag}(a-e) are corrected for propagation lag till the nose of the terrestrial bow shock. To compare the variations of these parameters with the magnetospheric and ionospheric parameters, additional time lags that account for the magnetosheath transit time and the Alfv\'en transit time are calculated \citep{Chakrabarty2005}. Therefore, the solar wind parameters presented in Figures~\ref{IntNovgeomag}(a-e) are corrected for the propagation lag, point by point, till ionosphere.

\begin{figure}[!htb]
\begin{center}
\includegraphics[width=28pc]{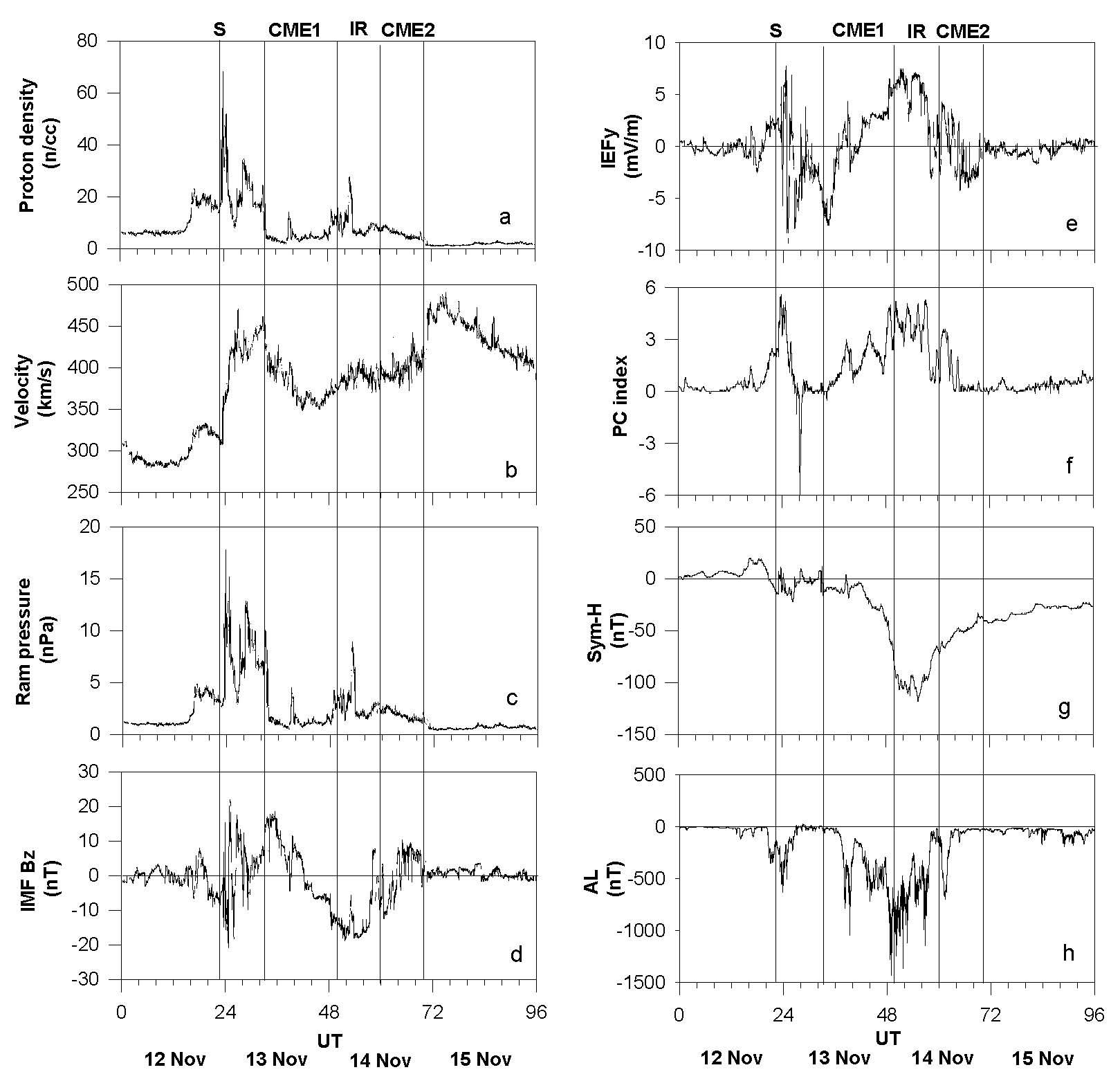}
\caption[Variations in geomagnetic field and in situ parameters during the interval of 2012 November 12-15]{Variations in geomagnetic field and in situ measured CMEs parameters during the interval 2012 November 12-15: (a) Proton density (cm$^{-3}$), (b) velocity (km s$^{-1}$), (c) ram pressure (nPa), (d) B$_{z}$ component (nT), (e) interplanetary electric field's $y$ -component IEF$_{y}$ (mV m$^{-1}$), (f) polar cap (PC) index, (g) Sym-H (nT), and (h) AL index (nT). The vertical lines and their labels are as defined in Figure~\ref{IntNovinsitu}}.
\label{IntNovgeomag}
\end{center} 
\end{figure}

Figure~\ref{IntNovgeomag}(f) represents the variation in the polar cap (PC) index. The PC index is shown \citep{Troshichev2000} to capture the variations in the ionospheric electric field over the polar region efficiently. Figure~\ref{IntNovgeomag}(g) shows the variation in the Sym-H (in nT) index which primarily represents the variation in the magnetospheric ring current \citep{Iyemori1996}. Figure~\ref{IntNovgeomag}(h) shows variations in the westward (midnight sector) auroral electrojet (AL) current (in nT) which captures auroral substorm processes reasonably well. Further, in Figure~\ref{IntNovgeomag}, the arrival of the shock is denoted by S; different vertical lines mark the arrival of different features of CME1 and CME2 and are labeled in the same manner as in Figure~\ref{IntNovinsitu}.

Figure~\ref{IntNovgeomag}(a) reveals the arrival of two distinctly enhanced density structures; first enhancement occurred at the arrival of the shock-sheath region before CME1 LE (on November 12 at 20:00 UT to November 13 at 04:00 UT), and the second enhancement occurred at the arrival of the CME1 TE-IR region (November 13 at 12:00 UT to November 14 at 12:00 UT). During the arrival of the shock-sheath region before CME1 LE, the peak density reached $\approx$ 68 cm$^{-3}$ which is more than two times of the corresponding peak density ($\approx$ 30 cm$^{-3}$) observed at the arrival of CME1 TE-IR. Sharp enhancement in the solar wind velocity was also observed vis-a-vis the sharp density enhancement at the arrival of the shock-sheath region (Figure~\ref{IntNovgeomag}b) when the velocity increased from $\approx$ 300 km s$^{-1}$ to $\approx$ 470 km s$^{-1}$. However, the solar wind velocity did not change sharply and significantly at the arrival of CME1 TE-IR. 

The changes in the density and the velocity resulted in changes in the solar wind ram pressure shown in Figure~\ref{IntNovgeomag}(c). The peak ram pressure in the shock-sheath region before CME LE ($\approx$ 18 nPa) was almost twice than that observed during CME1 TE-IR ($\approx$ 9 nPa). Figure~\ref{IntNovgeomag}(d) reveals that IMF B$_{z}$ was predominantly southward in both density enhanced intervals, although fast fluctuations were observed corresponding to the arrival of the shock-sheath region. The first peak in IMF 
B$_{z}$ is $\approx$ -8 nT before the arrival of the shock. Thereafter, IMF B$_{z}$ fluctuated sharply between $\approx$ -20 nT to $\approx$ +20 nT during the passage of the shock-sheath region before the arrival of CME1 LE. During CME1 TE-IR, IMF B$_{z}$ reached a peak value of $\approx$ -18 nT. No significant change in IMF B$_{z}$ was observed during the passage of the CME2 on November 14-15,  when its magnitude hovered around the zero line. 

Figure~\ref{IntNovgeomag}(e) elicits the variation in IEF$_{y}$ during 2012 November 12-15. It is observed that the peak value of 
IEF$_{y}$ before the arrival of the shock was $\approx$ 2 mV m$^{-1}$. However, during the shock event, IEF$_{y}$ fluctuated between $\pm$ $\sim$8 mV m$^{-1}$. At the arrival of CME1 TE-IR, the peak value of IEF$_{y}$ reached $\approx$ 7.5 mV m$^{-1}$. In fact, similar to IMF B$_{z}$, a sharp polarity change in IEF$_{y}$ was noticed at $\approx$ 09:40 UT on November 14. No significant change in IEF$_{y}$ was observed during the passage of the CME2 on  November 14-15, when its magnitude hovered around the zero line. Figure~\ref{IntNovgeomag}(f) shows that the PC index increased during both density enhanced intervals, and the peak value was $\approx$ 5 for both intervals. Figure~\ref{IntNovgeomag}(g) shows variation in the Sym-H index, which revealed the development of a geomagnetic storm during the CME1 TE region. The Sym-H index reached a value of $\approx$ -115 nT during the passage of IR on November 14. The ring current activity was not significant at the arrival of the shock-sheath region before CME1. Lastly, Figure~\ref{IntNovgeomag}(h) elicits the variation in the AL index. It is seen that AL reached  $\approx$ -600 nT during the arrival of the shock-sheath before CME1 and $\approx$ -1400 nT at the arrival of CME1 TE-IR. Therefore, significant intensification of westward auroral electrojet occurred at the arrival of CME1 TE-IR. 

The above observations (from Figure~\ref{IntNovgeomag}) reveal several interesting points. First, the magnitude of AL seems to remain unaffected by large amplitudes of fluctuations in IMF B$_{z}$ and IEF$_{y}$ in the shock-sheath region. The second and the most important point is that the duration of occurrence of the AL intensification in the shock-sheath before CME1 LE and in the CME1 TE-IR region was nearly identical to the duration of the southward IMF B$_{z}$ and positive IEF$_{y}$ phases. However, the AL amplitudes seem to be more closely correlated with the IEF$_{y}$ amplitudes in the CME1 TE-IR region compared to the shock-sheath region. Third, although the peak amplitudes of the PC index were nearly the same in the shock-sheath region before CME1 LE and CME1 TE-IR regions, the peak amplitudes of the AL index were significantly different during these two intervals; fourth, the substorm activity seems to be over during the passage of the CME2; therefore, CME2 did not have any bearing on the triggering of substorms. Figure~\ref{IntNovgeomag} demonstrates the direct role played by the positive IEF$_{y}$ (or southward IMF B$_{z}$) in the storm-time AL intensification, particularly when the terrestrial magnetosphere encountered the CME1 TE and IR region. Further, the arrival of CME2 did not affect the terrestrial magnetosphere-ionosphere system. 

As aforementioned, a major geomagnetic response was noticed at the arrival of the trailing edge of the preceding CME (CME1) and the IR of the two CMEs near the Earth. We also conclude that the following CME (CME2) failed to cause a significant geomagnetic activity, possibly due to the fact that the spacecraft encountered the flank of this CME as mentioned in Section~\ref{IntNovInsitu}. We understand that due to the interaction and collision between the trailing edge of the preceding CME1  and the leading edge of the following CME2, as revealed in imaging observations (described in Section~\ref{IntNovRecnsHI}), the parameters responsible for geomagnetic activity were significantly intensified at the CME1 rear edge and in the interaction region found between CME1 and CME2. These results bring out the importance of CME-CME interaction in the formation of the IR and its role in the significant development of geomagnetic disturbances.

\subsection{Results and Discussion on 2012 November 9-10 CMEs}
\label{IntNovResults}
The CMEs of November 9 and 10 provide us a rare opportunity to investigate the consequences of CME-CME interaction. A combination of heliospheric imaging and in situ observations are used for improving our understanding of CME kinematics, post-collision characteristics, and the nature of the collision. From the estimated kinematics, the site of collision of CMEs could be located at a distance of approximately 35 \textit{$R_{\odot}$} at 12:00 UT on November 10, which is at least 85 \textit{$R_{\odot}$} before and 15 h earlier than as predicted by using the initial kinematics of CMEs in COR2 FOV. 

As described earlier, CME1 LE is found to be propagating with speed higher than CME1 TE. Therefore, CME1 LE (green track in Figure~\ref{IntNovkinematics}) did not collide with CME2 LE (blue track). With the start of the collision phase, the LE of CME1 is also found to accelerate, and after the collision phase, significant acceleration of CME1 LE is noticed. This may occur either because of sudden impact (push) from CME1 TE to CME1 LE during a collision or due to the passage of a shock driven by the CME2 or a combined effect of both. Beyond the collision phase, up to an estimated distance of nearly 100 \textit{R$_{\odot}$}, LE of CME2 was found moving behind the TE of CME1, and both these features propagated together, decelerating slowly. For a few hours after the observed collision phase, it was noticed that the speed of the LE and TE of CME1 was slightly higher than the LE of CME2, which could have increased the observed separation of these structures. The \textit{J}-maps show that both these features could be tracked for further elongations after the interaction. However, due to the limitation of the HM method for higher elongations \citep{Lugaz2009}, we restricted our measurements on the tracked features up to 100 \textit{R$_{\odot}$}. Our analyses show evidence that after the interaction, the two features did not merge. It appears that LE of CME2 interacted with TE of CME1 and continued to propagate with a reduced speed of 500 km s$^{-1}$.

During the collision between CME1 and CME2, we noticed a large deceleration of fast CME2 while relatively less acceleration of slow CME1 till both approached an equal speed. This is expected to occur if the mass of CME1 is larger than that of CME2, which is indeed the case. In Section~\ref{IntNovMomentum} we have shown that the mass of CME1 is $\approx$ 2.0 times larger than the mass of CME2. This result is an important finding and is in agreement with the second scenario of interaction described in \citet{Lugaz2009}. It also must be noted that we have estimated the mass of CMEs in COR2 FOV while interaction takes place in the HI FOV. Further, we cannot ignore the possibility of an increase in the mass of CME due to mass accretion at its front via the snowplough effect in the solar wind beyond COR FOV \citep{DeForest2013}. We must also mention that in our calculation of momentum and energy transfer, although we use the total mass of CME1 but only a part of CME1 (\textit{i.e.} TE of CME1) took part in a collision with CME2.
However, considering the uncertainties in the derived speeds and mass of the CMEs, our analysis reveals that the nature of collision remains close to perfectly inelastic.

At the last point of measurement in the HI FOV, around 120 \textit{R$_{\odot}$}, the estimated speed of TE of CME1 and LE of CME2 are approximately equal ($\approx$ 470 km s$^{-1}$). Therefore, both features are expected to arrive at L1 at approximately the same time. Compared to in situ measured actual arrival time, the delayed arrival of LE of CME2 is possible due to higher drag force acting on it, resulting in its deceleration. However, the propagation direction of CME2 in the southern hemisphere, as noticed in HI1-A images and also estimated in COR2 FOV using 3D reconstruction, can also account for the flank encounter of CME2 and thus its delayed sampling by the \textit{WIND} spacecraft at L1. This is consistent with our interpretation in Section~\ref{IntNovInsitu}. The estimated speed using the single spacecraft HM method can also lead to some errors and can result in the delayed arrival of LE of CME2. Keeping all these issues in mind, we believe that it is quite probable that the tracked feature corresponding to LE of CME2 (blue track in \textit{J}-map) is not sampled by in situ spacecraft. Therefore, the erroneous predicted arrival time for LE of CME2 is related to its incorrect identification in in situ data at L1.

The association of HI observations with in situ measurements from \textit{WIND} leads to many interesting results. Despite two CMEs launched from the Sun in succession in the Earthward direction, we observe only one shock in in situ data which may suggest the merging of shocks driven by CME1 and CME2 if both CMEs would have driven shocks. However, such a claim cannot be made unless we are familiar with the in situ signatures of merged shock and plasma structure following it. The sweeping of plasma to high density at the front of CME1 and its compressed heating is most likely due to the passage of CME2-driven shock through the MC associated with CME1. Based on the predicted arrival time of the tracked features, it seems that the CME2-driven shock and CME1 sheath region is tracked as CME1 LE in the \textit{J}-map. Therefore, we infer that CME1 LE propagated probably into an unperturbed solar wind. Our study also provides a possibility of the formation of the interaction region (IR) at the junction of the trailing edge of the preceding CME and the leading edge of the following CME. We show that during the collision of the CMEs, kinetic energy exchange up to 50\% and momentum exchange between 23\% to 30\% took place. Our study also demonstrates that the arrival time prediction was significantly improved by using HI compared to COR2 observations and also emphasizes the importance of understanding post-collision kinematics in further improving the arrival time prediction for a reliable space weather prediction scheme.

Our study reveals clear signatures of the interaction of these CMEs in remote and in situ observations and also helps in the identification of separate structures corresponding to these CMEs. Despite the interaction of the two CMEs in the interplanetary medium, which generally results in complex structures as suggested by \citet{Burlaga1987} and \citet{Burlaga2002}, in this case, we could identify interacting CMEs as distinguished structures in the \textit{WIND} spacecraft data. Even after the collision of these CMEs, they did not merge, which may be possible because of the strong magnetic field and higher density of CME1 than CME2. This needs further confirmation, and therefore, it is worth investigating what decides the formation of merged CME structure or complex ejecta during CME-CME interaction. Due to single point in situ observations of CME, we acknowledge the possibility of ambiguity in marking the boundaries of CMEs. In the present case, the boundaries for CME1 and IR are distinctly evident. Also, slight ambiguity in the boundary of CME2 (flank encounter) will not change our interpretation because the main geomagnetic response is caused due to enhanced negative B$_{z}$ in the trailing portion of CME1 and its extension in IR. Here, we also point out that the temperature in IR was lower than the temperature in the CME2 region, but temperatures in both regions were elevated as compared to a normal non-interacting CME. The observations of unexpected larger temperature in the CME2 region than IR region may be due to the possibility that the sheath region of CME2 was intersected by in situ spacecraft, as we have interpreted.

The association of geomagnetic storms with isolated single CMEs has been carried out extensively for a long time \citep{Tsurutani1988, Gosling1991,Gosling1993, Gonzalez1989,Gonzalez1994, Echer2008,Richardson2011}. Our study is essential as it focuses on the role of interacting CMEs in the generation of geomagnetic storms and substorms. Our study suggests that the trailing edge of the preceding CME (CME1) and IR formed between the two interacting CMEs are efficient candidates for intense geomagnetic storms. In the context of substorms, our study highlights that the persistence of IMF B$_{z}$ in the southward direction is more important than the amplitude in driving the substorm activity as manifested by the AL intensification. Using the WINDMI model, \citet{Mays2007} have shown that the interplanetary shock and sheath features for CMEs contribute significantly to the development of storms and substorms. However, in this case of 2012 November 9-10 CMEs, sharp and large southward excursions in the midst of fluctuating IMF $B_{z}$ associated with the shock (the shock-sheath region before CME1) were found to be less effective in producing strong substorm activity. Therefore, further investigations are required regarding the characteristics (geometry, intensity) of shock and the preceding CME in the context of triggering of substorms, as has been shown in earlier studies \citep{Jurac2002, Wang2003a}.

Another interesting aspect regarding substorms noticed in our study is that the nearly equal amplitude responses of the PC index corresponding to the shock-sheath region (IMF $B_{z}$ sharply fluctuating between southward and northward directions) preceding the CME1 interval vis-\`a-vis the CME1 interval (IMF $B_{z}$ steadily turning southward). This is interesting as the responses of Sym-H and AL during these two intervals are quite different in terms of amplitudes of variations. 

\section{Conclusion}
\label{IntFebConclu}

Based on our analysis of interacting CMEs of 2011, February 13-15 and 2012 November 9-10 by combining the wide-angle imaging and in situ observations, the main results derived are given in Table~\ref{Concompint}. The two studies highlight the following:
\begin{enumerate}

\item{The collision between February 14 and 15 is observed at around 25 \textit{R}$_\odot$ while the collision between November 9 and 10 took place at 35 \textit{R$_{\odot}$}. These collision sites are much closer to the Sun than obtained by using the estimated kinematics in the COR2 FOV. This highlights that heliospheric imaging is important to observe the collision of CMEs and to estimate their post-collision dynamics.}

\item{We find that the observed collision of February 14 and 15 is in an inelastic regime reaching close to elastic, while for November 9 and 10 CMEs the collision is close to perfectly inelastic. These findings are in contrast to an earlier study by \citet{Shen2012} who reported a case of interacting CMEs in a super-elastic regime. Therefore, it is worth investigating further what decides the nature of the collision and which process is responsible for magnetic and thermal energy conversion to kinetic energy to make a collision super-elastic. Further in-depth study is required to examine the role of duration of collision phase and impact velocity of CMEs in deciding the nature of the collision.

We acknowledge that the present analysis for collision dynamics may have small uncertainties due to the adopted boundary for the start and end of the collision phase. In this context, we emphasize that it is often difficult to define the start of collision as the following CME starts to decelerate (due to its interaction with the preceding CME) and the preceding CME starts to accelerate before (most possibly due to shock driven by the following CME) they actually merge as observed in HI FOV. Hence, different timing and large time-interval of acceleration of one CME and deceleration of other prevent us from pinpointing the exact start and end of the collision phase. Further, we have not considered the expansion velocity and propagation direction of the centroid of CMEs, which may be different before and after the collision.}

\item{The total kinetic energy of February 14 and 15 CMEs after the observed collision is reduced by 1.3\%, and for November 9 and 10 CMEs, it decreased by 6.7\% to its value before the collision. We also found that speeds and momentum of 2011 February CMEs changed from 35\% to 68\%, and for 2012 November CMEs, it changed from 23\% to 30\% compared to their values before the collision.

Our analysis used the total mass of CMEs to study their collision dynamics, but as the CME is not a solid body, its total mass is not expected to participate in the collision. Keeping in mind various limitations of the present study, we believe that more detailed work, by incorporating multiple plasma processes, is required to understand the CME-CME interaction. Also, the assumption that there is no mass transfer between CME2 and CME3 during the collision may result in some uncertainties.}

\item{The in situ measurements of these CMEs near 1 AU show that interacting CMEs are accelerated or decelerated, compressed, and heated. Our analysis also highlights that the estimated heliospheric kinematics, particularly after collisions, is important to combine with DBM for improving the estimation of the arrival times of different features of CME that experience different drag forces during their propagation through the heliosphere.} 

\item{For 2011 February CMEs, our results do not favor the possibility of strengthening the geomagnetic response as a consequence of the arrival of two or more interacting CMEs near the Earth. However, for 2012 November CMEs, the interaction region (IR), formed due to the collision between November 9 CME TE and November 10 LE, has been found to be associated with intensified plasma and magnetic field parameters, which are responsible for major geomagnetic activity.}

\end{enumerate}

\begin{table}
  \centering

 \begin{tabular}{|p{4.0cm}| p{4.1cm}| p{4.1cm}|}
    \hline
Characteristics &  2011 February 13-15 CMEs & 2012 November 9-10 CMEs  \\ \hline

\vspace{0.2cm}Interaction distance & CME2-CME3 at 25 \textit{R}$_{\odot}$ (expected at 37 \textit{R}$_{\odot}$ using speed in COR2 ) &  CME1-CME2 at 35 \textit{R}$_{\odot}$ \newline (expected at 130 \textit{R}$_{\odot}$ using speed in COR2)\\  \hline

Momentum exchange & 35\% to 68\%  & 23\% to 30\%  \\  \hline

Total kinetic energy & Reduced by 1.3\% & Reduced by 6.7\% \\  \hline

Nature of collision & Close to elastic & Close to perfectly \newline inelastic \\  \hline

Geomagnetic response & Minor storm \newline (Dst = -30 nT), Strong SSC & Major storm \newline (Dst = -108 nT) \\  \hline

 \end{tabular}

\caption{A comparison of estimated characteristics for the interacting CMEs of the 2011 February 13-15 and 2012 November 9-10.}
\label{Concompint}
\end{table}

Our study of interacting CMEs highlight the importance of \textit{STEREO's} Heliospheric Imager (HI) observations and their association with in situ observations to understand the nature of CME-CME interaction in detail and for improved prediction of their arrival time using their post-interaction kinematics. Our study reveals that tracking different features of CMEs is necessary for a better understanding of CME-CME interaction. This study concludes that CMEs cannot be treated as completely isolated magnetized plasma blobs, especially when they are launched in quick succession. We have also highlighted the difficulties inherent in reliably understanding the kinematics, arrival time, nature of collision, and morphological evolution of CMEs. Further statistical study of such interacting CMEs is required to understand their nature of the collision and to investigate the characteristics of the CMEs (mass, strength, and orientation of magnetic field, speed and direction of propagation, and duration of collision phase), which are responsible for their interaction.

\chapter{Conclusions and Future Work}
\label{Chap9:ConFut}
\rhead{Chapter~\ref{Chap9:ConFut}. Conclusions and Future Work}

In this thesis, I have attempted to understand the evolution and consequences of CMEs in the heliosphere. For this purpose, we have exploited the remote sensing observations of CMEs mainly from the \textit{STEREO} spacecraft and in situ observation from \textit{ACE} and \textit{WIND} spacecraft. Our study is crucial to achieving a better scheme for arrival time prediction of CMEs at 1AU, which is a prime concern for any solar-terrestrial physicist. The principal conclusions of my thesis are given in Section~\ref{Conclulast}. Some of the limitations of the current study undertaken for this thesis will be addressed in the future to  enhance our understanding of the heliospheric evolution and consequences of CMEs. These are described in 
Section~\ref{Futulast}.    

\section{Conclusions}
\label{Conclulast}

Based on our study, we conclude that by exploiting white light images from twin vantage points of \textit{STEREO}, the uncertainties in the morphology and kinematics of CMEs due to projection effects in the white light images from a single viewpoint can be resolved. This can be done by implementing appropriate 3D reconstruction techniques on the CME images near and far from the Sun. Our study highlights that by constructing the \textit{J}-maps using the running difference white light images of CMEs from \textit{STEREO}, a CME can be unambiguously tracked continuously from its liftoff in the inner corona to almost the Earth. By continuously tracking different features of a CME and implementing 3D reconstruction techniques to estimate their kinematics, we found that characteristics (speed, direction, and morphology) of CMEs change as they propagate in the heliosphere. Such findings, particularly that the CMEs accelerate or decelerate until they obtain the speed of ambient solar wind medium, have been inferred before. However,  our analysis, using the reconstruction techniques on the SECCHI/HI observations of several Earth-directed CMEs, helps to show directly.

We have shown that the 3D speed (estimated using 3D reconstruction techniques) of CMEs near the Sun (in COR field of view (FOV)), and assuming that it remains constant for the remaining distance, i.e., up to 1 AU, is not sufficient to predict the arrival time accurately for a majority of CMEs at the Earth. This is true, especially for a fast speed CME traveling in the slow solar wind environment or a slow speed CME traveling in the high-speed stream. We show that a small error in the estimated speed of CMEs in the COR2 FOV may result in a large variation in the predicted arrival time near 1 AU. This variation may be due to the large distance between the COR2 FOV and L1. However, for fast CMEs, this variation in arrival time using slightly different estimated 3D speeds will be minimized. We also conclude that a fast-speed stream, from a coronal hole hitting the back of a CME, can significantly change its dynamics. Our study shows that if the estimated 3D kinematics of CMEs in the heliosphere are used as inputs in drag-based model (DBM), the arrival time prediction of these at 1 AU is improved compared to using only 3D speed estimated close to the Sun. Based on this, we conclude that the role of drag forces in the dynamics of CMEs, is effective farther out (few thirties of solar radii) from the Sun. From estimating the propagation direction of CMEs, we conclude that a CME may undergo the non-radial motion even far from the Sun.

We assessed the relative performance of 10 reconstruction techniques by applying them to SECCHI/HI observations for estimating the arrival time of CMEs at the Earth. For the first time, such a detailed analysis pointing out the limitations and advantages of each reconstruction technique was carried out. Recently, \citet{Mostl2014} also assessed the accuracy and limitations of three fitting methods. Based on our study, we conclude that although single spacecraft reconstruction techniques are capable of estimating the 3D kinematics of CMEs in the heliosphere, relative performance of such techniques is inferior than double spacecraft reconstruction techniques. Among the double spacecraft reconstruction techniques, the TAS method performs the best for predicting the arrival time and transit speed of CMEs at 1 AU, while GT method performs the least accurately. This is true, irrespective of the characteristics of the CMEs taken in our study. We also conclude that the HM method best estimates the arrival time among the single spacecraft techniques that we applied. Among the single spacecraft fitting methods, our study concludes that  HMF and SSEF methods are always superior to FPF. We also show that these techniques can estimate different kinematics (speed and direction) of CMEs. Therefore, anyone needs to be careful before using these estimates directly.

We also attempted to track two density structures at the front and the rear edge of a CME. We found that density structure at the rear edge corresponds most probably to cool and dense filament identified after the magnetic cloud, while density structure at the front corresponds to the sheath before the leading edge of CME in the in situ observations taken by \textit{ACE} and \textit{WIND} spacecraft. This is also one of the first attempts to associate optical and in situ observations, and only a few studies have been carried out so far with \textit{STEREO} observations \citep{Howard2012a}. We highlight the importance of \textit{J}-maps constructed from SECCHI/HI observations for associating the three-part structure of CME in remote and in situ observations.

Our study shows that a CME can interact/collide with another CME under favorable conditions. We also show that tracking different features of a CME in the HI observations is important to witness the CME-CME interaction. In this thesis, we have attempted to investigate the nature of the collision of two CMEs. There has been only one case study by \citet{Shen2012} who showed that collision was super-elastic. We conclude that there is a significant exchange of momentum and kinetic energy during the collision of the CMEs. Hence, the consequences of CMEs on other CMEs can be noted; therefore, post-collision kinematics of CMEs must be used for their improved arrival time prediction at 1 AU. Our analysis reveals that CMEs collide at much earlier than that derived from the estimated 3D kinematics in the COR2 FOV. Our study found that the nature of CME-CME collision can range from elastic to perfectly inelastic. This finding raises a question as to what decides the nature of collision of magnetized plasma blobs, which are expected to be different than gaseous blobs or solid balls. We also conclude that such collision of CMEs has significant effects on the magnetic and plasma parameters of both preceding and following CMEs. Our study shows the formation of interaction region and magnetic hole (probably the signatures of magnetic reconnection) by the collision of CMEs. These interaction regions are seen to be responsible for significant geomagnetic activity. Our study also shows that the morphology (angular width) of a fast CME can change significantly during its propagation in the heliosphere, especially when it traverses in a dense solar wind environment which may be formed due to remnants of a preceding CME.

Our study is a dedicated attempt to understand and implement the several reconstruction techniques applicable to COR and HI observations of CMEs for estimating their 3D kinematics, arrival time at 1 AU, morphological evolution, and CME-CME interaction geomagnetic consequences of interacting CMEs. We conclude that tracking the CMEs out to longer elongation in the heliosphere using HI is necessary for an improved understanding of their evolution and consequences.

\section{Future Work}
\label{Futulast}
In the study undertaken for this thesis, we have analyzed only a few CMEs. However, a large number of CMEs need to be studied in order to answer the questions regarding the heliospheric propagation and consequences of CMEs. In the thesis, we have focused mainly on the Earth-directed CMEs; however, CMEs launched in different directions from the Sun need to be included to assess the performance of reconstruction techniques. In our study of the collision of CMEs, the actual mass of the CMEs estimated in the COR FOV is assumed to be constant up to the HI FOV, which may not always be true. Therefore, the mass of CMEs in the HI FOV must be estimated. We have not compared the obtained kinematics of the CMEs to the kinematics profiles of the CMEs derived from the theoretical models. Such comparison must be made for a better understanding of the propagation of CMEs. Based on our study of interacting CMEs, we need to examine what decides the nature of collision of CMEs and why only some interacting CMEs cause significant geomagnetic activity. We plan to pursue studies that can remove some of the uncertainties in our current approach to understanding the evolution and consequences of different features of CMEs.

\subsection{Dynamic evolution of the CMEs}
We plan to understand how the physical parameters of CMEs vary with heliocentric distance, particularly the polytropic index of plasma of CMEs, Lorentz force and thermal energy inside the CMEs, and their role in the dynamic evolution of the CMEs. The contribution of Lorentz forces in driving the CMEs in the heliosphere will also be investigated because in several CME propagation models, the effect of Lorentz force is considered negligible after a few solar radii from the Sun. One also needs to incorporate the interaction of CMEs in the solar wind in the drag-based model of propagation of CMEs. We also plan to investigate how the magnetic energy stored in the CMEs is used in driving, expanding, and heating the CME material during their heliospheric propagation. Based on tracking of different features of a CME in remote and in situ observations, we can attempt to understand the forces acting on them as they travel in the heliosphere.

\subsection{Consequences of interacting CMEs}
 The possibility of the formation of complex ejecta on the interaction of CMEs and their geomagnetic response will be investigated. This will be attempted to understand how differently complex ejecta affects the Earth’s magnetosphere than an isolated CME or magnetic cloud. The dependence of the geoeffectiveness of the preceding CME on the intensity of the following shock will also be examined, which will help us improve the status of predicting major geomagnetic storms. In a few earlier studies, the enhancement in long-wavelength radio emission is presented as evidence for interacting CMEs \citep{Gopalswamy2001apj, Martinez-Oliveros2012}. We plan to investigate the favorable conditions or the mechanism for enhanced radio emission during CME-CME interaction.

\subsection{Assessing the performance of reconstruction methods using the off-ecliptic CMEs}   
We will track different features of CMEs in the \textit{J}-maps along with different position angles to derive their time-elongation profiles and estimate their kinematics by implementing reconstruction techniques. Such an analysis of the selected CMEs will better assess the relative performance of several reconstruction techniques. Further, the estimated kinematics at different PAs will help to understand the global evolution, i.e. at the different latitudinal extent of CMEs, especially in the case of CME-CME interaction. 

The future investigation described above can improve our understanding of the evolution and consequences of CMEs and to interact CMEs in the heliosphere. Some of these studies will better understand and resolve the critical issues of space weather, viz. propagation of CMEs and estimation of their arrival time.


\newpage


\begin{singlespacing}

\rhead{Bibliography}                

\setstretch{0.75}



\end{singlespacing}




%

\end{document}